\RequirePackage{fix-cm}

\documentclass[twocolumn,epjc3]{svjour3}  

\smartqed 
\RequirePackage[numbers,sort&compress]{natbib}
\journalname{Eur. Phys. J. A}
\usepackage{color}
\usepackage{amssymb}
\usepackage{verbatim}

\usepackage{blochsphere}

\usepackage{makecell}

\usepackage[T1]{fontenc}
\usepackage[utf8]{inputenc}
\usepackage[english]{babel}
\usepackage[unicode=true,pdfusetitle,bookmarks=true,bookmarksnumbered=false,bookmarksopen=false,
 breaklinks=true,pdfborder={0 0 0},pdfborderstyle={},backref=false,colorlinks=true]{hyperref}
\usepackage{braket}
\usepackage{stmaryrd}
\usepackage{mathtools}
\usepackage{bm}
\usepackage{soul}
\usepackage{ulem}
\usepackage{array}
\usepackage{tabularx}

\numberwithin{equation}{section} % to label equation by section

\usepackage{chngcntr}
\counterwithin{figure}{section} % to label figures by section

\begin{document}
\sloppy %break words at end of line
\title{Quantum computing with and for many-body physics}

\author{Thomas Ayral \thanksref{e1,addr1}  
\and 
Pauline Besserve
\thanksref{e2, addr1, addr3}
\and
Denis Lacroix
\thanksref{e3,addr2}
\and
Edgar Andr\'es Ruiz Guzman
\thanksref{e4,addr2} 
}

\thankstext{e1}{e-mail: thomas.ayral@atos.net}
\thankstext{e2}{e-mail: pauline.besserve@atos.net}
\thankstext{e3}{e-mail: denis.lacroix@ijclab.in2p3.fr}
\thankstext{e4}{e-mail: ruiz-guzman@ijclab.in2p3.fr}

\institute{Atos Quantum Lab, 78340 Les Clayes-sous-Bois, France \label{addr1}           
\and
Universit\'e Paris-Saclay, CNRS/IN2P3, IJCLab, 91405 Orsay, France \label{addr2}
\and
Centre de Physique Th\'eorique, 91120 Palaiseau, France
\label{addr3}
}

\maketitle

\date{Received: date / Accepted: date}

\begin{abstract}
Quantum computing technologies are making steady progress.
This has opened new opportunities for tackling problems whose complexity prevents their description on classical computers.
A prototypical example of these complex problems are interacting quantum many-body systems:
on the one hand, these systems are known to become rapidly prohibitive to describe using classical computers when their size increases.
On the other hand, these systems are precisely those which are used in the laboratory to build quantum computing platforms.
This arguably makes them one of the most promising early use cases of quantum computing. 

In this review, we explain how quantum many-body systems are used to build quantum processors, and how, in turn, current and future quantum processors can be used to describe large many-body systems of fermions such as electrons and nucleons.
The review includes an introduction to analog and digital quantum devices, the mapping of Fermi systems and their Hamiltonians onto qubit registers, as well as an overview of methods to access their static and dynamical properties.
We also highlight some aspects related to entanglement, and touch on the description, influence and processing of decoherence in quantum devices.
\end{abstract}

\section{Introduction}

For decades since they were envisioned by Richard Feynman in the 1980s \cite{feynman2018}, quantum computers have been imagined as futuristic objects  overcoming the limitations of classical devices.
Today, with significant progress in the manipulations of various quantum systems \cite{Acin2018}, the possibility of using them as a computational unit is becoming a reality \cite{GYONGYOSI2019,Meter2016}.
The race towards proving "quantum advantage" is now underway \cite{Harrow2017,Arute2019,Zhong2020}. 
The ultimate challenge of this race is to provide one or several reliable quantum processing units (QPUs) with unprecedented capabilities in terms of hard memory storage or the ability to solve specific complex problems in record time \cite{Arora2009,Nielsen2010}.

Quantum computing is now at a turning point in its practical development thanks to the growing availability of quantum machines \cite{SOEPARNO2021}.
Yet, these machines have a limited quality:
The noise that degrades each operation of a quantum circuit strongly constrains the complexity and types of algorithms that can be used today, and requires specific denoising methods \cite{Devitt2013,Gadioli2020}.
This had led to the notion of \textit{Noisy Intermediate Scale Quantum} (NISQ) \cite{Preskill2018,Bharti2021,Endo2021} processors to describe this intermediate stage of development.
Despite the limitations of the NISQ era, the possibility to experiment with actual quantum devices has led to intensive scientific emulation to test the capacity of current computers, propose new quantum algorithms, and ultimately prepare for the coming second quantum revolution and surpass the limitations of classical algorithms. 

Quantum many-body systems---formed by a set of particles interacting with one another---appear as natural test benches for quantum platforms \cite{McClean2016,McArdle2018a,Cao2019,Bauer2020,Tilly2021a,Claudino2022}.
This class of problems is characterized by a Hilbert space size that increases steeply when the number of one-body degrees of freedom (the number of particles or accessible single-particle space or both) increases.
This large size leads to severe restrictions in the class of many-body systems that one can solve exactly on classical computers.  

This increase in complexity is common to all fields of physics or chemistry.
In particular, it applies to quantum devices themselves:
assemblies of quantum bits or "qubits" are also characterized by an exponential growth of the spanned Hilbert space, as it is of size $2^{n_q}$ where $n_q$ is the number of qubits.
What is a hindrance for the classical description of many-body systems may thus become a blessing: the fact that they contain an exponential complexity a priori makes quantum computers suitable for tackling many-body (and thus also exponentially complex) systems, provided such systems can be accurately encoded and probed via a precise manipulation of qubits.
This is facilitated by the fact that particles treated in the formalism of second quantization share many formal aspects with qubits (as will be described in section \ref{sec:mapping_to_qc}).
Many-body systems such as those encountered in quantum chemistry \cite{McArdle2018a,Cao2019,Bauer2020,Claudino2022,Bassman2021,Fedorov2021a}, condensed-matter physics \cite{Zeng2019}, atomic physics \cite{Saffman2016,Bian2021,Graham2022,Bluvstein2022}, astrophysics \cite{KerenLi2019,mielczarek2019,Czelusta2021,Mielczarek2021,Mielczarek2018}, or nuclear physics \cite{Zhang2021,Stetcu2022,Roggero2020,Hobday2022,Romero2022}, have thus become key application domains of quantum computing.

The primary goal of the present article is to introduce quantum computing from a {\it double} many-body physics perspective:
quantum computers are many-body systems that can help understand (among others) many-body problems. 
We start by highlighting some specific aspects of selected many-body systems one might find in nature and underline their common features (section \ref{sec:Quantum-Many-body-problems}) with a focus on the complexity of treating them using classical computers.
Section \ref{sec:Quantum-computers:-artificial} introduces different types of quantum computing devices, namely, analog and digital, to the nonexpert reader.
This section is not only a discussion of the basic concepts in quantum computation but also an opportunity to underline the fact that quantum computers are built upon many-body interacting systems.  
In section \ref{sec:mapping_to_qc}, we discuss various aspects related to 
the solution of quantum many-body problems with quantum computers.
We introduce selected quantum algorithms to solve these problems, be it with post-NISQ (section \ref{sec:idealalgorithm}) or NISQ (section \ref{sec:nisq}) processors. 
In section \ref{sec:entanglement}, we briefly discuss how entanglement in many-body systems can be described on a quantum computer.
Finally, in section \ref{sec:noise}, we discuss the modeling of the noise impacting quantum hardware, how its effect can be mitigated on current machines as well as the main principles of error correction, which could bring about fault tolerance in the long term. 

\begin{figure}
\begin{center}
\includegraphics[width=0.8\columnwidth]{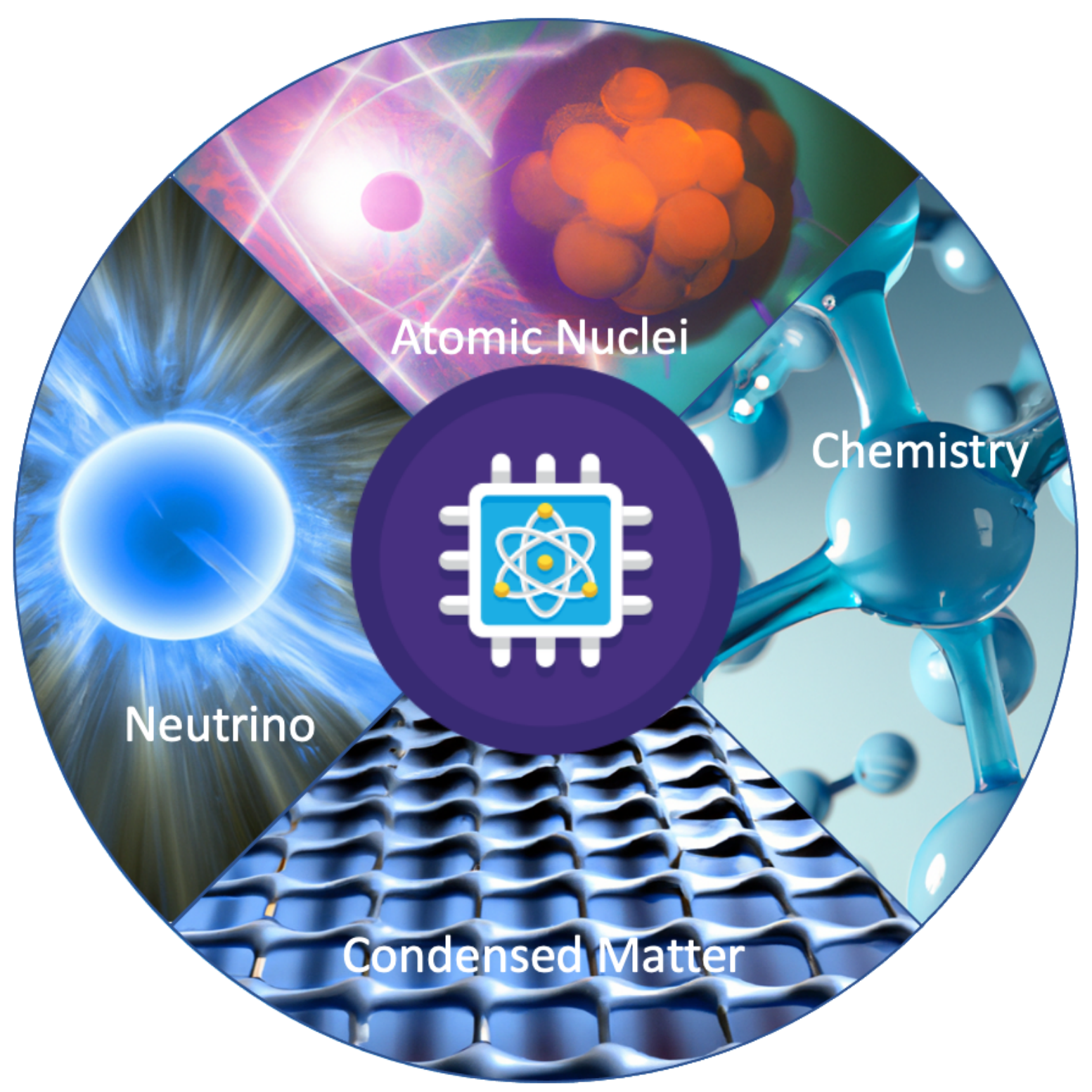}
\end{center}
    \caption{Illustration of some of the many-body systems where quantum computing is now being explored as a disruptive technique compared to classical computing: quantum chemistry, condensed matter, nuclear and neutrino physics.
    } 
\label{fig:manybodypic}
\end{figure}

\section{Quantum Many-body problems\label{sec:Quantum-Many-body-problems}}

In this section, we discuss various types of many-body systems where quantum computing could be of help.
The reader interested mainly in quantum computing aspects can skip this section and directly go to section  
\ref{sec:Quantum-computers:-artificial}.

Many-body systems are encountered in many fields of physics. 
They are ruled in all generality by a Hamiltonian that reads, in a second-quantized form:
\begin{eqnarray}
    H &=& H_{\rm 1-body}+ H_{\rm 2-body} + H_{\rm 3-body} + \cdots  \nonumber \\
      &=& \sum_{\alpha \beta} h_{\alpha \beta} c_{\alpha}^\dagger c_{\beta} 
    + \frac{1}{2} \sum_{\alpha \beta \gamma \delta} v_{\alpha \beta \gamma \delta} c^\dagger_{\alpha} c^\dagger_{\beta} c_{\gamma} c_{\delta} + \cdots, 
    \label{eq:manybodyhamiltonian}
\end{eqnarray}
where $\alpha, \beta, \gamma, \delta$ are multi-indices whose components depend on the specific many-body system at hand.
As we will see, depending on the system, one may not need terms involving more than two particles. 

The defining feature of many-body problems is that they cannot be solved within the mean-field approximation. This approximation is obtained, at the level of Hamiltonian (up to constant terms), by replacing all operators but one by its average value in each term.
In other words, in a many-body system, the correlations between its key constituents---be they electrons, nucleons, or spins---must be handled with sophisticated methods, hence their other name of many-body systems: \textit{strongly correlated systems}.

In the following subsections, we introduce a few fields---illustrated in Fig. \ref{fig:manybodypic}---where many-body problems are found, with a focus on the form of the Hamiltonians that describe them, and the typical classical methods that are used to investigate their properties.

\subsection{The electronic structure problem: electrons in solids and quantum chemistry}
\label{sec:electron}

Electrons in solids or molecules are interacting particles.
In many solids, the Coulomb interaction can be dealt with in an averaged fashion 
because electrons do not come too close to each other due to the Pauli principle. 
{Due to this, Hartree-Fock (HF) theory, where electrons are treated as independent particles in an average field, is often used as the building block for more complex theoretical methods to treat many-body effects. Another  powerful technique that allows treating a priori all correlations while keeping the independent particle or quasi-particle picture is Density Functional Theory (DFT)~\cite{Lipparini2008}.  }
%\sout{There, electrons merely become quasi-particles without radically changing their behavior. This observation is at the heart of mean-field methods; two prominent representatives of them are the Hartree-Fock (HF) method and Density Functional Theory (DFT)~\cite{Lipparini2008}.}

However, in some solids and most molecules, such an averaged description of Coulomb interactions leads to wrong predictions. For instance, in solids where valence electrons are of $d$ or $f$ character (namely very localized), neither HF nor DFT can predict the transition from a metallic \textit{Fermi liquid} to a \textit{Mott} insulator. In such insulators, Coulomb interactions freeze the charge degree of freedom. 
This type of solids has received much attention in the past forty years since the high-temperature superconductors discovered in the mid-1980s are believed to be doped Mott insulators. Despite prolonged efforts to crack this problem, a complete theoretical account of the origin of high-Tc superconductivity---and numerous other phenomena, especially when driving these systems out of equilibrium---is still missing. This lack of explanation is due to the complexity of dealing with such systems beyond mean field.

The simplest model describing such strongly-correlated solid-state systems is the so-called (Fermi) Hubbard model, which, in its single-band version, reads:
\begin{equation}
    H = - t \sum_{\langle ij\rangle} c_{i\sigma}^\dagger c_{j\sigma} +   U \sum_i n_{i\uparrow} n_{i \downarrow} - \mu \sum_{i \sigma} n_{i\sigma}, 
        \label{eq:Hubbard_solids}
\end{equation}
where $i, j = 1 \dots N$ denote lattice sites ($N$ is typically infinite in a solid; $\langle ij\rangle$ denote nearest neighbors on the lattice), $t$ denotes a tunneling or hopping term, $U$ the on-site Coulomb interaction and $\mu$ the chemical potential. The fermionic creation and annihilation operators $c^\dagger_{i\sigma}$ and $c_{i\sigma}$ typically create and annihilate electrons in localized orbitals $\phi_i(r)$ (sometimes called \textit{Wannier orbitals}). $n_{i\sigma}$ is a density operator which reads $n_{i\sigma} = c^{\dagger}_{i\sigma} c_{i\sigma}$. $\mu = U/2$ corresponds to half-filling (undoped case), namely, one electron per site.
Despite its apparent simplicity, this "spherical cow" of strongly-correlated electronic systems is difficult to solve on a classical computer in physically "interesting" regimes, like the doped, low-temperature regime in two spatial dimensions and the thermodynamical limit ($N\rightarrow\infty)$.

The Hubbard model can either be tackled "directly" via exact diagonalization (ED), quantum Monte-Carlo (MC) or tensor-network methods, or "indirectly" via embedding methods, which map the model to a smaller many-body problem as will be described in subsection \ref{subsec:Reducing-the-number}.
The exponentially growing size of the Hilbert space with $N$ quickly makes ED prohibitive, although advanced versions like Krylov or Lanczos methods can help reach relatively large sizes \cite{Saad2011}.
An exponential \textit{sign problem} typically plagues MC methods; namely, the statistical error bar of MC methods grows exponentially with system size and decreasing temperature, requiring an exponential number of samples and hence exponential run time.
Tensor-network methods (like Matrix Product States, MPS) \cite{schollwock_2011, Ran2020,Cirac2021} can also be used but usually require a space complexity (memory usage) that scales exponentially with the so-called \textit{entanglement entropy} $S$ (see section \ref{sec:entanglement} for a more in-depth discussion).
These methods are potent when $S$ is constrained to small values, as happens e.g in one dimension. However, in two and more dimensions, $S$ usually grows in a way that makes these methods difficult to apply.

The Hubbard model proves insufficient to describe molecules in quantum chemistry, as chemical systems are less prone to screening and thus cannot be described by on-site interactions only. Thus, one typically deals with the following more general Hamiltonian:
\begin{equation}
    H = \sum_{pq, \sigma} h_{pq} c_{p\sigma}^\dagger c_{q\sigma} 
    + \frac{1}{2} \sum_{pqrs} \sum_{\sigma \sigma'} v_{pqrs} c^\dagger_{p \sigma} c^\dagger_{q \sigma'} c_{r \sigma'} c_{s \sigma} 
    \label{eq:el_structure_h}
\end{equation}
where $p,q,r,s = 1 \dots N$ denote orbitals $\phi_q(r)$ (typically molecular orbitals obtained after a first Hartree-Fock computation).
The one- and two-body matrix elements $h_{pq}$ and $v_{pqrs}$ are computed as integrals over the molecular orbitals. Typically, $N$ is of the order of $10-100$. 

%\sout{The exact diagonalization of this Hamiltonian (ED, usually called \textit{Full Configuration Interaction} 
% -- FCI -- in a quantum chemical context) is limited to a small number $N$ of orbitals.
%Variational methods based on a perturbative expansion on the HF wave function are less costly than FCI. Typically, the \textit{Coupled Cluster} (CC) method, based on the variational state (here limited to single and double excitations)}

{The exact diagonalization of this Hamiltonian (ED, usually called \textit{Full Configuration Interaction} 
 -- FCI -- in a quantum chemical context \cite{David1999}) is limited to a small number $N$ of orbitals.
Variational methods that optimize parameterized wave-functions to minimize the energy are less costly than FCI. One simple example of such variational approaches is the HF method itself. Another illustration of a widely used variational approach in chemistry is the multi-configurational self-consistent field (MCSCF) method \cite{Hinze67,schmidt1998,jensen2007,cramer2013}.
Alternatively, non-variational methods based on systematic   
expansions on top of the HF wave-function, like many-body perturbation theory (MBPT) \cite{shavitt2009,Werner2009}, are also less costly than FCI.
A prominent example of such methods is the \textit{Coupled Cluster} (CC) method, based on the parametric many-body state (here limited to single and double excitations)} \cite{Bartlett2007,shavitt2009,Evangelista2018}
\begin{align}
    |\Psi(\vec{\theta}) \rangle &= e^{T_1 + T_2}\ket{\Psi_\mathrm{HF}} \nonumber \\
    &=e^{ \sum \limits_{ia, \sigma} \theta_{i}^{a} c_{i\sigma}^\dagger c_{a\sigma} + \sum \limits_{ijab, \sigma\sigma'} \theta_{ij}^{ab} c^\dagger_{i\sigma} c^\dagger_{j\sigma'} c_{a\sigma'} c_{b\sigma} }\ket{\Psi_\mathrm{HF}}, \label{eq:cc_ansatz}
\end{align}
with $i,j$ (resp. $a, b$) empty (resp. occupied) orbitals. {The non-variational nature of the CC method comes from the non-unitary character of the so-called \textit{excitation operator} $e^{T_1 + T_2}$, which implies that the variational principle does not hold. A unitary, variational variant of the CC method, called unitary coupled cluster (UCC), has been investigated for quantum computing and is tackled in \ref{subsubsec:ucc}.} This method is regarded as one of the most advanced methods in chemistry (for systems with dynamical correlations, as opposed to systems with large static correlations, where methods based on Matrix Product States [see section \ref{sec:entanglement}] are among the most advanced \cite{Baiardi2020}).
It can be combined with active-space methods, whose goal is to reduce the number of relevant (or \textit{correlated} or \textit{active}) degrees of freedom, similar to embedding methods for solids. This selection of degrees of freedom will be further discussed in subsection \ref{subsec:Reducing-the-number}.

\subsection{Nuclear physics}
\label{sec:nucleon}
Atomic nuclei are self-bound, strongly interacting systems with 
a wide range of numbers of particles, from very few ($2$ for the deuteron) to several hundreds for the heaviest nuclear systems existing in nature.
Nucleons organize themselves to form quantum droplets with a large variety of static and dynamical physical phenomena \cite{Bohr1998}.
These phenomena can be observed in the laboratory through the use of accelerators. 
The many-body treatment of nuclei is particularly complex due to the non-perturbative nature of the two-body interaction with a strong repulsion at short distances between particles.

The nuclear Hamiltonian is of the form \eqref{eq:manybodyhamiltonian} where the multi-indices $(\alpha, \beta, \gamma, \delta) =1,..., N$ 
label single-particle states characterized by the usual quantum numbers $n, l, m, \sigma$, as well as an isospin component $\tau$.  
These states can for instance be 3-dimensional harmonic oscillators (HO) states with $\phi_\alpha({\bf r}) = \phi_{nlm\sigma \tau}({\bf r})$ ($\tau = -1/2$ and $+1/2$ for neutrons and protons respectively), where the strength of the HO is optimized to reproduce nuclei sizes \cite{Ring80}.
Due to the presence of spin-orbit coupling, it is usual to introduce the angular momentum $\vec j = \vec l + \vec s$, and relabel the state with $(nljm \sigma \tau)$.    
Since there is no external field, the one-body term only contains the kinetic component. The two-body interaction contains nuclear (short-range) and Coulomb (long-range) interactions.
The Coulomb part acts only on protons, while the nuclear part, which depends on the spin of the particles, acts on all nucleons.
Note that the latter is almost the same for all particles, a property know as isospin symmetry of the nuclear force. 
Altogether, $v_{\alpha \beta \gamma \delta}$ is both spin and isospin-dependent.
In addition, the existence of 3-body and, more generally, multi-body interactions was recognized only recently with the recent advances in Effective-Field-Theory to construct nuclear interactions \cite{Epelbaum2009,Machleidt2011, Hammer2013,Machleidt2016,Hammer2020}.
The presence of multi-body interactions is an extra complication compared with the electronic structure Hamiltonian \eqref{eq:el_structure_h}, and even in the most advanced many-body techniques, such interactions are usually treated only approximately.

Despite this complexity, a variety of simplified Hamiltonians have been proposed to understand specific properties of nuclei.
A typical example is the Lipkin-Meshkov-Glick model \cite{Lipkin1965}, which is often used to understand the concept of spontaneous symmetry breaking in finite systems.
Another example is the pairing, also called Richardson \cite{Vondelft2001}, Hamiltonian that is often used to understand superconducting effects.
This Hamiltonian can be justified by (i) assuming that only a set of particles, typically close to the Fermi energy are active, (ii) the interaction between them is constant, and (iii) it is non-negligible only when pairs of time-reversed states, denoted by $(i,\bar i)$, are involved.
These states are often taken as opposite spin particles or as particles with spin projection $j_z = m$ and $-m$ when a single $j$-shell is considered as active.
The Hamiltonian then reduces to:
\begin{eqnarray}
H &=& \sum_i \varepsilon_i (c^\dagger_i c_i + c^\dagger_{\bar i} c_{\bar i}) - g \sum_{ij} c^\dagger_i c^\dagger_{\bar i} c_{\bar j} c_j . 
\label{eq:pairing}     
\end{eqnarray}
These simple, schematic Hamiltonians are studied today on quantum computers as first steps towards future applications.  

An overview of the microscopic approaches used to describe atomic nuclei can be found in Ref. \cite{Lacroix2010}.
The only approach able to describe the large variety of phenomena ranging from static (nuclear structure), dynamical (nuclear dynamics), and thermodynamical properties is nuclear Density Functional Theory, often referred to as Energy Density Functional (EDF) theory \cite{Bender2003,Nakatsukasa2016,Colo2020}.
Another powerful approach, restricted to studying nuclear structure properties, consists of performing a direct CI method in a restricted subspace of single-particle states forming the valence space \cite{Caurier2005,Nowacki2021}.
In this approach, often referred to as the \textit{Shell Model}, the effective interaction is fine-tuned to account for the truncation of the model space.
One of the main difficulties that forces the restriction to a set of active valence particles is the size of the many-body Hilbert space when the number of single-particle states increases.
The current scope of restricted CI approaches is the treatment of eigenvalue problems in spaces with $10^{11}-10^{12}$ states.
These values are still far from the requirements to treat the whole nuclear chart with all single-particle active states. 

Another breakthrough in the most recent description of interactions lies in the possibility of getting rid of the hard core and using \textit{soft interactions} for low-energy nuclear physics problems \cite{Bogner2010}.
Such interactions have opened the way to the so-called \textit{ab-initio method} that aims at treating the nuclear many-body problem directly, starting from the bare Hamiltonian (\ref{eq:manybodyhamiltonian}).
This advance has led to a significant boost in applying several many-body techniques, some already mentioned in section \ref{sec:electron}.
One can, in particular, mention the use of 
the full CI technique---known, in this context, as the no-core shell-model \cite{Quaglioni2008,Navratil2009},
the Green's Function Monte-Carlo method \cite{Pudliner1997,Wiringa1998,Wiringa2000},
the Self-Consistent Green Function method \cite{Dickhoff2004,Soma2020},
the Coupled-Cluster method \cite{Bartlett2007,Hagen2014},
or Many-Body Perturbation Theory \cite{Tichai2020}, among others.
One specific aspect of atomic nuclei is the necessity to generalize some of these theories to allow for possible spontaneous symmetry breaking like particle number or rotational symmetries (see, for instance, a few examples of extensions in Refs.~\cite{Tichai2011,Soma2011,Henderson2014,Duguet2016,Demol2021}).
This generalization is fundamental to describing open-shell nuclei. 

Although significant progress has been made in recent years, ab-initio methods are still applied to study the nuclear structure effects in a limited region of the nuclear chart or to nuclear reactions involving only very light systems.
This limitation is due to the increment in the complexity of the problem when the number of particles increases. 
For now, very few quantum computing pilot applications on real devices have been made, and these have been limited to rather simplistic nuclear Hamiltonians \cite{Dumitrescu2018,Lu2019,Kiss2022}.
However, the use of quantum computers for nuclear many-body problems has recently gained momentum \cite{Roggero2020,Roggero2020b,Lacroix2020,Dimatteo2021,Siwach2021,Du2021,Siwach2021b,Cervia2021,ruiz-guzman2021,Hlatshwayo2022,Perezfernandez2022,RuizGuzman2022,Stetcu2022,Siwach2022,Romero2022,Hodbay2022,Perez2023,Robin2023,Beck2023}.

\subsection{Common difficulties}
\label{sec:commondifficulty}

A defining property of many-body systems is that they are exponentially difficult to solve on classical computers.
As we will see in this section, this exponential difficulty may trivially arise from the size of the Hilbert space.
However, most classical methods strive to circumvent this difficulty by either using the structure of the problem to merely reduce the number of relevant degrees of freedom or by adopting another representation---shifting the exponential difficulty from the size of the Hilbert space to other parameters, like, for instance, the severity of the Monte-Carlo sign problem or the internal dimension of a tensor network.

One standard strategy to circumvent the exponential complexity is to use parameterized variational wave function ansätze.
Another strategy---based on so-called \textit{reduced density matrices}---relies on the assumption that some degrees of freedom contain more information than others. 
A typical starting point is to assume that one-body degrees of freedom are the most relevant.
The information on them is contained in the one-body reduced density matrix (1-RDM). Reducing the information to these DoFs leads to  Hartree-Fock (HF) or mean-field theory. 
Usually, such simplified approaches miss significant correlation effects, and approximations beyond the mean-field are necessary to describe many-body systems accurately. For instance, this description can be done by truncating the Bogolyubov-Born-Green-Kirkwood-Yvon (BBGKY) hierarchy and treating 2-body or higher DoFs explicitly through the two-body reduced density matrix (2-RDM) or higher-order reduced density matrices \cite{Bonitz2016} (see also the discussion of embedding or active-space methods in section \ref{sec:dofreduction}).

Despite the dimensionality reduction that the aforementioned approximate methods afford, they always, at some point, reach a computational limit on classical computers. Quantum computers---provided fermionic problems can be turned into spin or qubit problems (see section \ref{subsec:From-fermions-to})---can a priori overcome these limitations provided enough qubits can be efficiently manipulated. Interestingly, quantum computers do not start from a blank page: many quantum algorithms are strongly guided by the accumulated expertise and methods gained on classical devices. An illustration of that will be discussed in section \ref{sec:mapping_to_qc}.   

Besides the above general considerations on many-body systems, each physical system has specificities that will render its encoding on quantum computers more or less difficult. For instance, electrons can have two spin components (spin up and down), while nucleons can have both spin and isospin (neutron and proton components). Some physical systems, such as solids or atomic ones, can be suitably described on a lattice, sometimes with only nearest neighbor two-body interaction. This case is advantageous when encoding such a problem on a set of qubits with limited connectivity. Some other systems might be more complex, like atomic nuclei, with long-range two-body or, more generally, multi-body interactions or the necessity to describe unbound states.
A prerequisite to the success of future applications is the efficient transposition of a given many-body problem with good mapping of its characteristics into an analog or a digital quantum computer (see section \ref{sec:mapping_to_qc}).
The complexity of this transposition will strongly depend on the problem itself, but many-body problems will undoubtedly be among the first stringent benchmarks for current and future quantum technologies. 

The exponential size of the Hilbert space is not a sufficient condition for making a problem computationally hard from classical computers.
For instance, the ground state of quadratic fermionic Hamiltonians---namely Hamiltonians that are bilinear in the creation and annihilation operators---can be found in polynomial time on a classical computer (see e.g \cite{Bruus2002}). The problem becomes complicated only when terms beyond quadratic terms are added. This result explains why certain classes of time evolutions---quantum circuits (see section \ref{subsubsec:digital_qc} below for a definition) stemming from a quadratic Hamiltonian, with so-called \textit{matchgates}---are also simulatable classically in polynomial time.
Interestingly, other circuits, this time unrelated to "uncorrelated systems", are efficiently simulatable. 
For instance, Clifford circuits, with gates belonging only to the Clifford group \cite{Gottesman1998}, are simulatable with time and space complexity $\mathcal{O}(n^2)$ (with $n$ the number of qubits), a result known as the Gottesman-Knill theorem \cite{Gottesman1998a, Aaronson2004}. This result holds even though these circuits can generate highly entangled states. The key idea behind the efficiency of the simulation of such circuits is that the states generated by Clifford circuits can be represented in a compact (polynomial) fashion.

Leveraging a compact representation is also very common in many-body classical methods. For instance, some Monte-Carlo methods decouple quartic (interaction) terms in the Hamiltonian to eliminate this term at the expense of an additional auxiliary field \cite{BAEURLE2004}. Then, the exponential difficulty is shifted from the size of the Hilbert space to the Monte-Carlo sign problem.
In tensor network methods, like Matrix Product States \cite{schollwock_2011}, the accuracy of the representation is tuned by the size of the internal indices (often called the \textit{bond dimension}), which is related to the degree of entanglement of the state at stake (see section \ref{sec:entMPS}).

\subsection{A few quantum complexity considerations}

In this section, we comment on the formal expectations regarding speedups that quantum computers can bring when solving many-body problems. For general reviews on the topic, we refer the reader to e.g \cite{Watrous2012a, Nielsen2010}.

Classical computational problems with a yes/no answer (decision problems) are classified using complexity classes.
For instance, finding the ground state energy of a Hamiltonian $H$ acting on $n$ qubits (fermions) can be formulated as a decision problem by picking numbers $a$ and $b$ such that $\epsilon \equiv b - a > 1/\mathrm{poly}(n)$ defines a region where it is promised that the ground state energy does not lie in, and asking whether $E_0 < a$ or $E_0 > b$ (see e.g., \cite{Watson2022} and references therein).

\begin{figure}
\includegraphics[width=\columnwidth]{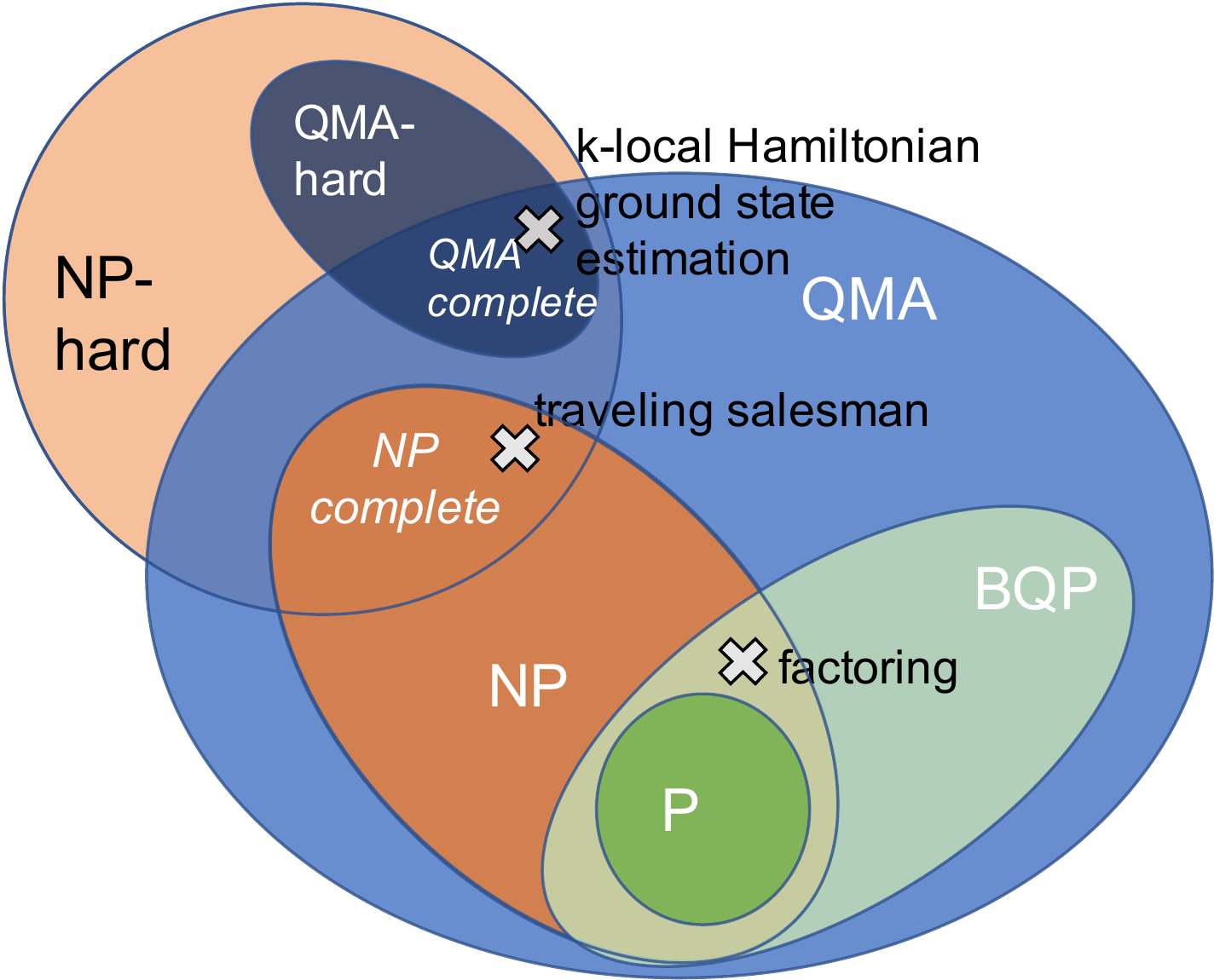}
    \caption{Computational complexity classes and selected problems. The three crosses indicate three examples of complex problems. The "factoring" and "traveling salesman" are common problems often quoted in complexity theories. The "k-local Hamiltonian ground state estimation" is a common problem in many-body systems (see text).} 
\label{fig:complexity_classes}
\end{figure}

The two main classes of classical complexity are P and NP, which describe problems that can be solved in polynomial time or whose solution can be verified in polynomial time. NP-hard problems are problems at least as hard as any problem in NP, and NP-complete problems are the NP-hard problems that belong to NP. NP-complete problems are very likely (that is unless P=NP) not to have a polynomial-time solution, i.e., they are colloquially referred to as \textit{exponentially hard} problems.

One central question is whether quantum computers can reduce the complexity of solving many-body problems \cite{Bernstein1997}.
To answer this question systematically, two quantum complexity classes, {BQP (bounded-error quantum polynomial time) and QMA (quantum Merlin–Arthur)}, have been designed as quantum counterparts to P and NP, respectively: the computer that solves the problem or, respectively, certifies the solution of the problem is a quantum computer.
Fig~\ref{fig:complexity_classes} illustrates these classes and the connection between quantum and classical complexity classifications. 

A significant result is that NP-complete problems are very likely not in BQP.
In other words, it is improbable that quantum computers can solve exponentially hard classical problems in polynomial time.
The fact that the factoring problem can be solved using Shor's algorithm \cite{Shor1994} with an exponential speedup is possible because factoring is not a NP-complete problem. Importantly, this also does not mean that quantum computers are not helpful for NP-complete problems: quantum heuristics (polynomial algorithms with no quality guarantee) can still reach better solutions than classical heuristics. 

A natural question to know what to expect from quantum computers for many-body problems is to which complexity class they belong.
The ground state estimation problem is QMA-complete for $k$-local (spin) Hamiltonians (Hamiltonians whose terms act on at most $k$ qubits, see Eq.~\eqref{eq:pauli_decomp} below) as long as $k \geq 2$  \cite{Kempe2008}.
QMA-completeness is retained for the ground state estimation problem of geometrically 2-local Hamiltonians -- that is to say 2-local Hamiltonians only involving pairs of adjacent qubits -- in a square lattice qubit layout (aka a quantum spin glass) \cite{Oliveira2008}.
These statements hold for general values of the couplings of the spin Hamiltonians mentioned above.
Restrictions on the coefficients can lead to a reduction in computational complexity.
For instance, the transverse Ising model with negative transverse field and ferromagnetic interactions (i.e., Eq.~\eqref{eq:Ising} below with $\Omega < 0$ and $C < 0$ [which is not the case for Rydberg atoms, where $C>0$]) is a so-called \textit{stoquastic} Hamiltonian. Its ground state energy can be approximated polynomially in the system size $n$ and $1/\epsilon$ \cite{Bravyi2015}.

As for fermionic and bosonic models, \cite{Schuch2009} showed that the Fermi-Hubbard model with local magnetic fields (with an additional $\sum_i \bf{\sigma}_i \cdot \bf{B}_i$ term in Eq.~\eqref{eq:Hubbard_solids}) is QMA-complete, while \cite{Childs2013} showed that the Bose-Hubbard model is QMA-complete.

These formal considerations call for two comments. First, despite these hardness results, the ground state energy of the models cited above can be determined with great accuracy on classical computers in special regimes, i.e., for certain classes of parameters.
For instance, this is the case of the Fermi-Hubbard model on a bipartite lattice, whose solution with quantum Monte-Carlo does not suffer from a sign problem at half-filling (see, e.g., \cite{Leblanc2015}).
The existence of these special regimes means that one must be careful to look for truly hard classical computational regimes to identify a useful application of quantum computers.
In other words, classical methods make the most of any symmetry or structure of the problem to overcome the underlying exponential complexity of the many-body problem so that truly hard regimes are hard to come by.
In these regimes, quantum computers ought to also leverage these symmetries and structures in order to outperform classical computers. 

Second, even in those regimes where classical computers fail to reach an accurate enough result, the QMA-completeness of the problem is a strong indication that the problem will also be hard to solve on a quantum computer. In other words, classical and quantum algorithms likely end up running into an exponential wall (see, e.g., \cite{Lee2022} for a concrete example). This limitation is not necessarily a showstopper: what matters is whether quantum computers can reach regimes inaccessible to classical computers before they run into said wall. 

Lastly, complexity theory can also be used to appraise, at least formally, the feasibility of hybrid quantum-classical algorithms like the \textit{Variational Quantum Eigensolver} (see section \ref{subsec:vqe} below).
For instance, the classical optimization procedure of the energy $E(\bf{\theta})$ in VQE is generically a challenging computational problem \cite{Bittel2021}.
In practice, this does not exclude the existence of heuristics for finding accurate enough variational parameters for concrete (as opposed to generic) problems. What is more, the classical counterpart---an entirely classical variational algorithm---also suffers from the same problems.
Ultimately, what matters is whether quantum processors can accelerate parts of the computation {\it relative to} the best classical algorithm.

\section{Quantum computers: artificial many-body systems\label{sec:Quantum-computers:-artificial} }

This section explains how to investigate the many-body problems mentioned above by building an artificial many-body system with a similar Hamiltonian (aka analog quantum computers or quantum simulators) or, still starting from a many-body system, but with individual control over the particles/degrees of freedom, by building a gate-based (digital) quantum computer (subsection \ref{subsec:From-analog-to}).
We then explain the basic building blocks and rules of ideal quantum computers.

\subsection{From analog to digital\label{subsec:From-analog-to}}

Quantum computers are essentially synthetic many-body systems whose state can be manipulated according to some predefined plan---{\it aka} a quantum program---and measured to learn something. 

Depending on the level of control of this quantum system, one speaks of quantum \textit{simulators} (or analog quantum computers) or quantum \textit{computers} (or digital quantum computers).
While (analog) simulators offer only a limited and specific set of controls, (digital) computers offer controls---usually called \textit{gates}---that are universal. This universality allows them to reach, in principle, any state of the Hilbert space by performing any unitary operation. 

\subsubsection{Analog quantum computers ({\it aka} quantum simulators)}

The term \textit{analog quantum computer} refers to any synthetic many-body system with a certain amount of control over its degrees of freedom. 
Each analog computer is characterized (in the absence of defects) by a many-body Hamiltonian $H(t)$ whose time-dependence is "programmed" more or less at will, depending on the experimental constraints.
This Hamiltonian is usually chosen as close as possible to the "real-life" Hamiltonians introduced in the previous section.
Thus, by measuring the properties of the analog simulator, one hopes to get insights into the physics of real-life systems. We highlight below some illustrations of physical systems used as analog simulators.

{\it Ultracold atoms} can be described as implementing a Fermi- or Bose-Hubbard model, depending on the atomic isotopes used \cite{Bloch2008}.
For instance, the Bose-Hubbard model reads:
\begin{equation}
    H(t) = \frac{U(t)}{2} \sum_i n_{i} (n_{i} - 1) - J (t) \sum_{\langle ij\rangle} b_i^\dagger b_j .
        \label{eq:Hubbard}
\end{equation}
Here, the creation and annihilation operators $b^\dagger_i$ and $b_i$ create and annihilate (bosonic) atoms in orbitals $\phi_i(r)$ and $n_i = b^\dagger_i b_i$; $J(t)$ is the tunneling between two neighboring "sites" $\langle i j \rangle $ of the optical lattice, and $U(t)$ is the on-site repulsion between two atoms.
Both $U$ and $J$ can be temporally modulated by changing the amplitudes of the lasers creating the lattice. In addition, the interaction $U(t)$ can also be tuned by changing the background magnetic field using a phenomenon known as the Feshbach resonance \cite{Bloch2008}.

{\it Spin qubits}, which are essentially electrons trapped in quantum dots, can also be described by a Fermi-Hubbard model or, when neglecting charge fluctuations, by a Heisenberg model \cite{Loss1998}:
\begin{equation}
    H(t) =  J_\mathrm{ex}(t) \sum_{\langle ij \rangle }  \left( X_i X_j + Y_i Y_j + Z_i Z_j \right) + \sum_i H_\mathrm{loc}^{(i)},
    \label{eq:Heisenberg}
\end{equation}
with $(X_i,Y_i,Z_i)$ denoting the Pauli matrices acting on the $i^{\mathrm{th}}$ spin, and with   
\begin{eqnarray}
    H_\mathrm{loc}^{(i)}=\frac{\omega_{0}^{(i)}-\delta_i(t)}{2}Z_i +
    \Omega_{i}(t)\cos\left(\omega_{c}^{(i)}t+\phi_{i}(t)\right)X_i.
    \label{eq:H_loc_spin}
\end{eqnarray}

The exchange constant $J_\mathrm{ex}\sim 4 J(t)^2/ U$ can be turned on and off via the tuning of the tunneling term $J(t)$ between two dots using a gate voltage. 
The local term $H_\mathrm{loc}$ can, for instance, come from a magnetic field with a static ($Z_i$ term) and a rotating ($X_i$ term) component.

Depending on which atomic levels they target, platforms of Rydberg atoms (see, e.g., \cite{Browaeys2020, Henriet2020a}) may implement an Ising Hamiltonian:
\begin{align}
    H(t) &=  \sum_{ij, i\neq j} \frac{C}{|r_i - r_j|^6} n_i n_j  
    \nonumber \\
    &
    + \frac{\Omega(t)}{2} \sum_i X_i - \delta(t) \sum_i Z_i , 
    \label{eq:Ising}
\end{align}
with $n_i = (1 - Z_i)/2$, or a $XY$ Hamiltonian:
\begin{align}
    H(t) &=  2 \sum_{ij, i\neq j} \frac{C}{|r_i - r_j|^3} \left( X_i X_j + Y_i Y_j \right) \nonumber \\
    &+ \Omega(t) \sum_i X_i - \frac{\delta(t)}{2} \sum_i Z_i.
    \label{eq:XY_model}
\end{align}

{\it Superconducting qubits}, which are usually thought of as (digital) computers, can also be seen as analog computers realizing a Bose-Hubbard model (see Eq. (\ref{eq:Hubbard})). 
There, the creation and annihilation operators refer to bosonic excitations relative to the charge and flux variables inside Josephson junctions.

All these Hamiltonians are of many-body nature owing to the coupling terms (first term of each equation). Thanks to their closeness to the many-body Hamiltonians encountered when studying quantum matter (see previous section, \ref{sec:Quantum-Many-body-problems}), these Hamiltonians have long been used as proxies (or \textit{simulators}) for the many-body systems one wants to understand.

By nature, quantum simulators are very specific in that
(i) they implement (or "simulate") only one class of Hamiltonians, and
(ii) they usually implement partial (often only global) control of their degrees of freedom.
This limitation is both temporal and spatial: e.g., in Eq. (\ref{eq:Ising}), the coupling term cannot be made time-dependent as it corresponds to a van der Waals interaction that cannot be switched off or decreased, except by moving the atoms, a very slow operation.
Also, the second (Rabi) and third (detuning) terms can most often not be controlled at an individual site level (i.e., $\Omega$ and $\delta$ are the same for all atoms).

Analog platforms have advantages and drawbacks. Their primary advantage is that the limited degree of control usually allows them to work with significantly more degrees of freedom (atoms, spins, ions, junctions, and other building blocks). The major drawback is that {their scope is limited according to the specific resource Hamiltonian at hand, whose terms might not be tunable at will.}
%they are not "universal" or "all-purpose".
{
Experimental platforms 
%must reach a reasonably good temporal and spatial degree of control to become "universal"  (in a sense that will be made more explicit later)
endowed with reasonably good levels of temporal and spatial control earn the qualification of \textit{gate-based quantum computers.}}

%This performance level is required to have gate-based or digital computers.

\begin{figure}
\includegraphics[width=\columnwidth]{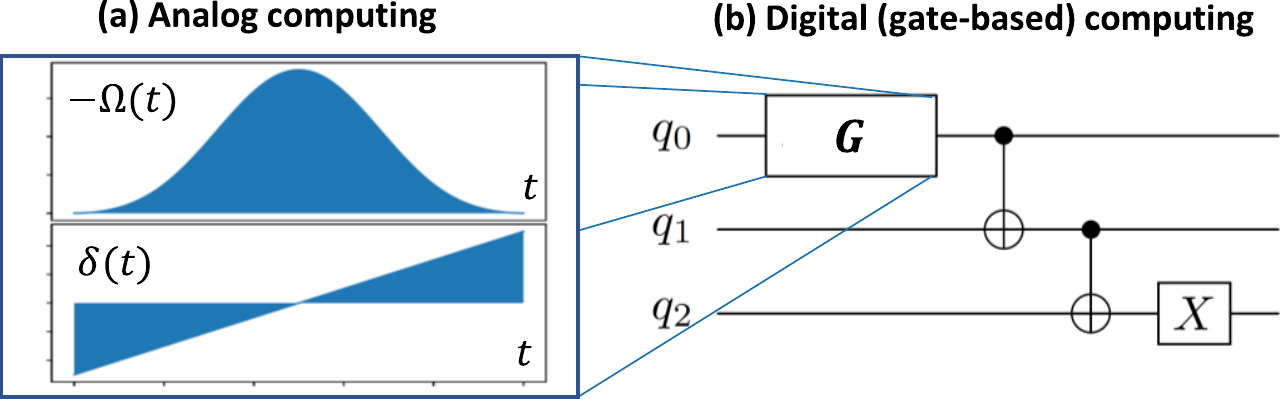}
    \caption{Analog computation (a) vs. digital computation (b). In analog computation, one directly specifies the (analog) parameters (here $\Omega(t)$ and $\delta(t)$, see, e.g., \eqref{eq:H_loc_spin}, \eqref{eq:Ising}, or \eqref{eq:XY_model}). These controls are not necessarily local (as in \eqref{eq:Ising} or \eqref{eq:XY_model}). In digital computations, the user discretely describes the sought-after evolution with quantum gates, which are usually local, i.e., act only on a few qubits (lines in the diagram). Internally, each gate is performed using an analog description.
    }
\label{fig:analog_vs_digital}
\end{figure}

\subsubsection{Digital ({\it aka} gate-based) quantum computers}\label{subsubsec:digital_qc}

Digital, or gate-based quantum computers, refer to physical setups (i) whose description can be narrowed to an assembly of interacting 
%two-level 
{$d$-level} quantum systems and (ii) whose Hamiltonian can be controlled at a local level.\\

{In this review, we focus on the \textit{qubit} case $d=2$, but note that \textit{qudit} approaches, where $d>2$, were also developed \cite{Wang2020,Jiang2020,Ringbauer2022}.}

\paragraph{Two-level quantum systems: qubits.}
Let us focus on criterion (i). For instance, among the systems cited above, Rydberg atoms or spin qubits are already naturally described as two-level systems: two atomic levels for Rydberg atoms and two spin levels for spin qubits. In photonic platforms, the photon's two polarizations can play the role of the two levels. 
Superconducting platforms, which are naturally described with bosonic variables, can be restricted to a two-level subspace by tuning their parameters so that "leakage" out of the two lowest levels---called \textit{computational subspace}---is very improbable.
The two levels of the computational subspace are usually denoted as $\ket{0}$ and $\ket{1}$. Hence, the wavefunction of a single two-level system, or \textit{qubit}, is, in general, the superposition:
\begin{equation}
    \ket{\psi} = a_0 \ket{0} + a_1 \ket{1},
\end{equation}
with $a_i\in \mathbb{C}$ and $|a_0|^2+|a_1|^2 = 1$.
More generally, a $n$-qubit wavefunction $\ket{\Psi}$ is the superposition of $2^n$ \textit{computational basis states} $|00\dots 0\rangle, |00\dots 01\rangle, \dots |11\dots 1\rangle$.
We see that all states can be written as $| b_{n - 1}, \cdots , b_0 \rangle = \bigotimes_{i=0}^{n -1} | b_i \rangle$  where $|b_i \rangle = | 0_i \rangle$ or $| 1_i \rangle$ refers to the state of the $i^{\mathrm{th}}$ qubit.
Below, we will use the same convention as in Eq. (\ref{eq:Ising}), and operators that act on this qubit will be labeled by $i$ like, for instance, the Pauli operators $(X_i,Y_i,Z_i)$. 
Each state $| b_{n - 1}, \cdots , b_0 \rangle$ can also be labeled by a single integer $k=\sum_{i} b_i 2^{i}$.

\paragraph{Manipulating qubits: gates.}\label{par:gates}
Criterion (ii) ensures that one can reach any state of this Hilbert space using operations called quantum gates. Mathematically, these gates are unitary operations $U$ acting on the wavefunction $\ket{\Psi}$: $\ket{\Psi'} = U \ket{\Psi}$.
Such operations are performed by letting the system evolve under a given Hamiltonian. For instance, let us consider a $n$-qubit system described by the (non-interacting) Hamiltonian
\begin{equation}
    H=\sum_{i=1}^{n} H_\mathrm{loc}^{(i)},
    \label{eq:digital_ham}
\end{equation}
with $H_\mathrm{loc}^{(i)}$ defined in Eq.~\eqref{eq:H_loc_spin}. 
$\omega_0^{(i)}$ is the $i^{\mathrm{th}}$ qubit's frequency (energy difference between the two levels), and $\omega_{c}^{(i)}$ is the drive frequency. $\delta_i(t)$, $g_i(t)$ and $\phi_i(t)$ in Eq.~\eqref{eq:H_loc_spin} are controllable fields.
If one switches off all but the $i^{\mathrm{th}}$ qubit's field, one goes to the frame rotating at frequency $\omega_0^{(i)}$, one drives at resonance (namely $\omega_{c}^{(i)}=\omega_{0}^{(i)}$), and one neglects terms oscillating as $2 \omega_0^{(i)}$ (so-called \textit{rotating wave approximation}), the Hamiltonian reads, up to terms acting on the other qubits, as
\begin{equation}
    H_{\omega_0^{(i)}} = - \frac{\delta_i(t)}{2} Z_i 
    + \frac{\Omega_i(t)}{2} \left[\cos(\phi_i)X_i 
    +\sin(\phi_i)Y_i\right]. \label{eq:rabi}
\end{equation}
If we turn off the Rabi term $\Omega_i$, solving the Schr\"odinger equation yields a wavefunction $\ket{\psi(t)} = R_z^{(i)}(\theta(t)) \ket{\psi(0)}$ with $R_z^{(i)}(\theta) \equiv e^{- i \frac{\theta}{2} Z_i}$ and $\theta(t) = -\int_0^t \delta_i(\tau) d\tau$.
Thus, we have operated a rotation of angle $\theta(t)$ around axis $z$ for the $i$th qubit.
Similarly, if we turn off the "detuning" term $\delta_i$, we are going to effect a rotation along axes $x$ ($\phi = 0$) and $y$ ($\phi = \pi /2$). 

\begin{figure}[htbp]  
\centering
\includegraphics[width=0.8\columnwidth]{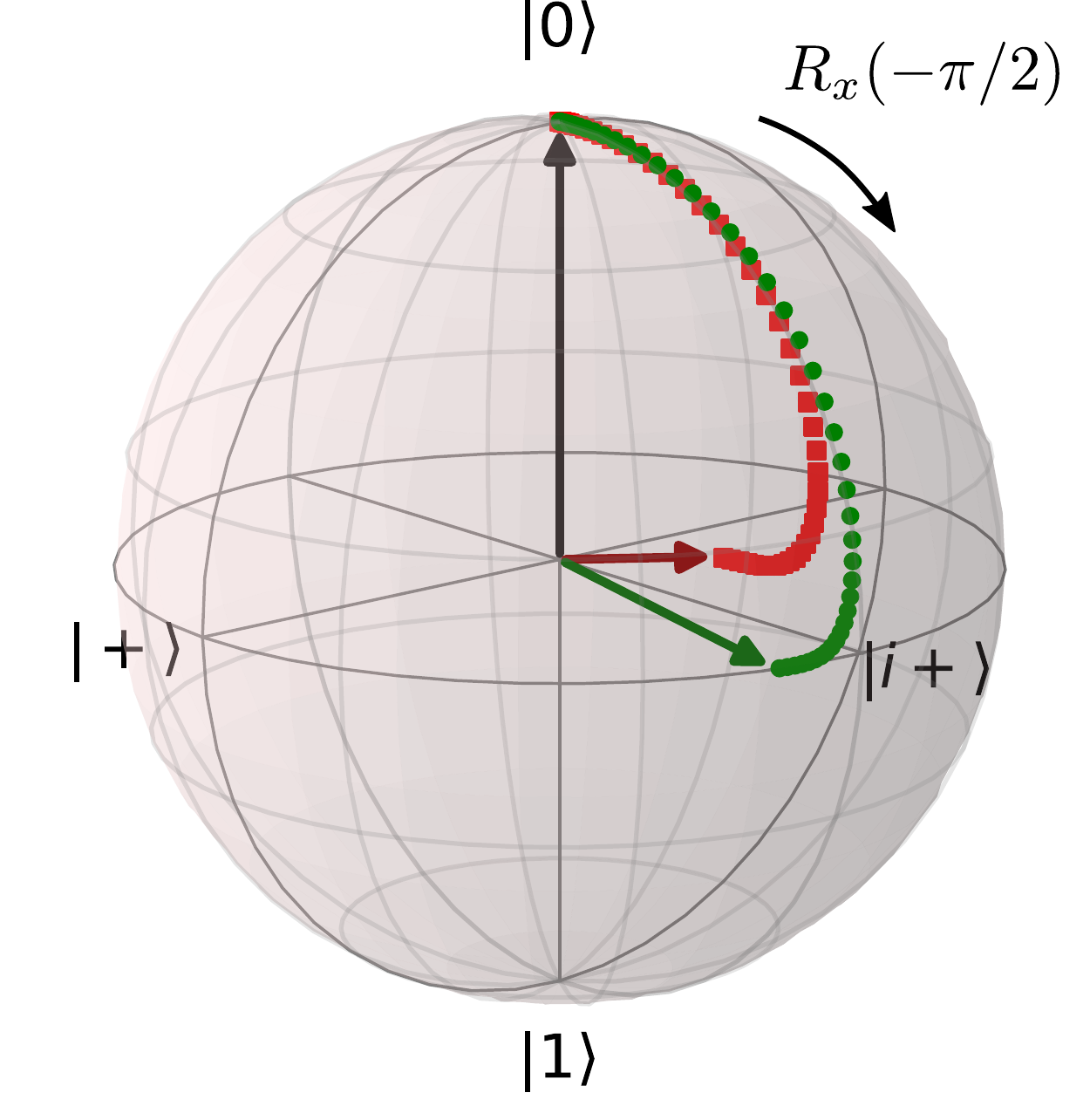}
\caption{Bloch sphere with the North and South pole corresponding to $|0\rangle $ and $| 1 \rangle$, respectively. The other poles shown in the figure are decomposed in terms of $|0\rangle$ and $|1\rangle$ as $|+\rangle = \frac{1}{\sqrt{2}}\left(|0\rangle+|1\rangle\right)$ and $|i+\rangle = \frac{1}{\sqrt{2}}\left(|0\rangle+i|1\rangle\right)$. The green and red trajectories represent the evolution of a one-qubit state, starting from $|0\rangle$, when subject to a $R_x(-\pi/2)$ rotation.
Green curve: noiseless (pure state) evolution. This evolution approximately realizes a $R_x(-\pi/2)$ rotation (it would be exactly such a rotation if the detuning drive $\delta(t)$ were rigorously 0). Red curve: noisy evolution under dephasing and relaxation noise, see section~\ref{sec:noise}.}

\label{fig:bloch_sphere}
\end{figure}

Such a time evolution is illustrated in Fig.~\ref{fig:bloch_sphere}, using the standard Bloch sphere representation \cite{Nielsen2010}. We show the evolution of a one-qubit state under a Rabi drive $\Omega(t)$ and a detuning drive $\delta(t)$ (see Fig.~\ref{fig:analog_vs_digital}), with $\phi = 0$ (green trajectory).
The area under the Rabi curve $-\Omega(t)$ is chosen to effect a $-\pi/2$ rotation, but as a consequence of the detuning being not strictly zero, a small $z$-axis rotation is effected in addition to the $x$-rotation.
We will explain what happens when the qubit is affected by decoherence (red trajectory) in a later section (section~\ref{sec:noise}).

\begin{table*}
\centering
\includegraphics[width=1.0\linewidth]{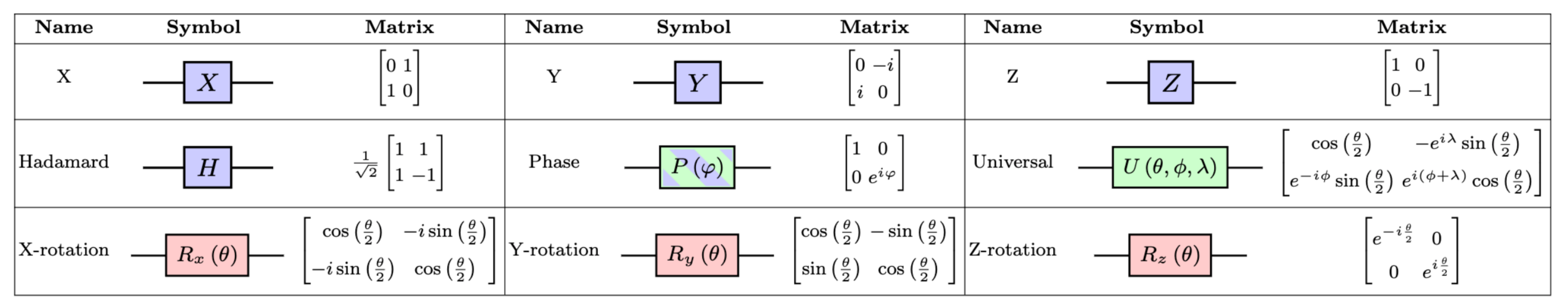}
 \caption{Summary of some standard single-qubit quantum gates. The Clifford group mentioned in section \ref{sec:commondifficulty} is depicted in purple and can be generated using the {\it H}, $S=P\left(\pi/2\right)$ and CNOT gates (see Table \ref{tab:double_gates}). The purple and green hatches on the phase gate $P\left(\varphi\right)$ reflect that it is Clifford only for $\varphi=\pm \pi/2$. Table adapted from \cite{Motta2022}.}
\label{tab:single_gates}
\end{table*}

\begin{table*}
\centering
\includegraphics[width=1.0\linewidth]{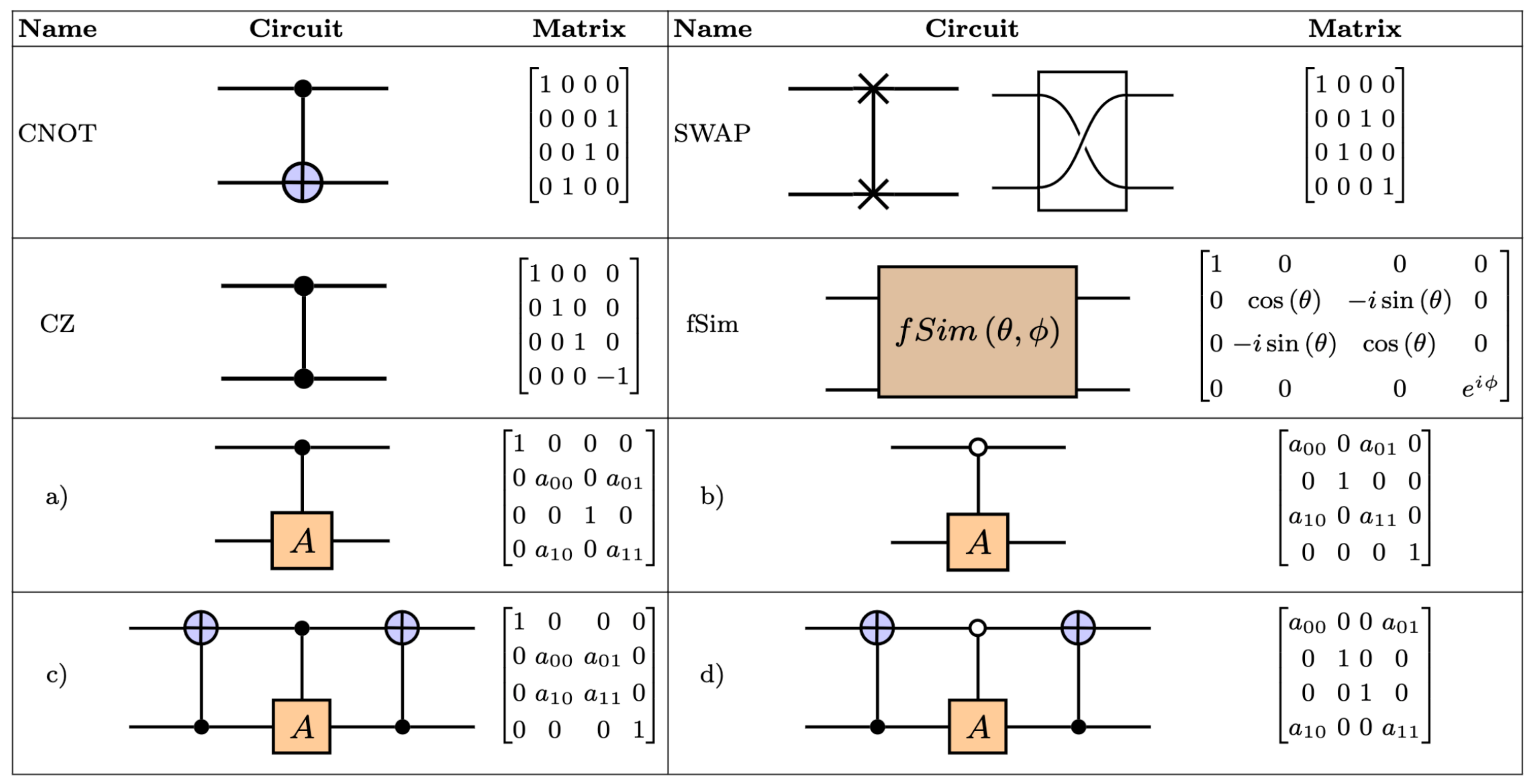}    
\caption{Summary of some common two-qubit quantum gates. We use the convention where the uppermost qubit line corresponds to the least significant bit in the binary representation of the computational state. {This convention corresponds to the so-called "little endian" convention used for bit/qubit storing. We will use this convention systematically throughout the article for circuit representation.} The {\it A} matrix presented from a) to d) in the Table is equal to 
    $\big(\begin{smallmatrix}
    a_{00} & a_{01}\\
    a_{10} & a_{11}
    \end{smallmatrix}\big)$. Some circuits are adapted from \cite{Seki2022}.}
\label{tab:double_gates}
\end{table*}

To produce entanglement, we need a Hamiltonian with interacting spins. For instance, if we can switch on a term $J Z_i Z_j$ in Hamiltonian \eqref{eq:digital_ham} (a term similar to the van der Waals term in Eq.~\eqref{eq:Ising}), we can perform operations of the type $e^{- i \frac{\theta}{2} Z_i Z_j}$. Such operations can create entanglement between qubits $i$ and $j$.

\paragraph{Universal quantum gates} 
It turns out that, with the one-qubit rotations $R_{x}\left(\theta\right)$, $R_{y}\left(\theta\right)$, and $R_{z}\left(\theta\right)$ presented in Table \ref{tab:single_gates}, together with, for instance, a two-qubit gate called ``CNOT'' presented in Table \ref{tab:double_gates}, one can achieve any unitary operation $U$ acting on $n$-qubits as a finite sequence of these gates (a result known as the Solovay-Kitaev theorem \cite{Kitaev1997,Nielsen2010}).
Therefore, one calls this gate set a {\it universal gate set}.
The Clifford group (see section \ref{sec:commondifficulty}) can become universal if a $T$ gate is added (with $T=P\left(\pi/4\right)$, see Table~\ref{tab:single_gates}).
Below, we briefly discuss the fundamentals of digital quantum computation; for more advanced considerations, we refer to different textbooks \cite{Nielsen2010,Mcmahon2007,Kaye2006,deLima2019}.

\paragraph{Quantum circuits} 
A standard digital quantum computation is a sequence of simple manipulations of the system's Hamiltonian; each described as a quantum gate. The computation usually starts from an initial state corresponding to all qubits in state $\ket{0}$. In other words, the "quantum register" is in state  $|0\rangle ^{\otimes n}$. One then applies gates $U_1, U_2, \dots, U_m$. This sequence of gates is usually represented as a so-called \textit{quantum circuit}, where each line stands for a qubit (time flowing from left to right) and each gate is pictured by a symbol that acts only on a subset of these lines.
Table \ref{tab:single_gates} and \ref{tab:double_gates} respectively give examples of the most common gates acting on one or two qubits.

\paragraph{Quantum measurements}

After applying the gates, the register is in its final state $\ket{\Psi} = U_m U_{m-1} \cdots U_1 |0\rangle ^{\otimes n}$ and one can measure some observable.
In most platforms, one can only measure the observable $Z_i$ (or a tensor product $Z_{i_1} \otimes \cdots \otimes Z_{i_k}$ if one "measures" $k$ qubits). One can translate a measurement in the $X$ or $Y$ basis into a measurement in the $Z$ basis using the insertion of one or two gates before the measurement, as pictured in Table \ref{fig:meas_XY}.

The outcome of a measurement of $Z_{i_1} \otimes \cdots \otimes Z_{i_k}$ is a bitstring $b_{i_1} \dots b_{i_k}$, with $b_i \in \lbrace 0,1 \rbrace$. This bitstring is obtained with a probability given by Born's rule,
\begin{equation}
    p(b_{i_1} \dots b_{i_k}) = \langle \Psi | P_{b_{i_1} \dots b_{i_k}} | \Psi \rangle,
\end{equation}
with $P_{b_{i_1} \dots b_{i_k}} = |b_{i_1}\rangle \langle b_{i_1} | \otimes  \cdots \otimes| b_{i_k} \rangle \langle b_{i_k}|$ (we do not explicitly write identities for qubits that are not being measured).
The measurement projects the register to the state $P_{b_{i_1} \dots b_{i_k}} | \Psi \rangle/\sqrt{p(b_{i_1} \dots b_{i_k})}$, so that if one wants to measure another observable that does not commute with $Z_{i_1} \otimes \cdots \otimes Z_{i_k}$, one needs to rerun the circuit.

The estimation of the expectation value of an observable (hermitian operator), like $\langle O \rangle = \bra{\Psi} {O} \ket{\Psi}$, is typically done by measuring the given observable a number $n_\mathrm{shots}$ of times, resulting in values $\{o_k\}_{k=1, n_\mathrm{shots}}$ that can be averaged to yield the estimator
\begin{equation}
    \overline{O} = \frac{1}{n_\mathrm{shots}} \sum_{k=1}^{n_\mathrm{shots}} o_k.    
\end{equation}
In the limit $n_\mathrm{shots}\rightarrow \infty$, this estimate converges to $\langle O \rangle$.
Due to the central limit theorem, $\mathcal{O}(1/\varepsilon^2)$ samples are needed to reach an accuracy $\varepsilon$.

Quantum circuits can also be used to compute the average value of any unitary operator, as shown in Table \ref{fig:meas_circ}. 
This table also shows how to measure time-dependent correlation functions of the form $\langle P_l(t) P_k(t') \rangle$ (using fermion-spin transforms, this type of circuit can be used to compute fermionic correlation functions).

\begin{table}
\centering
\includegraphics[width=1.0\linewidth]{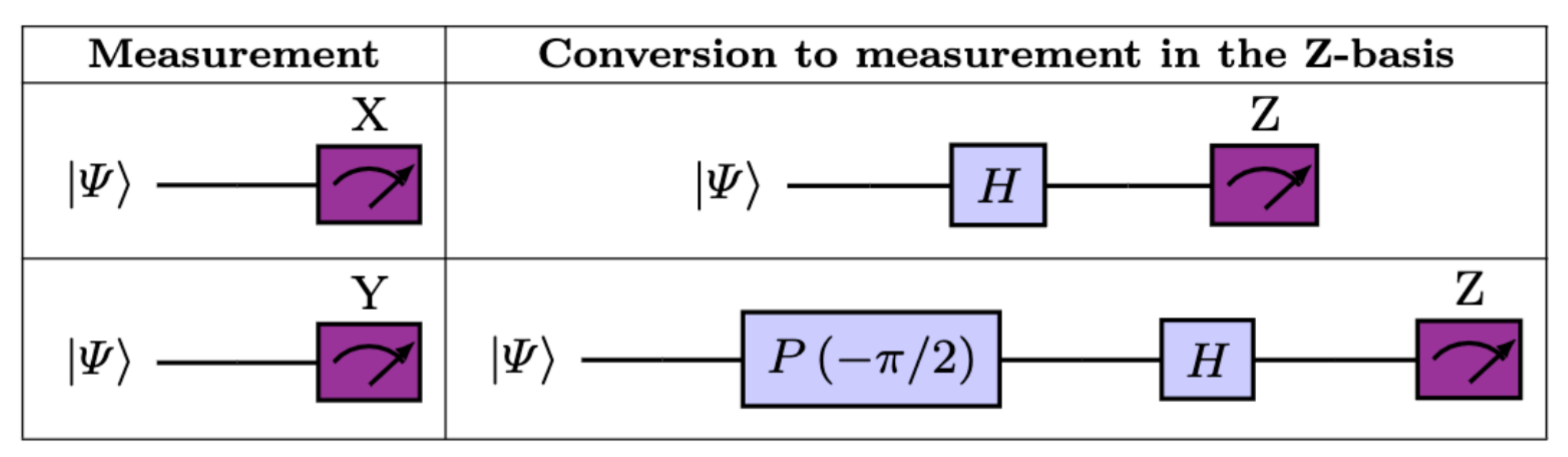}    
 \caption{Conversion of measurements in $X$ or $Y$ bases into measurements in the $Z$ basis.}
\label{fig:meas_XY}
\end{table}

\begin{table*}
\centering
\includegraphics[width=1.0\linewidth]{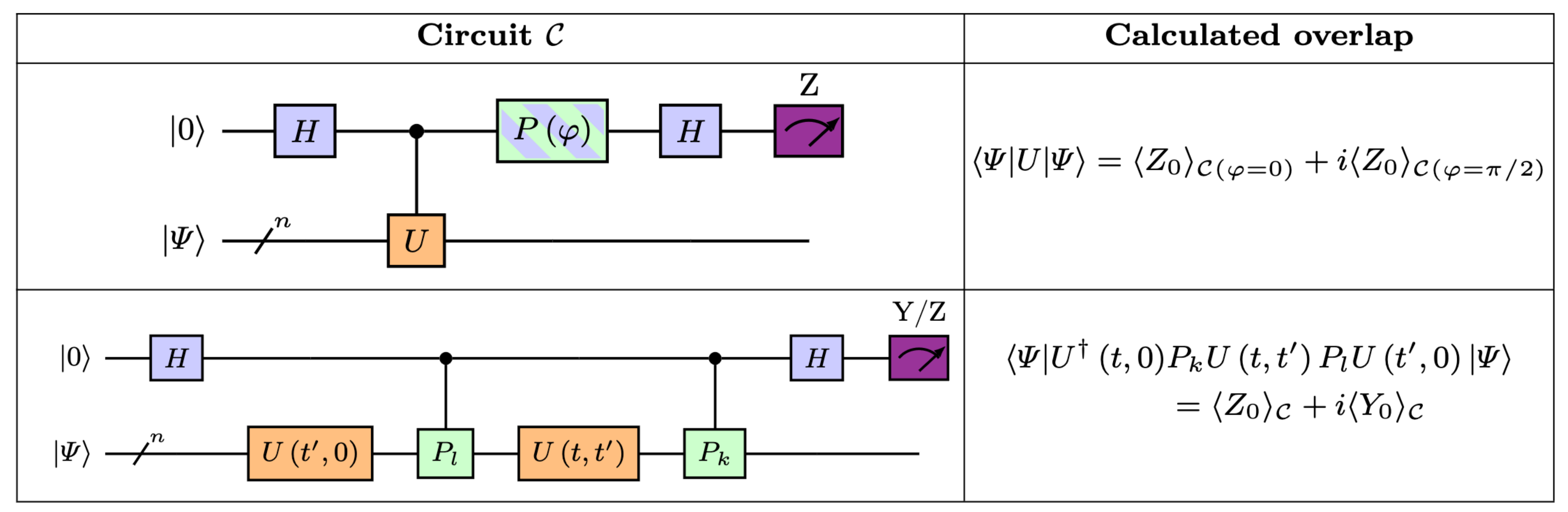}    
 \caption{Here we present some standard interferometry circuits which allow computing overlaps by making measurements on an ancillary qubit. The first circuit evaluates quantities of the form $\langle \Psi | U | \Psi \rangle$ where $UU^{\dagger}=\mathbb{I}$, e.g., the overlap between a state and its time-evolved counterpart. The second circuit enables to retrieve two-time correlators of the form $\langle \Psi | A(t) B(t') |\Psi \rangle$, with $A(t) = U^\dagger(t,0)P_kU(t,0)$ and $B(t')=U^\dagger(t',0) P_lU(t',0)$ (where $P_i$ denotes a Pauli operator). On the right, $\langle {O}_0 \rangle_{\cal C}$ indicates that the ancillary qubit (labeled by convention as the "$0^{\mathrm{th}}$" qubit) is to be measured in the $O \in \{ X, Y, Z\}$ basis after the execution of the circuit $\mathcal{C}$ drawn on the left.}
\label{fig:meas_circ}
\end{table*}

\paragraph{DiVincenzo criteria}

The principles introduced above have been gathered in a list of five criteria known as the DiVincenzo criteria~\cite{divincenzo_physical_2000}:
\begin{enumerate}
\item The ability to work with a scalable number of two-level systems (qubits) without "leakage" out of the computational subspace.
\item The capacity to initialize and reset qubits in a reliable (usually fast enough) fashion.
\item  A long coherence time (compared to the typical time scales of gates, measurements, and resets, i.e., compared to the "clock time" of the processor). \item  A universal set of gates (with the possibility to parallelize operations on disjoint sets of qubits).
\item  Reliable qubit-wise measurements.
\end{enumerate}

Current quantum processors are strongly impacted by decoherence effects, as will be explicited in section \ref{sec:noise}.
As a result, they do not meet all five criteria.
However, they help develop and test algorithms in real-life conditions.

Before reviewing these algorithms, we describe how many-body problems can be translated to forms amenable to quantum computing.

\section{Mapping a many-body problem to a quantum computer}
\label{sec:mapping_to_qc}

This section explains how to go from the many-body problem at hand to the one that the quantum computer models.
In particular, if we focus on digital (gate-based) quantum platforms, such devices usually have constraints:
(i) they have qubits (two-level systems), not fermions/bosons; (section \ref{subsec:From-fermions-to}), 
(ii) they have a limited number of qubits (section \ref{subsec:Reducing-the-number}), and
(iii) they have a limited coherence (section \ref{subsec:NISQ-Noisy-quantum}).
Thus, one needs to transform, reduce or/and map the original problem so that the quantum computer can give insights into the properties of the original many-body problem.

\subsection{From fermions to qubits\label{subsec:From-fermions-to}}

The treatment of fermions on a quantum computer can be made starting from the first or from the second quantization \cite{McArdle2018a}.
{While we mainly focus in this review on the second option, the two strategies, i.e., first or second quantization as a starting point, 
are briefly discussed below.}  

\subsubsection{Encoding fermions as qubits in second quantization} 
\label{sec:second}

To perform any Hamiltonian simulation written in second quantized form on a quantum computer, one must map the fermion Hamiltonian to a spin Hamiltonian.
This mapping is not unique. The most standard mapping techniques are the Jordan-Wigner (JW) transformation  \cite{JordanWigner1928}, the Bravyi-Kitaev (BK) \cite{Bravyi2002} or the parity mapping \cite{Seeley2012} (for a comprehensive discussion see \cite{Bauer2016,Fano2019,McArdle2018a}).
We illustrate the JW case that is often retained for many-body applications due to its relative simplicity. 

Let us consider a set of fermions associated with the creation/annihilation operators $(a^\dagger_p, a_p)$ where $p$ labels a complete basis of single-particle states $\phi_p(r)$. These operators act on the many-body vacuum by changing the occupation of orbital $p$. 
In the JW fermion-to-qubit mapping, the occupation (resp. vacancy) of a state is usually encoded as the state $| 1\rangle_p$ (resp. $| 0 \rangle_p$).
Then, the operator $Q^+_p = |1\rangle_p \langle 0|_p = \frac{1}{2}\left(X_p - i Y_p\right)$ and its hermitian conjugate $Q^-_p$ can be seen as the qubit equivalents of the creation/annihilation operators.
The difficulty is that these sets of operators commute between each other for different qubits while they should anticommute for fermions.
One solution to this issue is to choose a specific ordering for the one-to-one correspondence between the single-particle state and the qubits and use the following prescription: 
\begin{eqnarray}
    a^{\dagger}_p &\longleftrightarrow&  \bigotimes_{k=1}^{p-1} Z_k  \otimes Q_{p}^{+}, ~ 
    a_p \longleftrightarrow  \bigotimes_{k=1}^{p-1} Z_k  \otimes Q^{-}_{p}. 
\end{eqnarray}
In this transformation, the fermionic sign (which comes from the anticommutation rules of fermions) is kept track of via the string $\bigotimes_{k=1}^{p-1} Z_k$ of Pauli-$Z$ operators.
At the circuit level, this means that operations that are one-body at a fermionic level (like $\exp(-i t \{c_0^\dagger c_1 + \mathrm{h.c}\})$) might become a multi-qubit operation. For instance, $a^\dagger_3 a_1$ leads to  a term $Q^\dagger_3 Z_2 Q_1$ that acts on the three qubits $(1,2,3)$. 

{The fermion--to--qubit mapping corresponds to the mapping of a set of $\tilde{n}$ single-particle orbitals obeying fermionic anticommutation rules to a set of $n$ qubits. The mapping is not unique, and depending on which choice is made the number of qubits  $n$ may not be identical to the number of spin-orbitals $\tilde{n}$.} Furthermore, the locality $d$ (a $d$-local Hamiltonian can be expressed as the sum of Hamiltonian terms acting upon at most $d$ qubits) is usually not conserved upon encoding.
For instance, in the Jordan-Wigner mapping, $\tilde{n} = n$ and  $\tilde{d} = O(\tilde{n})$, while another transform called the Bravyi-Kitaev transformation that also has $\tilde{n}=n$ achieves a better locality, namely $\tilde{d} = O(\log_2{\tilde{n}})$.
{Note that  in general it is advantageous to use the Bravyi-Kitaev encoding compared to the Jordan-Wigner encoding in terms of circuit depth \cite{Tranter2018}. Despite this, the Jordan-Wigner mapping is often used in practice due to its relative simplicity. }
Another example is the so-called superfast fermionic encoding \cite{Setia2018}, which requires $\tilde{n} = O(m d)$ and achieves $\tilde{d} = O(d)$. {Here $m$ represents the number of Fermi modes, such that each mode can be either empty or occupied by a fermionic particle.}

{Note that we have implicitly referred to $SU(2)$ fermions only, but $SU(N)$ fermions can also be handled. For instance, the JW mapping scheme presented above readily generalizes to $SU(N)$ fermions \cite{Consiglio_2022}.}

\subsubsection{Encoding fermions in first quantization}
\label{sec:first}

{
One major drawback of the encoding based on second quantization is that the many-body problem is encoding directly in the entire Fock space, whose 
size grows exponentially with the number of considered orbitals $n$ as $2^n$. On the other hand, we know that a problem of $A$ interacting particles
described on $n$ single-particle states requires at most $C^A_n$ many-body states, just by invoking the particle number conservation. Additional 
symmetries can further reduce the size of the relevant Hilbert space. Said differently, second quantization applied to many-body problems 
leads to a significant dark sector that might ultimately reduce quantum advantage. First quantization, which was the original formulation of quantum mechanics, has been explored 
as a possible solution to this problem. An exhaustive discussion of the particles--to--qubit encoding using first quantization 
is out of the scope of the present review and we only give some brief guidelines. For more details, see for instance Ref. \cite{McArdle2018a} and 
the historical overview given in Fig. 5 of Ref. \cite{Cao2019}.  
}

{
First quantization directly relies on the wave-function representation of the problem of interacting 
particles without invoking the machinery associated with Fock space and creation/annihilation operators. 
Some anticipated advantages of this formulation are underlined below. We start from a Hamiltonian describing a set of particles $i=1,\dots,N$ 
and written in first quantization as:
\begin{eqnarray}
    H = \sum_{i=1}^{N} h(i) + \sum_{i<j=1}^{N} V(i,j), \label{eq:hamfirst}
\end{eqnarray}
where $h(i)$ is a one-body Hamiltonian acting on the $i$th particle, and $V(i,j)$ denotes the interaction between particles $i$ and $j$. Two main strategies have been used to encode the problem directly starting from  Hamiltonian (\ref{eq:hamfirst}):
\begin{enumerate}    
\item {\it Grid-based method:} Let us assume that the particle is described by a certain set of coordinates $\{\bf q \}$, for instance, its position and spin, i.e., ${\bf q} = (x,y,z,\sigma)$. Then, the many-body wave-function, denoted by $|\Psi_N \rangle$, can be decomposed as 
    \begin{eqnarray}
      | \Psi_N \rangle &=& \int d^{3N} \Omega \Psi_N({\bf q}_1, \cdots ,{\bf q}_N) | {\bf q}_1, \cdots {\bf q}_N \rangle. \label{eq:wffirst}   
    \end{eqnarray}
    Here $\int d^{3N} \Omega$ denotes the integral over the $3N$ dimensional space and the sum over the $2N$ spin components. One advantage of this representation is 
    that the state $| {\bf q}_1, \cdots {\bf q}_N \rangle$ can already contain the antisymmetry property \cite{Blaizot86}. Continuous variables like position or momentum are generally not bounded and application requires specific discretization schemes on a grid. 
    The possibility to describe quantum systems on a grid using quantum computers has been explored already in pioneering articles
    \cite{Zalka1998,Lloyd1996}. 
    The encoding can be made by assuming that 
    each $(x,y,z)$ component is discretized on a mesh with $L$ points. As discussed in detail in Ref. \cite{McArdle2018a}, the description of a wave-function given by (\ref{eq:wffirst}) requires storing $L^{3N} \times 2^{N}$ amplitudes and, therefore, becomes rapidly prohibitive on a classical computer. Brute-force few-body problems treated on a classical computer are usually restricted to small number of particles $N$. Assuming that $L=2^{m-1}$, such amplitudes can be stored on $(3m+1)N$ qubits. Such encoding becomes more compact and automatically treats the problem in the proper space of wave-function having $N$ particles. 
    Using grid-based methods, the different operators entering in (\ref{eq:hamfirst}) can be encoded in the basis by rewriting them as:
    \begin{eqnarray}
\left\{ 
\begin{array}{l}
\displaystyle h(i)   = \sum_{{\bf p}, {\bf q}} \langle {\bf p} | h | {\bf q} \rangle \left(| {\bf p} \rangle \langle {\bf q} | \right)_i       \\
%\\
\displaystyle  V(i,j) = \sum_{{\bf p}, {\bf q}, {\bf r}, {\bf s}} \langle {\bf p} , {\bf r}| V | {\bf q} , {\bf s} \rangle 
        \left(| {\bf p}\rangle \langle {\bf q}| \right)_i  \left(| {\bf r}\rangle \langle {\bf s}| \right)_j ,  
\end{array} \label{eq:hivijfirst} 
\right.
    \end{eqnarray}
    where we use the notation \cite{Babbush2019} 
    \begin{eqnarray}
        \left(| {\bf p} \rangle \langle {\bf q} | \right)_i  \equiv | {\bf p} \rangle_i \langle {\bf q} |_i  \bigotimes_{k \neq i} I_k .\nonumber
    \end{eqnarray}
    With these expressions, we see that each term of the one- and two-body components will only couple one state of the computational basis to
    a single state of the same basis. Following this discretization technique, a number of quantum algorithms eventually taking  advantage of the sparsity of the 
    Hamiltonian have been proposed \cite{Kivlichan2017,Babbush2018,Babbush2019,Su2021} (see also \cite{McArdle2018a,Cao2019}). 
    \item {\it Basis set method or restricted CI method:} the direct solution of a many-body problem directly on a mesh is rather challenging and requires 
    a good control of the precision both on the quantum algorithm and on the discretization scheme itself \cite{Babbush2019}. As discussed in the introduction 
    and further elaborated in section \ref{sec:dofreduction}, classical algorithms based on CI techniques are built using a set of $M_a$ "active" 
    single-particle states $\{| \phi_l \rangle_{l=1,M_a} \}$ where a number $M_p$ of particles can be distributed to form Slater determinants. If $M_p$ identifies with the total number of particles $N$ and a sufficient number of single-particle states are considered the CI calculation is said unrestricted; 
    otherwise, it is referred to as restricted CI. One can then write the Hamiltonian components in a similar way as in Eq. (\ref{eq:hivijfirst}) replacing the states $| {\bf p} \rangle$ by the new single-particles states. The number of many-body states to consider is $C^{M_p}_{M_a}$ and the Hamiltonian is sparse in this basis. As proposed in Ref. \cite{Toloui2013}, one can label a given state as $|m_1, \cdots , m_N \rangle$ with $m_l =0,M_a - 1$ denoting the orbital that the particle $l$ is occupying. Let us assume simply that $M_a = 2^{n_a}$, then, a set of integers $m_l = 0,M_a - 1$ can be 
    encoded on $n_a$ qubits \cite{Abrams1997}. Accordingly, the set of states $|m_1, \cdots , m_N \rangle$ can be encoded on $N \log_2(M_a)$ qubits. Illustrations 
    of quantum algorithms using this technique can be found in \cite{Toloui2013,McArdle2018a}.     
\end{enumerate}
}

\subsection{Reducing the number of degrees of freedom\label{subsec:Reducing-the-number}}
\label{sec:dofreduction}

\begin{figure} 
\centering

\includegraphics[width=\columnwidth]{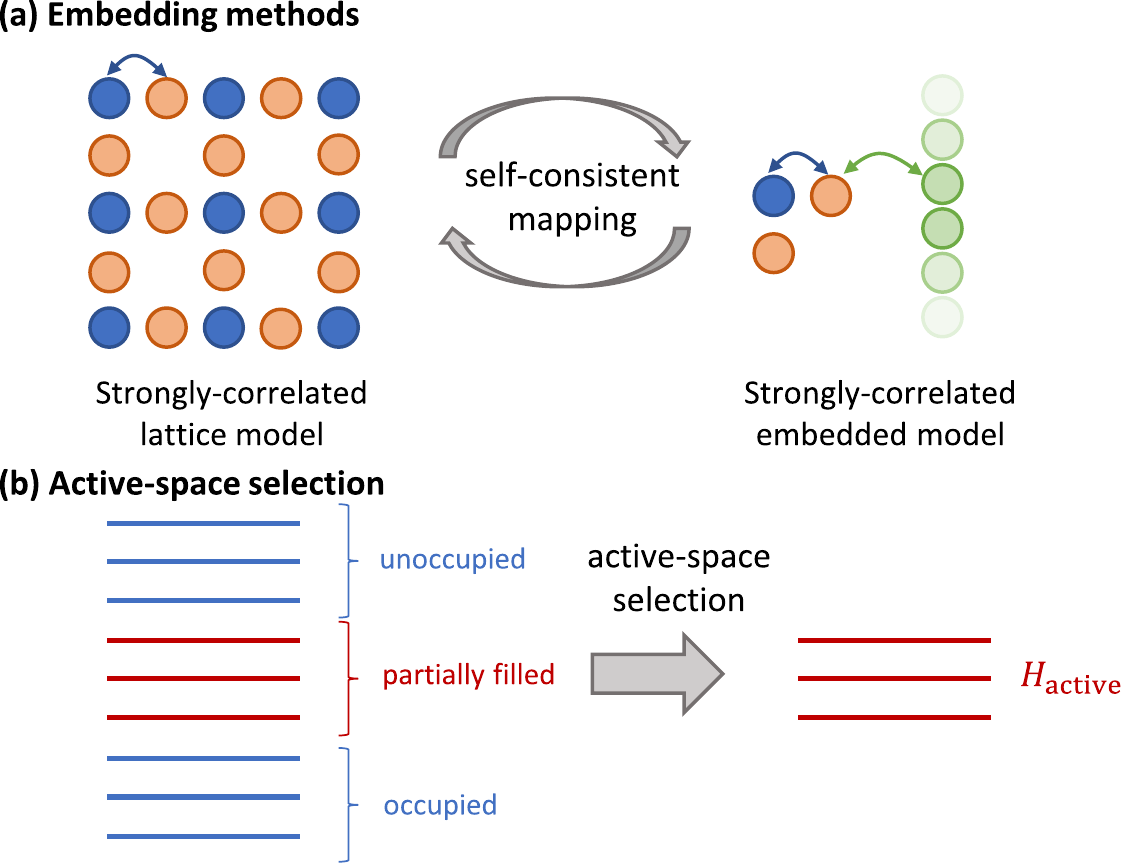}

    \caption{Embedding methods (a) and active-space selection (b). In embedding methods, an extended (lattice) model is self-consistently mapped to a local (impurity or embedded) model. In active-space methods, a subset of orbitals (usually partially filled ones) is selected to construct the active space Hamiltonian, while the other orbitals are treated at the mean-field (Hartree-Fock) level. Due to their smaller size, the embedded or active-space models are better suited for a solution with today's quantum computers.}

\label{fig:dof_reduction}
\end{figure}

NISQ devices come with a limited number of qubits and limited coherence. These constraints limit the number of degrees of freedom (typically orbitals) of the system one wants to study. 
Nevertheless, many-body condensed matter, quantum chemistry, or nuclear problems typically comprise tens or hundreds of orbitals.
Directly tackling these large model spaces with a quantum processor appears unfeasible, if not ill-advised.
Indeed, over the last century, numerous classical many-body methods have been devised to reduce the number of truly correlated---sometimes called "ultra quantum"---degrees of freedom.
These classical methods then rely on advanced algorithms to solve the "reduced model". Despite its reduced complexity, this model is usually hard to tackle in some physically relevant regimes due to its strongly-correlated character. This is where quantum coprocessors could be used to extend the power of classical algorithms. 
We highlight below a few classical-inspired methods that were used to reduce the number of qubits for many-body systems treated on quantum computers. These methods are illustrated in Fig.~\ref{fig:dof_reduction}. 

\subsubsection{Embedding methods}\label{subsubsec:embedding}

Typical condensed-matter problems are formulated on a lattice of atomic sites (e.g., the Hubbard model introduced in Eq.~\eqref{eq:Hubbard_solids}). The number of sites needed to observe collective phenomena like phase transitions typically exceeds the capacity of exact diagonalization or Monte-Carlo methods. Classical methods collected under the term \textit{embedding methods} have been developed to overcome this limitation. They draw inspiration from mean-field methods in that they self-consistently map the original, extended problem onto a smaller, more local many-body problem (sometimes called \textit{fragment}) "embedded" in a (usually) non-interacting environment (also called \textit{bath}).
One can then leverage the fact that this embedded problem has fewer correlated degrees of freedom to tackle it with classical or, if need be, quantum methods \cite{Bauer2016}. An illustration for these methods is provided in Figure~\ref{fig:dof_reduction} (a).

Examples of embedding methods include, but are not limited to, Dynamical Mean Field Theory (DMFT)\cite{Georges1996}, the Gutzwiller or Rotationally-Invariant Slave Boson (RISB)\cite{Lechermann2007} method, and the Density-Matrix Embedding Theory (DMET)\cite{Knizia2012} method.

Generically, the embedded problem has the form: 
\begin{align}
  H &= \sum_{i=1}^{N_c} U n^c_{i\uparrow} n^c_{i\downarrow} - \mu \sum_{i=1}^{N_c} \sum_{\sigma}n^c_{i\sigma} \\
    &+ \sum_{p=1}^{N_b} \sum_{i=1}^{N_c} \sum_{\sigma}\left( V_p c^\dagger_{i\sigma} a_{p\sigma} + \mathrm{h.c} \right)   
    +\sum_{p=1}^{N_b} \sum_{\sigma} \varepsilon_p a^\dagger_{p\sigma} a_{p\sigma},
    \label{eq:H_AIM}
\end{align}
where $a^\dagger_{p\sigma}$ creates electrons in the bath (of size $N_b$), while $c^\dagger_{i\sigma}$ creates electrons in the correlated orbitals ($n_{i\sigma}^c = c^\dagger_{i\sigma} c_{i\sigma}$). 
Compared to the Hubbard model, Eq.~\eqref{eq:H_AIM} has fewer ($N_c$) interacting sites (the $N_b$ bath sites are uncorrelated). However, it is still a complicated many-body problem. Typically, $N_c$ is adapted to the spatial resolution one wants. For regimes with considerable correlation lengths, it can exceed the reach of advanced Monte-Carlo methods.

The embedding methods mentioned above differ by the number of bath sites, the observables that need to be computed (generally, Green's functions on the impurity or reduced density matrices), and the way the self-consistent parameters ($\varepsilon_p$ and $V_p$) are updated.
For instance, within RISB and DMET, $N_b = N_c$; this results in the embedded model being much smaller than the original Hubbard model. 

Recent works have used quantum processors to tackle the embedded model within an embedding method \cite{Rungger2019,Keen2019,Tilly2021,Yao2020, besserve_ayral_2022, Backes2023}. They are limited so far to small sizes ($N_c \leq 2$) due to NISQ limitations. Until now, classical methods still outperform quantum methods in solving these problems.

\subsubsection{Active space methods}\label{subsubsec:active_space}

In quantum chemistry, like in condensed matter or nuclear physics problems, the number of degrees of freedom can be reduced to the genuinely complicated degrees of freedom. These are usually called active orbitals. Instead of handling all orbitals at the same level of theory, orbitals are divided into active ones---which require an advanced many-body method---and inactive ones---for which mean-field (Hartree-Fock) methods will be sufficient.
The active space selection can be based on the occupation level of molecular orbitals (the orbitals resulting from a Hartree-Fock optimization). Empty and occupied orbitals  (with occupation numbers close to $0$ or $1$, respectively) are inactive, while partially-filled orbitals are considered active. The active space size is adjusted according to the sought-after accuracy, available computational capacity, or both.
The so-obtained active space Hamiltonian has the same form as the original Hamiltonian, which is given by Eq.~\eqref{eq:el_structure_h}. However, it usually has a much smaller number of orbitals: $N$ is reduced to $N_a$. Then, with a classical computer, one can tackle this reduced problem with advanced methods like FCI (if $N_a$ is very small) or CC otherwise. {The simplification of a many-body problem's solution by selecting some active space without degrading the quality of an approach is a complex and delicate issue where one can take advantage of quantum information concepts to make the selection of relevant states \cite{Stein2016,Stein2019}.}

With a quantum coprocessor, the reduction from $N$ to $N_a$ orbitals directly translates, via fermion-spin transforms (see section \ref{subsec:From-fermions-to}, to a reduced number of required qubits (namely $N_a$). An illustration for these methods is provided in Figure~\ref{fig:dof_reduction} (b).

Many recent works use this active space selection to reduce the number of required qubits, see, e.g., \cite{McCaskey2019}, or to explore the resource requirements on future quantum computers \cite{Reiher2016, Li2018a}. 

\section{Ideal algorithms}
\label{sec:idealalgorithm}

Here, we discuss some textbook methods to simulate a quantum system and solve the eigenvalue problems on a quantum computer \cite{Nielsen2010}.  

\subsection{Quantum Phase Estimation for the eigenvalue problem}
\label{subsec:qpe}

\paragraph{Description}

Quantum Phase Estimation (QPE), also called Phase Estimation Algorithm, is a generic method to shed light on the spectrum of a unitary operator with a quantum computer \cite{Nielsen2010,deLima2019}. It is already well documented, and we only give here the key ingredients of the approach. 

Suppose that the operator $U$ has a set of eigenvalues $\{ e^{2 \pi i \phi^\alpha} \}$. We assume that for all $\alpha$, we have $0 \le \phi^\alpha < 1$ and denote by $\{ | \alpha \rangle \}$ the corresponding eigenvectors. The system's initial state can be 
decomposed as:
\begin{eqnarray}
|\Psi \rangle &=& \sum_\alpha c_\alpha | \alpha \rangle. \label{eq:initdec}
\end{eqnarray}

\begin{figure}[htbp]  
\centering
\includegraphics[width=1.0\linewidth]{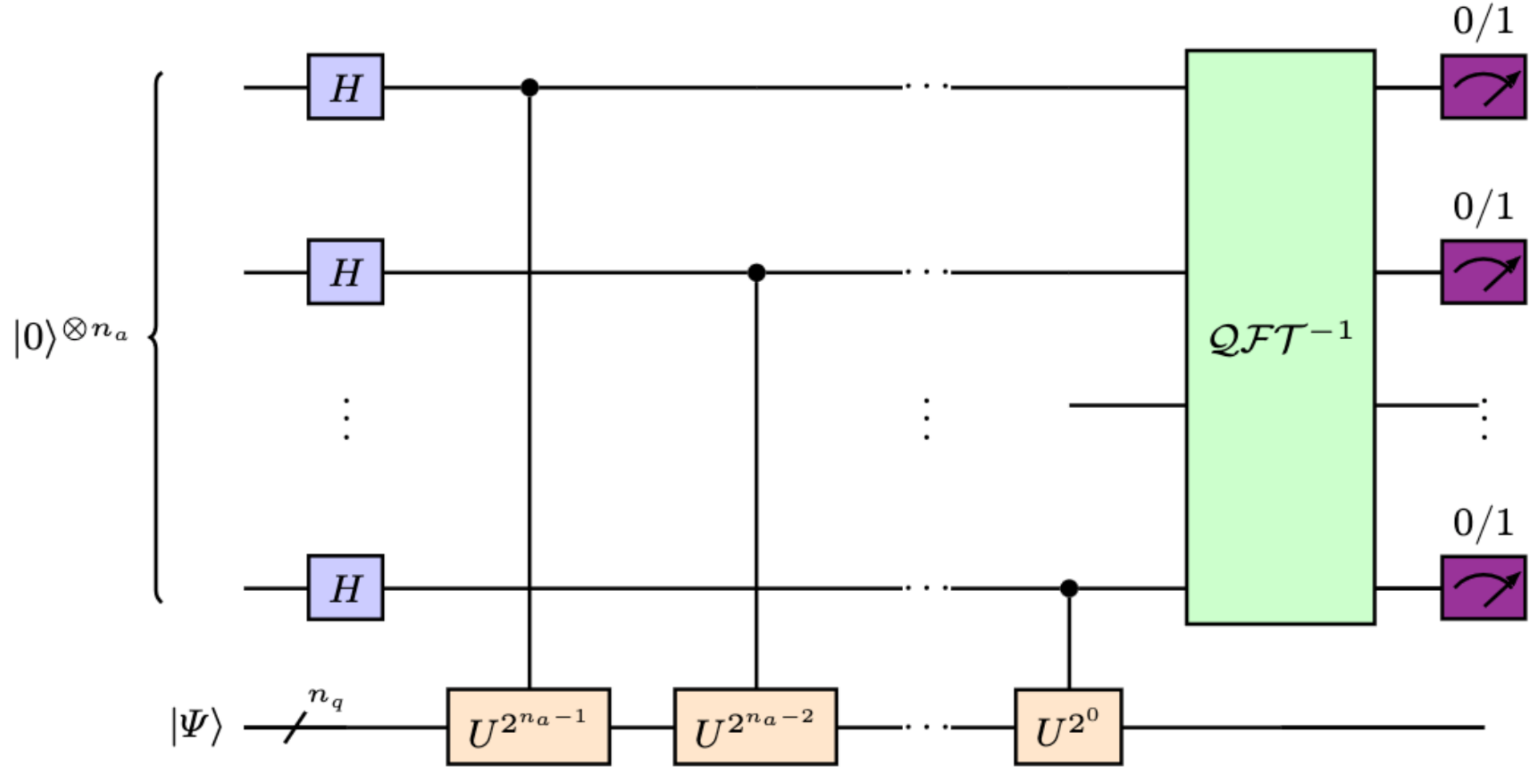}    
 \caption{Illustration of the QPE circuit used to get the eigenvalues of an arbitrary unitary operator $U$. The QPE circuit uses the inverse Quantum Fourier Transform (QFT$^{-1}$) \cite{Nielsen2010}. The QPE requires $n_a$ additional ancillary qubits. The precision of the method will directly depend on $n_a$.} 
\label{fig:qpe}
\end{figure}

The QPE method consists of the following schematic sequence:
\begin{eqnarray}
| \Psi \rangle  \xrightarrow[]{\text{QPE}} \sum_\alpha c_\alpha |\alpha \rangle \otimes | \widetilde{\phi}^\alpha \rangle  \xrightarrow[]{\text{Measure} } 
|\alpha \rangle \otimes | \widetilde{\phi}^\alpha \rangle, \label{eq:secqpe} 
\end{eqnarray}
where the bitstring  $\widetilde{\phi}^\alpha = \phi^\alpha_{n_a - 1} \dots \phi^\alpha_0$ ($\phi^\alpha_i \in \{0,1\}$ being the result of the measurement of the ancillary qubit $i$) encodes an estimation of the phase $\phi^\alpha$ as the binary fraction $0.\phi^\alpha_0 \cdots \phi^\alpha_{n_a - 1} \equiv \sum \limits_{j=0}^{n_{a}-1} \frac{\phi^\alpha_j}{2^{j+1}}$.
We will use the notation $0.\widetilde{\phi}^\alpha$ for this number, although the order of the bits is reversed. A schematic view of the QPE circuit is shown in Fig. \ref{fig:qpe}.

The effect of QPE is twofold: the initial state is projected into one eigenstate (or a set of degenerate states) having non-vanishing overlap with the initial state, and the associated eigenvalue is retrieved with a precision $|{\phi}^\alpha - 0.\widetilde{\phi}^\alpha| \le 1/2^{n_a}$, which improves exponentially with $n_a$. Note that the projection is only approximate unless the binary fraction of ${\phi}^\alpha$ is finite and the number $n_a$ is sufficient to have ${\phi}^\alpha = 0.\widetilde{\phi}^\alpha$.
This projection effect was used recently in a many-body system to restore broken symmetries (see \cite{Lacroix2020,RuizGuzman2022}).

For many-body problems, QPE can be seen as a gold standard to solve the eigenvalue problem for the Hamiltonian $H$ in a large Hilbert space.
In this case, the operator $U$ can be chosen as the propagator itself, with
\begin{eqnarray}
U(\tau) = e^{-2\pi i\tau (H-E_0)}, 
\end{eqnarray}
where $\tau$ and $E_0$ are parameters chosen to map the spectrum of $H$ into elements of $[0,1[$.
The circuit to prepare the unitary $U$ starting from a given $H$ is usually obtained via trotterization, a method summarized in section \ref{subsubsec:trotter}.

\begin{figure}[htbp]  
\includegraphics[scale=0.58, trim = 0.6cm 0.65cm 0.5cm 0.5cm]{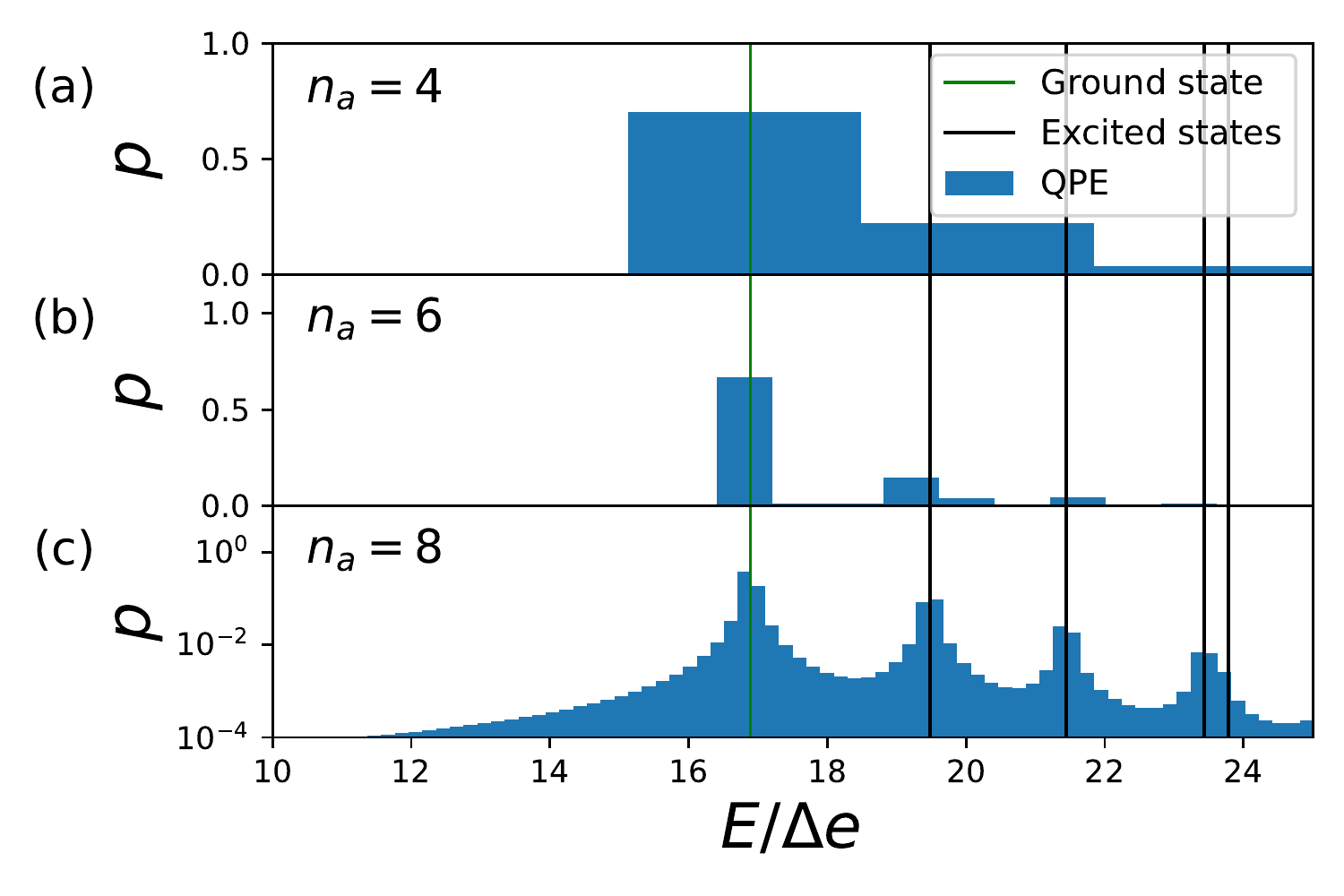}
    \caption{Illustration of the QPE algorithm applied to the Hartree-Fock state $|\psi\rangle = |00001111\rangle$ of a Pairing Hamiltonian, Eq. (\ref{eq:pairing}), with one body energies at levels {\it j}, $\varepsilon_{j=0,1,\dots,n_q-1} = j
    \Delta e$, $g=0.5\Delta e$ and $\Delta e = 1$ \cite{RuizGuzman2022}. {\it p} is the probability of measuring the energy value {\it E} in the ancillary register. The ancillary qubits $n_a$ used for the QPE were  4 (a), 6 (b), and 8 (c). The vertical green and black lines correspond to the ground and excited energies of the Hamiltonian, respectively. The figure has been adapted from \cite{RuizGuzman2022}. Note that panel (c) is shown in linear-log scale.}
\label{fig:qpe_aplication}
\end{figure}

{ A key feature of the QPE algorithm is that its complexity for finding the phase $ \phi^\alpha $ associated with the eigenstate $ |\alpha\rangle $ with additive error $ \epsilon $ scales as:
\begin{equation}
\mathcal{O}\left(\frac{{\text{poly}(n_q)}}{\epsilon S^2}\right).
\label{eq:complexity_qpe}
\end{equation}
Here, $ S $ denotes the overlap between the initial state $ |\Psi\rangle $ and $ |\alpha\rangle $, i.e., $ S = |\langle \alpha | \Psi \rangle|$. For large systems, $S$ is generically small because it becomes exponentially difficult to attain a high overlap value due to the exponential expansion of the Hilbert space, a phenomenon known as the \textit{orthogonality catastrophe} \cite{Lee2022,Louvet2023}. It is important to note that the cost of preparing the initial state $ |\Psi \rangle $ has not been included in this analysis, as we assume that a cost-efficient approach is used: the term $ 1/S^2 $ simply reflects the scaling of the number of measurements required to identify $ \phi^\alpha $. In the general case, however, preparing an arbitrary initial state has exponential scaling with respect to $ n_q $. 
%The term $ 1/S^2 $ relates to the number of measurements required to identify $ \phi^\alpha $.

%It is known that
Since $ \epsilon \sim 1/2^{n_a} $, 
%linking 
the term $ 1/\epsilon $ is linked both to the number of ancilla qubits required for a specified precision and to the circuit depth, given that there are $ n_a $ controlled-$ U $ gates and $ O(n_a^2) $ operations are needed for the inverse Quantum Fourier Transform. Typically, the $ U $ gates exhibit polynomial scaling with respect to the number of system qubits $ n_q $, i.e., $ O(\text{poly}(n_q)) $, and for electronic structure problems, this scaling is commonly $ O(n_q^5) $. Furthermore, as highlighted in a later section (\ref{subsubsec:trotter}) and due to the \textit{no fast-forwarding theorem} \cite{berry_2006}, the complexity of the $ U $ gates generally scales linearly with the evolution time $ t $. This implies that the application of a sequence of controlled-$ U^{2^k} $ gates will usually result in exponential complexity, given that the total evolution time needed to execute the algorithm is $ \tau(2^{n_a} - 1) $.
}

These scalings illustrate the advantages and shortcomings of the ``perfect'' QPE (i.e., performed on a fault-tolerant quantum computer).
On the flip side, it means that QPE requires an input state that reasonably overlaps the actual eigenstate. One could, for instance, resort to \textit{adiabatic state preparation} \cite{Farhi2000} to rotate some simple initial state into a state exhibiting a significant overlap with the eigenstate of interest. 
This requirement over the overlap becomes problematic in high dimension for many-body problems due to the 
{orthogonality catastrophe referred to above.}
%\textit{orthogonality catastrophe} (see, e.g., \cite{Lee2022}).
Note that this initial state problem can be cleverly handled in some cases. A case in point is Shor's factoring algorithm \cite{Shor1994}, which uses QPE as the main ingredient after a clever state preparation, providing an exponential speedup over the classical version of the factoring algorithm. 

On the bright side, the $1/\epsilon$ scaling is much better than the $1/\epsilon^2$ typical of classical Monte-Carlo methods. This advantage is used in "quantum Monte-Carlo" methods. These techniques use QPE as a critical building block within another algorithm called quantum amplitude estimation \cite{Brassard2000}. We point out that these techniques should not be confused with the many-body quantum Monte-Carlo methods, which refer to purely classical methods to solve quantum many-body problems.
We also note in passing that recent proposals have been made to hybridize "classical" quantum Monte-Carlo methods such as Auxiliary-Field Quantum Monte-Carlo (AFQMC) or Full-Configuration-Interaction-Quantum-Monte-Carlo (FCIQMC) with quantum algorithms \cite{Huggins2021, Mazzola2022, Zhang2022a}.

On NISQ processors, QPE is hardly applicable owing to the number of operations required (whether the previously mentioned $2^{n_a}$ depth or those of the QFT). Estimates for chemical problems yield tremendous numbers \cite{Wecker2014a}, not compatible with the number of qubits and error rates of current and near-term machines.

Today, intensive efforts are being made to provide less costly methods for Hamiltonian eigenvalue problems. Some of them will be further reviewed in section \ref{sec:nisq}.

\subsection{Trotterization}\label{subsubsec:trotter}

\paragraph{Description}

Trotterization is a technique to implement a time evolution $U=e^{-iHt}$ on digital quantum hardware \cite{Lloyd1996}, i.e., as a sequence of few-qubit gates. 

Most many-body Hamiltonians are $k$-local, meaning they can be decomposed as a sum of terms acting on at most $k$ qubits:
\begin{equation}
    H = \sum_{j=1}^m \lambda_j P_j, \label{eq:pauli_decomp}
\end{equation}
with $P_j$ a product of at most $k$ Pauli operators.

Trotterization approximates the exponential of a sum $e^{-iHt}$ as the product of the individual terms.
The so-called first-order \textit{Trotter-Suzuki} formula \cite{Lloyd1996} reads
\begin{eqnarray}
    \label{eq:trotter}
    e^{-iHt} &= \left( \prod_{j=1}^{m} e^{-i \lambda_j P_j \frac{t}{n_t} }\right)^{n_t} + O\left(\frac{m^2t^2}{n_t} \right)
\end{eqnarray}
where $n_t$ is the number of \textit{Trotter steps}. 
The rationale behind formula (\ref{eq:trotter}) is that the whole Hamiltonian evolution is carried out in the form of repeated sequences of step-wise evolutions $e^{-i \lambda_j P_j \frac{t}{n_t}}$. Each such evolution can be simplified as a sequence of one- and two-qubit gates.

Notably, the number $n_t$ of Trotter steps must be increased when $t$ increases: the circuit structure is not fixed with $t$; instead, its complexity grows linearly with $t$. In some cases, a sublinear scaling can be proposed, but this is not generally the case because of the \textit{no fast-forwarding theorem} \cite{berry_2006}. In the QPE described in the previous section, this implies that a unitary operator of the form $U^{2^n} = \exp(-i H 2^n t)$ has depth $O(2^n)$, explaining the exponential scalings of QPE.

\paragraph{Beyond standard trotterization} 
The possibility of reducing the significant scaling associated with the Trotter-Suzuki methods is an active field of research. Several alternative methods have been proposed: the Variational Fast-Forwarding \cite{Cirstoiu2020,Gibbs2021}, Incremental Structure Learning \cite{Jaderberg2020}, the Adaptive Product Formula \cite{Zhang2020}, and the Variational Time Dependent Phase Estimation \cite{Klymko2021}, to quote some of them.

\section{NISQ Algorithms}
\label{sec:nisq}

Quantum algorithms like the QPE presented above require, in general, a large number of qubits or gates, or both.
Because of the limitations of current quantum platforms in qubit and gate counts, these algorithms cannot be used in the presence of noise.
Specific algorithms have been designed to circumvent those limitations and study the applicability of today's quantum processors to concrete problems.
These algorithms are designed, for instance, to reduce the circuit depth by allocating only a specific task to the quantum computer or, 
as when using variational methods, to allow better control of these states.

\subsection{Variational algorithms}

Variational methods are standard tools for many-body physicists using classical computers \cite{Blaizot86}. In recent years, they have emerged as a tool of choice for applications on quantum platforms, and are an important part of the broader class of hybrid quantum-classical methods \cite{Bharti2021, Endo2021}.

\subsubsection{Variational Quantum Eigensolver (VQE)(and VQS~\cite{Kokail2018})}
\label{subsec:vqe}

The VQE algorithm, first introduced in \cite{Peruzzo2014}, aims at finding the approximate ground-state energy and wavefunction of a Hamiltonian $H$ by minimizing the energy over a parameterized trial space (also called \textit{ansatz}), i.e.:
\begin{equation}
    E_{\rm VQE}= {\min}_{\bm{ \theta}} \left[ \langle \Psi \left ( {\bm{ \theta}} \right ) |H | \Psi \left ( {\bm{\theta}}\right ) \rangle\right] \equiv {\min}_{\bm{ \theta}}\left[E(\bm{\theta})\right],
    \label{e_vqe}
\end{equation}
where $\bm{ \theta} \equiv \{\theta_p\}$ is a set of parameters that defines the trial state $|\Psi ({\bm{ \theta}} )\rangle$. In most applications, the preparation of the trial state vector is made using a unitary transformation $U({\bm{ \theta}})$ of the qubit vacuum, denoted hereafter by $|\bm{0}\rangle \equiv | 0, \dots, 0 \rangle$ with:
\begin{eqnarray}
|\Psi ({\bm{ \theta}} )\rangle &=& U({\bm{ \theta}}) | \bm{0} \rangle. \label{eq:utheta}
\end{eqnarray}
Some illustrative examples of trial state vectors and associated unitary transformation will be given in section \ref{sec:varansatz}. 

The parameterized observable $E(\bm{\theta})$ is further decomposed into observables that are directly measurable on the quantum device. To do so, being provided a fermion-to-qubit mapping (see paragraph \ref{subsec:From-fermions-to}), one can write $H$ as qubit operator
\begin{equation}
    H = \sum_k \alpha_k P_k, 
    \label{d_h}
\end{equation}
where each $P_k$ corresponds to a string of Pauli operators. The expectation value of each operator over the trial state can be evaluated via sampling in the computational basis ($Z$ measurements) with negligible gate overhead (as illustrated on Table \ref{fig:meas_XY}). 
$E(\bm{\theta})$ is then obtained by classically aggregating the expectation value of each of the Pauli strings:
\begin{equation}
    E(\bm{\theta}) = \sum_k \alpha_k \langle P_k \rangle_{\bm{\theta}}
\end{equation}
where we use the shorthand notation $ \langle. \rangle_{\bm{\theta}} \equiv \langle \Psi \left ( {\bm{ \theta}} \right ) | . | \Psi \left ( {\bm{\theta}}\right ) \rangle$.

The expectation value $\langle  P_k \rangle$ is computed by sampling many instances of the parameterized quantum circuit to gather enough statistics to curb statistical or "shot" noise.
Numerous strategies to limit the sampling overhead incurred by VQE have been put forward (like term-grouping strategies, see, e.g., \cite{Huggins2019}).
Note that the associated time burden of the algorithm can also be partly alleviated by running these circuits in parallel on several quantum devices or parts of a large chip.

In VQE, the optimization of parameters is delegated to classical computers using standard optimization methods, either gradient-free (like the Nelder-Mead or COBYLA method) or gradient-based (like gradient descent). 
{In order to compute the gradients within gradient-based methods, one can apply either finite difference methods, utilizing the generating function from Eq.~\eqref{eq:generating} (notably, this method might be challenging for NISQ devices due to the requisite precision), or 'analytical' gradients.}
%To obtain gradients in gradient-based approaches, one can use either finite difference methods (with the generating function of Eq.~\eqref{eq:generating}) or compute "analytical" gradients.
The computation of the first derivative needed in those methods can be done either with a Hadamard-test-like circuit
or with the parameter shift-rule technique \cite{Mitarai2018,Schuld2019,izmaylov_analytic_2021}. 
{The latter method tends to be more suitable for noisy devices as it typically necessitates running the original circuit twice while adjusting a single parameter. However, in more general cases, the employment of a gate conditioned on an ancilla qubit may be needed.}
%The latter is more suitable for noisy devices as it requires fewer qubits and fewer multi-qubit gates.
Some optimization methods are particularly well-suited in that they are more robust to the shot noise mentioned above, like the SPSA method or the {parameter-shift-rule-based} "rotosolve" method \cite{Nakanishi2019, Ostaszewski2021}.

\begin{figure}[htbp]  
\centering
\includegraphics[width=\linewidth]{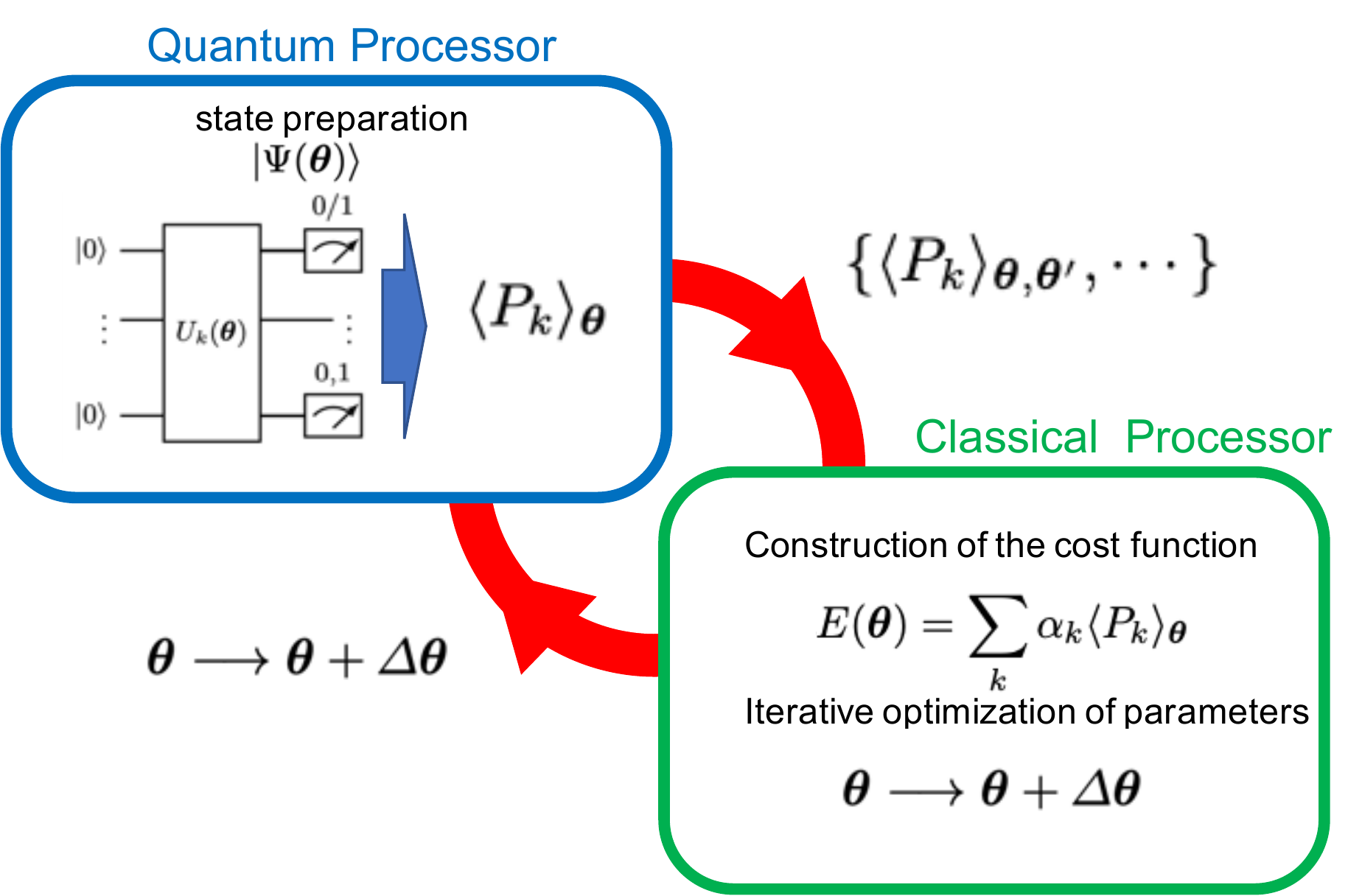}
    \caption{Schematic illustration of the VQE approach. A set of expectation values of Pauli strings  $\{  P_k \}$ is obtained upon measurements on a quantum processor, while the cost function reconstruction and the parameter optimization are made through classical processing. The unitary $U_k(\bm{\theta})$ comprises both the ansatz circuit instance $U(\bm{\theta})$ as well as the additional basis rotation gates necessary to access the expectation value of operator $P_k$ as described in Table \ref{fig:meas_XY}.}
    \label{fig:vqe}
\end{figure}

A schematic view of the VQE hybrid method is given in Fig. \ref{fig:vqe}. 

The implementation of the VQE algorithm on current noisy devices faces several challenges. To mention some important ones: the preparation of the initial state (in the presence of noise), the accurate estimation of expectation values (Eq. \eqref{e_vqe}) (despite shot noise), and the (classical) optimization of the set of parameters $\bm{ \theta}$ to find the minimum of $E_{\rm VQE}$.
VQE's main advantage is the flexibility in the choice of the unitary transformation. 

Variational methods like VQE are not limited to digital quantum processors. 
Elementary operations $\{U_{R_j}(\theta_j)\}$ of the "resource" Hamiltonian of an analog quantum computer can also be assembled into a parameterized ansatz circuit  $U_{R_1}(\theta_1) U_{R_2}(\theta_2) \cdots U_{R_k}(\theta_k)$ in order to minimize the energy of the target Hamiltonian in the final state. This method is called generically \textit{Variational Quantum Simulation} (VQS, \cite{Kokail2018}). It has been applied e.g to the Schwinger model on trapped-ion processors \cite{Kokail2018} or to simple molecules on Rydberg processors \cite{Michel2023}.

Some refinements of VQE are highlighted below. 

\subsubsection{\label{sec:advanced_vqe} Advanced VQE schemes}

Various refined algorithmic schemes have been proposed to either extend the scope of or overcome some limitations of plain-vanilla VQE. 

\paragraph{Penalty methods}
Other types of states than the ground state can be valuable to prepare, such as excited states. VQE schemes with an alternative cost function have thus been proposed to enforce the exploration of specific subspaces of interest. For instance, leveraging the orthogonality of the eigenvectors of the Hamiltonian, one can prepare a sequence of eigenstates by supplementing the cost function with penalty terms proportional to the overlaps (inner products) between the trial state $\Psi(\bm{\theta})\rangle$ and the eigenstates already prepared $\{|\Psi(\bm{\theta}^{*}_{j})\rangle\}$: $\mathcal{C}(\bm{\theta}) = E(\bm{\theta}) + \sum_j \lambda_j \langle \Psi(\bm{\theta}^{*}_{j}) | \Psi(\bm{\theta})\rangle$.
This approach was dubbed the \textit{Variational Quantum Deflation} scheme \cite{higgott_wang_brierley_2019}. More generally, penalty terms chosen accordingly can focus the variational search on, e.g., a specific spin sector \cite{ryabinkin_genin_izmaylov_2018}. 

\paragraph{Subspace-search VQE}
A significant inconvenience of the penalty methods listed above is the need to evaluate inner products. Hence, the proposal in \cite{nakanishi_mitarai_fujii_2019} to leverage the preservation of the inner product under unitary transformation to look for the circuit that best maps a set of orthogonal states to the Hamiltonian's eigenstates.

\paragraph{ctrl-VQE}
VQE can be applied more fundamentally by optimizing the control pulses which underlie quantum operations (see Fig.~\ref{fig:analog_vs_digital}) \cite{meitei_2021}. This low-level optimization allows to address the time-limited character of coherence in NISQ devices. 

\paragraph{Orbital-rotating VQE schemes}
To use shorter circuits, one had better work in a suitable one-particle orbital basis, depending on the target state. Some VQE schemes were proposed to tailor the basis to the problem. We can distinguish two different approaches:
(i) a "classical dressing" of the Hamiltonian observable with a general orbital rotation whose parameters must be determined along with the circuit's parameters (Orbital Optimized-VQE \cite{sokolov_2020,koridon_2021}), and
(ii) iterative basis updates: a converged variational state in the current basis (starting from the usual site-spin basis in lattice models for instance) is leveraged to extract ground state features, setting forth advantageous updates to the single-particle basis.
The terms of the Hamiltonian are transformed accordingly before a new VQE optimization is run.
The latter strategy was applied in two different settings.
The first one---dubbed permVQE \cite{tkachenko_2021}---consists in mere single-particle basis permutations guided by the form of the current converged state's \textit{mutual information matrix}. This matrix measures the information shared by pairs of qubits. For a nearest-neighbor qubit topology, one aims at concentrating its high magnitude coefficients around the diagonal so as to lower the count of entangling gates required.
The second one---called NOization \cite{besserve_ayral_2022}---performs general basis updates aimed at iteratively reaching a specific, state-dependent basis known as the Natural Orbitals basis. This basis is associated with a compact state representation, namely the one encompassing the lowest number of computational basis states (Slater determinants). In this case, the quantity being monitored is the 1-RDM, whose eigenbasis provides the new basis for the subsequent VQE run. 

\paragraph{Projective Quantum Eigensolver (PQE)}
The PQE method \cite{Stair2021a} minimizes the residuals (that measure the non-orthogonality of excited states to the ground state manifold) instead of the energy, yielding accuracies on a par with VQE's using fewer resources, and with less size-dependence.

%\sout{A possible method to access excited states that might also be very useful in case of degenerate state is to use the quantum equivalent of the state-averaged complete active space self-consistent field used in classical computers \cite{Siegbahn1981}. This method, combined with the VQE approach provides an interesting way to obtain excited states in a reduced Hilbert space that can be used even in the case of level crossings, using a modified version of the VQE method, called State-Averaged VQE (SA-VQE) \cite{Yalouz2021}.}

{\paragraph{State-Average VQE}: A possible way to access excited states as well as to handle degeneracies is to use the quantum equivalent of the state-averaged complete active space self-consistent field used in classical computers \cite{Siegbahn1981}. This method, dubbed State-Average-VQE (SA-VQE) \cite{Yalouz2021} provides a way to obtain excited states in a reduced Hilbert space that can be used even in presence of degeneracies. }
\subsubsection{On the use of variational principles in quantum computers}
Although it is slightly out of the scope of the present review, we want to mention broader applications of variational principles in quantum computing. 

One can, for instance, use McLachlan’s variational principle (MVP) \cite{McLachlan64} to obtain approximate quantum systems' unitary evolutions. In that case, the MVP takes the form: 
\begin{eqnarray}
 \delta \left\| {(i \hbar \partial / \partial t - H )\left| {\Psi (t)} \right\rangle } \right\| = 0,
\end{eqnarray}
where $\left\| | \Psi \rangle \right\| \equiv \sqrt{\langle \Psi | \Psi\rangle }$. This variational principle can be interpreted as a cost function which, given a parametric form for the trial state, minimizes the deviation of the approximate evolution $i \hbar \partial_t | \Psi(t) \rangle$ from the true evolution $H | \Psi \rangle$. This principle can be connected with other variational principles generally used in many-body problems \cite{Blaizot86}. A complete discussion of the MVP and its manipulation is out of the scope of the present article but can be found in Ref. \cite{yuan_theory_2019} in the quantum computing context. Still, it is interesting to mention that the MVP is not restricted to real-time unitary evolution but can be adapted to other problems involving non-unitary motion or mixed-state evolution. 

{\it Mixed state evolution:} The approximate solution of a density matrix evolution that could differ from that of a pure state density can be obtained from the MVP:
\begin{eqnarray}
\delta \left\| i \hbar d\rho / dt - {\cal L}(\rho) \right\|^2 = 0,
\end{eqnarray} 
where $\rho$ is the density matrix (see section~\ref{subsec:NISQ-Noisy-quantum} for a definition), and the Liouvillian ${\cal L}(\rho)$ is a general functional of the density.
This variational principle can be used to find an approximation of $i \hbar \dot \rho = {\cal L}(\rho)$. The pure state Hamiltonian evolution can be recovered by setting $\rho = | \Psi \rangle \langle \Psi |$ and ${\cal L}(\rho) = [H , \rho]$.
Besides this case, it can also be used to simulate dissipative processes using a Lindblad-type equation (see section~\ref{sec:noise}) for ${\cal L}(\rho) $ \cite{Endo2020}.

{\it Imaginary-time propagation} When the real-time evolution is replaced by an imaginary-time evolution ($t \rightarrow i\tau$), an initial state with a good enough overlap with the exact ground state is projected to the exact ground state by the evolution operator $U(\tau) = e^{-\tau H}$.
This method is a standard practical one to obtain the lowest energies and associated eigenstates of the Hamiltonian on a classical computer.
The operator $U(\tau)$ is non-unitary and cannot a priori be directly implemented on a quantum computer.
This problem was overcome in Ref. \cite{mcardle_variational_2019,McArdle2018a} using the following variant of the MVP:
\begin{eqnarray}
    \delta \left\| {(\partial_\tau + \left[ H - \langle H \rangle_\tau \right] )\left| {\Psi (\tau )} \right\rangle } \right\| = 0.
\end{eqnarray}
This approach is called Quantum Imaginary-Time Evolution (QITE). 
It can be variational if $\ket{\Psi}$ is parameterized by variational parameters, or not \cite{Motta2019}.  

It is interesting to mention that variational techniques, that are fundamental tools in many-body systems, have also been exported to general learning problems. Such techniques are a growing field of interest today \cite{Mitarai2018,Schuld2018,Benedetti2019,Biamonte2021,Schuld2021,Goto2021,Cerezo2021}.

\subsection{Variational ans\"atze for many-body problems}
\label{sec:varansatz}

\subsubsection{Adiabatic state preparation vs. variational state preparation}

\begin{figure}
    \centering
    \includegraphics[width=\columnwidth]{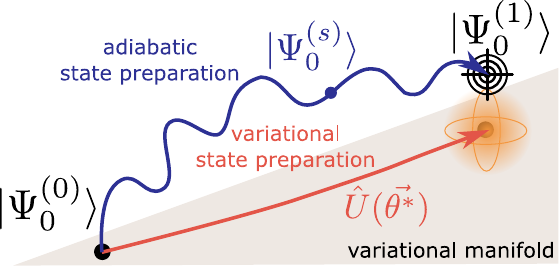}
    \caption{Ground state preparation of $\ket{\Psi_0^{(1)}}$ from a simple-to-prepare ground state $\ket{\Psi_0^{(0)}}$, either via adiabatic state preparation or variational state preparation. The optimal variational state is represented with a dot in the variational manifold. It is the one with the highest overlap with $\ket{\Psi_0^{(1)}}$. Its distance to the target state $\ket{\Psi_0^{(1)}}$ measures the \textit{expressivity} of the ansatz. The orange sphere materializes the variability on the state obtained upon optimization of the variational ansatz, due to \textit{(i)} the optimization being done with regards to an energetic criterion rather than the overlap \textit{(ii)} the presence of local minima in the optimization landscape, \textit{(iii)} shot noise (all of which incur solely errors making the state move in the variational manifold), and \textit{(iv)} hardware noise (possibly also incurring errors kicking the state out of the variational manifold).}
    \label{fig:asp_vs_vqe}
\end{figure}

The adiabatic state preparation method is an essential inspiration to many variational state preparation techniques that we will describe.
Its goal is reach the ground state $\ket{\psi_0^{(1)}}$ of a Hamiltonian $H_1$ by starting from the (easy-to-prepare) ground state $\ket{\psi_0^{(0)}}$ of an initial Hamiltonian $H_0$.

This is achieved by designing a time-dependent Hamiltonian:
\begin{equation}
    \label{eq:asp_schedule}
    H(t) = (1-s(t))H_0 + s(t) H_1,
\end{equation}
where $s$ denotes a "schedule" such that $s(t=0) = 0$ and $s(t=T)=1$.
Typically, $\ket{\psi_0^{(0)}}$ can be a Slater determinant.

The adiabatic theorem guarantees that provided one proceeds slowly enough (adiabatically) with regards to the spectral gap along the path, the system remains in the ground state of the instantaneous Hamiltonian $H(t)$.
This property is linked to the so-called Gell-Mann and Low theorem \cite{Gell-Mann1951,Fetter2012}.
Upon trotterizing the evolution under perturbed Hamiltonians, one thus has a general---but costly---recipe for ground state preparation: adiabatic state preparation (ASP) \cite{Farhi2000}.
This method was used in, e.g., \cite{Bauer2016}, which furthermore makes use of QPE (see paragraph \ref{subsec:qpe} for a discussion on the limitations of QPE for state preparation) to pin the state into the ground state of the perturbed Hamiltonian.

By construction, ASP gives rise to long time evolutions and hence deep circuits, which is a problem for NISQ processors.
It can nevertheless be used as a formal inspiration to design variational states to be used in the variational methods described above: these can be regarded as a way to find unitary "shortcuts" (sometimes called diabatic evolution) to go from a simple initial state $\ket{\psi_0^{(0)}}$ to a target ground state $\ket{\psi_0^{(1)}}$, as pictured on Fig.~\ref{fig:asp_vs_vqe}. 

The general strategy to borrow from ASP to design VQE states is the following: Hamiltonian \eqref{eq:asp_schedule} induces a unitary evolution that can be trotterized as $\prod_{k=1}^{n_t} e^{-i (1-s(t_k))H_0 \delta t} e^{-i s(t_k) H_1 \delta t}$, with $\delta_t = T/n_t$ and $t_k = k \delta_t$. This unitary evolution can be transformed into a variational circuit 
\begin{equation}
    U(\bm{\theta}) = \prod_{k} e^{-i \theta_{2k} H_0} e^{-i \theta_{2k+1} H_1}.
\end{equation}

In the following subsections, we consider ans\"atze with increasing complexity starting from a rather academic methodology (Hartree-Fock approach, Bogolyubov transformation) and then present the recent efforts to craft more minimal ans\"atze so that they avoid running into the usual limitations of VQE (limited coherence time of NISQ devices, or the barren plateaus encountered in numerous-parameter ansatz optimizations \cite{Romero2018}).

\subsubsection{Uncorrelated ansätze}

\paragraph{Hartree-Fock theory and Slater determinants}
\label{sec:hf}

In many-body problems, a complete basis of the Fock space is given by the set of Slater determinants:
\begin{eqnarray}
|\delta_{n_q-1}, \dots , \delta_{0} \rangle = \prod_k \left[ a^\dagger_k
\right]^{\delta_k} | \bm{0} \rangle , \label{eq:slater}
\end{eqnarray}
where $\{ a^\dagger_k \}_{k=0, n_q -1}$ correspond to the creation operators associated with a complete set of single-particle states $\{ | k \rangle \}$. Here it is assumed that the one-body Hilbert space is finite with dimension $n_q$.
For a set of $A$ particles, the Hartree-Fock procedure consists in estimating the ground state energy of a (possibly correlated) Hamiltonian ${H}$ as $E_{\rm HF} = \langle \Psi_{\rm HF} |  H | \Psi_{\rm HF} \rangle$ assuming that the trial wave-function is a Slater determinant given by 
\begin{eqnarray}
| \Psi_{\rm HF} \rangle &=& \prod_{\alpha} \left[  b^\dagger_{\alpha}\right]^{\gamma_\alpha} | \bm{0} \rangle.   \label{eq:hfstate}  
\end{eqnarray}
In this equation, only $A$ coefficients  $\gamma_\alpha$ are equal  to $1$ (corresponding to particles below the Fermi level) while $n_q - A$ of them are equal to zero (particle states).
The creations operators $\{ b^\dagger_{\alpha} \} $ are associated to a complete basis $\{ |\alpha \rangle \}$, with the relationship:
\begin{eqnarray}
| \alpha \rangle &=& \sum_k | k \rangle \langle k | \alpha \rangle ~\longrightarrow b^\dagger_\alpha = \sum_k a^\dagger_k U_{k \alpha} ,
\label{eq:unitsingle}      
\end{eqnarray}
with $U_{k \alpha} = \langle k | \alpha \rangle$ a unitary transformation to be variationally determined.
In other words, the HF procedure consists in variationally finding the best Slater determinant approximation to the target ground state. It is a very simple mean-field approach to the original problem: it replaces the many-body problem with a set of particles influencing one another through a self-consistent, average one-body potential (or mean field), instead of actual interactions.
Practical aspects of finding the HF solution, which consists in minimizing the energy with respect to the variations of unitary matrix $U$, are standard.

Below, we focus on the implementation of HF on a quantum computer because the preparation of arbitrary Slater determinants is the starting point of many more advanced methods (indeed, implementing HF on a quantum computer is per se not useful as HF can be implemented efficiently---i.e in polynomial time---on a classical computer).

When the Jordan-Wigner transformation is used to map the Fock space to qubit space (see section \ref{subsec:From-fermions-to}), the set of Slater determinants, given by Eq. (\ref{eq:slater}), directly identifies with the computational basis $\{ | \delta_{n_q-1}, \dots, \delta_{0} \rangle \}$ with $\{\delta_k = 0,1\}_{k=0,n_q-1}$, that we can rewrite as:
\begin{eqnarray}
| \delta_{n_q-1}, \dots , \delta_{0} \rangle = \prod_k \left[ X_k
\right]^{\delta_k} | \bm{0} \rangle . \label{eq:slaterqubit}
\end{eqnarray}
The key ingredient to implementing the HF theory on a quantum computer is to be able to realize a general unitary transformation $U$ on the fermionic modes, as defined in Equation (\ref{eq:unitsingle}), using a parametric circuit and the possibility to prepare a state equivalent to (\ref{eq:hfstate}) with exactly $A$ particles.
The technique that has been employed, for instance, in Ref. \cite{Dumitrescu2018, Arute2020}, is based on the use of the Thouless theorem \cite{Thouless1960} (see also Appendix E of Ref. \cite{Ring80}).
Starting from one of the Slater determinants given in (\ref{eq:slater}) and denoted generically as $| \Psi_0 \rangle$, one can generate an ensemble of new Slater determinants given by 
\begin{eqnarray}
| \Psi ({Z}) \rangle &\equiv& e^{i \sum_{ij} Z_{ij} a^\dagger_i a_j} |\Psi_0 \rangle, \label{eq:thouless}
\end{eqnarray}
where $Z$ is supposed to be hermitian. 
These states identify with the form (\ref{eq:hfstate}) with $b^\dagger_\alpha({Z}) = \sum_k (e^{i{Z}})_{\alpha k} a^\dagger_k$, i.e. $U^* = e^{iZ}$.
The circuit that performs the Thouless transformation starting from a product state as given by Eq.  (\ref{eq:slaterqubit}) is discussed in detail in Ref.  \cite{Arute2020}. Additional discussion on the application of HF can be found in Refs. \cite{Wecker2014a,McArdle2018a}.

\paragraph{General quasi-particle vacuum}\label{subsubsec:hfb}
The transformation (\ref{eq:unitsingle}) and the Thouless method can be generalized to a larger class of trial states known as quasi-particle vacua (also known as Gaussian states in other contexts).
Below, such vacuum states are written generically as $| \Psi_\beta \rangle \propto \prod \beta_\alpha | \bm{0} \rangle$, where $\{\beta_\alpha, \beta^\dagger_\alpha \}$ denotes a complete set of quasi-particles creation/annihilation operators. These operators can be connected through a generalization of Eq. (\ref{eq:unitsingle}) given by \cite{Ring80,Blaizot86}:
\begin{eqnarray}
\beta^\dagger_k &=& \sum_k  \left[ a^\dagger_k U_{k \alpha} + a_k V_{k \alpha}  \right]. \label{gen:bogo}
\end{eqnarray}
Using general quasi-particle vacua instead of restricting to Slater determinants leads to the Hartree-Fock-Bogolyubov (HFB) theory, where the $U(1)$ symmetry, associated with particle number conservation, is broken.
The advantage of breaking this symmetry is the possibility to describe superfluid systems \cite{brink_broglia_2005}.

The quantum state preparation of a general quasi-particle vacuum using the Thouless transformation, as done at the HF level in \cite{Arute2020}, was addressed in, e.g., \cite{Dallaire-Demers2018} and relies on two main arguments.
First, the mapping between Thouless' transformation $\mathcal{U}(Z) = e^{i \sum_{ij} Z_{ij} \gamma_i \gamma_j}$ (where the $\gamma_i$ denote Majorana modes, i.e., $\gamma_{2k}= a_k + a_k^{\dagger}$ and $\gamma_{2k + 1}= -i(a_k - a_k^{\dagger})$) and quantum gates can be found by leveraging the decomposition of $R = e^{iZ}$ as a product of elementary Givens rotations \cite{hoffman_1972}. Let $M$ be the number of fermionic modes (and hence, qubits), then $R = \prod \limits_{k=1}^{2M \choose 2} r_k(\theta_k)$, where the Givens rotations consist of $M$ local phase rotations which can be implemented as $R_z$ gates and $2M(M-1)$ $SO(4)$ rotations acting non-trivially only on two modes.
The $SO(4)$ rotations can be rendered in a quantum circuit by means of matchgates \cite{valiant_2001, jozsa_miyake_2008} and SWAP gates. All in all, Thouless' transformation can be implemented as a nearest-neighbour matchgate circuit with depth $O(M)$.

Several applications have also been explored in quantum computers using a Bardeen–Cooper–Schrieffer (BCS)-like ansatz given by:
\begin{eqnarray}
| \Psi \rangle &=& \prod_k \left(u_k + v_k a^\dagger_k a^\dagger_{\bar{k}} \right)| \bm{0} \rangle   , \label{eq:bcssimple}
\end{eqnarray}
where $(k, \bar{k})$ refers to two single-particle states forming a pair of time-reversed states, and where $u^2_k + v^2_k = 1$.
The encoding of such a state on a qubit register is not unique. If the brute force JWT is used to make a direct mapping between single-particle states and qubits, the trial state can be written as:
\begin{eqnarray}
| \Psi(\bm{\theta}) \rangle &=& \bigotimes_k \left[ \sin(\theta_k) |00 \rangle_k + \cos(\theta_k) | 11 \rangle_k \right],  \label{eq:bitsimple}
\end{eqnarray}
where we made the identification $u_k = \sin(\theta_k)$ and $v_k = \cos(\theta_k)$.
To map Eq. (\ref{eq:bcssimple}) into Eq. (\ref{eq:bitsimple}), time-reversed states are assumed to be represented by adjacent qubits, and $|. \rangle_k$ denotes the two qubits associated with these states. We recognize a generalized Bell state that can be obtained by performing a $R_y(\theta_k)$ rotation on one of the qubits followed by a CNOT operation with the second qubit. Such encoding was used, for instance, in \cite{Verstraete2000,ovrum2007quantum,Jiang2018,Lacroix2020,Khamoshi2020,Khamoshi2022}.

This encoding is general and allows treatment of systems where one or several pairs are broken (usually referred to as nonzero seniority~\cite{brink_broglia_2005}) as illustrated in Ref. \cite{Lacroix2020} for instance, to treat odd systems. If we restrict to the situation with seniority $0$, i.e., when no pairs are broken, one can reduce the number of qubits by directly encoding the occupation of the two adjacent time-reversed states onto one qubit. In this case, the state $| 1\rangle_k$ or $|0\rangle_k$ represents the simultaneous occupation or not of the two time-reversed particles $(k,\bar k)$. 
This technique, used in Refs.~\cite{Khamoshi2020,ruiz-guzman2021,RuizGuzman2022}, has the advantage of reducing by a factor of $2$ the required number of qubits compared to the case where one particle is encoded on one qubit. It also avoids the use of controlled operations since we have, for this encoding scheme: 
\begin{eqnarray}
\label{eq:bcs_ansatz}
| \Psi(\bm{\theta}) \rangle &=& \bigotimes_{k} \left[ \sin{\theta_k}|0\rangle_k + \cos{\theta_k}|1\rangle_k \right]
\nonumber \\
&=& \prod_{k} R^{(k)}_{y}\left(\pi - 2\theta_k\right) |\bm{0}\rangle.
\end{eqnarray}
Quasiparticle-like states have been extensively explored in quantum computers \cite{Wecker2015a, Kivlichan2018} (see also \cite{Dallaire-Demers2018}), as well as their experimental implementation, \cite{Arute2020}.

We focused here on relatively standard quasi-particle vacuum states that play a particular role in many-body systems and lead to the HF and HFB frameworks. Below is a selection of other ans\"atze that are widely discussed today in the literature.

\subsubsection{Hamiltonian Variational Ansatz}
Inspired both from adiabatic state preparation (ASP, see Fig.~\ref{fig:asp_vs_vqe}) and the Quantum Approximate Optimization Algorithm circuit (QAOA \cite{Farhi_2014}), which is ubiquitous in quantum combinatorial optimization, the Hamiltonian Variational Ansatz (HVA \cite{wecker_2015}) state reads
\begin{equation}
    \ket{\Psi(\bm{\theta})} = \prod_{l = 1}^L \left( \prod_k e^{-i\theta_k^l H_k}\right) \ket{\Psi_0},
\end{equation}
where the terms $H_k$ come from decomposing $H$ as $H = \sum_k H_k$, with $[H_k, H_{k'}] \neq 0 $ for $k \neq k'$.
The dimension $L$ of index $l$ is referred to as the \textit{depth} of the ansatz. The initial state $\ket{\Psi_0}$ is the ground state of some $H_{k_0}$ that is taken not to act first on $\ket{\Psi_0}$.
Optimizing the HVA parameters amounts to optimizing the Hamiltonian schedule $s$ of ASP (as introduced in Eq. \eqref{eq:asp_schedule}). HVA was applied, {\it e.g.} to the study of the 1-D Hubbard model in \cite{anselme_martin_2022}.

\subsubsection{Hardware-Efficient Ansatz (HEA)}
Today's quantum computers are not uniformly accurate in performing different operations. Knowing the strength or weaknesses of a given platform, one might adapt the ansatz to the operations most efficiently realized. The HEA technique consists in writing the trial state from a set of operations that are "native" in the quantum processor.
This heuristic approach can optimize the trial state construction with respect to the specific hardware, but also restricts the type of trial states that can be constructed \cite{Kandala2017,Barkoutsos2018}.
One issue with the HEA is that the classical optimization of the variational parameters can become difficult due to gradients that vanish exponentially with the number of qubits, a phenomenon dubbed the \textit{barren plateau problem} \cite{McClean2018}.
{Interestingly, it was observed on the other hand that beyond a certain depth at a given number of qubits, gradients increased again for e.g. the transverse-field Ising model as well as for the Sachdev-Ye-Kitaev model \cite{Kim_2021}. This phenomenon is general, and is easier to understand with the help of the concept of \textit{overparametrization} \cite{Larocca_2023}.}
{Moreover, HEA are typically defined with gates taken from the native gateset of the specific technology that is used, and are therefore strongly hardware-driven. This might in practice lead to quantum circuits that do not respect some of the symmetries of the physical problem of interest, leading to specific additional difficulties.}

\subsubsection{Unitary coupled cluster (UCC)} \label{subsubsec:ucc}
The UCC offers a framework that naturally extends the HF method based on the Thouless approach described in section \ref{sec:hf} (see also section \ref{sec:electron}).
The trial wavefunction is written in a generalized form \cite{Romero2018}:
\begin{eqnarray}
        |\Psi_{\rm  UCC}(\bm \theta) \rangle &=& e^{T(\bm \theta)-T^\dagger(\bm \theta)} | \Psi_0 \rangle \equiv U(\bm{\theta})  | \Psi_0 \rangle,  \label{eq:stateucc}
\end{eqnarray}
where $T$ can be expanded as a set of operators of increasing complexity with $T= T_1+ T_2 + \dots$. Here, $T_1$, $T_2$, ... stands for single, double, ... particle-hole excitation operators with respect to the state $| \Psi_0 \rangle$, with
\begin{eqnarray}
        T_1 &=& \sum_{i,j} T^{(1)}_{ij} a^\dagger_i a_j, \label{eq:single} \\
        T_2 &=& \sum_{i,j} T^{(2)}_{ij, kl} a^\dagger_i a^\dagger_j a_l a_k, \label{eq:doubles} \\
        &\cdots& \nonumber
\end{eqnarray}
We see, in particular, that the HF is recovered by using a Slater determinant and restricting $T$ to single excitations. After truncation, the state prepared using Eq. (\ref{eq:stateucc}) is used as a trial state in the VQE approach discussed in section \ref{subsec:vqe}.
This technique is currently widely applied in quantum chemistry (see the recent review~\cite{Abhinav2022} and references therein) {, but mostly in noise-free simulations. The reason is that whereas it has the capacity to yield highly accurate results, the very large depth of the UCC ansatz circuit leads to degraded performances on real quantum platforms. Ways around this prohibitive gate count include circuit compilation---tailoring the implementation to the subspace under scrutiny, as in Ref.~\cite{Dumitrescu2018}, as well as adaptive approaches, for instance as applied to atomic nuclei in Ref.~\cite{Kiss2022}.}
%It was also used in most applications to atomic nuclei on real quantum platforms~\cite{Dumitrescu2018,Kiss2022}.
Finally, the possibility of combining such an approach with $U(1)$ symmetry that is relevant for strongly interacting systems like nuclei is of current interest \cite{Henderson2014,Qiu2019} and is currently explored on quantum computers too \cite{Khamoshi2020,Khamoshi2022}.   

\subsubsection{The Low-Depth Circuit Ansatz (LDCA)}
\label{subsubsec:ldca}

Elaborating on the general quasiparticle vacua preparation routine reviewed in paragraph \ref{subsubsec:hfb}, the LDCA circuit \cite{Dallaire-Demers2018} (standing for \textit{Low-Depth Circuit Ansatz}) possibly allows reaching any state due to the insertion of $R_{zz}$ rotation gates into to the Hartree-Fock-Bogolyubov circuit. It also allows the replication of similar layers in the ansatz to increase its representability systematically.
Intuitively, $R_{zz}(\theta) \equiv e^{-i\theta/2 {Z}_p {Z}_q}$ gates generate correlated states as these gates correspond {to density-density interactions within a Jordan-Wigner encoding:} the density operator of orbital $p$, $n_p \equiv a^{\dagger}_p a_p$, maps to the qubit operator $\frac{I-{Z}_p}{2}$; therefore $n_q n_p$ interactions translate into $Z_p Z_q$ terms.

%The main drawback of LDCA is that despite its gentle scaling {compared with the UCC ansatz}, it still incurs a prohibitive gate count with 
%respect to NISQ capacities, e.g., \cite{besserve_ayral_2022}. 

{The LDCA is competitive compared to the UCC approach in terms of circuit depth, especially when double or higher excitations are considered.  Despite its gentle scaling in terms of the number of quantum operations, it still incurs a prohibitive gate count with respect to NISQ capacities, e.g., \cite{besserve_ayral_2022}.}

\subsubsection{Projected Ans\"atze}
An important cornerstone for future applications, especially in nuclear systems, is the possibility of making, for instance, symmetry restoration after symmetry breaking. A discussion for the particle number symmetry was made in section \ref{subsubsec:hfb}. Assume, for instance, that a state $| \Psi(\bm{\theta}) \rangle$ can be prepared on a quantum computer and that such state does not respect the symmetry of the physical problem that is encoded in the Hamiltonian $ H$. Instead of using $| \Psi(\bm{\theta}) \rangle$ in the variational principle, one can use the projected wavefunction: 
\begin{eqnarray}
        | \Psi_{\rm P}'(\bm{\theta})\rangle &=&
        \frac{1}{\sqrt{\langle  \Psi(\bm{\theta})  |  P_S |  \Psi(\bm{\theta} ) \rangle } }  P_{S} |  \Psi(\bm{\theta}) \rangle.
\end{eqnarray}
Here, $P_S$ is a projector onto the subspace of the Hilbert space containing states with the desired property (for instance, the proper symmetries).
Such a strategy is used in many-body physics to grasp specific correlations between particles or strongly entangled states that are hard to describe otherwise \cite{Ring80,Blaizot86}.
One difficulty is that the projector is a non-unitary operation and cannot be directly implemented on a quantum computer. Several methods have been proposed recently to construct projected states \cite{Lacroix2020,Lacroix2022}. These methods have been combined with VQE in Ref. \cite{RuizGuzman2022}, leading to the Quantum-Variation After Projection (Q-VAP) framework.

\subsubsection{Adapt-VQE}
Plain-vanilla VQE takes a fixed variational state as an input to the computation. This incurs the risk of overfitting the target state if the variational manifold is too large.
The ADAPT-VQE method iteratively constructs this ansatz instead. It relies on a predefined operator pool from which operators are drawn adaptively along the optimization procedure. Typically one selects the operator maximizing the gradient at the current step so that its addition to the circuit has the most significant effect on the variational energy. The aim is to reduce the number of parameters of the ansatz at the expense of an increased measurement overhead due to the gradients. 
In the initial proposal (ADAPT-VQE \cite{grimsley_economou_barnes_mayhall_2019})), the operators in the pool were fermionic, but the large gate overheads resulting from long Jordan-Wigner strings can be avoided by directly using qubit operators instead (qubit-ADAPT VQE \cite{tang_shkolnikov_2021}). 
This method can reach chemical accuracy at a relatively low gate count, at least with noiseless computers (see, e.g., \cite{Haidar2022} for an example with large molecules).

\subsubsection{Tensor-network-inspired quantum circuits}
\label{subsec:TN_ansatz}

Tensor-network states refer to widely used representations of the wave function $|\Psi \rangle$ of a many-body problem.
Instead of storing the information contained in a $n$-qubit wavefunction in a multi-array $a_{b_1, b_2, \dots b_n}$ (with storage cost $2^n$ complex floating-point numbers), one assumes a particular factorization of this multi-array, with the hope that storing the different factors will be less costly.
The Matrix Product State (MPS) class is a widespread subclass of tensor networks. It consists in factorizing the multi-array as a product of matrices $[A^{(k)}]^{b_k}$:
\begin{equation}
    a_{b_1, b_2, \dots b_n} = \sum_{\alpha_1, \dots \alpha_{n-1}} [A^{(1)}]^{b_1}_{\alpha_1} [A^{(2)}]^{b_2}_{\alpha_1, \alpha_2} \cdots [A^{(n)}]^{b_n}_{\alpha_{n-1}} \label{eq:MPScoef}
\end{equation}
The internal indices $\alpha_k$ have a dimension called the bond dimension, which is usually denoted as $\chi$. As we will see in section \ref{sec:entMPS}, this parameter is closely connected to the degree of entanglement one can access with such a state.  

MPS can be used as an inspiration or starting point for quantum computations. Specifically, methods have been proposed to convert a given MPS to a quantum circuit \cite{Ran2020, Rudolph2022}. Paragraph \ref{sec:mps_quantum} presents an outline of the quantum circuitry involved. An important outcome is that MPS with a given bond dimension requires a circuit with a logarithmic depth in $\chi$, resulting in a complexity gain and thus the possibility to use a quantum computer to generate states with an entanglement level inaccessible to classical computers due to too large a bond dimension. The conversion from MPS to a circuit can also be used to warm-start a variational quantum computation \cite{Rudolph2022a}.

Other tensor networks can also be used as a comparison point or inspiration to quantum circuit design, like Tree Tensor Networks \cite{Yuan2021a}, the Multiscale Entanglement Renormalization Ansatz \cite{Miao2021}, or Projected Entangled Pair States \cite{Cirac2021}.

\subsection{Beyond variational methods}

As described in section \ref{subsec:vqe}, variational methods usually target the search of approximate ground states. Here, we discuss several methods that can give access to excited states too. Previously (section \ref{subsec:qpe}), we saw that the QPE could be a tool of choice to obtain energy eigenvalues and associated eigenvectors. Unfortunately, it cannot be used in current devices, and probably, it will take some time before the fidelity of quantum machines becomes sufficiently high to apply it.
The search for alternative methods, less costly in terms of quantum resources, is therefore an intensive domain of activity today. 
We report below some of the methods that have been proposed to access excited state properties. Note that these methods sometimes only give access to the energies, not the states associated with them.      

\subsubsection{Quantum Subspace Expansion methods}
\label{sec:hse}

\begin{figure}[htbp]  
\centering
\includegraphics[width=0.8\columnwidth]{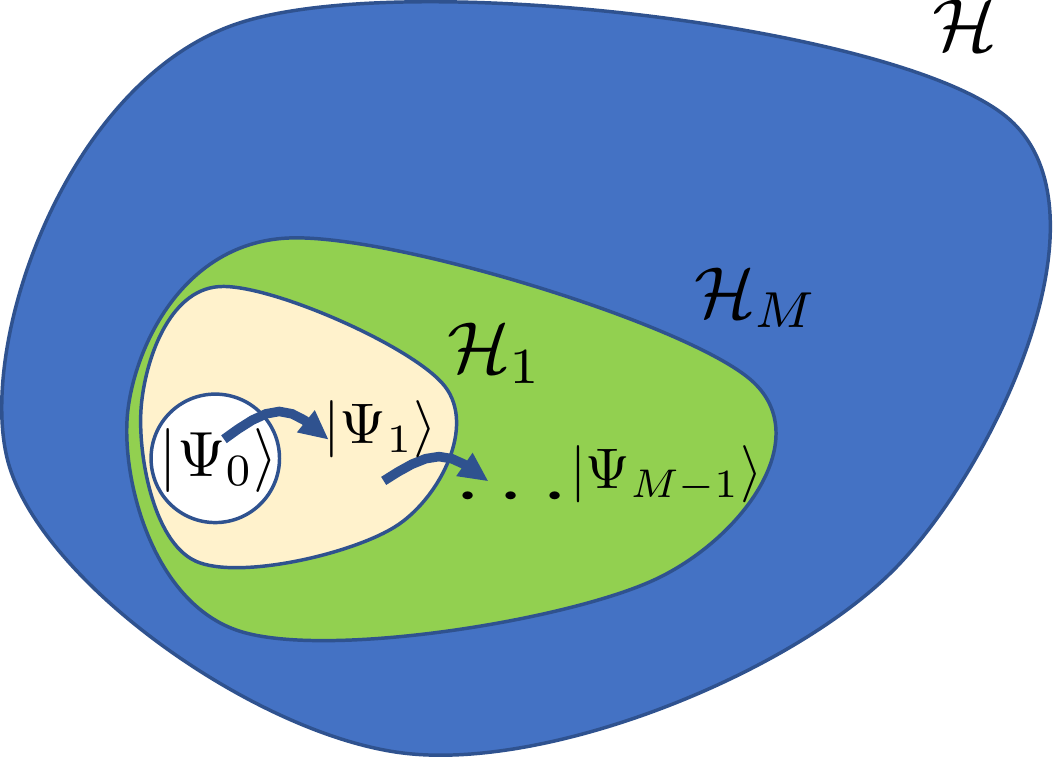}
    \caption{Illustration of QSE philosophy where the eigenvalue problem is considered 
    in a subspace with increasing dimension.}
    \label{fig:hse}
\end{figure}

A common strategy used in classical computers to obtain the approximate solution of a diagonalization problem when a complete CI solution is prohibitive consists of iteratively constructing subspaces of the total Hilbert space ${\cal H}$ of increasing complexity. 
In many cases, at a given level $M$ of complexity,  a subspace is generated by a set of states $\{ |\Psi_0\rangle, \dots, | \Psi_{M-1}\rangle \}$ that spans a subspace denoted as ${\cal H}_M$.
The method to obtain the states is usually iterative in the sense that $| \Psi_{k+1} \rangle$ is constructed from $| \Psi_{k} \rangle$ using specific operations.
Arnoldi or Lanczos methods are famous examples widely used on classical computers \cite{Saad2011}.
The generic strategy of using an increasing number of states to form a subspace of the Hilbert space will be called hereafter Quantum Subspace Expansion (QSE) \cite{McClean2017a}.
A schematic view of the QSE strategy is shown in Fig. \ref{fig:hse}. This strategy's success depends on its capacity to construct the relevant subspace for a given problem. Key ingredients are the seed state $| \Psi_0 \rangle$ and the rules for the iterative generation of states.
 
The generated states are usually not orthogonal with one another. 
Any state belonging to the reduced space  ${\cal H}_M$ can be written as $| \Psi \rangle = \sum \limits_{K=0}^{M-1} c_K | \Psi_K \rangle$. An approximate eigenvalue $E$ in this space can be obtained by solving the generalized eigenvalue problem written as a set of $K=0, M-1$ equations given by:
\begin{eqnarray}
\sum \limits_{K'=0}^{M-1} c_{K'} H_{KK'} &=& E \sum \limits_{K'=0}^{M-1} c_{K'} O_{KK'}, \label{eq:eigen}
\end{eqnarray} 
with $H_{KK'}= \langle \Psi_K |  H | \Psi_{K'} \rangle$ and $O_{KK'} =  \langle \Psi_K | \Psi_{K'} \rangle$ the overlap matrix. Such equations can be solved in a two-step process by first diagonalizing the overlap matrix prior to the Hamiltonian diagonalization. Potentially, in ${\cal H}_M$, $M$ approximate eigenstates can be obtained, which makes the method quite attractive. 

On classical computers, a typical choice for the states is the Krylov basis where $| \Psi_{k+1} \rangle =  H | \Psi_{k} \rangle $. Great effort is currently devoted to the possibility of extending the Krylov space technique to quantum computers. Then, the strategy is to compute the matrix elements of the two matrices $O$ and $H$ in the reduced space using the quantum computer. Subsequently, this matrix is diagonalized via classical methods \cite{Seki2021}. A discussion on the possibility of obtaining the strict equivalent of the Krylov basis using derivatives of the generating function $F(t)$ introduced below in Eq. (\ref{eq:generating}) was scrutinized in Ref. \cite{ruiz-guzman2021}. This analysis was made using the fact that $H_{KK'} = \langle \Psi_0| H^{K+K'+1}| \Psi_0 \rangle$ and that $F(t)$ is the generating function of the Hamiltonian moments. However, this method is exceptionally susceptible to numerical noise.

Alternatively, one can generate the states using unitary transformation of the seed state such that $| \Psi_K \rangle = U_K | \Psi_0 \rangle$. This approach is particularly well adapted to quantum computing, where circuits are automatically unitary. In the {\it Quantum Krylov} technique, the hamiltonian propagator itself is used such that $U_K \equiv e^{-i  H \tau_K}$ where a set  of times $\{ \tau_K \}_{K=0,\dots,M}$ has been assumed with the convention $\tau_0=0$.
The possibility of using Krylov-inspired techniques on quantum computers has more generally attracted much attention in recent years \cite{McClean2017a,Parrish2019,Stair2020,Seki2021,Bharti2021b,Bharti2021c,Bespalova2021,Jamet2021, Cortes2022,Seki2022,Haug2022,Wei2022,RuizGuzman2022} (see also the survey \cite{Aulicino2022}).

\subsubsection{QPE-inspired quantum algorithms} 

Different methods inspired by the QPE algorithm have been proposed to reduce the quantum resources in ancilla qubits or the number of operations in the quantum circuit. For instance, the methods we will discuss subsequently use only one ancilla qubit and, contrary to the QPE algorithm, do not require the inverse quantum Fourier Transform. Ref. \cite{McArdle2018a} discusses a comprehensive list of these methods.

As an alternative to the standard phase estimation, Kitaev's algorithm \cite{Kitaev1996} and the iterative QPE algorithm based on the semiclassical quantum Fourier transform \cite{Griffiths1996,Dobsicek2007} (see also \cite{Svore2013,Lin2022}) were proposed to find the eigenvalue of a single eigenstate.
More recently, further progress has been made with the {\it Rodeo} algorithm \cite{Choi2021} that appears as a practical tool in the NISQ context \cite{Qian2021,Bee-Lindgren2022}.  We briefly describe below how these iterative techniques can be implemented.

\begin{figure}[htbp]  
    \centering
    \includegraphics[width=1.0\columnwidth]{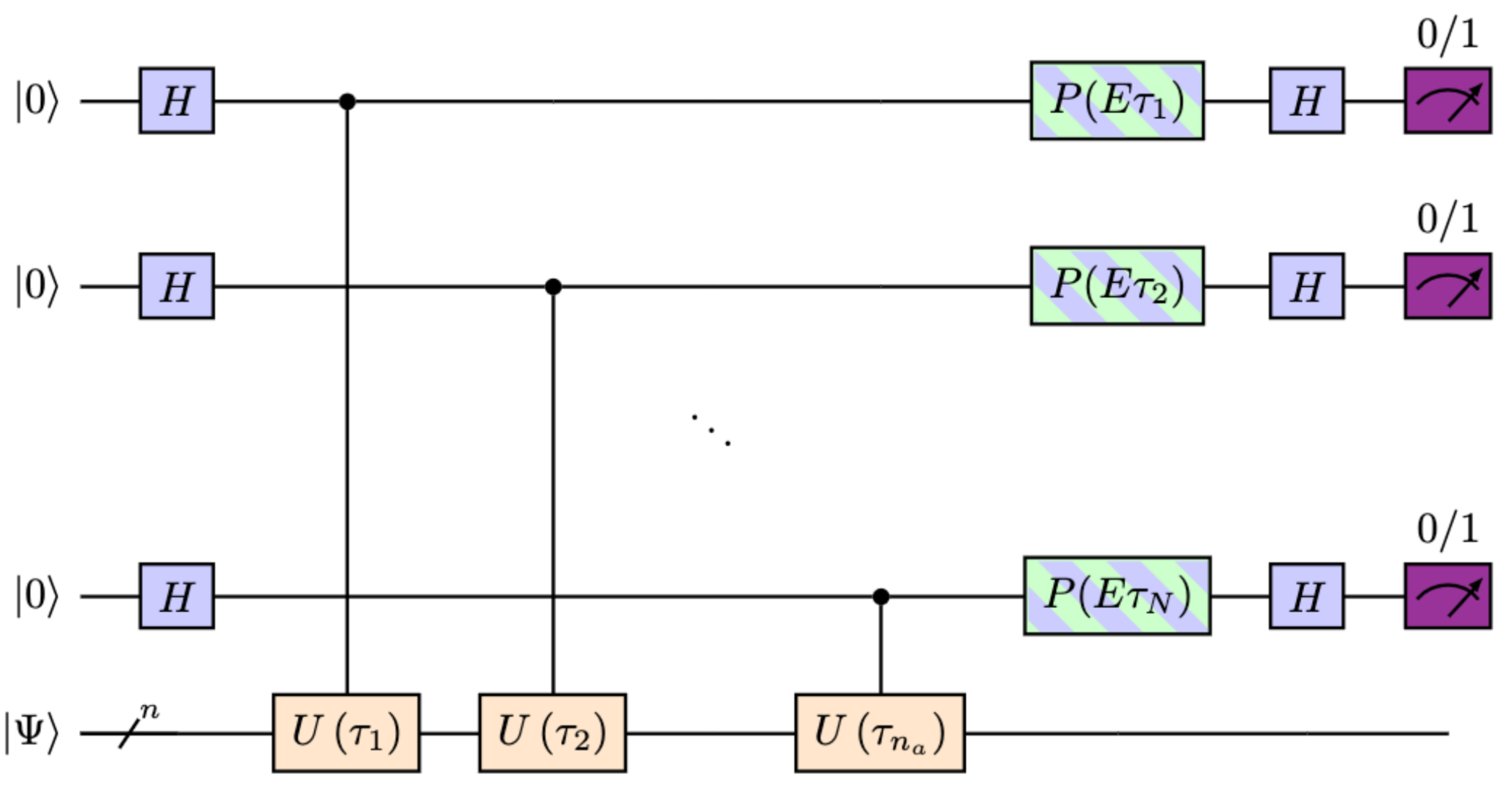}
    \caption{Circuit used for the Rodeo algorithm \cite{Choi2021}. Given that there are no entangling gates between the ancillary qubits, it is possible to use a procedure where only one is used and measured multiple times.}
    \label{fig:rodeo_qc}
\end{figure} 

We consider again an initial state $| \Psi \rangle$ that decomposes onto the Hamiltonian eigenstates $\{|\alpha \rangle\}$ (associated to a set of eigenvalues $\{ E_{\alpha} \}$) as $|\Psi\rangle = \sum_{\alpha} c_{\alpha} |\alpha \rangle$.
Provided that the initial state is prepared on a quantum register, the circuit depicted in Fig. \ref{fig:rodeo_qc} is applied, and the ensemble of ancilla qubits is measured. The parameter $E$ appearing as a scaling factor in the phase rotations shown in Fig. \ref{fig:rodeo_qc} can be freely varied. Given that there are no entangling gates between the ancilla wires in Fig. \ref{fig:rodeo_qc}, it is possible to see this circuit as a consecutive series of measurements on a single qubit, similar to the iterative QPE procedure \cite{Dobsicek2007}.
To be more specific, assuming $n_a$ indirect measurements, a set of times $(\tau_1, \dots , \tau_{n_a})$ are considered and the controlled operation of the $j^{\mathrm{th}}$ measurement is made using $U(\tau_j)=e^{-i{H}\tau_{j}}$ with ${H}$ the Hamiltonian. It can then be shown \cite{Choi2021,Qian2021} that the probability to obtain only the $|0\rangle$ state in all of the consecutive $n_a$ measurements of the ancilla qubits is:

\begin{eqnarray}
    p_{0^{n_a}} \left(E,\{\tau_{i}\}\right)&=&\sum_{\alpha}|c_{\alpha}|^{2}\prod_{i=1}^{n_a}
    \cos^{2}\left(\left(E_{\alpha}-E\right)\frac{\tau_{i}}{2}\right).
    \label{eq:prob_0_N_kitaev}
\end{eqnarray}

\begin{figure}[htbp]  
\centering
\includegraphics[width=1.0\columnwidth]{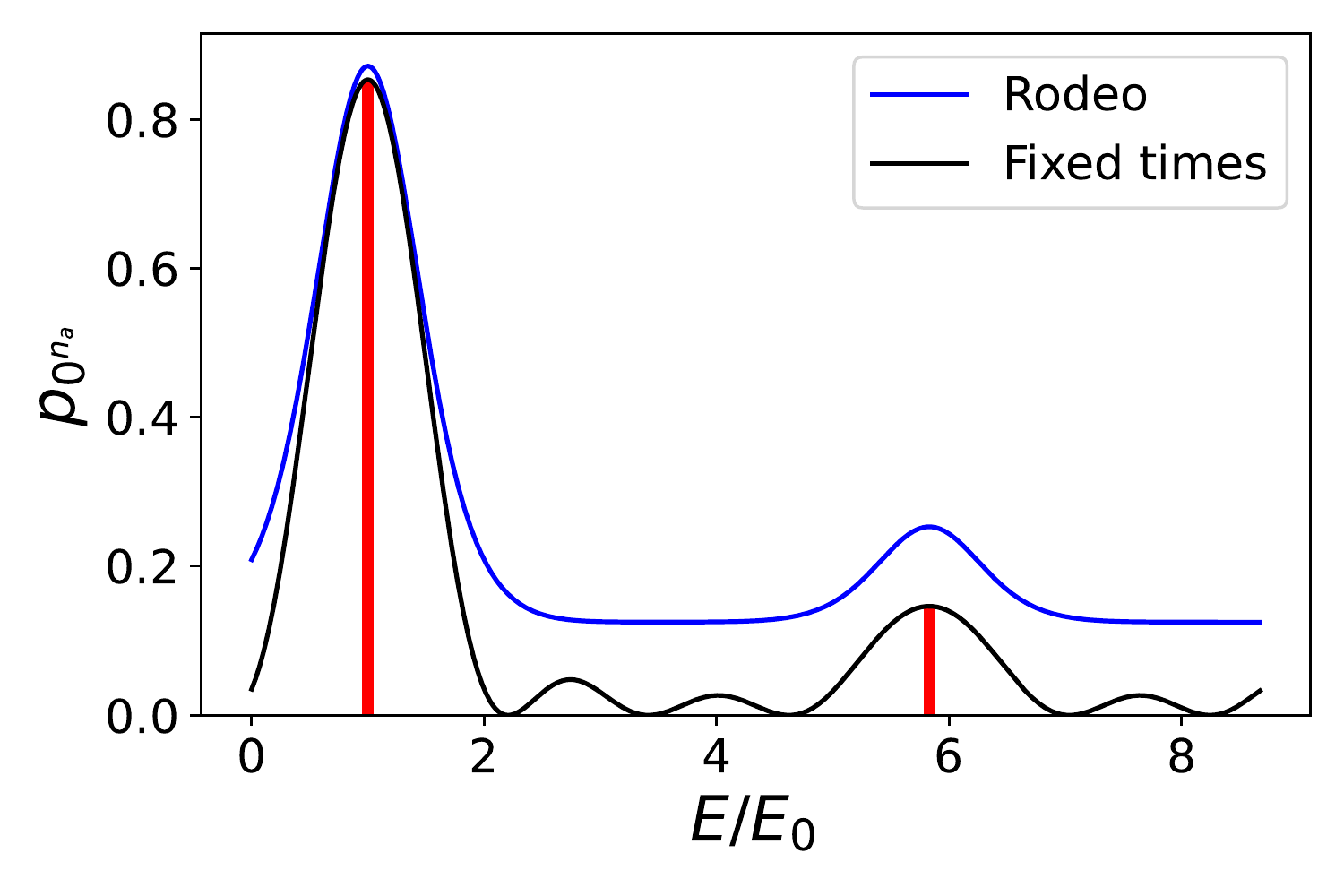}
    \caption{Illustration of the iterative methods using the fixed times (black line) and Rodeo (blue line) prescriptions discussed in the text. The red bars indicate the (exact) decomposition of the initial wave function in the eigenbasis of the Hamiltonian $\{|\alpha\rangle \}$, i.e., the points $\left( |c_{\alpha}|^2, E_{\alpha}\right)$ from $|\psi \rangle=\sum_{\alpha} c_{\alpha} |\alpha \rangle$. Here the system is assumed to have two eigenvalues, and $n_a = 3$ is used. Note that in the Rodeo case, each eigenstate $k$ contributes to a flat background proportional to $|c_{\alpha}|^2/2^{n_a}$ (see Eq. (\ref{eq:p0_N_rodeo})). Contrary to the fixed times case, this background being flat cannot be misinterpreted as an eigenstate contribution.}
    \label{fig:rodeo_results}
\end{figure}

As the number of repetitions $n_a$ increases, the above function of $E$ peaks around the $E_{\alpha}$ values. The flexibility in choosing the $\{ \tau_i\}$ values can be further used to improve the convergence. Below, we discuss two main options:

(i) {\it Fixed times prescription:} We can assume
that we have $\tau_i = \tau/2^{i-1}$ and $\tau = \frac{\pi 2^{n_a-2}}{|E_{up} - E_{low}|}$ where $E_{up}$ (resp. $E_{low}$) is an upper bound (resp. lower bound) on the spectrum of the Hamiltonian. An example of the resulting probability $p_{0^{n_a}}$ given by Eq. (\ref{eq:prob_0_N_kitaev}) at various $E$ is shown in Fig. \ref{fig:rodeo_results}. From the positions of the peaks in the distribution, as well as their amplitude, we 
can extract approximate eigenenergies and the weights of the associated eigenstates in the decomposition of $\ket{\Psi}$, $|c_{\alpha}|^2$.   
(ii) {\it Rodeo prescription:} The key idea behind the Rodeo method is to assume a Gaussian statistical ensemble of times $\{ \tau_i\}$ with an adjustable Gaussian width $\sigma$. Averaging over the statistical ensemble gives the probability:
\begin{equation}
\centering
p_{0^{n_a}}\left(E\right)=\sum_{\alpha}|c_{\alpha}|^{2}\left[\frac{1+e^{-\left(E_{\alpha}-E\right)^{2}\sigma^{2}/2}}{2}\right]^{n_a},
\label{eq:p0_N_rodeo}
\end{equation}
that is also strongly peaked around the eigenenergies. An illustration of the Rodeo prescription is also given in Fig.~\ref{fig:rodeo_results}.
{
The Rodeo method has two advantages compared to the fixed times' (original Kitaev method) approach. First, the probabilities are flattened away from the energies, which helps to identify peaks. Second, the extra parameter $\sigma$ can be used as a resolution to rapidly scan a given energy range (see \cite{Qian2021}).}

\paragraph{Accelerated-VQE}
Finally, let us mention a proposal to take the "best of both worlds" of QPE and VQE by interpolating between these two regimes \cite{wang_2019}. The idea is to tune an interpolation parameter $\alpha \in [0, 1[$ to achieve an optimal trade-off between measurement variance and circuit depth.

\subsubsection{Response and Green's function methods}

Starting from the decomposition \eqref{eq:initdec}, a direct 
method on a classical computer to get both the amplitude $|c_{\alpha}|^2$ and the eigenvalues $E_{\alpha}$ would be to compute the function:
 \begin{equation}
     F(t) = \langle \Psi | e^{-it{H}} | \Psi \rangle, \label{eq:generating}
 \end{equation}
and perform its classical Fourier transform, leading to
 \begin{equation}
    \widetilde F(E) \propto \sum_k |c_{\alpha}|^2 \delta(E - E_{\alpha}).  \label{eq:response}
 \end{equation}
The function $F(t)$ is called a \textit{generating function} for reasons that will become apparent hereafter. In contrast, $\widetilde F(E)$ is named \textit{response function} in analogy to the response of a system to an external field. Two essential conditions are necessary to extract the energies from this technique accurately. The first one is the possibility of computing the propagator entering in Eq. (\ref{eq:generating}). For complex systems like many-body systems, quantum computers seem appropriate platforms. For these reasons, a hybrid method where (\ref{eq:generating}) is estimated on a quantum device while the Fourier is performed classically has been advertised in Ref. \cite{ruiz-guzman2021} (see also discussion in \cite{Somma2019}). One clear advantage is that the real and imaginary parts of $F$ at a given time $t$ can be obtained using standard techniques with a single Hadamard-like test, as pictured in Table \ref{fig:meas_circ}. An illustration of such a function was given in Ref. \cite{ruiz-guzman2021} for superfluid systems and the Hubbard model. 

A second constraint is that the energy resolution achieved in Eq. (\ref{eq:response}) will strongly depend on the maximal time $\tau_{\rm max}$ over which $F$ is known due to the Heisenberg uncertainty relation between time and energy. Such a long-time evolution requirement prevents using the response function technique in the NISQ period.  

Along the same line, with performant quantum computers, one can also imagine a priori to access Green's function in many-body systems without approximation. 
For instance, the one-body Green's function matrix elements can be defined as \cite{Fetter2012}:
\begin{eqnarray}
G_{ij}(t,t') &=& \langle \Psi(0) |{\rm T} \left[a^\dagger_j(t) a_i (t') \right] | \Psi(0) \rangle.     
\end{eqnarray}
Here, $| \Psi(0) \rangle$ is the initial state we suppose normalized. ${\rm T}$  is the time-ordering operator, and we use the Heisenberg interaction representation, i.e., $a^{\dagger}_i(t) = U^\dagger(t) a^\dagger_i U(t)$, with $U(t)= e^{-iHt}$.
Provided that the propagator can be efficiently implemented on a digital quantum platform, the Green's function matrix elements  
can be obtained using, for instance, a circuit similar to the one shown in the bottom part of Table \ref{fig:meas_circ}.
The possibility of computing Green's functions on quantum computers is now being explored \cite{Endo2020b,Baker2021,Tong2021,Stenger2022,Rizzo2022}.

\section{Entanglement and quantum entropy}
\label{sec:entanglement}

One promise of quantum computing is the possibility to construct quantum states that include complex internal correlations between particles. Hence, a question of utmost importance underpinning the design of quantum circuits is their ability to generate entanglement beyond classical correlations.
Here, we describe some figures of merit to measure the level of entanglement exhibited by a state and how to connect this degree of entanglement to requirements on the depth of a state-preparation circuit or the complexity and expressive power of a quantum ansatz.  

\subsection{Basic aspects of entanglement and some measures of it}

Entanglement is one branch of quantum information theory that is a vast subject of research \cite{Nielsen2010,Wilde2017}.
This subsection briefly introduces how to measure entanglement between two systems, starting from the von Neumann entropy concept. 

\subsubsection{Measures of entanglement} 
The entanglement degree is relative to a partition of a quantum system into subsystems, denoted by $A$ and $B$. 
It measures how far the state of the entire system $\lbrace A+B \rbrace$ is from being factorized into a product of the states of its subparts $A$ and $B$.

\paragraph{Von Neumann entanglement entropy}

Let us assume that the total system is described by a density matrix $ \rho_{AB}$ (see section \ref{sec:noise} for more details about density matrices); the densities of the two subsystems can be obtained by performing partial traces: 
\begin{eqnarray}
 \rho_A &=& {\rm Tr}_B ( \rho_{AB}), ~~ \rho_B = {\rm Tr}_A ( \rho_{AB}),
\end{eqnarray}
where $ \rho_A$ (resp. $ \rho_B$) is the density of the system $A$ (resp. $B$). If the two subsystems are not entangled, then we have the simple property:
\begin{eqnarray}
 \rho_{AB} &=&  \rho_A \otimes  \rho_B. \label{eq:rhotensor}
\end{eqnarray}
One key aspect of quantum computing is the possibility to control the degree of entanglement between two subsets of the complete qubit register. Entanglement is a specific feature of quantum mechanics that does not exist in classical mechanics. Quantum algorithms, as opposed to classical algorithms, generally use entanglement as a tool. This is, for instance, the case of most algorithms discussed in section \ref{sec:idealalgorithm}.
In the many-body context, when one qubit represents one orbital, the possibility of a given ansatz to produce entanglement between qubits should also be linked to the onset of correlation between particles. Therefore, the possibility of generating and controlling entanglement is essential.  
A possible measure of the entanglement between two subsystems is based on the von Neumann entropy, defined for a given density $ \rho_x$ as:
\begin{eqnarray}
S_x &=& - {\rm Tr}( \rho_x \log_2 \rho_x). \label{eq:entropy_x}
\end{eqnarray}
Using $x= AB$, $A$, or $B$, we obtain three entropies $S_{AB}$, $S_A$, and $S_B$ associated with the total system or with either subsystem ($S_A$ and $S_B$ are called bipartition entropies). These entropies are real positive numbers. They can quantify the complexity, disorder or entanglement in a system.
An interesting property of the entanglement entropy is the so-called subadditivity condition \cite{Breuer2002}:
\begin{eqnarray}
S_{AB} \le S_A + S_B ,
\end{eqnarray}
where the equality holds strictly when Eq. (\ref{eq:rhotensor}) is verified, i.e., when the two systems are not entangled.

One can also define the \textit{entanglement entropy} $S_{\rm max}$ of a system as the maximum bipartition entropy over all the possible bipartitions of the system.

\paragraph{Mutual information}
Another measure of entanglement is given by the so-called \textit{mutual information} $M_{AB}$, defined as:
\begin{eqnarray}
    M_{AB} = S_A + S_B - S_{AB}.   
\end{eqnarray}
This quantity is the crux of the algorithm developed in Ref.~\cite{tkachenko_2021} to limit circuit depth by adapting the qubit order to the chip's topology according to the leading correlations among qubit pairs (there, $A$ is chosen to be the Hilbert space of individual qubits $A = \lbrace i \rbrace$ or qubit pairs $A = \lbrace i, j \rbrace$).
It is also used in different fields of physics and chemistry to characterize the entanglement between particles (see, for instance, \cite{Rissler2006,Boguslawski2015,Robin2021,Lacroix2022b}). {We also mention that the concepts of entanglement and mutual information can be used to reduce the complexity of solving a many-body problem, like in the case  of active space selection \cite{Stein2016,Stein2019}}.

\subsubsection{Schmidt decomposition}
Here, we restrict ourselves to the specific case where the total system is a pure state $|\Psi \rangle$, which is the case for all ans\"atze discussed in section \ref{sec:varansatz}.
In this case, we have $\rho_{AB} = | \Psi \rangle \langle \Psi |$ and 
$S_{AB} = 0$.
Splitting the system into two subsystems and introducing the two bases $\{ |\alpha\rangle \}_{\alpha=1,{\cal N}_A}$ and $\{ |\beta \rangle \}_{\beta=1,{\cal N}_B}$ of subsystem $A$ and $B$ respectively, one can decompose the total state as:
\begin{eqnarray}
| \Psi \rangle &=& \sum \limits_{\alpha, \beta} c_{\alpha \beta} \ket{\alpha} \otimes \ket{\beta}. 
\end{eqnarray}
One can then interpret $c_{\alpha\beta}$ the matrix element of a ${\cal N}_A \times {\cal N}_B$ matrix ${\cal C}$ and use the Singular Value Decomposition (SVD) to rewrite is as
\begin{equation}
   c_{\alpha \beta} = \sum \limits_{k=1}^{\chi} U_{\alpha k} s_k V^{\dagger}_{k\beta} 
\end{equation}
with $s_k > 0$ and $\chi \leq \mathrm{min}(\mathcal{N}_A, \mathcal{N}_B$). The normalization of state $\ket{\Psi}$ ensures that $\sum \limits_{k=1}^{\chi} s_k^2 = 1$.
$\chi$, the number of nonzero singular values, and is called, in this context, the \textit{Schmidt rank}. The $s_k$ are the Schmidt coefficients. They define the \textit{entanglement spectrum} of the state, from which the entropy of each subsystem can be obtained. For a total pure state, we have:
\begin{eqnarray}
    S_{A} &=& S_B = -\sum \limits_{k=1}^{\chi} s_k^2 \log_2(s_k^2).
\end{eqnarray}
Such a decomposition is useful to provide upper limits to the subsystems' entropies.
For instance, in the case of a factorized state (Eq. (\ref{eq:rhotensor})), there is only one coefficient $s_k$ in the Schmidt decomposition, therefore $\chi=1$ and $S_A=S_B=0$.

Perhaps more importantly, the upper bound on the entanglement entropy at a fixed number of Schmidt coefficients $\chi$ corresponds to a flat singular value spectrum ($s_k =1/\sqrt{\chi}$ for all $k$). In this case, $S_{A/B} = \log_2(\chi)$.
Thus, in general,
\begin{equation}
    S_{A/B}  \leq \log_2(\chi) \leq \log_2\left[ \mathrm{min}(\mathcal{N}_A, \mathcal{N}_B) \right].\label{eq:sboundary}
\end{equation}

\subsection{Gaussian qubit states}
\label{sec:Gaussqubit}

In section \ref{sec:varansatz}, we discussed the case of uncorrelated ans\"atze like HF states or, more generally, Gaussian states. Disregarding the extra complexity induced by the Pauli principle, we consider here such a state. More precisely, taking inspiration from a many-body density obtained usually for a set of non-interacting particles at thermal equilibrium, we consider a qubit register whose density matrix is given by:
\begin{eqnarray}
 \rho &=& \frac{1}{Z} \exp\left(-\sum_{i=0}^{n-1} \alpha_i Q^+_i Q^-_i  \right),  \label{eq:densthermal}
\end{eqnarray}
where $Z$ is a normalization factor ensuring that ${\rm Tr}(\rho) = 1$.
Here $Q^+_i$ (resp. $Q^-_i$) are the operators acting on qubit $i$.
Using a technique similar to the one used in the Fock space to treat non-interacting fermions in the grand canonical ensemble, we deduce that the density can be rewritten as: 
\begin{eqnarray}
 \rho &=& \bigotimes_{i=0}^{n-1} \left[ (1-p_i) |0_i \rangle \langle 0_i | +  p_i |1_i \rangle \langle 1_i |  \right],  \label{eq:Sindtensor}
\end{eqnarray}  
with $p_i = (1+e^{\alpha_i})^{-1}$.
The mixed-state equivalent of the HF pure state case is obtained when all the $p_i$ are either equal to $0$ or $1$.  

\begin{figure}
    \centering
    \includegraphics[width=\columnwidth]{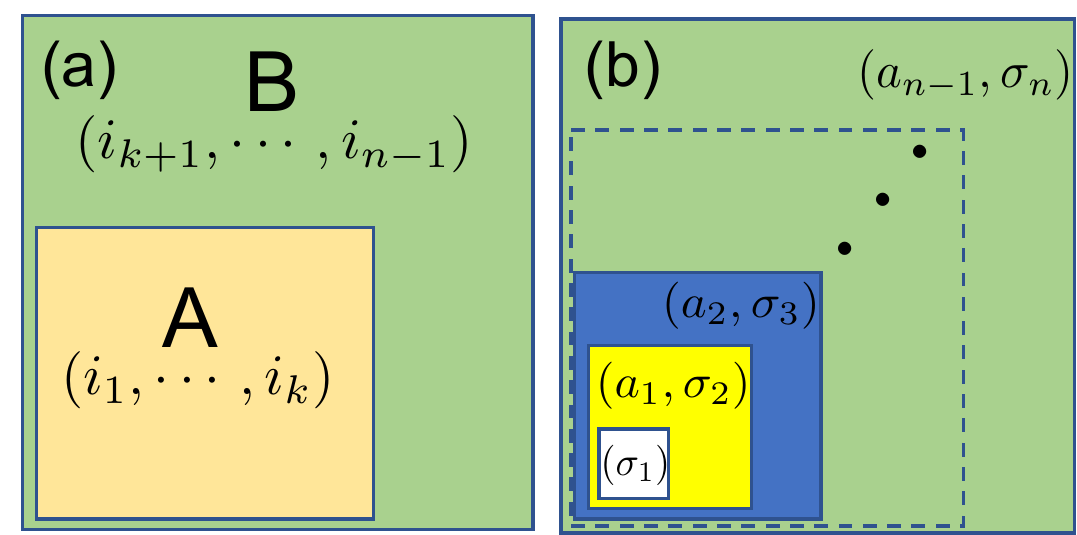}
    \caption{(a) Schematic illustration of a separation of a qubit register into two subsystems where a set of $k$ qubits $(i_1, \cdots i_k)$ from a subsystem $A$, while $B$ contains all other qubits of the total register (A+B). (b) schematic view of how a general tensor is decomposed to give an MPS form.}
    \label{fig:entregister}
\end{figure}

We then consider that the total register is separated into two sets of qubits as depicted schematically in Fig.~\ref{fig:entregister}-a forming the subsystems $A$ and $B$ discussed previously.
We then denote by $S_{(i_1, \dots, i_k)}$ the entropy associated to the subsystem containing the qubits $(i_1, \dots , i_k)$.
Because of the tensor structure of the total density, Eq. (\ref{eq:Sindtensor}), this entropy verifies:
\begin{eqnarray}
S_{(i_1, \dots, i_k)} &=& \sum_{m=1,k} S_{(i_m)}, \label{eq:SNto1}
\end{eqnarray}
where $S_{(i)}$ denotes the entropy of a subsystem formed by the single qubit $i$. This entropy is given by:
\begin{eqnarray}
S_{(i)} &=& - \left[ p_i \ln p_i  + (1-p_i) \ln (1-p_i) \right]. \label{eq:S1}
\end{eqnarray}
Eq. (\ref{eq:SNto1}) implies that the mutual information $M_{AB}$ of any partition of the total register is zero independently of the number of qubits or which qubits are included in each subsystem.
Said differently, there is no entanglement when a density like (\ref{eq:densthermal}) is considered.

\subsection{Understanding entanglement generation with the Matrix Product State representation}
\label{sec:entMPS}

Information flow along a circuit can be described as causal, `light' cones relating a local action on a subset of qubits to its effects on other qubits at a later stage. Lieb-Robinson bounds limit the speed at which quantum correlations, aka entanglement, can be generated \cite{nachtergaele_2010}. When translated into the language of digital quantum circuits, these bounds prescribe that a certain depth is required to reach a certain amount of entanglement. A natural framework to better understand this is the Matrix Product State (MPS) representation \cite{schollwock_2011, orus_2014} and the associated quantum circuits \cite{ran_2020} (see section 
\ref{subsec:TN_ansatz}).

\subsubsection{Constructing the MPS representation of any state $\ket{\Psi}$}

In this section, we briefly review a standard derivation (see also \cite{schollwock_2011}), that of the MPS representation starting from any pure state, to shed light on the link between the MPS representation and entanglement entropy.

Let us consider a general state of $n$ qubits $| \Psi \rangle = \sum_{\sigma_i = 0,1} c_{\sigma_1 \dots \sigma_n} | \sigma_1 ,\dots ,\sigma_n \rangle$.
The complex amplitudes $c_{\sigma_1 \dots \sigma_n}$, understood as elements of a tensor of rank $2^n$, can be written as a MPS as given by Eq. (\ref{eq:MPScoef}). The proof briefly recalled below uses a strategy schematically represented in Fig. \ref{fig:entregister}-b. It corresponds to an iterative set of separations of the full register in two subsystems together with applications of singular value decompositions (SVDs). This proof can be summarized as follows: 

(i)  Consider $c_{\sigma_1 \cdots \sigma_n}$ as a $2 \times 2^{n-1}$
 matrix $c_{\sigma_1, (\sigma_2 \cdots \sigma_n)}$ and perform a SVD of it:
\begin{eqnarray}
c_{\sigma_1, (\sigma_2 \dots \sigma_n)} &=& \sum_{a_1 = 0}^{r_1 - 1} U_{\sigma_1 a_1} s_{a_1} V^\dagger_{a_1, (\sigma_2 \dots \sigma_n)},
\end{eqnarray}
where $s_{a_1}$ are the nonzero eigenvalues whose number, i.e., the Schmidt rank, is denoted by $r_1$. It verifies $r_1 \le 2$. One can then introduce the notation $[A^{(1)}]^{\sigma_1}_{1,a_1} = U_{\sigma_1, a_1}$ and absorb the $s_{a_1}$ in $V$ to give:
\begin{eqnarray}
c_{\sigma_1 \cdots \sigma_n} &\equiv& \sum_{a_1} [A^{(1)}]^{\sigma_1}_{1,a_1} G_{(a_1, \sigma_2) , (\sigma_3, \dots , \sigma_n)}.
\end{eqnarray}

(ii) The matrix $G$ has the dimension $(2 r_1) \times 2^{n-2}$. We can then redo an SVD on the matrix $G$ giving a number $r_2$ of nonzero eigenvalues with $r_2 \le {\rm min}(2r_1, 2^{n-2})$. The process is then iterated until the amplitudes get rewritten as contractions over a \textit{tensor train} comprising $n$ tensors with rank one (vectors) or two (matrices):
\begin{eqnarray}
\label{eq:mps_without_truncation}
c_{\sigma_1 \cdots \sigma_n} &\equiv& \left( [A^{(1)}]^{\sigma_1}\right)\left( [A^{(2)}]^{\sigma_2}\right)\cdots \left( [A^{(n)}]^{\sigma_n}\right) \nonumber \\
&=& \sum_{\substack{\{a_i=0,\\\dots,r_{i}-1\}}} [A^{(1)}]^{\sigma_1}_{1,a_1} [A^{(2)}]^{\sigma_2}_{a_1,a_2}  \cdots  [A^{(n)}]^{\sigma_n}_{a_{n-1},1}. 
\end{eqnarray} 
The indices not summed over (the $\sigma_j$) are referred to as the \textit{physical indices}. Here, each can take two different values.
On the other hand, internal indices that are summed over (the $a_i$) correspond to so-called \textit{virtual indices}. The different ranks $\{ r_i \}$ verify $r_{i+1} \le {\rm min} (2 r_i, 2^{n-i-1}) \le 2^{n/2}$ and $\chi = {\rm max}_i(r_i)$ is nothing but the bond dimension (BD) discussed in section \ref{subsec:TN_ansatz}.

A corollary of the above demonstration is the following inequality:
\begin{eqnarray}
2^{S_{\rm max}} \le \chi \le 2^{n/2},
\end{eqnarray}
that is a consequence of Eq. (\ref{eq:sboundary}).

\begin{figure}
    \centering
    \includegraphics[width=\columnwidth]{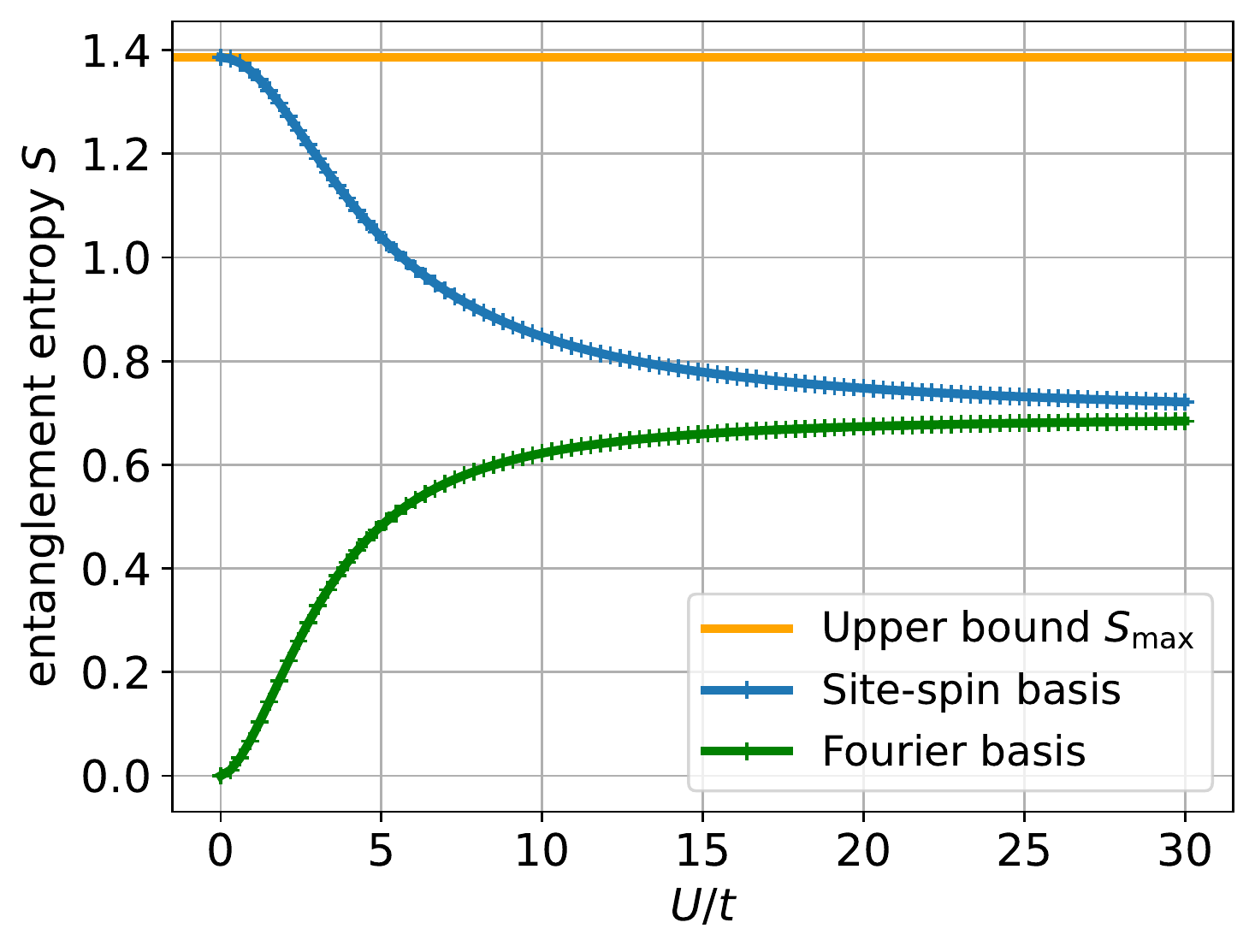}
    \caption{Entanglement entropy displayed by the ground state of the half-filled ($\mu=U/2$) Hubbard dimer, i.e., with two doubly degenerated sites, as a function of the ratio $U/t$ of the Hamiltonian defined in Eq. \eqref{eq:Hubbard_solids}. In the site-spin basis, the entanglement entropy saturates the upper bound at $U/t=0$. At high $U/t$, local charge fluctuations are suppressed, pinning the entanglement entropy to a nonzero asymptotic value. Turning to the reciprocal Fourier basis -- the diagonalization basis of the quadratic/single-particle part of the Hamiltonian --, one sees that the ground state exhibits no entanglement at $U/t=0$ and monotonically increases to the asymptotic value. This fact illustrates the strong dependence of the entanglement entropy on the single-particle basis used. Note that here, we have calculated $S$ using the natural logarithm $\ln$ rather than $\log_2$.}
    \label{fig:Sent_Hubbard}
\end{figure}

\subsubsection{Entanglement in various systems}
In some systems, such as the ground state of gapped, local Hamiltonians, the bond dimension scales favorably with system size $n$ in virtue of the \textit{area law} \cite{Srednicki_1993}: the bipartition entropy increases as the area $\propto n^{d-1}$ of the bipartition ($d$ refers to the dimension) rather than the volumes $\propto n^{d}$ of the subsystems. 
For such systems, the MPS representation thus offers a tractable way to store the wave function. Conversely, other systems require exponentially-big BDs, and it may be advantageous to turn to a quantum computer to represent them, for instance, for ground states of 2D local Hamiltonians.
A counterexample is the time-evolving state of a quenched many-body problem, which typically displays a ballistic growth of entanglement with time $t$, $S \propto t$: then, $\chi$ needs to scale exponentially with $t$.

Due to their properties, MPS can be used to simulate quantum computers that generate weakly entangled states \cite{Vidal2003} or that are plagued with a finite fidelity \cite{Zhou2020,Ayral2022}. 
MPS are also more and more used to study quantum chemical system despite the nonlocal character of the Coulomb interaction tensor \cite{Baiardi2020}.
 
Let us also mention that the entanglement entropy, and thus the size of the MPS representation, is heavily basis-dependent. This dependence is illustrated in Figure \ref{fig:Sent_Hubbard} where the ground state entanglement entropy of the half-filled Hubbard dimer is plotted as a function of the ratio $U/t$, which is a measure of correlations in the system. In the original basis, the Hubbard Hamiltonian is written (denoted here as the site-spin basis), the entanglement entropy is maximal (and saturates the bound) at $U/t=0$ and decreases to a non-vanishing asymptotic value as $U/t \rightarrow \infty$.
Conversely, in the Fourier-transformed basis, the entanglement entropy vanishes at $U/t=0$ and increases monotonically to the asymptotic value as $U/t \rightarrow \infty$. This asymptotic value can be understood from the form of the ground state, which tends to the superposition $\frac{1}{\sqrt{2}}\left( \ket{\uparrow \downarrow} + \ket{\downarrow \uparrow} \right)$ as $U/t$ increases.

\subsubsection{Generating an MPS with a quantum computer}
\label{sec:mps_quantum}
Above a specific entanglement entropy and correspondingly a certain MPS bond dimension, Matrix Product States become impractical to store on a classical computer. This subsection explains how MPS can be generated using a quantum computer.

\begin{figure}[htbp]  
    \centering
    \includegraphics[width=\linewidth]{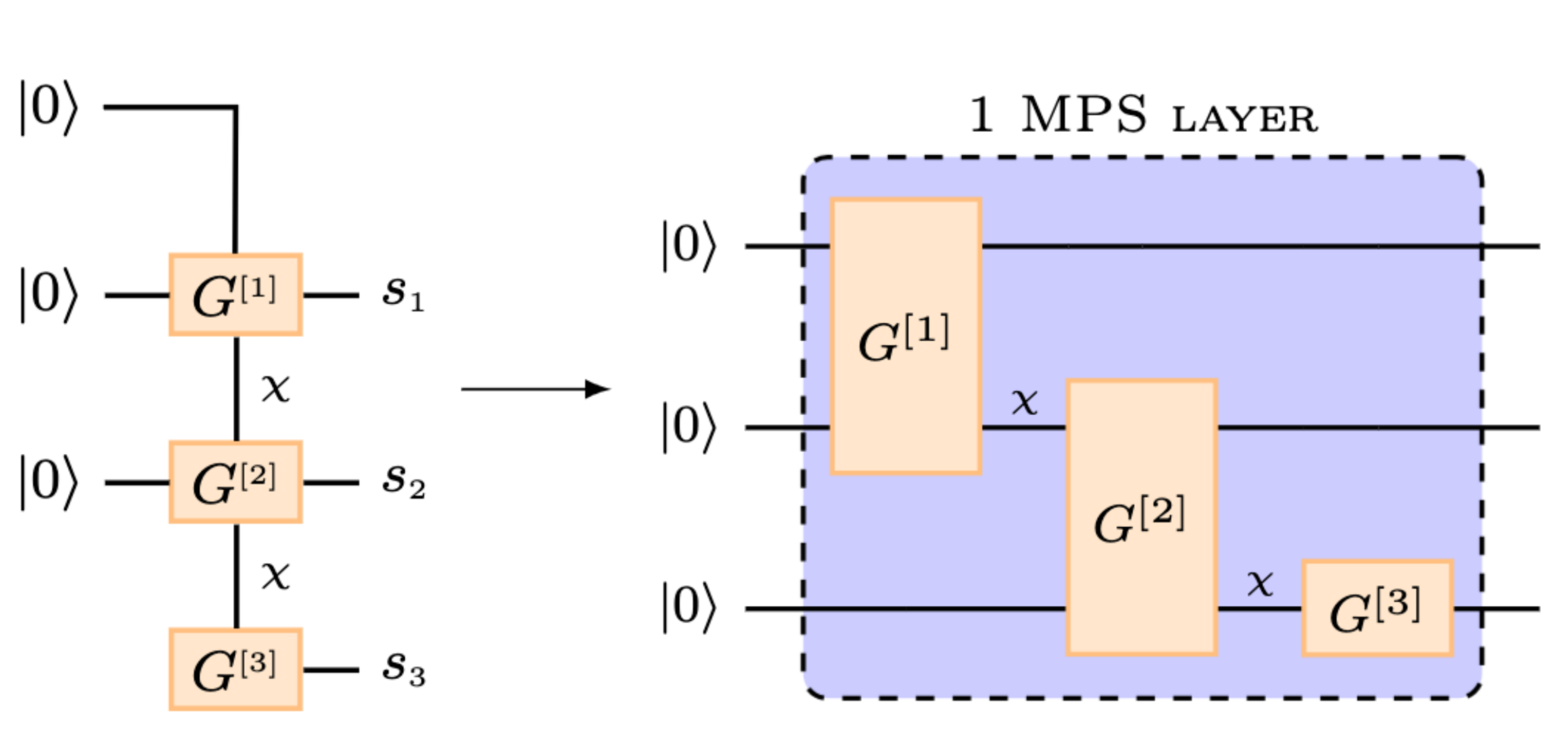}
    \caption{Illustration of a one-layer set of quantum gates used to create an MPS circuit (adapted from \cite{ran_2020}). The ensemble of gates $G^{\left[i\right]}$ has been constructed by truncating the SVD decomposition of the tensors in Eq. (\ref{eq:mps_without_truncation}) following the method described in \cite{Paeckel2019}. This MPS has $\chi=2$. Higher $\chi$ can be constructed by repeating the sequence as a set of layers.}
    \label{fig:mps_circuit}
\end{figure}

The MPS is entirely characterized by the set of tensors $[A^{(i)}]^{\sigma_i}_{a_i, a_i+1}$.
In Fig. \ref{fig:mps_circuit}, we show a simple circuit to create a MPS with a uniform $\chi=2$. MPS with such a bond dimension are, for instance, Greenberger–Horne–Zeilinger (GHZ) states and so-called $| W\rangle$ states \cite{Cabello2002,Greenberger2007}.   

\begin{figure} \centering
    \includegraphics[width=\linewidth]{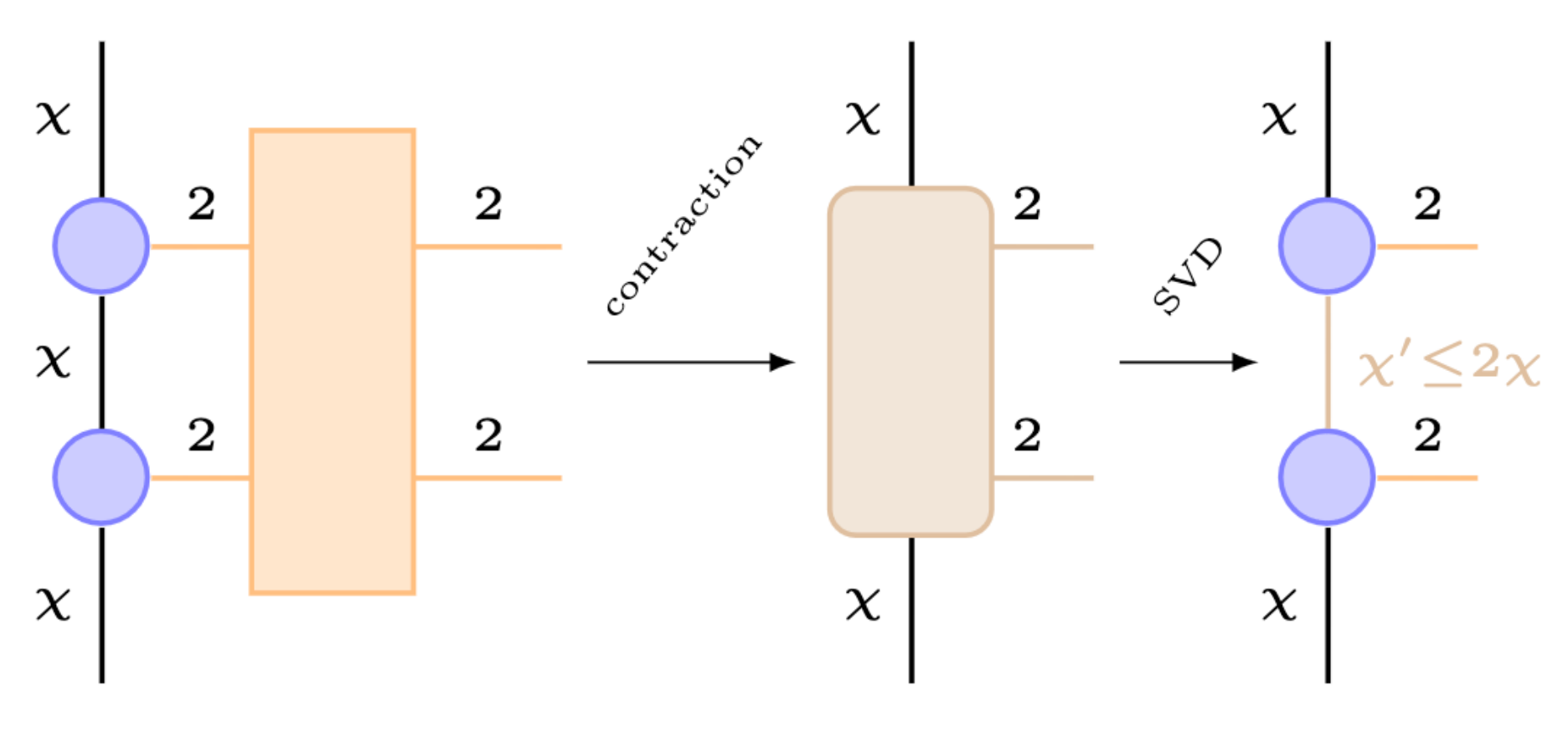}
    \caption{Schematic representation of the effect of a two-qubit gate on the associated local bond dimension of an MPS. The tensor network's bonds (horizontal edges, standing for the summation over virtual indices) and legs (vertical edges, representing physical indices) are represented by their dimensionality. After contraction over the internal indices, an MPS form is retrieved with an SVD.}
    \label{fig:2qb_gate_on_MPS}
\end{figure}

The MPS example shows that the bond dimension $\chi$ controls the entanglement entropy 
and prescribes a certain depth for the quantum circuit preparing the MPS on a linearly-connected chip.
Indeed, applying a two-qubit gate on qubits with local BD $\chi_k$ yields an MPS with local BD $\chi'_k \leq 2\chi_k$. This result is illustrated in Figure \ref{fig:2qb_gate_on_MPS} with tensor network formalism. As a consequence, to prepare a MPS with BD $2^n$, one has to resort to a circuit of depth $n$.

The possibility of designing complex trial states and controlling the degree of entanglement is an active field of research today.
The tensor network discussed here is beneficial to understand the link between the gate structure used in a circuit and the achieved complexity in entangling particles in many-body systems.
The capability of layered ansätze to encompass some physical Hamiltonian ground states, as well as the BD of the converged wavefunction they yield, was studied in, e.g., Ref \cite{bravo_2020}.

\section{Noise in quantum processors}
\label{sec:noise}

Quantum computers are imperfect and noisy. Performing calculations with quantum devices today means being able to accommodate these imperfections. Enormous efforts are being made today to understand/correct the different sources of noise. In the meantime, methods are developed to obtain acceptable results despite the various noise sources.
Here, we present a brief discussion on (i) how imperfect quantum computing can be understood and might affect the evolution of a quantum system and on (ii) some methods that are used today to, at least partially, get rid of the effects of noise.
The discussion below is not explicitly dedicated to application in the many-body sector but applies to any quantum computing problem.

We refer the reader to e.g \cite{Schlosshauer2019} for an more in-depth review of decoherence, and to \cite{Watrous2018} for mathematical aspects of noisy quantum computations.

\subsection{Decoherence in NISQ processors\label{subsec:NISQ-Noisy-quantum}}

Imperfections on current quantum processors can be broken down into two categories: coherent and incoherent errors.
Coherent errors are systematic errors like calibration errors. For instance, if the qubit's frequency is not known precisely (say it is $\omega_0 + \epsilon$ instead of $\omega_0$), executing a $z$-rotation gate as described in section~\ref{subsubsec:digital_qc} with a drive frequency $\omega_c = \omega_0$ will result in an over $z$-rotation of angle $\epsilon t$. Coherent errors can thus be described as additional unwanted unitary operations. In theory, they are reversible since a unitary operation $U$ can be undone by applying the hermitian conjugate operator $U^\dagger$.

Incoherent errors, on the other hand, are stochastic. They come from the uncertainty on the quantum processor's state brought by its interaction with the outside world, often called the environment.
In principle, they cannot be undone and are thus irreversible. The only way to avoid decoherence induced by the environment is to isolate as much as possible the quantum computer from the rest of the world.

In this section, we focus on describing incoherent errors and their modeling in analog and digital quantum processors.

\subsubsection{Describing the state of a noisy quantum computer: the density matrix}\label{subsubsec:rho}

Thus far, we have described the state of a quantum processor, whether analog or digital, by its wavefunction $\ket{\Psi}$. Gates and measurements have been introduced as acting on this object.

In noisy computers, unwanted interactions with the environment lead to a loss of information on the system's state. To capture this uncertainty, the state of the quantum system can no longer be described as a single wavefunction $\ket{\Psi}$, but as a statistical mixture of wavefunctions: the system is said to be in states $\lbrace \ket{\Psi_i} \rbrace_i$ with probabilities $\lbrace p_i \rbrace_i$. Thus, the average of an observable is no longer $\langle O \rangle = \bra{\Psi} O \ket{\Psi}$ but $\langle O \rangle = \sum_i p_i \bra{\Psi_i} O \ket{\Psi_i} $.
A convenient object to manipulate this uncertain (or \textit{mixed}) state is the so-called density matrix $\rho$:
\begin{equation}
    \rho = \sum_i p_i \ket{\Psi_i} \bra{\Psi_i}.
    \label{eq:dens_mat}
\end{equation}
This object completely describes the state of a noisy quantum computer.
For instance, one can check that the expectation value $\langle O \rangle$ given above can be recovered as $\mathrm{Tr} [\rho O]$.

The density matrix has important properties: hermitian, positive semidefinite, and unit trace \cite{Breuer2002}. These properties ensure it can describe a statistical mixture. 
In the absence of noise, the state of the quantum processor becomes deterministic: $\rho$ is given by $\rho = \ket{\Psi}\bra{\Psi}$. The state is called "pure", and the Schrödinger equation describes its evolution. For a given $\rho$, one can tell whether it corresponds to a pure state or a mixed state by looking at the rank of the operator (rank one is a pure state) or at a quantity called purity, $\mathcal{P} = \mathrm{Tr} \rho^2$. The state is pure if $\mathcal{P} = 1$. Otherwise, $\mathcal{P} < 1$.

Let us now describe how (possibly noisy) operations act on a noisy processor's state $\rho$.

\subsubsection{Describing noise in analog processors: Lindblad master equation}
\label{sec:lindblad}

Schrödinger's equation describes the temporal evolution of perfect analog processors. In theory, one could describe the temporal evolution of noisy analog processors by describing the state of the processor and of the environment as a single wavefunction $\ket{\Psi_\mathrm{tot}}$.
Its evolution would be driven by a total Hamiltonian $H_\mathrm{tot} = H + H_\mathrm{env} + H_\mathrm{coupling}$ (where $H_\mathrm{env}$ is the Hamiltonian of the environment and $H_\mathrm{coupling}$ that of the coupling between the processor and the environment).
One could then recover, e.g., average values of the processor's observables by computing $\bra{\Psi_\mathrm{tot}} O \ket{\Psi_\mathrm{tot}}$, or, equivalently, $\mathrm{Tr} [\rho O]$ with $\rho$ defined by "eliminating" the environmental degrees of freedom via a partial trace operation. This operation is denoted as $\rho = \mathrm{Tr}_\mathrm{env} \ket{\Psi_\mathrm{tot}} \bra{\Psi_\mathrm{tot}}$.

However, this strategy is often impractical because the environment generically comprises many degrees of freedom that (i) one cannot describe individually and (ii) one cannot solve the corresponding Schrödinger equation because of the huge size of the total Hilbert space. One thus looks for time-evolution equations that directly focus on the minimal description of the noisy quantum processor, namely the reduced density matrix of the processor, $\rho$ (instead of $\ket{\Psi_\mathrm{tot}}$).
Such equations go under the name of "master equations".
One of them---the so-called Lindblad equation \cite{Breuer2002} (also known as Gorini–Kossakowski–Sudarshan–Lindblad equation)---is of particular interest since it guarantees that the time evolution of the density matrix will preserve the essential properties of $\rho$, namely its unit trace and its positive semidefinite character. It reads:
\begin{eqnarray}
    i\hbar \frac{d\rho}{dt}=\left[H(t),\rho\right]-\frac{i}{2}\sum_{m}\left[\left\{ L_{m}^{\dagger}L_{m},\rho\right\} -2L_{m}\rho L_{m}^{\dagger}\right]. \label{eq:Lindblad}
\end{eqnarray}
Here, the $L_m$ operators are known as Lindblad or "jump" operators. They are responsible for decoherence. In the absence of these operators, the system follows a unitary evolution (and the equation is called the Liouville - von Neumann equation).
In the presence of these operators, the density matrix evolution becomes non-unitary with dissipation induced by the second term in the right-hand side of Eq. (\ref{eq:Lindblad}).

For instance, for a one-qubit system with idle qubits ($H=0$) and $L = \sqrt{\gamma_\varphi/2} Z$, the density matrix evolves as 
\begin{equation}
    \rho(t)=\left[\begin{array}{cc}
\rho_{00}(t=0) & \rho_{01}(t=0) e^{-\gamma_\varphi t}\\
\rho_{01}(t=0)^{*} e^{-\gamma_\varphi t} & (1-\rho_{00}(t=0))
\end{array}\right] . 
\end{equation}
The off-diagonal elements of $\rho$ (sometimes called "coherences") become negligibly small with a characteristic "dephasing" time $T_\varphi = 1/ \gamma_\varphi$. For $t \gg T_\varphi$, the state of the quantum system becomes $\rho \approx \rho_{00} |0\rangle \langle 0| + \rho_{11} |1\rangle \langle 1|$. If one starts from a superposed pure state $\ket{\psi} = (\ket{0} + \ket{1})/\sqrt{2}$, i.e $\rho_{ij}(t=0) = 1/2$, for all $(i,j)$, one ends up in state $\rho = 1/2 |0\rangle \langle 0| + 1/2 |1\rangle \langle 1|$. 
In other words, under this dephasing noise, we went from a system in a (quantum) state $0$ {\it AND} $1$ to a (classical-like) state $0$ {\it OR} $1$.

Other Lindblad operators lead to different types of noise; for instance, $L = \sqrt{\gamma_1} Q^{-}$ leads to a kind of noise called relaxation (or "amplitude damping") noise, which causes the qubit to lose energy to its environment by "relaxing" to its "ground state" $\rho = |0\rangle \langle 0|$:
\begin{equation}
\rho(t)=\left[\begin{array}{cc}
1-\rho_{11}(t=0)e^{-\gamma_{1}t} & \rho_{01}(t=0)e^{-\gamma_{1}t/2}\\
\rho_{10}(t=0)e^{-\gamma_{1}t/2} & \rho_{11}(t=0)e^{-\gamma_{1}t}
\end{array}\right].
\end{equation}
The characteristic time is $T_1 = 1/\gamma_1$. 

Putting these two noise models together yields the time evolution:
\begin{equation}
    \rho(t)=\left[\begin{array}{cc}
1-\rho_{11}(t=0)e^{-t/T_{1}} & \rho_{01}(t=0)e^{-t/T_{2}}\\
\rho_{10}(t=0)e^{-t/T_{2}} & \rho_{11}(t=0)e^{-t/T_{1}}
\end{array}\right]\label{eq:T1T2_noise}
\end{equation}
with the characteristic times:
\begin{align}
    \frac{1}{T_1} &= \gamma_1,\\
    \frac{1}{T_2} &= \gamma_\varphi + \frac{\gamma_1}{2}= \frac{1}{T_\varphi} + \frac{1}{2 T_1}.
\end{align}
These times can be measured experimentally on real hardware by conducting Rabi experiments (for $T_1$) and Ramsey experiments (for $T_2$) (see e.g \cite{Oliver2013}). In real hardware, the $t$-dependence of the off-diagonal term is generally not as simple as an exponential decay because noise is usually not white (contrary to the assumptions leading to the Lindblad equation) \cite{Krantz2019}.

The effect of dephasing and relaxation noise is illustrated in Fig.~\ref{fig:bloch_sphere}: the red trajectory represents the evolution of $\rho(t)$ under a Rabi and detuning drive and Lindblad jump operators of the dephasing and relaxation type. Relaxation pushes states towards the North pole (since it tends to relax states to $|0\rangle$), while dephasing pushes states towards the vertical axis of the sphere (it destroys superposed states, which sit on the equator of the sphere). These effects are visible in the figure, where the red trajectory is deformed towards the vertical axis and the North pole of the Bloch sphere.

In practice, these two coherence times are handy to crudely assess the number of gates that can be executed on a given hardware platform. Since the quantum execution time $\tau_\mathrm{run} \propto \tau_\mathrm{gate}$ must be much shorter than the coherence time $T$, the allowed depth is $\ll T/\tau_\mathrm{run} \propto T/ \tau_\mathrm{gate}$. Thus a rough quality factor for a quantum algorithm is the ratio $T/ \tau_\mathrm{gate}$ (as opposed to the sole coherence time).

As already mentioned, the Lindblad master equation is itself an approximate evolution equation. It assumes that the coupling between the environment and the processor is weak and that the environment has no memory effect, a property called Markovianity. In other words, it can only describe "white" noise, i.e., noise without temporal correlations. This description may not be sufficient for some architectures. A prominent example is superconducting qubits, where dephasing noise is known to be "pink", i.e., its autocorrelation function decays as $1/f$ \cite{Paladino2014} (instead of being a constant in frequency for white noise).
Other more complex equations can be used to describe dissipative and decoherence effects, particularly non-Markovian effects \cite{Breuer2002,Breuer2016,deVega2017}. Such effects can be incorporated at the price of a significant increase in the numerical effort, the nature of the stochastic jumps, or both (see, for instance, \cite{Diosi1998,Breuer2004,Sargsyan2017,Lacroix2020_oqs}), and are in general not incorporated to describe noisy qubits.        

\subsubsection{Describing noise in digital processors: quantum channels.}\label{subsec:quantum_channels}

\paragraph{Noisy gates}
In digital quantum processors, the time evolution of the quantum state is specified by a sequence of gates. One usually does not have direct access to the underlying Hamiltonian $H(t)$: for each gate, the Hamiltonian is tuned by the hardware maker to reach a target unitary operator $U$. Because of this discrete description of the time evolution, the Lindblad equation introduced in the previous subsection is not the most convenient way to study the time evolution of the quantum state.

The most straightforward way to translate the perfect unitary evolution of the wavefunction $\ket{\Psi}$ induced by quantum gates into a noisy evolution is to describe each operation (gate) as a transformation of the density matrix $\rho$ introduced in subsection \ref{subsubsec:rho}.
Owing to the linear nature of the Schrödinger equation, this transformation---that we shall call $\mathcal{E}$---is linear.
It must preserve the critical properties of $\rho$, namely its unit trace ($\mathcal{E}$ is said to be "trace-preserving" (TP)) and positive semidefinite character ($\mathcal{E}$ is then said to be "positive"). The mapping must also be such that any extension $\mathcal{E}\otimes I$ to a larger space is positive, a property called "complete positivity". 
Thus, a noisy quantum gate is a completely positive, trace-preserving (CPTP) map acting on density matrices. It is also called a quantum channel.

Quantum channels have several equivalent representations that are used in different contexts. A widespread representation is the Kraus, or operator-sum representation \cite{Kraus1971, Kraus1983}:
\begin{equation}
    \mathcal{E}(\rho) = \sum_{k=1}^{K} E_k \rho E_k^\dagger.\label{eq:kraus_repr}
\end{equation}
The $E_k$ operators are called Kraus operators. $K$ is called the Kraus rank. Trace preservation imposes $\sum_k E_k^\dagger E_k = I$.
The $K=1$ case corresponds to a unitary evolution since, in this case, $E_1^\dagger E_1=I$ and the density matrix transforms as $\rho\rightarrow E_1 \rho E_1^\dagger$, i.e., a pure state $\ket{\Psi}$ is mapped to a pure state $E_1 \ket{\Psi}$.

Alternative representations that can be used include the Pauli transfer matrix (PTM, \cite{Chow2012}), the matrix representation of the linear map written on the basis of Pauli matrices. 
The matrix representation is sometimes called the superoperator (or $\mathcal{S}$-matrix) representation when expressed on the canonical matrix basis.
One can also mention the $\chi$-matrix (or process) representation \cite{Chuang1997}, the Choi-Jamiolkovski representation \cite{Choi1975a, Jamiolkowski1972}, and the Stinespring dilation \cite{Stinespring1955}. Graphical representations of these equivalent variants are given in \cite{Wood2015}.

\begin{figure}
    \centering
     \includegraphics[width=0.9\columnwidth]{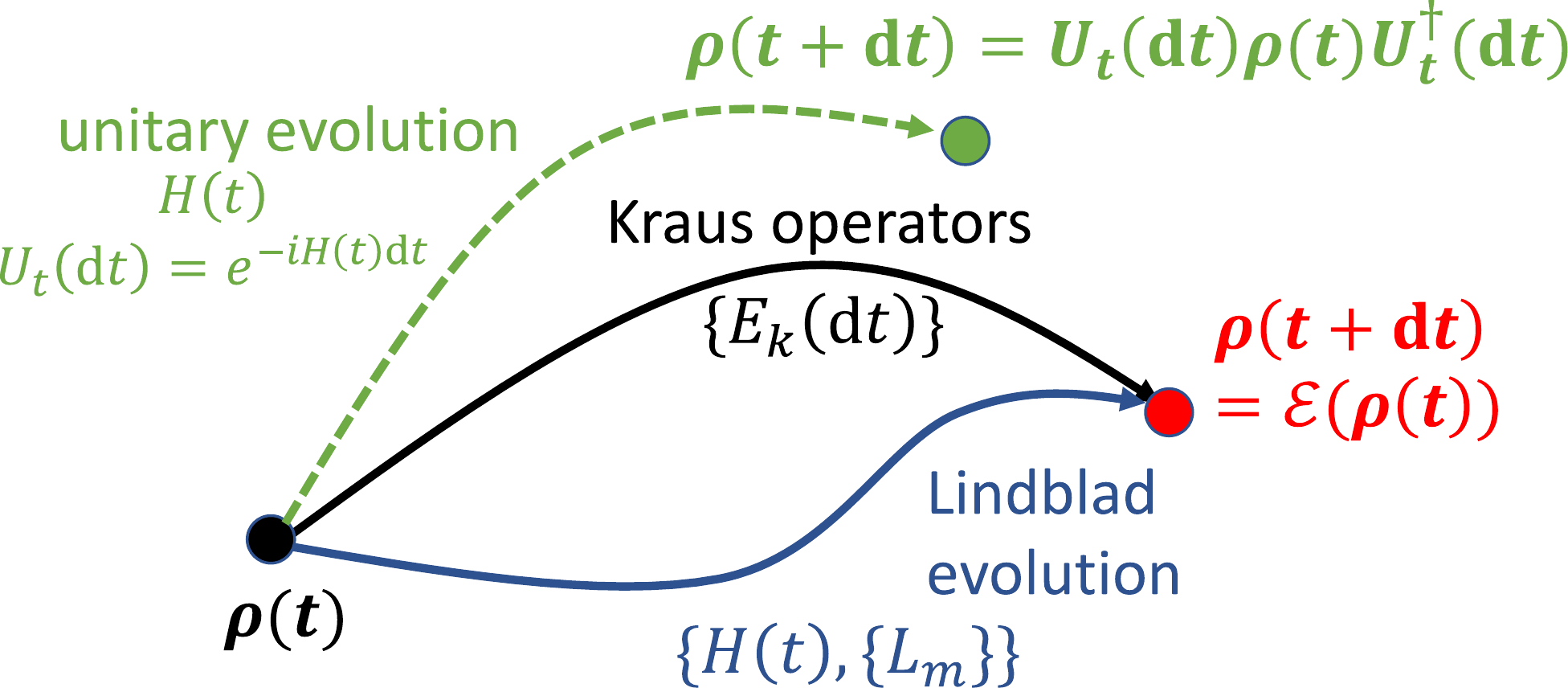}
    \caption{Two equivalent ways (via the Lindblad equation [solid blue arrow] or Kraus operators [solid black arrow]) to describe the evolution of the density matrix from time $t$ to time $t+dt$, compared to unitary evolution (dashed green arrow). 
    }
    \label{fig:krauslindblad}
\end{figure}

The time-dependent approach presented in section \ref{sec:lindblad} to describe the effect of noise and the one presented here are related. Assuming a certain density at time $t$, denoted by $\rho(t)$, the Lindblad equation makes the evolution in the presence of noise. Said differently, through the solution of the Lindblad equation, 
for a given time $\mathrm{d} t > 0$, we obtain $\rho(t+\mathrm{d} t)$.
For a given $\mathrm{d} t$, one can introduce a set of Kraus operators $\{ E_k (\mathrm{d} t) \}$ that can be related to the Hamiltonian and Lindblad operators (see, e.g., \cite{Preskill2015}).
A schematic view of the connection between the Lindblad and Kraus techniques is given in Fig. \ref{fig:krauslindblad}.

In a quantum computer, each noisy quantum gate is entirely described by its Kraus operators (or any other representation of the quantum channel).
This description also includes "idling noise", namely the noise that qubits incur when left idle between two gate applications: idling noise merely corresponds to a "noisy identity" map. In their simplest form, quantum channels act only on the qubits operated on by the gate at stake.
However, crosstalk effects---the fact that the gate acts on other qubits than the intended ones---can, in principle, be taken into account by extending the channel's support.
Finally, let us note that non-Markovian effects (temporal correlations) are not captured by such a discrete description of noise: the quantum channel that comes after a given noisy gate is not modified by the preceding quantum channels.

\begin{figure}
    \centering
    \includegraphics[width=\columnwidth]{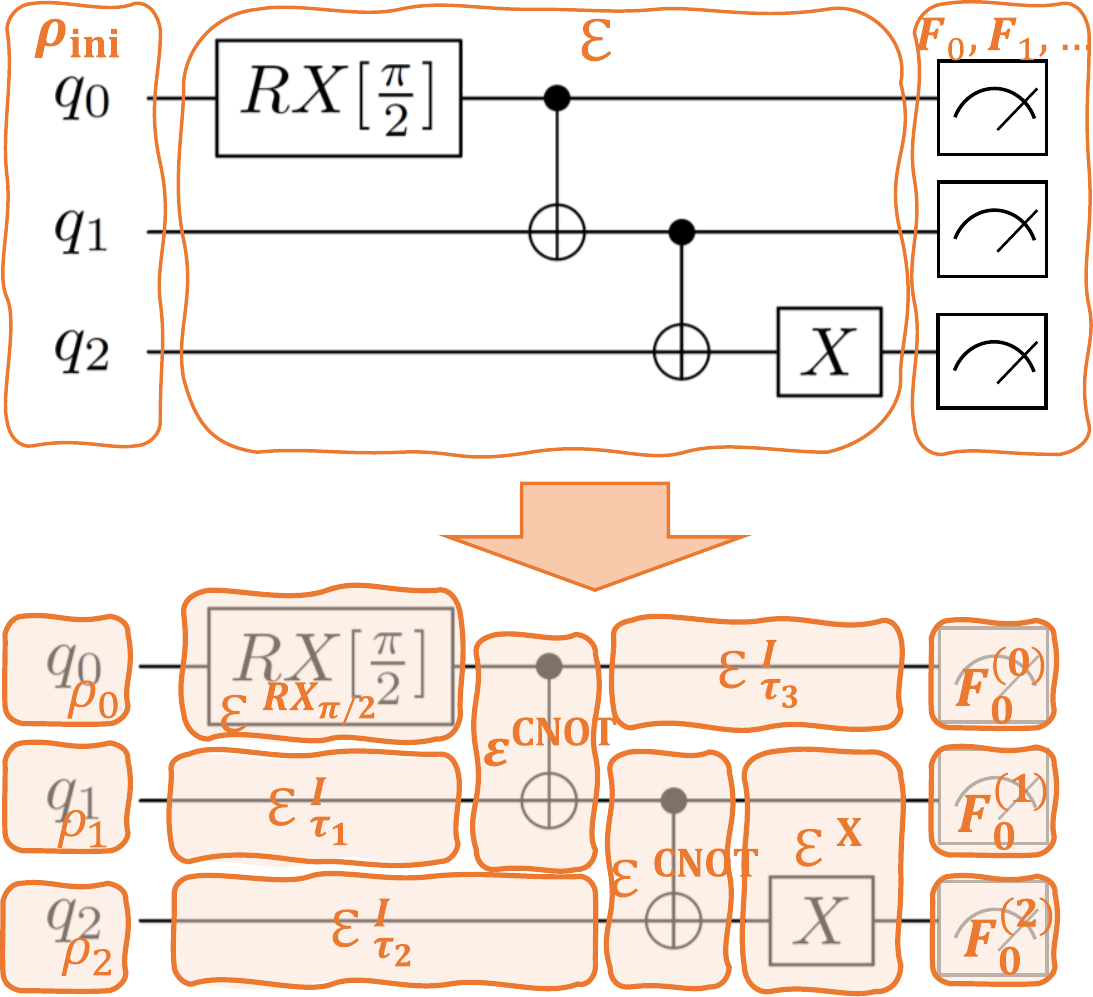}
    \caption{Schematic representation of a noisy circuit. Instead of a global quantum channel acting on all qubits of the initial state, followed by a global POVM, one can (approximately) break down the noisy evolution as a succession of more or less local quantum channels applied on a factorized initial state followed by local two-outcome POVMs.}
    \label{fig:noisy_circuit}
\end{figure}

\paragraph{Noisy circuits}
In all generality, a noisy quantum circuit is a $n$-qubit quantum channel $\mathcal{E}$ that turns an initial state $\boldsymbol{\rho}_\mathrm{ini}$ into a final state $\boldsymbol{\rho}_\mathrm{f} = \mathcal{E}(\boldsymbol{\rho}_\mathrm{ini})$. 
In practice, as illustrated in Fig.~\ref{fig:noisy_circuit}, one can approximate this "global" channel by a sequence of local channels ($\mathcal{E}^\mathrm{H}$, $\mathcal{E}^\mathrm{CNOT}$, etc in the figure) acting on an initial state that can be approximated as a product state: $\boldsymbol{\rho}_\mathrm{ini} = \rho_0 \otimes \rho_1 \otimes \rho_2$ \cite{Aharonov1998}.
If a gate is known to suffer from crosstalk (like the $X$ gate in the figure), one can take this into account by assuming that the corresponding channel acts on more qubits than expected from the ideal gate.
Noise also affects "idle" qubits: this is illustrated by the $\mathcal{E}^\mathrm{I}$ boxes in Fig.~\ref{fig:noisy_circuit}. Typically, if the "idling noise" is of dephasing and amplitude damping type, then the action of these CPTP maps is defined by the expression of Eq.~\eqref{eq:T1T2_noise}. 
Equivalently, this corresponds to the following Kraus operators:
\begin{align}
    E^{(\mathrm{PD})}_{0}&=\left[\begin{array}{cc}
1\\
 & \sqrt{1-p_{(\mathrm{PD})}}
\end{array}\right], E^{(\mathrm{PD})}_{1}=\left[\begin{array}{cc}
0\\
 & \sqrt{p_{(\mathrm{PD})}}
\end{array}\right],\\
E^{(\mathrm{AD})}_{0}&=\left[\begin{array}{cc}
1 & 0\\
0 & \sqrt{1-p_{(\mathrm{AD})}}
\end{array}\right], E^{(\mathrm{AD})}_{1}=\left[\begin{array}{cc}
0 & \sqrt{p_{(\mathrm{AD})}}\\
0 & 0
\end{array}\right]
\end{align}
with the pure dephasing and amplitude damping probabilities $p_{(\mathrm{PD})}(\tau)= 1 - e^{-2\tau/T_\varphi}$ and $p_{(\mathrm{AD})}(\tau)= 1 - e^{-\tau/T_1}$ (in a Markovian/white noise approximation).

The "local" Kraus operators corresponding to the local quantum channels can be determined by so-called quantum process tomography methods (a "process" is another name for transforming the density matrix). They are methods for experimentally characterizing the quantum channel by measuring the output distribution of a noisy gate for a well-chosen set of inputs. Since these inputs are prepared using a priori unknown noisy gates, one has to resort to self-consistent schemes to solve this chicken-and-egg problem. Such schemes go under the broad name of gate-set tomography (GST~\cite{Blume-Kohout2013,Merkel2013,greenbaum_2015}).

In the absence of tomography, one can also resort to generic quantum channels to study the effect of noise on the execution of quantum circuits. Such channels include the amplitude damping (or relaxation) mentioned above and pure-dephasing channels, as well as the depolarizing channel and the bit-flip channel \cite{Nielsen2010}. For instance, the depolarizing channel is defined by the expression:
\begin{equation}
    \mathcal{E}(\rho) = (1-p) \rho + p \frac{I}{2^n},\label{eq:depol}
\end{equation}
where $n$ is the number of qubits.
It leaves the density matrix unchanged with probability $1-p$ and turns it into the ``maximally mixed state'' $I/2^n$ with probability $p$.

\paragraph{Noisy measurements}

From a mathematical perspective, measurements are so-called positive operator-valued measures (POVM), defined as a set of so-called POVM elements $\{ F_i \}$, which are positive semi-definite matrices summing to identity ($\sum_i F_i = I$) and such that the probability of getting the outcome $i$ is given by Born's rule,
\begin{equation}
    P(i) = \mathrm{Tr} \left[ \rho F_i \right].
\end{equation}

In the description of perfect quantum computers (section \ref{subsubsec:digital_qc}), we introduced the measurement of observable "$Z$". In general, the measurement of an observable ${O}$ and the corresponding POVM is given by the decomposition ${O} = \sum_i o_i F_i$. For instance, for ${O} = {Z}$, we have $F_0 = |0\rangle \langle 0|$,  $F_1 = |1\rangle \langle 1|$ and $o_0 = 1$, $o_1=-1$. $Z$ is a particular example of two-outcome POVM. Generally, a noisy two-outcome POVM is completely determined by a matrix $F_0$ (the other given by $F_1 = I - F_0$).

Typically, one can suppose that the final measurements on each qubit are independent, and thus, since they are also two-outcome, completely determined by $\lbrace F^{(k)}_0 \rbrace_{k=0\dots n-1}$ (see Fig.~\ref{fig:noisy_circuit}).

\subsubsection{Decoherence and fidelity}

Noise in quantum circuits has a dramatic influence on the fidelity $F$ of the output states. $F$ measures the similarity of the state $\rho$ that is actually output by the (noisy) processor with the state $|\Psi\rangle$ that would have been output by a perfect computer, $F = \langle \Psi | \rho |\Psi \rangle$.

A heuristic law relates the average error rate $\varepsilon_k$ of individual operations (gates, measurements...) to the final fidelity \cite{Arute2019}:
\begin{equation}
    F = \prod_{k=1}^{N_\mathrm{ops}} \left(1 - \epsilon_k \right) \approx \exp(- \epsilon N_{\mathrm{ops}}),
\end{equation}
with $N_{\mathrm{ops}}$ the total number of operations, and we have assumed an identical error rate in deriving the approximate scaling. 

In other words, the output fidelity falls exponentially with the individual error rate and the number of operations.
This law is also heuristically observed when the errors come from compression algorithms in random circuits \cite{Zhou2020, Ayral2022}. It is exact in the case of depolarizing noise (see Eq. \eqref{eq:depol}). For other noise models or assumptions, more complex inequalities relate individual error characteristics and total errors (see, e.g., \cite{Carignan-dugas2019}).

This exponential decay puts strong constraints on quantum processors' capability to outperform classical processors without quantum error correction.

\subsection{Quantum Error Mitigation}

In the absence of an error-correcting scheme (see subsection \ref{subsec:qec}), one can try to limit the effect of the incoherent errors accumulated during the execution of the circuit on the estimation of observables: this is the scope of \textit{error mitigation}.
Error mitigation does not require more physical qubits but instead trades possibly large sampling overheads for enhanced accuracy. We review a few methods here and refer the reader to \cite{Cai2022}
for a more extensive review. 

\subsubsection{Post-selection and purification}
In most applications, the output state (or a related observable) respects some mathematical properties. 
For instance, using the JWT technique, for a problem where particle number is conserved, the number of $1$ measured is constant and equal to the particle number. Discarding measured states that do not respect the symmetries enforced by the circuit (in the example given, sampled states that have a number of 1 different from the total number of particles) provides a straightforward error mitigation scheme.    

In addition, it is sometimes possible to map a noisy quantity to the pure one it represents (or at least a close approximation), a procedure referred to as \textit{purification}. Purification can be based on so-called fermionic $N$-representability conditions \cite{rubin_2018}.  
For instance, a (well-conditioned) noisy density matrix $\rho$ can be mapped to a pure density matrix (satisfying the idempotency criterion $\rho^2=\rho$) through repeated application of the  McWeeny "purification" polynomial $P_{\mathrm{MW}}(\rho) = 3\rho^2 - 2\rho^3$ \cite{mcweeny_1960}.
However, as the number of qubits increases, full density matrix  tomography becomes cumbersome.  Part of this complexity can be bypassed by considering the marginals of the density matrix, namely the 1- and 2-RDM that contain all the information on one- and two-body observables. Then, the methods can also be tweaked by approximate $N$-representability conditions. This process requires, however, more advanced schemes than McWeeny purification. 
A notable exception is the preparation of a Slater determinant, e.g., within the Hartree-Fock procedure, where an idempotent 1-RDM is expected: McWeeny purification applies to the 1-RDM, providing dramatic increases in the accuracy \cite{Arute2020}.
However, the Hartree-Fock procedure is trivial on a classical computer, and the idempotency of the 1-RDM breaks down as soon as a non-Slater state is targeted.

\subsubsection{Zero-noise extrapolation (ZNE)}

\begin{figure}
    \centering
    \includegraphics[width=\columnwidth]{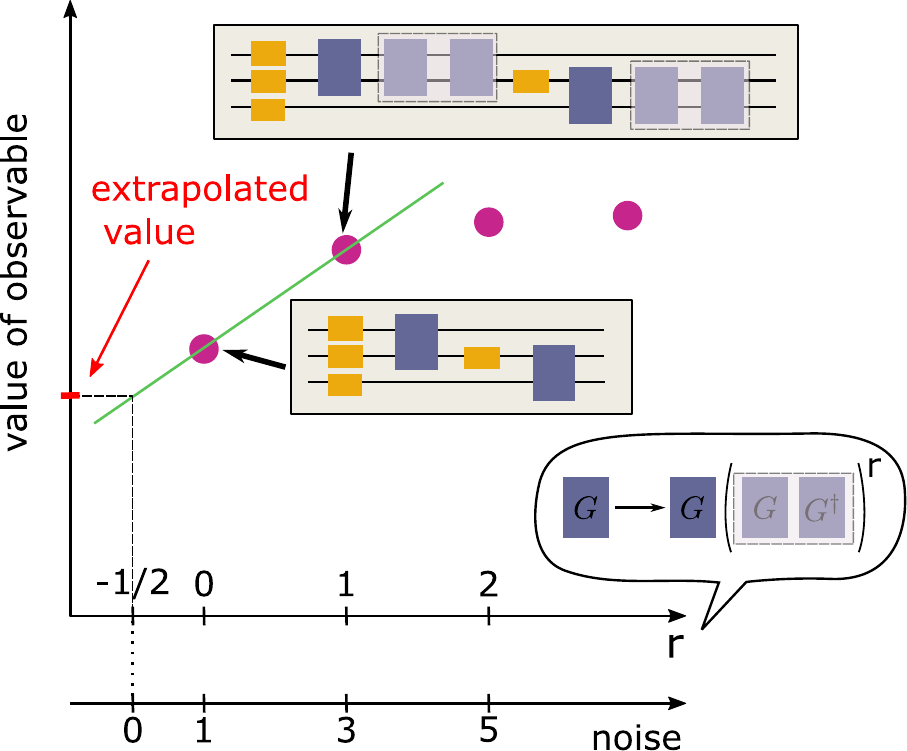} \\
    \caption{Principle of zero-noise extrapolation illustrated with a linear ansatz for inference. Occurrences of two-qubit gates $G$ are followed by a number $r$ of resolutions of the identity $I=GG^{\dagger}$ to scale the noise to a factor $(2r+1)$. A noiseless observable value can be inferred from the noisy observables measured on the original circuit and the circuit with $r=1$ by linearly extrapolating to $r=-1/2$.}
    \label{fig:ZNE_principle}
\end{figure}

Within ZNE, the departure of the observable as measured $\langle {O} \rangle_{\mathrm{meas}}$ from its noise-free counterpart $\langle {O} \rangle_{\mathrm{perfect}}$ is assumed to depend on a single parameter, an error rate $\epsilon_{\mathrm{phys}}$.
Assuming some ansatz for the precise form of how these two are related, one can infer an estimation of $\langle {O} \rangle_{\mathrm{perfect}}$ from a set of measurements corresponding to different effective error rates $\epsilon = f(\epsilon_{\mathrm{phys}}, r)$ where $r$ is a tunable parameter. 

A ZNE-specific challenge is to find a way to explore different error rates, which depend on the  noise processes at play. Typically, the noise to be mitigated is the one stemming from the two-qubit gate of the set, say $G$, and the 'rescaling' of the error rate is obtained by inserting decompositions of the identity under the form $GG^{\dagger}$ after each occurrence of $G$ \cite{he_zero-noise_2020}.
This process does not change the state encoded by the circuit. However, it makes it more error-prone: under the assumption that a depolarizing channel can model the two-qubit gates errors, $r$ insertions correspond to inflating the (two-qubit gate) error rate from its physical value $\epsilon^{(2)}_{\mathrm{phys}}$ to $\epsilon(r)  = (2r +1)\epsilon^{(2)}_{\mathrm{phys}}$, and a noise-free observable can subsequently be inferred by extrapolating to the $r=-1/2$ regime, see Figure \ref{fig:ZNE_principle}.
Alternatively, one can resort to pulse stretches rather than identity insertions to increase the noise picked along the execution of the circuit \cite{kandala_error_2019}: the only underlying assumption is that the noise is time-invariant.

\subsubsection{Clifford data regression}

\begin{figure}
    \centering
    \includegraphics[width=0.85\columnwidth]{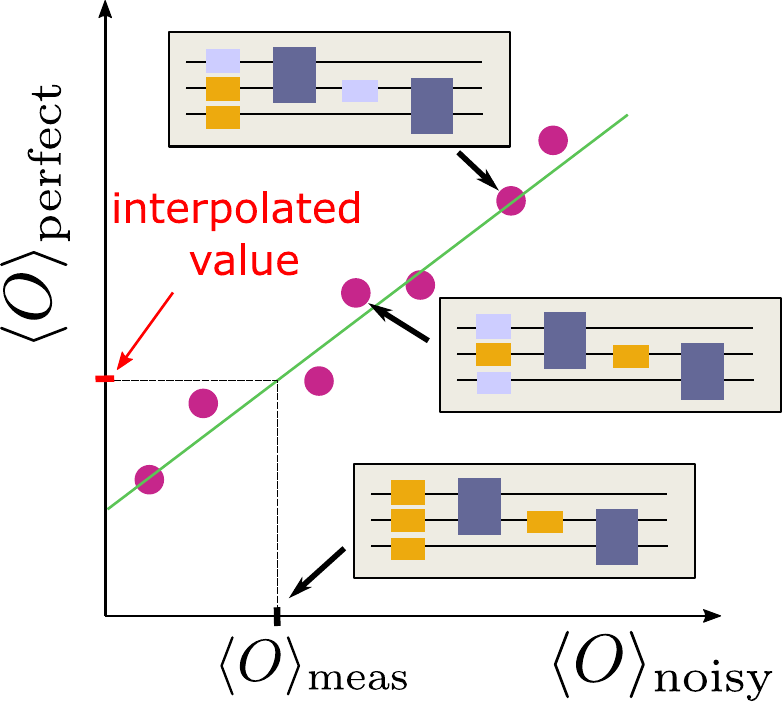} \\
    \caption{Principle of Clifford data regression illustrated with a linear ansatz for interpolation. A number $K$ (here $K=2$) of non-Clifford gates from the original circuit are replaced by Clifford gates (in light purple) to obtain circuits that can be simulated classically. A linear ansatz linking noisy observable measurements to noiseless values is then trained on the set of circuits obtained so. A noise-free observable can finally be inferred by interpolating from the noisy measurement result obtained on the original circuit.  
    }
    \label{fig:CDR_principle}
\end{figure}

Clifford data regression (CDR) \cite{czarnik_error_2021} is a learning-based method (and is thus sometimes referred to as Learning-Based Error Mitigation \cite{strikis_learning-based_2021}) where an ansatz is trained to map noisy values to noise-free ones. It applies only to digital quantum computers. 

For instance, one can look for a relation of the form
\begin{eqnarray}
    \langle O \rangle_{\mathrm{perfect}} = a \langle O \rangle_{\mathrm{noisy}} + b
    \label{eq:linear_cdr_ansatz}
\end{eqnarray}
by fitting on a set of tuples $(\langle O \rangle_{\mathrm{noisy}}^{\mathcal{C}_j},  \langle O \rangle_{\mathrm{perfect}}^{\mathcal{C}_j})$. 
The training set $\{ \mathcal{C}_j \}$ comprises circuits that are easy to simulate classically. In the original method, near-Clifford circuits were used to this end, and we will stick to this example in what follows. Alternatively, one can summon another class of easily-simulable circuits to study fermionic systems: gaussian circuits \cite{montanaro_2021}. 
To ensure the predictive character of Equation \eqref{eq:linear_cdr_ansatz} (namely, that coefficients $a$ and $b$ obtained by fitting over the training set give good predictions for the value of $\langle O \rangle_{\mathrm{noisy}}^{\mathcal{C}}$ for the circuit $\mathcal{C}$ of interest), the training set is obtained by replacing some of the non-Clifford gates in the original circuit with Clifford gates. Assume a universal gate set made of single-qubit rotations and the CNOT gate; this could be done by compiling the circuit replacing the $R_z(\theta)$ gates -- that are Clifford only for $\theta_n=n\pi/2, n \in [0, 1, 2, 3]$ because they correspond to the phase gate to the power of $n$, $S^n$ -- by some $R_z(\theta_n)$. 
The number of gates that are replaced acts as a refinement parameter. The non-Clifford gates to replace and their Clifford gates replacements are chosen out of a distance criterion. 
Alternatively, a Markov Chain Monte Carlo (MCMC) technique can be employed. 

A significant obstacle in successfully implementing CDR is that there is no known recipe for designing the training set optimally.
A method dubbed \textit{variable noise CDR} (vnCDR) was proposed, which mixes ZNE and CDR features. A training set's element in vnCDR is defined by both a circuit and a noise strength. The scheme consists in guiding the ZNE with CDR, cutting the need for precise knowledge of the noise strength. 
Note that CDR and ZNE, along with a third technique not reviewed here -- Virtual Distillation \cite{huggins_2021}, employing more qubits than the state fits in -- can be subsumed and combined in a unified framework \cite{bultrini_2021}.

\subsubsection{Quasiprobability method}

\begin{figure}
    \centering
    \includegraphics[width=\columnwidth]{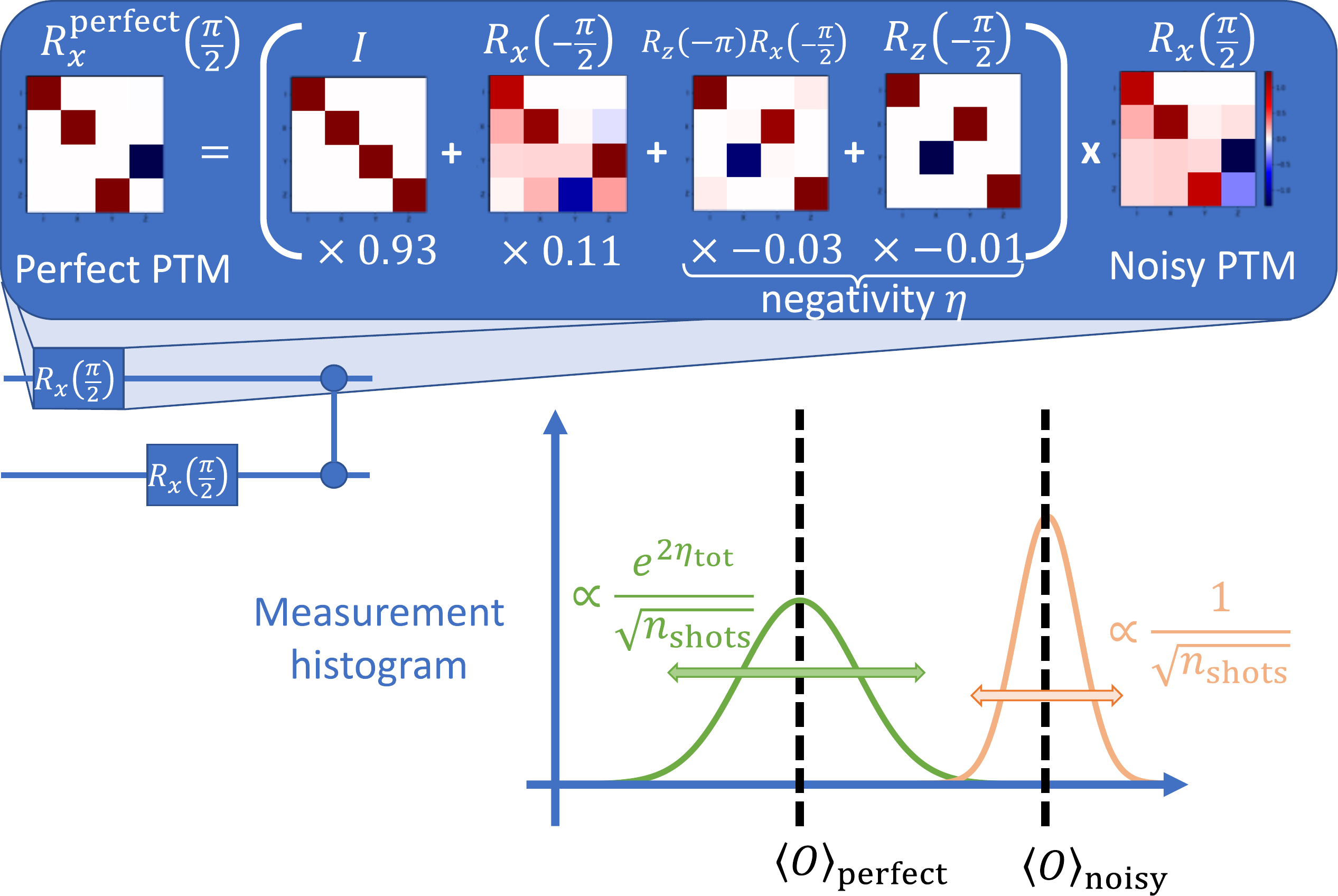} \\
    \caption{Principle of quasiprobability error mitigation. Each perfect gate (represented, e.g., as a Pauli transfer matrix [PTM]) is decomposed onto available (noisy) operations (see Eq.~\eqref{eq:QPD}). The perfect observable $\langle O \rangle_{\mathrm{perfect}}$ is then sampled, with a sampling overhead of $e^{2\eta_\mathrm{tot}}$ compared to the noisy observable, but with no bias.}
    \label{fig:qem_principle}
\end{figure}

The Quasi Probability Error Mitigation (QPEM) method, introduced in \cite{temme_error_2017}, originates from the so-called Quasi Probability Decomposition (QPD) of a perfect quantum channel $\mathcal{E}^\mathrm{perfect}$ onto a set of the noisy quantum channels $\{ \mathcal{E}_k\}$ that are implemented by the hardware:
\begin{equation}
    \label{eq:QPD}
    \mathcal{E}^\mathrm{perfect}(\rho) = \sum_{k \in \mathrm{available\;ops}} q_k \mathcal{E}_k(\rho).
\end{equation}
The set of coefficients $\{ q_k\}$ denote the "quasiprobabilities": trace preservation ensures $\sum_k q_k = 1$, but the $q_k$ may take negative values. The "negativity" of the channel is defined as $\eta = -\sum_{k, q_k < 0} q_k$.

Measuring the expectation value of an observable $O$ output by channel $\mathcal{E}_k$ picked with probability $ \frac{|q_k|}{\sum_k |q_k|}$ thus provides an unbiased estimator of $\langle O \rangle \equiv \mathrm{Tr}(O\mathcal{E}(\rho))$ as $ C \langle  \mathrm{sgn}(q_k) \mathrm{Tr}(O\mathcal{E}_k(\rho)) \rangle_k$. Here we have defined $C = \sum_k |q_k| = 1 + 2 \eta$. This factor measures the sampling overhead incurred by the QPEM procedure: to maintain a given variance, one needs $O(C^2)$ more shots to evaluate $\langle O \rangle$ with QPEM than would be required if one were able to implement the quantum channel $\mathcal{E}$ perfectly. 

Usually, the quasiprobability decomposition is obtained at the level of individual gates since performing the decomposition at the circuit level would be exponentially costly.
The resulting $C$-factor will be the product of the individual ones: $C_\mathrm{tot} = \prod_{l=1}^{N_g} C_l$. Since, on the other hand, $C_l = 1 + 2 \eta_l \approx e^{2 \eta_l}$ (assuming weak negativity), we see that the method incurs a cost exponential in $\eta_\mathrm{tot} = \sum_{l=1}^{N_g} \eta_l \approx N_g \eta$ if $\eta$ is uniform.
This fact is illustrated in Fig.~\ref{fig:qem_principle}.

The sampling overhead can be reduced by reintroducing some bias in the estimator of $\langle O \rangle$, using approximate QPDs \cite{piveteau_quasiprobability_2022,piveteau_2021}.

Similarly to the ZNE, it requires a good knowledge of the noise processes at work in the hardware.
The quantum channels (in the form, e.g., of Pauli transfer matrices) for each operation can be obtained via so-called \textit{gate set tomography} \cite{greenbaum_2015, nielsen_gamble_2021} as done in Ref. \cite{endo_practical_2018}.
This process assumes that noise is both local (meaning that crosstalk between qubits can be neglected) and Markovian (time-invariant); see discussion in section~\ref{subsec:quantum_channels}.

\subsection{Quantum error correction (QEC) and fault tolerance (FT) in a nutshell}\label{subsec:qec}

\subsubsection{Quantum error correction}

\begin{figure}
    \centering
    \includegraphics[width=\columnwidth]{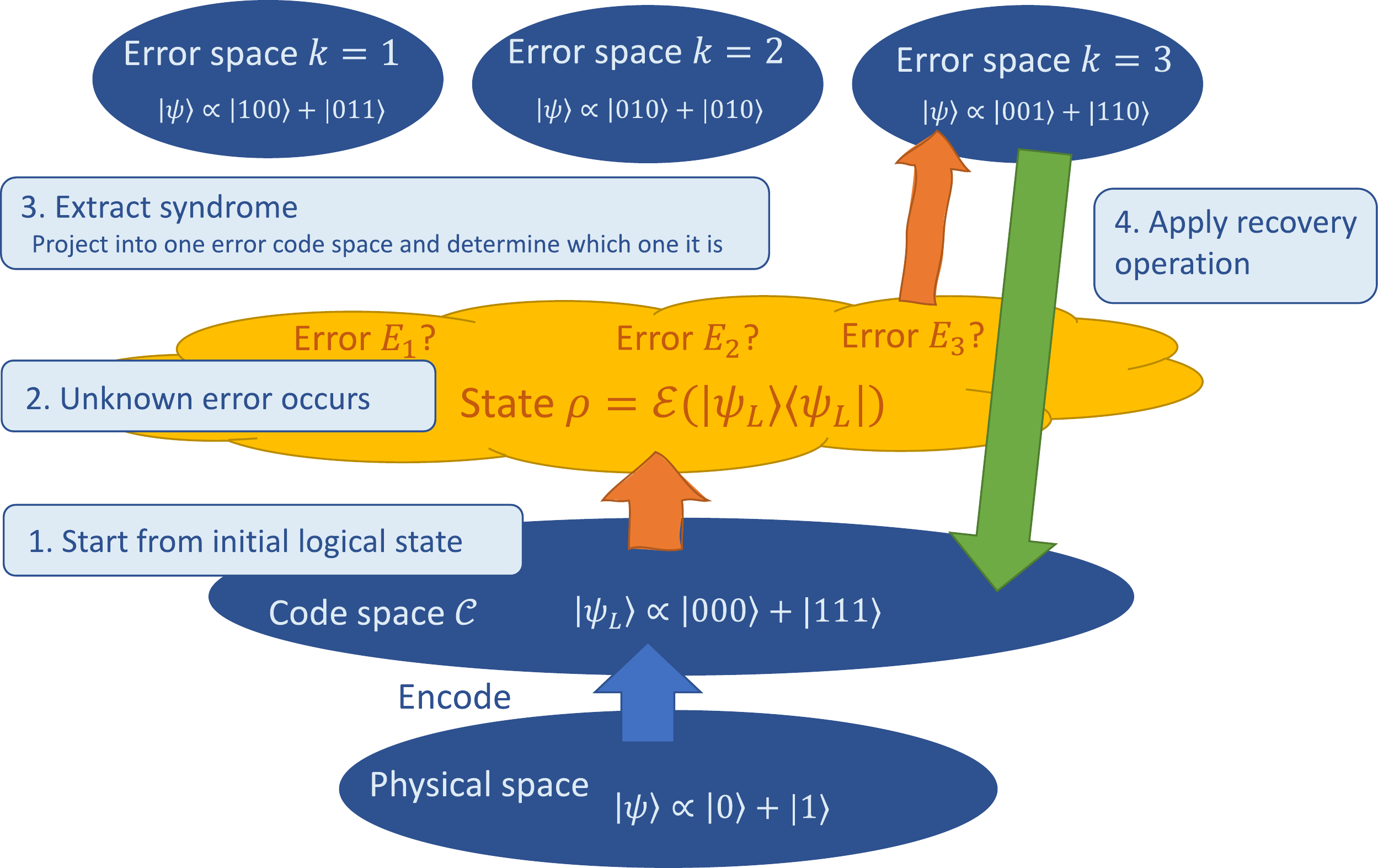} \\
    \caption{Schematic view of the principle behind the quantum error correction (see text for more details). }
    \label{fig:qec_principle}
\end{figure}

In analogy to classical computers, quantum computers can benefit from error correction by using redundancy, namely by encoding the information of one (quantum) bit, called a "logical qubit", into several physical qubits (see \cite{Girvin2022} for a recent reference).
Encoded states live in a subspace $\mathcal{C}$ (called codespace) of the physical Hilbert space designed in such a way that errors (described by a quantum channel $\mathcal{E}$, see subsection~\ref{subsubsec:rho} above) can be detected and then corrected using a recovery operation $\mathcal{R}$ such that $\mathcal{R} \circ \mathcal{E} (\rho) = \rho$, with $\rho \in \mathcal{C}$.
The encoding (or code) $\mathcal{C}$ is chosen based on the error model $\mathcal{E}$. The following necessary and sufficient conditions, known as QEC or Knill-Laflamme conditions \cite{Knill1997}, ensure the existence of a recovery operation:
\begin{equation}
    P_{\mathcal{C}}E_{k}^{\dagger}E_{l}P_{\mathcal{C}}=\beta_k \delta_{kl}P_{\mathcal{C}}, \;\;\; {\forall k,l}
\end{equation}
with $\lbrace E_k \rbrace$ the Kraus operators associated with $\mathcal{E}$, $P_\mathcal{C}$ the projector onto the codespace, and $\beta_k > 0$.

The principle of QEC is illustrated in Fig.~\ref{fig:qec_principle}. A state $\ket{\psi_L}$ of the codespace undergoes an error map $\mathcal{E}$ and thus becomes a mixed state $\rho = \mathcal{E}(|\psi_L \rangle \langle \psi_L|) = \sum_k E_k |\psi_L \rangle \langle \psi_L| E_k^\dagger$. In other words, there is uncertainty as to which error $E_k$ occurred. 
Measurements of so-called "syndromes" project the state into one of the error code spaces and also allow us to determine in which error code space the state was projected. The design of these measurements is subtle due to the wave function collapse: one wants to learn information about the error that occurred {\it without learning information on the quantum state that was corrupted} (lest information on the data qubit is lost).
Experimentally, this is achieved by measuring ancilla qubits entangled with the "data" qubits. 

As a last step, thanks to the syndrome information, the proper recovery operation is applied to recover the initial state $\ket{\psi_L}$.
Finding the recovery operation given a syndrome can be a complex (classical) computational task.
Designing good heuristics for finding the best recovery operation is an active research topic for advanced codes.

A consequence of the QEC conditions is that the recovery operation can correct any error that is a linear combination of the Kraus operators that satisfy the QEC condition. This fact implies that a code and recovery built to protect against one-qubit Pauli noise is enough for correcting any one-qubit noise (since its Kraus operators can be decomposed on the Pauli basis). This procedure is known as the "digitization" of errors.
Well-known codes include so-called stabilizer codes \cite{Gottesman1997}, which generalize classical linear codes and are particularly well-suited for one-qubit Pauli errors.

\subsubsection{Fault tolerance}

While QEC is meant to preserve quantum information, fault tolerance (FT) is the ability to perform quantum circuits without propagating errors.

\paragraph{The influence of errors during recovery}
One simple context where errors need to be considered is the recovery operation. If recovery were made with perfect gates, a code able to correct a number $t$ of errors would have an error per gate probability of $p_1 = c \epsilon^{t+1}$ after correction, for an error per gate before correction of $p_0 = \epsilon$. Thus, one could reach arbitrarily small error rates by choosing a large enough $t$. In practice, however, recovery is made with noisy gates. If recovery involves $O(t^\alpha)$ gates, then the error probability becomes $ \tilde{p}_1 = c(t^\alpha \epsilon)^{t+1}$. This function is monotonic for $t$: at a certain point, a larger $t$ means that recovery brings more errors than it corrects. There is thus an optimal $t$. At the optimal $t$, one finds a minimal error probability $\tilde{p}_1^\mathrm{min}(\epsilon, \alpha)$. This probability is an increasing function of the physical error rate $\epsilon$. To ensure that no error occurs for a circuit of length $N_g$, one must choose $N_g \tilde{p}_1^\mathrm{min}(\epsilon, \alpha) < 1$.
This in turn requires $\epsilon < \epsilon_0$, with
\begin{equation}
    \epsilon_0 \propto \frac{1}{\log(cN_g)^\alpha}.
\end{equation}
The rate $\epsilon_0$ is much more favorable than that one would have obtained in the absence of error correction, namely $\epsilon_0 \propto 1/N_g$.

\paragraph{Concatenation and the threshold theorem \cite{Shor1996, Aharonov2008}}

\begin{figure}
    \centering
    \includegraphics[width=\columnwidth]{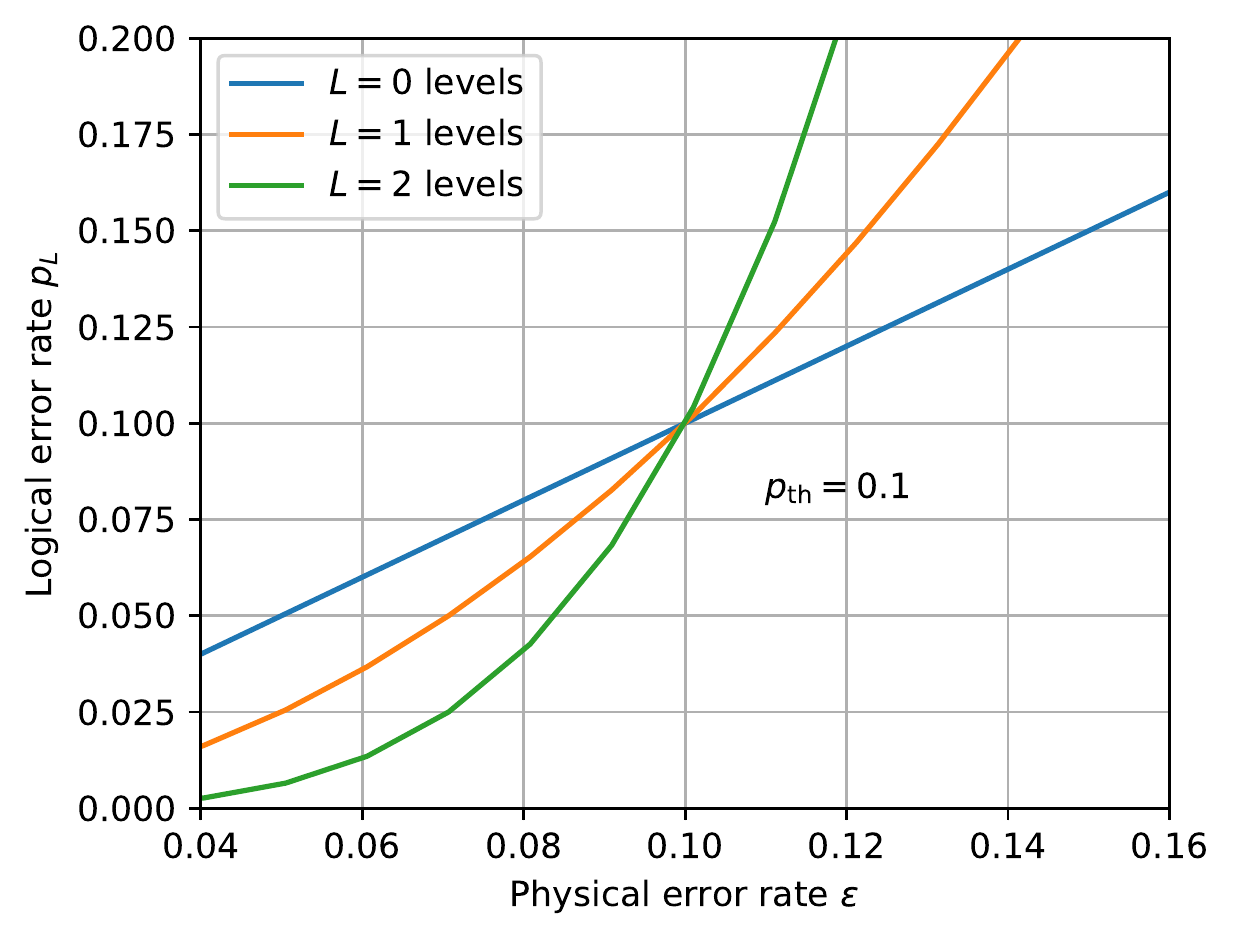} \\
    \caption{Logical error as a function of physical error in a concatenated code for a threshold $p_\mathrm{th} = 0.1$. }
    \label{fig:ft_threshold}
\end{figure}

In the reasoning above, with a fixed error $\epsilon$, one still reaches a limit in terms of the circuit length one can execute without error.
One way to solve this issue is concatenation, a method where the code is replicated several times, similar to a kind of renormalization group flow or fractal structure. Using $L$ nested levels of encoding, the error probability (with $t=1$) becomes $p_L = (c \epsilon)^{2^L}/c$. 
To reach an accuracy $\epsilon$ for a circuit with $N_g$ gates, i.e. an accuracy per gate $\epsilon/N_g$, we need to use $L$ such that $p_L \leq \epsilon/N_g$. Such a $L$ exists provided $\epsilon \leq p_\mathrm{th} \equiv 1/c$. Summarizing: after $L$ levels of concatenation, starting from a physical error rate $\epsilon$, one can achieve an error rate of 
\begin{equation}
    p_L = p_\mathrm{th} \left ( \frac{\epsilon} {p_\mathrm{th}} \right )^{2^L}.
\end{equation}
Provided the physical error rate is below a threshold value $p_\mathrm{th}$, the error rate after concatenation is reduced doubly exponentially.
This fact is illustrated in Fig.~\ref{fig:ft_threshold}.
At the same time, the circuit length is increased exponentially (we have a length $N_g^L$ after concatenation).

\paragraph{Surface and color codes}
In practice, concatenation often requires long-range two-qubit gates, which are unavailable in current and near-term hardware. 
This issue is addressed by, e.g., surface codes \cite{Kitaev2003, Dennis2002, Fowler2012}, and color codes \cite{Bombin2006, Landahl2011}, which are subclasses of stabilizer codes with good locality properties and are thus more suitable for actual hardware. A threshold theorem also holds for these codes, except one does not scale the number of concatenation levels but the size of the lattice in which the codes live. The error rate scales exponentially with the lattice size (instead of the double exponential of concatenated codes) for the surface code. In contrast, the circuit length scales linearly with the size (instead of exponentially).

First implementations of QEC with logical error rates close or slightly below the threshold have been demonstrated recently with surface codes on superconducting qubits \cite{Krinner2021, Acharya2022}, and color codes on trapped ions \cite{Ryan-Anderson2022}.

Interestingly, new types of superconductor-based hardware implementations are being developed specifically to perform more robust and/or economical quantum error correction \cite{Guillaud2022}.

\section{Conclusions}

Many-body systems are among the hardest problems to solve with classical computers. Their defining property is an exponential difficulty that can appear in many guises: the sheer size of the relevant portion of the Hilbert space, the Monte-Carlo sign problem, or a bond dimension exponential in the entanglement.

In the last century, scientists have accumulated deep expertise in solving this problem in many areas of physics and chemistry on classical computers despite the aforementioned exponential wall:
a plethora of heuristic methods to gain insights into the exotic physics of these systems has been designed.
These methods usually resort to the most advanced numerical techniques and are thus a difficult target for nascent quantum processors.
However, there are some regimes where the complexity of the problems at stake still prevents these classical methods from uncovering the critical physical mechanisms at play.

Quantum computers can be regarded as promising complementary tools to tackle such problems in difficult regimes. 
Indeed, as illustrated in this review, quantum processors are physical many-body systems, contrary to classical computers. Therefore, at least on paper, they appear as an ideal tool for understanding many-body phenomena. 
We presented several methods proposed in recent years to leverage this many-body nature with many different computational paradigms. 
In theory, these methods offer interesting solutions to the exponential wall.
Yet, contrary to classical computers, today's quantum processors must also reckon with decoherence.
This hurdle is also generically, in the absence of error correction, of exponential nature.
Therefore, the practical gain of using quantum processors needs to be carefully assessed by factoring in the advantages of exploiting inherent many-body phenomena together with the corresponding weaknesses.

Today's efforts have not yet converged to an example of a many-body problem that can be solved more efficiently with a quantum processor's (maybe partial) help. However, steady experimental and theoretical progress gives reasonable hope that such examples will emerge. 
More importantly, the fact that natural many-body systems can exhibit large-scale entangled states (like high-temperature superconductivity, superfluidity, etc.) despite decoherence is a quite robust indication that quantum processors---that is, synthetic many-body systems---can also be engineered to generate, and therefore gain insights into, phenomena with large-scale entanglement.

While the possibility of performing accurate enough computations with quantum computers or proving quantum advantage is still uncertain \cite{Lee2022}, there is no doubt that this domain is progressing very fast both in terms of technology and in terms of algorithms.

\begin{acknowledgements}
This project has received financial support from the CNRS through
the 80Prime program and the AIQI-IN2P3 project. This work is part of HQI initiative (www.hqi.fr) and is supported by France 2030 under the
French National Research Agency award number “ANR-22-PNCQ-0002".
This project has received funding from the European Union’s Horizon 2020 research and innovation programme under grant agreement No 951821. 
This project has received funding from the European High-Performance Computing Joint Undertaking (JU) under grant agreement No 101018180.

\end{acknowledgements}

\bibliographystyle{elsarticle-num}

\bibliography{abbreviated}

\begin{thebibliography}{100}
\expandafter\ifx\csname url\endcsname\relax
  \def\url#1{\texttt{#1}}\fi
\expandafter\ifx\csname urlprefix\endcsname\relax\def\urlprefix{URL }\fi
\expandafter\ifx\csname href\endcsname\relax
  \def\href#1#2{#2} \def\path#1{#1}\fi

\bibitem{feynman2018}
R.~P. Feynman, {\it Simulating physics with computers}, InterNatl. J. (Wash.)
  of Theoretical Physics 21 (1982) 467.
\newblock \href {https://doi.org/10.1007/BF02650179}
  {\path{doi:10.1007/BF02650179}}.

\bibitem{Acin2018}
A.~Acín, I.~Bloch, H.~Buhrman, T.~Calarco, C.~Eichler, J.~Eisert, D.~Esteve,
  N.~Gisin, S.~J. Glaser, F.~Jelezko, S.~Kuhr, M.~Lewenstein, M.~F. Riedel,
  P.~O. Schmidt, R.~Thew, A.~Wallraff, I.~Walmsley, F.~K. Wilhelm, {\it The
  quantum technologies roadmap: a European community view}, New J. Phys. 20
  (2018) 080201.
\newblock \href {https://doi.org/10.1088/1367-2630/aad1ea}
  {\path{doi:10.1088/1367-2630/aad1ea}}.

\bibitem{GYONGYOSI2019}
L.~Gyongyosi, S.~Imre, {\it A Survey on quantum computing technology}, Computer
  Science Review 31 (2019) 51.
\newblock \href {https://doi.org/10.1016/j.cosrev.2018.11.002}
  {\path{doi:10.1016/j.cosrev.2018.11.002}}.

\bibitem{Meter2016}
R.~Van~Meter, S.~J. Devitt, {\it The path to scalable distributed quantum
  computing}, Computer 49 (2016) 31.
\newblock \href {https://doi.org/10.1109/mc.2016.291}
  {\path{doi:10.1109/mc.2016.291}}.

\bibitem{Harrow2017}
A.~W. Harrow, A.~Montanaro, {\it Quantum computational supremacy}, Nature 549
  (2017) 203.
\newblock \href {https://doi.org/10.1038/nature23458}
  {\path{doi:10.1038/nature23458}}.

\bibitem{Arute2019}
F.~Arute, K.~Arya, R.~Babbush, D.~Bacon, J.~C. Bardin, R.~Barends, R.~Biswas,
  S.~Boixo, F.~G. Brandao, D.~A. Buell, et~al., {\it Quantum supremacy using a
  programmable superconducting processor}, Nature 574 (2019) 505.
\newblock \href {https://doi.org/https://doi.org/10.1038/s41586-019-1666-5}
  {\path{doi:https://doi.org/10.1038/s41586-019-1666-5}}.

\bibitem{Zhong2020}
H.-S. Zhong, H.~Wang, Y.-H. Deng, M.-C. Chen, L.-C. Peng, Y.-H. Luo, J.~Qin,
  D.~Wu, X.~Ding, Y.~Hu, P.~Hu, X.-Y. Yang, W.-J. Zhang, H.~Li, Y.~Li,
  X.~Jiang, L.~Gan, G.~Yang, L.~You, Z.~Wang, L.~Li, N.-L. Liu, C.-Y. Lu, J.-W.
  Pan, {\it Quantum computational advantage using photons}, Science 370 (2020)
  1460.
\newblock \href {https://doi.org/10.1126/science.abe8770}
  {\path{doi:10.1126/science.abe8770}}.

\bibitem{Arora2009}
S.~Arora, B.~Barak, {\it Computational complexity: a modern approach},
  Cambridge University Press, 2009.

\bibitem{Nielsen2010}
M.~A. Nielsen, I.~L. Chuang, {\it Quantum computation and quantum information},
  2010.
\newblock \href {https://doi.org/10.1017/CBO9780511976667}
  {\path{doi:10.1017/CBO9780511976667}}.

\bibitem{SOEPARNO2021}
H.~Soeparno, A.~S. Perbangsa, {\it Cloud Quantum Computing Concept and
  Development: A Systematic Literature Review}, Procedia Comput. Sci. 179
  (2021) 944.
\newblock \href {https://doi.org/10.1016/j.procs.2021.01.084}
  {\path{doi:10.1016/j.procs.2021.01.084}}.

\bibitem{Devitt2013}
S.~J. Devitt, W.~J. Munro, K.~Nemoto, {\it Quantum error correction for
  beginners}, Rep. Prog. Phys. 76 (2013) 076001.
\newblock \href {https://doi.org/10.1088/0034-4885/76/7/076001}
  {\path{doi:10.1088/0034-4885/76/7/076001}}.

\bibitem{Gadioli2020}
G.~G.~L. Guardia, {\it Quantum Error Correction}, Springer International
  Publishing, 2020.
\newblock \href {https://doi.org/10.1007/978-3-030-48551-1}
  {\path{doi:10.1007/978-3-030-48551-1}}.

\bibitem{Preskill2018}
J.~Preskill, {\it Quantum Computing in the NISQ era and beyond}, Quantum 2
  (2018) 79.
\newblock \href {https://doi.org/10.22331/q-2018-08-06-79}
  {\path{doi:10.22331/q-2018-08-06-79}}.

\bibitem{Bharti2021}
K.~Bharti, A.~Cervera-Lierta, T.~H. Kyaw, T.~Haug, S.~Alperin-Lea, A.~Anand,
  M.~Degroote, H.~Heimonen, J.~S. Kottmann, T.~Menke, W.-K. Mok, S.~Sim, L.-C.
  Kwek, A.~Aspuru-Guzik, {\it Noisy intermediate-scale quantum algorithms},
  Rev. Mod. Phys. 94 (2022) 015004.
\newblock \href {https://doi.org/10.1103/RevModPhys.94.015004}
  {\path{doi:10.1103/RevModPhys.94.015004}}.

\bibitem{Endo2021}
S.~Endo, Z.~Cai, S.~C. Benjamin, X.~Yuan, {\it Hybrid Quantum-Classical
  Algorithms and Quantum Error Mitigation}, J. Phys. Soc. Jpn. 90 (2021)
  032001.
\newblock \href {https://doi.org/10.7566/JPSJ.90.032001}
  {\path{doi:10.7566/JPSJ.90.032001}}.

\bibitem{McClean2016}
J.~R. McClean, J.~Romero, R.~Babbush, A.~Aspuru-Guzik, {\it The theory of
  variational hybrid quantum-classical algorithms}, New J. Phys. 18 (2016)
  023023.
\newblock \href {https://doi.org/10.1088/1367-2630/18/2/023023}
  {\path{doi:10.1088/1367-2630/18/2/023023}}.

\bibitem{McArdle2018a}
S.~McArdle, S.~Endo, A.~Aspuru-Guzik, S.~C. Benjamin, X.~Yuan, {\it Quantum
  computational chemistry}, Rev. Mod. Phys. 92 (2020) 015003.
\newblock \href {https://doi.org/10.1103/RevModPhys.92.015003}
  {\path{doi:10.1103/RevModPhys.92.015003}}.

\bibitem{Cao2019}
Y.~Cao, J.~Romero, J.~P. Olson, M.~Degroote, P.~D. Johnson, M.~Kieferov\'a,
  I.~D. Kivlichan, T.~Menke, B.~Peropadre, N.~P.~D. Sawaya, S.~Sim, L.~Veis,
  A.~Aspuru-Guzik, {\it Quantum Chemistry in the Age of Quantum Computing},
  Chem. Rev. 119 (2019) 10856.
\newblock \href {https://doi.org/10.1021/acs.chemrev.8b00803}
  {\path{doi:10.1021/acs.chemrev.8b00803}}.

\bibitem{Bauer2020}
B.~Bauer, S.~Bravyi, M.~Motta, G.~Kin-Lic~Chan, {\it Quantum Algorithms for
  Quantum Chemistry and Quantum Mater. Sci.}, Chem. Rev. 120 (2020) 12685.
\newblock \href {https://doi.org/10.1021/acs.chemrev.9b00829}
  {\path{doi:10.1021/acs.chemrev.9b00829}}.

\bibitem{Tilly2021a}
J.~Tilly, H.~Chen, S.~Cao, D.~Picozzi, K.~Setia, Y.~Li, E.~Grant, L.~Wossnig,
  I.~Rungger, G.~H. Booth, J.~Tennyson, Phys. Rep. 986 (2022) 1.
\newblock \href {https://doi.org/10.1016/j.physrep.2022.08.003}
  {\path{doi:10.1016/j.physrep.2022.08.003}}.

\bibitem{Claudino2022}
D.~Claudino, {\it The Basics of Quantum Computing for Chemists} (2022).
\newblock \href {http://arxiv.org/abs/2203.15063} {\path{arXiv:2203.15063}},
  \href {https://doi.org/10.48550/arxiv.2203.15063}
  {\path{doi:10.48550/arxiv.2203.15063}}.

\bibitem{Bassman2021}
L.~Bassman, M.~Urbanek, M.~Metcalf, J.~Carter, A.~F. Kemper, W.~A. de~Jong,
  {\it Simulating quantum materials with digital quantum computers}, Quantum
  Science and Technology 6~(4) (2021) 043002.
\newblock \href {http://arxiv.org/abs/2101.08836} {\path{arXiv:2101.08836}},
  \href {https://doi.org/10.1088/2058-9565/ac1ca6}
  {\path{doi:10.1088/2058-9565/ac1ca6}}.

\bibitem{Fedorov2021a}
D.~A. Fedorov, B.~Peng, N.~Govind, Y.~Alexeev, {\it VQE Method: A Short Survey
  and Recent Developments} (2021).
\newblock \href {http://arxiv.org/abs/2103.08505} {\path{arXiv:2103.08505}},
  \href {https://doi.org/10.48550/arxiv.2103.08505}
  {\path{doi:10.48550/arxiv.2103.08505}}.

\bibitem{Zeng2019}
B.~Zeng, X.~Chen, D.-L. Zhou, X.-G. Wen, {\it Quantum Information Meets Quantum
  Matter}, Springer New York, 2019.
\newblock \href {https://doi.org/10.1007/978-1-4939-9084-9}
  {\path{doi:10.1007/978-1-4939-9084-9}}.

\bibitem{Saffman2016}
M.~Saffman, {\it Quantum computing with atomic qubits and Rydberg interactions:
  progress and challenges}, J. Phys. B: At., Mol. Opt. Phys. 49~(20) (2016)
  202001.
\newblock \href {https://doi.org/10.1088/0953-4075/49/20/202001}
  {\path{doi:10.1088/0953-4075/49/20/202001}}.

\bibitem{Bian2021}
T.~Bian, S.~Kais, {\it Quantum computing for atomic and molecular resonances},
  J. Chem. Phys 154 (2021) 194107.
\newblock \href {https://doi.org/10.1063/5.0040477}
  {\path{doi:10.1063/5.0040477}}.

\bibitem{Graham2022}
T.~M. Graham, Y.~Song, J.~Scott, C.~Poole, L.~Phuttitarn, K.~Jooya, P.~Eichler,
  X.~Jiang, A.~Marra, B.~Grinkemeyer, M.~Kwon, M.~Ebert, J.~Cherek, M.~T.
  Lichtman, M.~Gillette, J.~Gilbert, D.~Bowman, T.~Ballance, C.~Campbell, E.~D.
  Dahl, O.~Crawford, N.~S. Blunt, B.~Rogers, T.~Noel, .~M. Saffman, {\it
  Multi-qubit entanglement and algorithms on a neutral-atom quantum computer},
  Nature 604 (2022) 457.
\newblock \href {https://doi.org/10.1038/s41586-022-04603-6}
  {\path{doi:10.1038/s41586-022-04603-6}}.

\bibitem{Bluvstein2022}
D.~Bluvstein, H.~Levine, G.~Semeghini, T.~T. Wang, S.~Ebadi, M.~Kalinowski,
  A.~Keesling, N.~Maskara, H.~Pichler, M.~Greiner, V.~Vuletić, M.~D. Lukin,
  {\it A quantum processor based on coherent transport of entangled atom arrays
  Entanglement transport in atom arrays}, Nature 604 (2022) 451.
\newblock \href {https://doi.org/10.1038/s41586-022-04592-6}
  {\path{doi:10.1038/s41586-022-04592-6}}.

\bibitem{KerenLi2019}
K.~Li, Y.~Li, M.~Han, S.~Lu, J.~Zhou, D.~Ruan, G.~Long, Y.~Wan, D.~Lu, B.~Zeng,
  R.~Laflamme, {\it Quantum spacetime on a quantum simulator}, Communications
  Physics 2~(1) (2019) 122.
\newblock \href {http://arxiv.org/abs/1712.08711} {\path{arXiv:1712.08711}},
  \href {https://doi.org/10.1038/s42005-019-0218-5}
  {\path{doi:10.1038/s42005-019-0218-5}}.

\bibitem{mielczarek2019}
J.~Mielczarek, {\it Spin Foam Vertex Amplitudes on Quantum
  Computer—Preliminary Results}, Universe 5 (2019) 179.
\newblock \href {https://doi.org/10.3390/universe5080179}
  {\path{doi:10.3390/universe5080179}}.

\bibitem{Czelusta2021}
G.~Czelusta, J.~Mielczarek, {\it Quantum simulations of a qubit of space},
  Phys. Rev. D 103 (2021) 046001.
\newblock \href {https://doi.org/10.1103/PhysRevD.103.046001}
  {\path{doi:10.1103/PhysRevD.103.046001}}.

\bibitem{Mielczarek2021}
J.~Mielczarek, {\it Prelude to Simulations of Loop Quantum Gravity on Adiabatic
  Quantum Computers}, Frontiers in Astronomy and Space Sciences 8 (2021).
\newblock \href {https://doi.org/10.3389/fspas.2021.571282}
  {\path{doi:10.3389/fspas.2021.571282}}.

\bibitem{Mielczarek2018}
J.~Mielczarek, {\it Quantum Gravity on a Quantum Chip}~(March) (2018).
\newblock \href {http://arxiv.org/abs/1803.10592} {\path{arXiv:1803.10592}},
  \href {https://doi.org/10.48550/arXiv.1803.10592}
  {\path{doi:10.48550/arXiv.1803.10592}}.

\bibitem{Zhang2021}
D.-B. Zhang, H.~Xing, H.~Yan, E.~Wang, S.-L. Zhu, {\it Selected topics of
  quantum computing for nuclear physics*}, Chin. Phys. B 30 (2021) 020306.
\newblock \href {https://doi.org/10.1088/1674-1056/abd761}
  {\path{doi:10.1088/1674-1056/abd761}}.

\bibitem{Stetcu2022}
I.~Stetcu, A.~Baroni, J.~Carlson, {\it Variational approaches to constructing
  the many-body nuclear ground state for quantum computing}, Phys. Rev. C 105
  (2022) 064308.
\newblock \href {https://doi.org/10.1103/PhysRevC.105.064308}
  {\path{doi:10.1103/PhysRevC.105.064308}}.

\bibitem{Roggero2020}
A.~Roggero, C.~Gu, A.~Baroni, T.~Papenbrock, {\it Preparation of excited states
  for nuclear dynamics on a quantum computer}, Phys. Rev. C 102 (2020) 064624.
\newblock \href {https://doi.org/10.1103/PhysRevC.102.064624}
  {\path{doi:10.1103/PhysRevC.102.064624}}.

\bibitem{Hobday2022}
I.~A. Hobday, P.~Stevenson, J.~Benstead, {\it Quantum computing calculations
  for nuclear structure and nuclear data}, in: S.~Ducci, E.~Diamanti, N.~Treps,
  S.~Whitlock (Eds.), Quantum Technologies 2022, SPIE, 2022, p.~61.
\newblock \href {http://arxiv.org/abs/2205.05576} {\path{arXiv:2205.05576}},
  \href {https://doi.org/10.1117/12.2632782} {\path{doi:10.1117/12.2632782}}.

\bibitem{Romero2022}
A.~M. Romero, J.~Engel, H.~L. Tang, S.~E. Economou, {\it Solving nuclear
  structure problems with the adaptive variational quantum algorithm}, Phys.
  Rev. C 105 (2022) 064317.
\newblock \href {https://doi.org/10.1103/PhysRevC.105.064317}
  {\path{doi:10.1103/PhysRevC.105.064317}}.

\bibitem{Lipparini2008}
E.~Lipparini, {\it Modern many-particle physics: atomic gases, nanostructures
  and quantum liquids}, World Scientific Publishing Company, 2008.

\bibitem{Saad2011}
Y.~Saad, {\it Numerical methods for large eigenvalue problems: revised
  edition}, SIAM, 2011.

\bibitem{schollwock_2011}
U.~Schollwöck, {\it The density-matrix renormalization group in the age of
  matrix product states}, Ann. Phys. 326 (2011) 96.
\newblock \href {https://doi.org/10.1016/j.aop.2010.09.012}
  {\path{doi:10.1016/j.aop.2010.09.012}}.

\bibitem{Ran2020}
S.-J. Ran, E.~Tirrito, C.~Peng, C.~Xi, L.~Tagliacozzo, G.~Su, M.~Lewenstein,
  {\it Tensor Network Contractions: Methods and Applications to Quantum
  Many-Body Systems}, 2020.
\newblock \href {https://doi.org/10.1007/978-3-030-34489-4}
  {\path{doi:10.1007/978-3-030-34489-4}}.

\bibitem{Cirac2021}
J.~I. Cirac, D.~P\'erez-Garc\'{\i}a, N.~Schuch, F.~Verstraete, {\it Matrix
  product states and projected entangled pair states: Concepts, symmetries,
  theorems}, Rev. Mod. Phys. 93 (2021) 045003.
\newblock \href {https://doi.org/10.1103/RevModPhys.93.045003}
  {\path{doi:10.1103/RevModPhys.93.045003}}.

\bibitem{David1999}
C.~{David Sherrill}, H.~F. Schaefer,
  \href{https://www.sciencedirect.com/science/article/pii/S0065327608605328}{The
  configuration interaction method: Advances in highly correlated approaches},
  Vol.~34 of Advances in Quantum Chemistry, Academic Press, 1999, pp. 143--269.
\newblock \href {https://doi.org/https://doi.org/10.1016/S0065-3276(08)60532-8}
  {\path{doi:https://doi.org/10.1016/S0065-3276(08)60532-8}}.
\newline\urlprefix\url{https://www.sciencedirect.com/science/article/pii/S0065327608605328}

\bibitem{Hinze67}
J.~Hinze, C.~C.~J. Roothaan,
  \href{https://doi.org/10.1143/PTPS.40.37}{{Multi-Configuration
  Self-Consistent-Field Theory*)}}, Progress of Theoretical Physics Supplement
  40 (1967) 37.
\newblock \href
  {http://arxiv.org/abs/https://academic.oup.com/ptps/article-pdf/doi/10.1143/PTPS.40.37/5289273/40-37.pdf}
  {\path{arXiv:https://academic.oup.com/ptps/article-pdf/doi/10.1143/PTPS.40.37/5289273/40-37.pdf}},
  \href {https://doi.org/10.1143/PTPS.40.37} {\path{doi:10.1143/PTPS.40.37}}.
\newline\urlprefix\url{https://doi.org/10.1143/PTPS.40.37}

\bibitem{schmidt1998}
M.~W. Schmidt, M.~S. Gordon, The construction and interpretation of mcscf
  wavefunctions, Annual review of physical chemistry 49~(1) (1998) 233.

\bibitem{jensen2007}
F.~Jensen, Introduction to computational chemistry, willey, Chichester, UK
  (2007).

\bibitem{cramer2013}
C.~J. Cramer, Essentials of computational chemistry: theories and models, John
  Wiley \& Sons, 2013.

\bibitem{shavitt2009}
I.~Shavitt, R.~J. Bartlett, Many-body methods in chemistry and physics: MBPT
  and coupled-cluster theory, Cambridge university press, 2009.

\bibitem{Werner2009}
W.~Kutzelnigg,
  \href{https://onlinelibrary.wiley.com/doi/abs/10.1002/qua.22384}{How
  many-body perturbation theory (mbpt) has changed quantum chemistry},
  International Journal of Quantum Chemistry 109~(15) (2009) 3858.
\newblock \href
  {http://arxiv.org/abs/https://onlinelibrary.wiley.com/doi/pdf/10.1002/qua.22384}
  {\path{arXiv:https://onlinelibrary.wiley.com/doi/pdf/10.1002/qua.22384}},
  \href {https://doi.org/https://doi.org/10.1002/qua.22384}
  {\path{doi:https://doi.org/10.1002/qua.22384}}.
\newline\urlprefix\url{https://onlinelibrary.wiley.com/doi/abs/10.1002/qua.22384}

\bibitem{Bartlett2007}
R.~J. Bartlett, M.~Musia\l, {\it Coupled-cluster theory in quantum chemistry},
  Rev. Mod. Phys. 79 (2007) 291.
\newblock \href {https://doi.org/10.1103/RevModPhys.79.291}
  {\path{doi:10.1103/RevModPhys.79.291}}.

\bibitem{Evangelista2018}
F.~A. Evangelista, \href{https://doi.org/10.1063/1.5039496}{{Perspective:
  Multireference coupled cluster theories of dynamical electron correlation}},
  The Journal of Chemical Physics 149~(3) (2018) 030901.
\newblock \href
  {http://arxiv.org/abs/https://pubs.aip.org/aip/jcp/article-pdf/doi/10.1063/1.5039496/15545193/030901\_1\_online.pdf}
  {\path{arXiv:https://pubs.aip.org/aip/jcp/article-pdf/doi/10.1063/1.5039496/15545193/030901\_1\_online.pdf}},
  \href {https://doi.org/10.1063/1.5039496} {\path{doi:10.1063/1.5039496}}.
\newline\urlprefix\url{https://doi.org/10.1063/1.5039496}

\bibitem{Baiardi2020}
A.~Baiardi, M.~Reiher, { \it The density matrix renormalization group in
  chemistry and Mol. Phys.: Recent developments and new challenges}, The J.
  Chem. Phys. 152~(4) (2020) 040903.
\newblock \href {http://arxiv.org/abs/1910.00137} {\path{arXiv:1910.00137}},
  \href {https://doi.org/10.1063/1.5129672} {\path{doi:10.1063/1.5129672}}.

\bibitem{Bohr1998}
A.~N. Bohr, B.~R. Mottelson, {\it Nuclear Structure (in 2 volumes)}, World
  Scientific Publishing Company, 1998.

\bibitem{Ring80}
P.~Ring, P.~Schuck, {\it The nuclear many-body problem}, Springer-Verlag, 1980.

\bibitem{Epelbaum2009}
E.~Epelbaum, H.-W. Hammer, U.-G. Mei\ss{}ner, {\it Modern theory of nuclear
  forces}, Rev. Mod. Phys. 81 (2009) 1773.
\newblock \href {https://doi.org/10.1103/RevModPhys.81.1773}
  {\path{doi:10.1103/RevModPhys.81.1773}}.

\bibitem{Machleidt2011}
R.~Machleidt, D.~Entem, {\it Chiral effective field theory and nuclear forces},
  Phys. Rep. 503 (2011) 1.
\newblock \href {https://doi.org/10.1016/j.physrep.2011.02.001}
  {\path{doi:10.1016/j.physrep.2011.02.001}}.

\bibitem{Hammer2013}
H.-W. Hammer, A.~Nogga, A.~Schwenk, {\it Colloquium: Three-body forces: From
  cold atoms to nuclei}, Rev. Mod. Phys. 85 (2013) 197.
\newblock \href {https://doi.org/10.1103/RevModPhys.85.197}
  {\path{doi:10.1103/RevModPhys.85.197}}.

\bibitem{Machleidt2016}
R.~Machleidt, F.~Sammarruca, {\it Chiral EFT based nuclear forces: achievements
  and challenges}, Phys. Scr. 91 (2016) 083007.
\newblock \href {https://doi.org/10.1088/0031-8949/91/8/083007}
  {\path{doi:10.1088/0031-8949/91/8/083007}}.

\bibitem{Hammer2020}
H.-W. Hammer, S.~K\"onig, U.~van Kolck, {\it Nuclear effective field theory:
  Status and perspectives}, Rev. Mod. Phys. 92 (2020) 025004.
\newblock \href {https://doi.org/10.1103/RevModPhys.92.025004}
  {\path{doi:10.1103/RevModPhys.92.025004}}.

\bibitem{Lipkin1965}
H.~Lipkin, N.~Meshkov, A.~Glick, {\it Validity of many-body approximation
  methods for a solvable model: (I). Exact solutions and perturbation theory},
  Nuclear Physics 62 (1965) 188.
\newblock \href {https://doi.org/10.1016/0029-5582(65)90862-X}
  {\path{doi:10.1016/0029-5582(65)90862-X}}.

\bibitem{Vondelft2001}
J.~von Delft, D.~Ralph, {\it Spectroscopy of discrete energy levels in
  ultrasmall metallic grains}, Phys. Rep. 345 (2001) 61.
\newblock \href {https://doi.org/10.1016/S0370-1573(00)00099-5}
  {\path{doi:10.1016/S0370-1573(00)00099-5}}.

\bibitem{Lacroix2010}
D.~Lacroix, {\it Introduction - Strong interaction in the nuclear medium: new
  trends} (2010).
\newblock \href {https://doi.org/10.48550/ARXIV.1001.5001}
  {\path{doi:10.48550/ARXIV.1001.5001}}.

\bibitem{Bender2003}
M.~Bender, P.-H. Heenen, P.-G. Reinhard, {\it Self-consistent mean-field models
  for nuclear structure}, Rev. Mod. Phys. 75 (2003) 121.
\newblock \href {https://doi.org/10.1103/RevModPhys.75.121}
  {\path{doi:10.1103/RevModPhys.75.121}}.

\bibitem{Nakatsukasa2016}
T.~Nakatsukasa, K.~Matsuyanagi, M.~Matsuo, K.~Yabana, {\it Time-dependent
  density-functional description of nuclear dynamics}, Rev. Mod. Phys. 88
  (2016) 045004.
\newblock \href {https://doi.org/10.1103/RevModPhys.88.045004}
  {\path{doi:10.1103/RevModPhys.88.045004}}.

\bibitem{Colo2020}
G.~Col\'o, {\it Nuclear density functional theory}, Adv. Phys.: X 5 (2020)
  1740061.
\newblock \href {https://doi.org/10.1080/23746149.2020.1740061}
  {\path{doi:10.1080/23746149.2020.1740061}}.

\bibitem{Caurier2005}
E.~Caurier, G.~Mart\'{\i}nez-Pinedo, F.~Nowacki, A.~Poves, A.~P. Zuker, {\it
  The shell model as a unified view of nuclear structure}, Rev. Mod. Phys. 77
  (2005) 427.
\newblock \href {https://doi.org/10.1103/RevModPhys.77.427}
  {\path{doi:10.1103/RevModPhys.77.427}}.

\bibitem{Nowacki2021}
F.~Nowacki, A.~Obertelli, A.~Poves, {\it The neutron-rich edge of the nuclear
  landscape: Experiment and theory.}, Prog. Part. Nucl. Phys. 120 (2021)
  103866.
\newblock \href {https://doi.org/10.1016/j.ppnp.2021.103866}
  {\path{doi:10.1016/j.ppnp.2021.103866}}.

\bibitem{Bogner2010}
S.~Bogner, R.~Furnstahl, A.~Schwenk, {\it From low-momentum interactions to
  nuclear structure}, Prog. Part. Nucl. Phys. 65 (2010) 94.
\newblock \href {https://doi.org/10.1016/j.ppnp.2010.03.001}
  {\path{doi:10.1016/j.ppnp.2010.03.001}}.

\bibitem{Quaglioni2008}
S.~Quaglioni, P.~Navr\'atil, {\it Ab initio no-core shell model and microscopic
  reactions: Recent achievements}, Few-Body Syst. 44 (2008) 337--339.
\newblock \href {https://doi.org/10.1007/s00601-008-0322-7}
  {\path{doi:10.1007/s00601-008-0322-7}}.

\bibitem{Navratil2009}
P.~Navr\'atil, S.~Quaglioni, I.~Stetcu, B.~R. Barrett, {\it Recent developments
  in no-core shell-model calculations}, J. Phys. G: Nucl. Part. Phys. 36 (2009)
  083101.
\newblock \href {https://doi.org/10.1088/0954-3899/36/8/083101}
  {\path{doi:10.1088/0954-3899/36/8/083101}}.

\bibitem{Pudliner1997}
B.~S. Pudliner, V.~R. Pandharipande, J.~Carlson, S.~C. Pieper, R.~B. Wiringa,
  {\it Quantum Monte Carlo calculations of nuclei with $A\sim 7$}, Phys. Rev. C
  56 (1997) 1720.
\newblock \href {https://doi.org/10.1103/PhysRevC.56.1720}
  {\path{doi:10.1103/PhysRevC.56.1720}}.

\bibitem{Wiringa1998}
R.~Wiringa, {\it Quantum Monte Carlo calculations for light nuclei}, Nucl.
  Phys. A 631 (1998) 70.
\newblock \href {https://doi.org/10.1016/S0375-9474(98)00016-5}
  {\path{doi:10.1016/S0375-9474(98)00016-5}}.

\bibitem{Wiringa2000}
R.~B. Wiringa, S.~C. Pieper, J.~Carlson, V.~R. Pandharipande, {\it Quantum
  Monte Carlo calculations of $A=8$ nuclei}, Phys. Rev. C 62 (2000) 014001.
\newblock \href {https://doi.org/10.1103/PhysRevC.62.014001}
  {\path{doi:10.1103/PhysRevC.62.014001}}.

\bibitem{Dickhoff2004}
W.~Dickhoff, C.~Barbieri, {\it Self-consistent Green's function method for
  nuclei and nuclear matter}, Prog. in Part. and Nucl. Phys. 52 (2004) 377.
\newblock \href {https://doi.org/10.1016/j.ppnp.2004.02.038}
  {\path{doi:10.1016/j.ppnp.2004.02.038}}.

\bibitem{Soma2020}
V.~Som\`a, {\it Self-Consistent Green's Function Theory for Atomic Nuclei},
  Front. Phys. 8 (2020).
\newblock \href {https://doi.org/10.3389/fphy.2020.00340}
  {\path{doi:10.3389/fphy.2020.00340}}.

\bibitem{Hagen2014}
G.~Hagen, T.~Papenbrock, M.~Hjorth-Jensen, D.~J. Dean, {\it Coupled-cluster
  computations of atomic nuclei}, Rep. Prog. Phys. 77 (2014) 096302.
\newblock \href {https://doi.org/10.1088/0034-4885/77/9/096302}
  {\path{doi:10.1088/0034-4885/77/9/096302}}.

\bibitem{Tichai2020}
A.~Tichai, R.~Roth, T.~Duguet, {\it Many-Body Perturbation Theories for Finite
  Nuclei}, Front. Phys. 8 (2020).
\newblock \href {https://doi.org/10.3389/fphy.2020.00164}
  {\path{doi:10.3389/fphy.2020.00164}}.

\bibitem{Tichai2011}
A.~Tichai, P.~Arthuis, T.~Duguet, H.~Hergert, V.~Somà, R.~Roth, {\it
  Bogoliubov many-body perturbation theory for open-shell nuclei}, Phys. Lett.
  B 786 (2018) 195.
\newblock \href {https://doi.org/10.1016/j.physletb.2018.09.044}
  {\path{doi:10.1016/j.physletb.2018.09.044}}.

\bibitem{Soma2011}
V.~Som\`a, T.~Duguet, C.~Barbieri, {\it Ab initio self-consistent
  Gorkov-Green's function calculations of semimagic nuclei: Formalism at second
  order with a two-nucleon interaction}, Phys. Rev. C 84 (2011) 064317.
\newblock \href {https://doi.org/10.1103/PhysRevC.84.064317}
  {\path{doi:10.1103/PhysRevC.84.064317}}.

\bibitem{Henderson2014}
T.~M. Henderson, G.~E. Scuseria, J.~Dukelsky, A.~Signoracci, T.~Duguet, {\it
  Quasiparticle coupled cluster theory for pairing interactions}, Phys. Rev. C
  89 (2014) 054305.
\newblock \href {https://doi.org/10.1103/PhysRevC.89.054305}
  {\path{doi:10.1103/PhysRevC.89.054305}}.

\bibitem{Duguet2016}
T.~Duguet, A.~Signoracci, {\it Symmetry broken and restored coupled-cluster
  theory: II. Global gauge symmetry and particle number}, J. Phys. G: Nucl.
  Part. Phys. 44 (2016) 015103.
\newblock \href {https://doi.org/10.1088/0954-3899/44/1/015103}
  {\path{doi:10.1088/0954-3899/44/1/015103}}.

\bibitem{Demol2021}
P.~Demol, M.~Frosini, A.~Tichai, V.~Somà, T.~Duguet, {\it Bogoliubov many-body
  perturbation theory under constraint}, Ann. Phys. 424 (2021) 168358.
\newblock \href {https://doi.org/10.1016/j.aop.2020.168358}
  {\path{doi:10.1016/j.aop.2020.168358}}.

\bibitem{Dumitrescu2018}
E.~F. Dumitrescu, A.~J. McCaskey, G.~Hagen, G.~R. Jansen, T.~D. Morris,
  T.~Papenbrock, R.~C. Pooser, D.~J. Dean, P.~Lougovski, {\it Cloud Quantum
  Computing of an Atomic Nucleus}, Phys. Rev. Lett. 120 (2018) 210501.
\newblock \href {https://doi.org/10.1103/PhysRevLett.120.210501}
  {\path{doi:10.1103/PhysRevLett.120.210501}}.

\bibitem{Lu2019}
H.-H. Lu, N.~Klco, J.~M. Lukens, T.~D. Morris, A.~Bansal, A.~Ekstr\"om,
  G.~Hagen, T.~Papenbrock, A.~M. Weiner, M.~J. Savage, P.~Lougovski, {\it
  Simulations of subatomic many-body physics on a quantum frequency processor},
  Phys. Rev. A 100 (2019) 012320.
\newblock \href {https://doi.org/10.1103/PhysRevA.100.012320}
  {\path{doi:10.1103/PhysRevA.100.012320}}.

\bibitem{Kiss2022}
O.~Kiss, M.~Grossi, P.~Lougovski, F.~Sanchez, S.~Vallecorsa, T.~Papenbrock,
  {\it Quantum computing of the $^{6}\mathrm{Li}$ nucleus via ordered unitary
  coupled clusters}, Phys. Rev. C 106 (2022) 034325.
\newblock \href {https://doi.org/10.1103/PhysRevC.106.034325}
  {\path{doi:10.1103/PhysRevC.106.034325}}.

\bibitem{Roggero2020b}
A.~Roggero, A.~C.~Y. Li, J.~Carlson, R.~Gupta, G.~N. Perdue, {\it Quantum
  computing for neutrino-nucleus scattering}, Phys. Rev. D 101 (2020) 074038.
\newblock \href {https://doi.org/10.1103/PhysRevD.101.074038}
  {\path{doi:10.1103/PhysRevD.101.074038}}.

\bibitem{Lacroix2020}
D.~Lacroix, {\it Symmetry-assisted preparation of entangled many-body states on
  a quantum computer}, Phys. Rev. Lett. 125 (2020) 230502.
\newblock \href {https://doi.org/10.1103/PhysRevLett.125.230502}
  {\path{doi:10.1103/PhysRevLett.125.230502}}.

\bibitem{Dimatteo2021}
O.~Di~Matteo, A.~McCoy, P.~Gysbers, T.~Miyagi, R.~M. Woloshyn, P.~Navr\'atil,
  {\it Improving Hamiltonian encodings with the Gray code}, Phys. Rev. A 103
  (2021) 042405.
\newblock \href {https://doi.org/10.1103/PhysRevA.103.042405}
  {\path{doi:10.1103/PhysRevA.103.042405}}.

\bibitem{Siwach2021}
P.~Siwach, P.~Arumugam, {\it Quantum simulation of nuclear Hamiltonian with a
  generalized transformation for Gray code encoding}, Phys. Rev. C 104 (2021)
  034301.
\newblock \href {https://doi.org/10.1103/PhysRevC.104.034301}
  {\path{doi:10.1103/PhysRevC.104.034301}}.

\bibitem{Du2021}
W.~Du, J.~P. Vary, X.~Zhao, W.~Zuo, {\it Quantum simulation of nuclear
  inelastic scattering}, Phys. Rev. A 104 (2021) 012611.
\newblock \href {https://doi.org/10.1103/PhysRevA.104.012611}
  {\path{doi:10.1103/PhysRevA.104.012611}}.

\bibitem{Siwach2021b}
P.~Siwach, D.~Lacroix, {\it Filtering states with total spin on a quantum
  computer}, Phys. Rev. A 104 (2021) 062435.
\newblock \href {https://doi.org/10.1103/PhysRevA.104.062435}
  {\path{doi:10.1103/PhysRevA.104.062435}}.

\bibitem{Cervia2021}
M.~J. Cervia, A.~B. Balantekin, S.~N. Coppersmith, C.~W. Johnson, P.~J. Love,
  C.~Poole, K.~Robbins, M.~Saffman, {\it Lipkin model on a quantum computer},
  Phys. Rev. C 104 (2021) 024305.
\newblock \href {https://doi.org/10.1103/PhysRevC.104.024305}
  {\path{doi:10.1103/PhysRevC.104.024305}}.

\bibitem{ruiz-guzman2021}
E.~A.~R. Guzman, D.~Lacroix, {\it Calculation of generating function in
  many-body systems with quantum computers: technical challenges and use in
  hybrid quantum-classical methods}, arXiv:2104.08181 (2021).
\newblock \href {https://doi.org/10.48550/ARXIV.2104.08181}
  {\path{doi:10.48550/ARXIV.2104.08181}}.

\bibitem{Hlatshwayo2022}
M.~Q. Hlatshwayo, Y.~Zhang, H.~Wibowo, R.~LaRose, D.~Lacroix, E.~Litvinova,
  {\it Simulating excited states of the Lipkin model on a quantum computer},
  Phys. Rev. C 106 (2022) 024319.
\newblock \href {https://doi.org/10.1103/PhysRevC.106.024319}
  {\path{doi:10.1103/PhysRevC.106.024319}}.

\bibitem{Perezfernandez2022}
P.~Pérez-Fernández, J.-M. Arias, J.-E. García-Ramos, L.~Lamata, {\it A
  digital quantum simulation of the Agassi model}, Phys. Lett. B 829 (2022)
  137133.
\newblock \href {https://doi.org/10.1016/j.physletb.2022.137133}
  {\path{doi:10.1016/j.physletb.2022.137133}}.

\bibitem{RuizGuzman2022}
E.~A. Ruiz~Guzman, D.~Lacroix, {\it Accessing ground-state and excited-state
  energies in a many-body system after symmetry restoration using quantum
  computers}, Phys. Rev. C 105 (2022) 024324.
\newblock \href {https://doi.org/10.1103/PhysRevC.105.024324}
  {\path{doi:10.1103/PhysRevC.105.024324}}.

\bibitem{Siwach2022}
P.~Siwach, P.~Arumugam, {\it Quantum computation of nuclear observables
  involving linear combinations of unitary operators}, Phys. Rev. C 105 (2022)
  064318.
\newblock \href {https://doi.org/10.1103/PhysRevC.105.064318}
  {\path{doi:10.1103/PhysRevC.105.064318}}.

\bibitem{Hodbay2022}
I.~Hobday, P.~Stevenson, J.~Benstead, {\it Variance minimisation on a quantum
  computer for nuclear structure}, arxiv:2209.07820. (2022).
\newblock \href {https://doi.org/10.48550/ARXIV.2209.07820}
  {\path{doi:10.48550/ARXIV.2209.07820}}.

\bibitem{Perez2023}
A.~Pérez-Obiol, A.~M. Romero, J.~Menéndez, A.~Rios, A.~García-Sáez,
  B.~Juliá-Díaz, Nuclear shell-model simulation in digital quantum computers,
  arxiv:2302.03641. (2023).
\newblock \href {https://doi.org/10.48550/ARXIV.2302.03641}
  {\path{doi:10.48550/ARXIV.2302.03641}}.

\bibitem{Robin2023}
C.~E.~P. Robin, M.~J. Savage, Quantum simulations in effective model spaces
  (i): Hamiltonian learning-vqe using digital quantum computers and application
  to the lipkin-meshkov-glick model, arxiv:2301.05976 (2023).
\newblock \href {https://doi.org/10.48550/ARXIV.2301.05976}
  {\path{doi:10.48550/ARXIV.2301.05976}}.

\bibitem{Beck2023}
D.~Beck, et~al, Quantum information science and technology for nuclear physics.
  input into u.s. long-range planning, 2023, arxiv:2301.2303.00113. (2023).
\newblock \href {https://doi.org/10.48550/ARXIV.2303.00113}
  {\path{doi:10.48550/ARXIV.2303.00113}}.

\bibitem{Bonitz2016}
M.~Bonitz, {\it Quantum kinetic theory}, Vol. 412, Springer, 2016.
\newblock \href {https://doi.org/10.1007/978-3-319-24121-0}
  {\path{doi:10.1007/978-3-319-24121-0}}.

\bibitem{Bruus2002}
H.~Bruus, K.~Flensberg, {\it Introduction to many-body quantum theory in
  Condens. Matter Phys.}, Oxford Graduate Texts, 2004.

\bibitem{Gottesman1998}
D.~Gottesman, {\it Theory of fault-tolerant quantum computation}, Phys. Rev. A
  57 (1998) 127.
\newblock \href {https://doi.org/10.1103/PhysRevA.57.127}
  {\path{doi:10.1103/PhysRevA.57.127}}.

\bibitem{Gottesman1998a}
D.~Gottesman, {\it The Heisenberg representation of quantum computers},
  arXiv:9807006 (1998).
\newblock \href {https://doi.org/10.48550/ARXIV.QUANT-PH/9807006}
  {\path{doi:10.48550/ARXIV.QUANT-PH/9807006}}.

\bibitem{Aaronson2004}
S.~Aaronson, D.~Gottesman, {\it Improved simulation of stabilizer circuits},
  Phys. Rev. A 70 (2004) 052328.
\newblock \href {https://doi.org/10.1103/PhysRevA.70.052328}
  {\path{doi:10.1103/PhysRevA.70.052328}}.

\bibitem{BAEURLE2004}
S.~Baeurle, {\it Grand canonical auxiliary field Monte Carlo: a new technique
  for simulating open systems at high density}, Comput. Phys. Commun. 157
  (2004) 201.
\newblock \href {https://doi.org/10.1016/j.comphy.2003.11.001}
  {\path{doi:10.1016/j.comphy.2003.11.001}}.

\bibitem{Watrous2012a}
J.~Watrous, {\it Quantum computational complexity}, Springer New York, 2012, p.
  2361.
\newblock \href {https://doi.org/10.1007/978-1-4614-1800-9_147}
  {\path{doi:10.1007/978-1-4614-1800-9_147}}.

\bibitem{Watson2022}
J.~D. Watson, T.~S. Cubitt, {\it Computational complexity of the ground state
  energy density problem}, ACM, 2022, p. 764.
\newblock \href {https://doi.org/10.1145/3519935.3520052}
  {\path{doi:10.1145/3519935.3520052}}.

\bibitem{Bernstein1997}
E.~Bernstein, U.~Vazirani, {\it Quantum complexity theory}, SIAM J. Comput. 26
  (1997) 1411.
\newblock \href {https://doi.org/10.1137/S0097539796300921}
  {\path{doi:10.1137/S0097539796300921}}.

\bibitem{Shor1994}
P.~Shor, {\it Algorithms for quantum computation: discrete logarithms and
  factoring}, in: Proceedings 35th Annual Symposium on Foundations of Computer
  Science, 1994, p. 124.
\newblock \href {https://doi.org/10.1109/SFCS.1994.365700}
  {\path{doi:10.1109/SFCS.1994.365700}}.

\bibitem{Kempe2008}
J.~Kempe, A.~Kitaev, O.~Regev, {\it The complexity of the local hamiltonian
  problem}, arXiv:0406180 (2004) 372\href
  {https://doi.org/10.48550/ARXIV.QUANT-PH/0406180}
  {\path{doi:10.48550/ARXIV.QUANT-PH/0406180}}.

\bibitem{Oliveira2008}
R.~Oliveira, B.~M. Terhal, {\it The complexity of quantum spin systems on a
  two-dimensional square lattice}, Quant. Inf. Comp. 8 (2005) 900.
\newblock \href {https://doi.org/10.5555/2016985.2016987}
  {\path{doi:10.5555/2016985.2016987}}.

\bibitem{Bravyi2015}
S.~Bravyi, {\it Monte carlo simulation of stoquastic hamiltonians}, Quant. Inf.
  Comp. 15~(13/14) (2015) 1122.
\newblock \href {https://doi.org/10.48550/arXiv.1402.2295}
  {\path{doi:10.48550/arXiv.1402.2295}}.

\bibitem{Schuch2009}
N.~Schuch, F.~Verstraete, {\it Computational complexity of interacting
  electrons and fundamental limitations of density functional theory}, Nat.
  Phys. 5 (2009) 732.
\newblock \href {https://doi.org/10.1038/nphys1370}
  {\path{doi:10.1038/nphys1370}}.

\bibitem{Childs2013}
A.~M. Childs, D.~Gosset, Z.~Webb, {\it The Bose-Hubbard model is QMA-complete},
  in: J.~Esparza, P.~Fraigniaud, T.~Husfeldt, E.~Koutsoupias (Eds.), Automata,
  Languages, and Programming, Springer Berlin Heidelberg, 2014, p. 308.
\newblock \href {https://doi.org/10.1007/978-3-662-43948-7_26}
  {\path{doi:10.1007/978-3-662-43948-7_26}}.

\bibitem{Leblanc2015}
J.~P.~F. LeBlanc, A.~E. Antipov, F.~Becca, I.~W. Bulik, G.~K.-L. Chan, C.-m.
  Chung, Y.~Deng, M.~Ferrero, T.~M. Henderson, C.~A. Jim\'enez-Hoyos, E.~Kozik,
  X.-w. Liu, A.~J. Millis, N.~V. Prokof'ev, M.~Qin, G.~E. Scuseria, H.~Shi,
  B.~V. Svistunov, L.~F. Tocchio, I.~S. Tupitsyn, S.~R. White, S.~Zhang, B.-X.
  Zheng, Z.~Zhu, E.~Gull, {\it Solutions of the two-dimensional Hubbard model:
  benchmarks and results from a wide range of numerical algorithms}, Phys. Rev.
  X 5 (2015) 041041.
\newblock \href {https://doi.org/10.1103/PhysRevX.5.041041}
  {\path{doi:10.1103/PhysRevX.5.041041}}.

\bibitem{Lee2022}
S.~Lee, J.~Lee, H.~Zhai, Y.~Tong, A.~M. Dalzell, A.~Kumar, P.~Helms, J.~Gray,
  Z.-H. Cui, W.~Liu, M.~Kastoryano, R.~Babbush, J.~Preskill, D.~R. Reichman,
  E.~T. Campbell, E.~F. Valeev, L.~Lin, G.~K.-L. Chan, {\it Is there evidence
  for exponential quantum advantage in quantum chemistry?}, arXiv:2208.02199
  (2022).
\newblock \href {https://doi.org/10.48550/arXiv.2208.02199}
  {\path{doi:10.48550/arXiv.2208.02199}}.

\bibitem{Bittel2021}
L.~Bittel, M.~Kliesch, {\it Training variational quantum algorithms is NP-hard
  -- even for logarithmically many qubits and free fermionic systems}, Phys.
  Rev. Lett. 127 (2021) 150502.
\newblock \href {https://doi.org/10.1103/PhysRevLett.127.120502}
  {\path{doi:10.1103/PhysRevLett.127.120502}}.

\bibitem{Bloch2008}
I.~Bloch, J.~Dalibard, W.~Zwerger, {\it Many-body physics with ultracold
  gases}, Rev. Mod. Phys. 80 (2008) 885.
\newblock \href {https://doi.org/10.1103/RevModPhys.80.885}
  {\path{doi:10.1103/RevModPhys.80.885}}.

\bibitem{Loss1998}
D.~Loss, D.~P. DiVincenzo, {\it Quantum computation with quantum dots}, Phys.
  Rev. A 57~(1) (1998) 120.
\newblock \href {https://doi.org/10.1103/PhysRevA.57.120}
  {\path{doi:10.1103/PhysRevA.57.120}}.

\bibitem{Browaeys2020}
A.~Browaeys, T.~Lahaye, {\it Many-body physics with individually controlled
  Rydberg atoms}, Nat. Phys. 16 (2020) 132.
\newblock \href {https://doi.org/10.1038/s41567-019-0733-z}
  {\path{doi:10.1038/s41567-019-0733-z}}.

\bibitem{Henriet2020a}
L.~Henriet, L.~Beguin, A.~Signoles, T.~Lahaye, A.~Browaeys, G.-O. Reymond,
  C.~Jurczak, {\it Quantum computing with neutral atoms}, Quantum 4 (2020) 327.
\newblock \href {http://arxiv.org/abs/2006.12326} {\path{arXiv:2006.12326}},
  \href {https://doi.org/10.22331/q-2020-09-21-327}
  {\path{doi:10.22331/q-2020-09-21-327}}.

\bibitem{Wang2020}
Y.~Wang, Z.~Hu, B.~C. Sanders, S.~Kais,
  \href{https://doi.org/10.3389%2Ffphy.2020.589504}{Qudits and high-dimensional
  quantum computing}, Frontiers in Physics 8 (nov 2020).
\newblock \href {https://doi.org/10.3389/fphy.2020.589504}
  {\path{doi:10.3389/fphy.2020.589504}}.
\newline\urlprefix\url{https://doi.org/10.3389%2Ffphy.2020.589504}

\bibitem{Jiang2020}
P.~J. Low, B.~M. White, A.~A. Cox, M.~L. Day, C.~Senko,
  \href{https://link.aps.org/doi/10.1103/PhysRevResearch.2.033128}{Practical
  trapped-ion protocols for universal qudit-based quantum computing}, Phys.
  Rev. Res. 2 (2020) 033128.
\newblock \href {https://doi.org/10.1103/PhysRevResearch.2.033128}
  {\path{doi:10.1103/PhysRevResearch.2.033128}}.
\newline\urlprefix\url{https://link.aps.org/doi/10.1103/PhysRevResearch.2.033128}

\bibitem{Ringbauer2022}
M.~Ringbauer, M.~Meth, L.~Postler, R.~Stricker, R.~Blatt, P.~Schindler,
  T.~Monz, \href{https://doi.org/10.1038%2Fs41567-022-01658-0}{A universal
  qudit quantum processor with trapped ions}, Nature Physics 18~(9) (2022)
  1053--1057.
\newblock \href {https://doi.org/10.1038/s41567-022-01658-0}
  {\path{doi:10.1038/s41567-022-01658-0}}.
\newline\urlprefix\url{https://doi.org/10.1038%2Fs41567-022-01658-0}

\bibitem{Motta2022}
M.~Motta, J.~E. Rice, {\it Emerging quantum computing algorithms for quantum
  chemistry}, WIREs Comput. Mol. Sci. 12 (2022) e1580.
\newblock \href {https://doi.org/10.1002/wcms.1580}
  {\path{doi:10.1002/wcms.1580}}.

\bibitem{Seki2022}
K.~Seki, S.~Yunoki, {\it Spatial, spin, and charge symmetry projections for a
  Fermi-Hubbard model on a quantum computer}, Phys. Rev. A 105 (2022) 032419.
\newblock \href {https://doi.org/10.1103/PhysRevA.105.032419}
  {\path{doi:10.1103/PhysRevA.105.032419}}.

\bibitem{Kitaev1997}
A.~Y. Kitaev, {\it Quantum computations: algorithms and error correction},
  Russ. Math. Surv. 52 (1997) 1191.
\newblock \href {https://doi.org/10.1070/RM1997v052n06ABEH002155}
  {\path{doi:10.1070/RM1997v052n06ABEH002155}}.

\bibitem{Mcmahon2007}
D.~McMahon, {\it Quantum computing explained}, John Wiley \& Sons, 2007.

\bibitem{Kaye2006}
P.~Kaye, R.~Laflamme, M.~Mosca, {\it An introduction to quantum computing},
  Oxford University Press, 2006.

\bibitem{deLima2019}
F.~de~Lima~Marquezino, R.~Portugal, C.~Lavor, {\it A primer on quantum
  computing}, Springer, 2019.
\newblock \href {https://doi.org/10.1007/978-3-030-19066-8}
  {\path{doi:10.1007/978-3-030-19066-8}}.

\bibitem{divincenzo_physical_2000}
D.~P. DiVincenzo, {\it The physical implementation of quantum computation},
  Fortschr. Phys. 48 (2000) 771.
\newblock \href
  {https://doi.org/10.1002/1521-3978(200009)48:9/11<771::AID-PROP771>3.0.CO;2-E}
  {\path{doi:10.1002/1521-3978(200009)48:9/11<771::AID-PROP771>3.0.CO;2-E}}.

\bibitem{JordanWigner1928}
P.~Jordan, E.~Wigner, {\it Über das Paulische Äquivalenzverbot}, Z. Physik 47
  (1928) 631.
\newblock \href {https://doi.org/10.1007/BF01331938}
  {\path{doi:10.1007/BF01331938}}.

\bibitem{Bravyi2002}
S.~B. Bravyi, A.~Y. Kitaev, {\it Fermionic quantum computation}, Ann. Phys. 298
  (2002) 210.
\newblock \href {https://doi.org/10.1006/aphy.2002.6254}
  {\path{doi:10.1006/aphy.2002.6254}}.

\bibitem{Seeley2012}
J.~T. Seeley, M.~J. Richard, P.~J. Love, {\it The Bravyi-Kitaev transformation
  for quantum computation of electronic structure}, J. Chem. Phys. 137 (2012)
  224109.
\newblock \href {https://doi.org/10.1063/1.4768229}
  {\path{doi:10.1063/1.4768229}}.

\bibitem{Bauer2016}
B.~Bauer, D.~Wecker, A.~J. Millis, M.~B. Hastings, M.~Troyer, {\it Hybrid
  quantum-classical approach to correlated materials}, Phys. Rev. X 6 (2016)
  031045.
\newblock \href {https://doi.org/10.1103/PhysRevX.6.031045}
  {\path{doi:10.1103/PhysRevX.6.031045}}.

\bibitem{Fano2019}
G.~Fano, S.~Blinder, {\it Chapter 11 - Quantum chemistry on a quantum
  computer}, Elsevier, 2019, p. 377.
\newblock \href {https://doi.org/10.1016/B978-0-12-813651-5.00011-5}
  {\path{doi:10.1016/B978-0-12-813651-5.00011-5}}.

\bibitem{Tranter2018}
A.~Tranter, P.~J. Love, F.~Mintert, P.~V. Coveney,
  \href{https://doi.org/10.1021/acs.jctc.8b00450}{A comparison of the
  bravyi–kitaev and jordan–wigner transformations for the quantum
  simulation of quantum chemistry}, Journal of Chemical Theory and Computation
  14~(11) (2018) 5617, pMID: 30189144.
\newblock \href {http://arxiv.org/abs/https://doi.org/10.1021/acs.jctc.8b00450}
  {\path{arXiv:https://doi.org/10.1021/acs.jctc.8b00450}}, \href
  {https://doi.org/10.1021/acs.jctc.8b00450}
  {\path{doi:10.1021/acs.jctc.8b00450}}.
\newline\urlprefix\url{https://doi.org/10.1021/acs.jctc.8b00450}

\bibitem{Setia2018}
K.~Setia, S.~Bravyi, A.~Mezzacapo, J.~D. Whitfield, {\it Superfast encodings
  for fermionic quantum simulation}, Phys. Rev. Research 1~(3) (2019) 033033.
\newblock \href {https://doi.org/10.1103/PhysRevResearch.1.033033}
  {\path{doi:10.1103/PhysRevResearch.1.033033}}.

\bibitem{Consiglio_2022}
M.~Consiglio, W.~J. Chetcuti, C.~Bravo-Prieto, S.~Ramos-Calderer, A.~Minguzzi,
  J.~I. Latorre, L.~Amico, T.~J. Apollaro, Variational quantum eigensolver for
  su(n) fermions, Journal of Physics A: Mathematical and Theoretical 55~(26)
  (2022) 265301.
\newblock \href {https://doi.org/10.1088/1751-8121/ac7016}
  {\path{doi:10.1088/1751-8121/ac7016}}.

\bibitem{Blaizot86}
J.~P. Blaizot, G.~Ripka, {\it Quantum Theory of Finite Systems}, The MIT Press,
  1986.

\bibitem{Zalka1998}
C.~Zalka, \href{https://onlinelibrary.wiley.com/doi/abs/10.1002/}{Efficient
  simulation of quantum systems by quantum computers}, Fortschritte der Physik
  46~(6-8) (1998) 877.
\newblock \href
  {http://arxiv.org/abs/https://onlinelibrary.wiley.com/doi/pdf/10.1002/}
  {\path{arXiv:https://onlinelibrary.wiley.com/doi/pdf/10.1002/}}, \href
  {https://doi.org/https://doi.org/10.1002/(SICI)1521-3978(199811)46:6/8<877::AID-PROP877>3.0.CO;2-A}
  {\path{doi:https://doi.org/10.1002/(SICI)1521-3978(199811)46:6/8<877::AID-PROP877>3.0.CO;2-A}}.
\newline\urlprefix\url{https://onlinelibrary.wiley.com/doi/abs/10.1002/}

\bibitem{Lloyd1996}
S.~Lloyd, {\it Universal Quantum Simulators}, Science 273~(5278) (1996) 1073.
\newblock \href {https://doi.org/10.1126/science.273.5278.1073}
  {\path{doi:10.1126/science.273.5278.1073}}.

\bibitem{Babbush2019}
R.~Babbush, D.~W. Berry, J.~R. McClean, H.~Neven, Quantum simulation of
  chemistry with sublinear scaling in basis size, npj Quantum Information 5~(1)
  (2019) 92.
\newblock \href {https://doi.org/https://doi.org/10.1038/s41534-019-0199-y}
  {\path{doi:https://doi.org/10.1038/s41534-019-0199-y}}.

\bibitem{Kivlichan2017}
I.~D. Kivlichan, N.~Wiebe, R.~Babbush, A.~Aspuru-Guzik,
  \href{https://dx.doi.org/10.1088/1751-8121/aa77b8}{Bounding the costs of
  quantum simulation of many-body physics in real space}, Journal of Physics A:
  Mathematical and Theoretical 50~(30) (2017) 305301.
\newblock \href {https://doi.org/10.1088/1751-8121/aa77b8}
  {\path{doi:10.1088/1751-8121/aa77b8}}.
\newline\urlprefix\url{https://dx.doi.org/10.1088/1751-8121/aa77b8}

\bibitem{Babbush2018}
R.~Babbush, D.~W. Berry, Y.~R. Sanders, I.~D. Kivlichan, A.~Scherer, A.~Y. Wei,
  P.~J. Love, A.~Aspuru-Guzik,
  \href{https://dx.doi.org/10.1088/2058-9565/aa9463}{Exponentially more precise
  quantum simulation of fermions in the configuration interaction
  representation}, Quantum Science and Technology 3~(1) (2017) 015006.
\newblock \href {https://doi.org/10.1088/2058-9565/aa9463}
  {\path{doi:10.1088/2058-9565/aa9463}}.
\newline\urlprefix\url{https://dx.doi.org/10.1088/2058-9565/aa9463}

\bibitem{Su2021}
Y.~Su, D.~W. Berry, N.~Wiebe, N.~Rubin, R.~Babbush,
  \href{https://link.aps.org/doi/10.1103/PRXQuantum.2.040332}{Fault-tolerant
  quantum simulations of chemistry in first quantization}, PRX Quantum 2 (2021)
  040332.
\newblock \href {https://doi.org/10.1103/PRXQuantum.2.040332}
  {\path{doi:10.1103/PRXQuantum.2.040332}}.
\newline\urlprefix\url{https://link.aps.org/doi/10.1103/PRXQuantum.2.040332}

\bibitem{Toloui2013}
B.~Toloui, P.~J. Love, \href{https://doi.org/10.48550/arXiv.1312.2579}{Quantum
  algorithms for quantum chemistry based on the sparsity of the ci-matrix}
  (2013).
\newblock \href {http://arxiv.org/abs/1312.2579} {\path{arXiv:1312.2579}},
  \href {https://doi.org/https://doi.org/10.48550/arXiv.1312.2579}
  {\path{doi:https://doi.org/10.48550/arXiv.1312.2579}}.
\newline\urlprefix\url{https://doi.org/10.48550/arXiv.1312.2579}

\bibitem{Abrams1997}
D.~S. Abrams, S.~Lloyd,
  \href{https://link.aps.org/doi/10.1103/PhysRevLett.79.2586}{Simulation of
  many-body fermi systems on a universal quantum computer}, Phys. Rev. Lett. 79
  (1997) 2586.
\newblock \href {https://doi.org/10.1103/PhysRevLett.79.2586}
  {\path{doi:10.1103/PhysRevLett.79.2586}}.
\newline\urlprefix\url{https://link.aps.org/doi/10.1103/PhysRevLett.79.2586}

\bibitem{Georges1996}
A.~Georges, G.~Kotliar, W.~Krauth, M.~J. Rozenberg, {\it Dynamical mean-field
  theory of strongly correlated fermion systems and the limit of infinite
  dimensions}, Rev. Mod. Phys. 68 (1996) 13.
\newblock \href {https://doi.org/10.1103/RevModPhys.68.13}
  {\path{doi:10.1103/RevModPhys.68.13}}.

\bibitem{Lechermann2007}
F.~Lechermann, A.~Georges, G.~Kotliar, O.~Parcollet, {\it Rotationally
  invariant slave-boson formalism and momentum dependence of the quasiparticle
  weight}, Phys. Rev. B 76 (2007) 155102.
\newblock \href {https://doi.org/10.1103/PhysRevB.76.155102}
  {\path{doi:10.1103/PhysRevB.76.155102}}.

\bibitem{Knizia2012}
G.~Knizia, G.~K.-L. Chan, {\it Density Matrix Embedding: a simple alternative
  to Dynamical Mean-Field Theory}, Phys. Rev. Lett. 109 (2012) 186404.
\newblock \href {https://doi.org/10.1103/PhysRevLett.109.186404}
  {\path{doi:10.1103/PhysRevLett.109.186404}}.

\bibitem{Rungger2019}
I.~Rungger, N.~Fitzpatrick, H.~Chen, C.~H. Alderete, H.~Apel, A.~Cowtan,
  A.~Patterson, D.~M. Ramo, Y.~Zhu, N.~H. Nguyen, E.~Grant, S.~Chretien,
  L.~Wossnig, N.~M. Linke, R.~Duncan, {\it Dynamical mean field theory
  algorithm and experiment on quantum computers}, arXiv:1910.04735 (2019).
\newblock \href {https://doi.org/10.48550/arXiv.1910.04735}
  {\path{doi:10.48550/arXiv.1910.04735}}.

\bibitem{Keen2019}
T.~Keen, T.~Maier, S.~Johnston, P.~Lougovski, {\it Quantum-classical simulation
  of two-site dynamical mean-field theory on noisy quantum hardware}, Quantum
  Sci. Technol. 5 (2020) 035001.
\newblock \href {https://doi.org/10.1088/2058-9565/ab7d4c}
  {\path{doi:10.1088/2058-9565/ab7d4c}}.

\bibitem{Tilly2021}
J.~Tilly, P.~V. Sriluckshmy, A.~Patel, E.~Fontana, I.~Rungger, E.~Grant,
  R.~Anderson, J.~Tennyson, G.~H. Booth, {\it Reduced density matrix sampling:
  Self-consistent embedding and multiscale electronic structure on current
  generation quantum computers}, Phys. Rev. Research 3~(3) (2021) 033230.
\newblock \href {https://doi.org/10.1103/PhysRevResearch.3.033230}
  {\path{doi:10.1103/PhysRevResearch.3.033230}}.

\bibitem{Yao2020}
Y.~Yao, F.~Zhang, C.-Z. Wang, K.-M. Ho, P.~P. Orth, {\it Gutzwiller hybrid
  quantum-classical computing approach for correlated materials}, Phys. Rev.
  Research 3 (2021) 013184.
\newblock \href {https://doi.org/10.1103/PhysRevResearch.3.013184}
  {\path{doi:10.1103/PhysRevResearch.3.013184}}.

\bibitem{besserve_ayral_2022}
P.~Besserve, T.~Ayral, {\it Unraveling correlated material properties with
  noisy quantum computers: Natural orbitalized variational quantum eigensolving
  of extended impurity models within a slave-boson approach}, Phys. Rev. B 105
  (2022).
\newblock \href {https://doi.org/10.1103/physrevb.105.115108}
  {\path{doi:10.1103/physrevb.105.115108}}.

\bibitem{Backes2023}
S.~Backes, Y.~Murakami, S.~Sakai, R.~Arita, { \it Dynamical mean-field theory
  for the Hubbard-Holstein model on a quantum device}, arXiv:1910.04735 (2023).
\newblock \href {https://doi.org/10.48550/arXiv.2301.01860}
  {\path{doi:10.48550/arXiv.2301.01860}}.

\bibitem{Stein2016}
C.~J. Stein, M.~Reiher,
  \href{https://doi.org/10.1021/acs.jctc.6b00156}{Automated selection of active
  orbital spaces}, Journal of Chemical Theory and Computation 12~(4) (2016)
  1760, pMID: 26959891.
\newblock \href {https://doi.org/10.1021/acs.jctc.6b00156}
  {\path{doi:10.1021/acs.jctc.6b00156}}.
\newline\urlprefix\url{https://doi.org/10.1021/acs.jctc.6b00156}

\bibitem{Stein2019}
C.~J. Stein, M.~Reiher,
  \href{https://onlinelibrary.wiley.com/doi/abs/10.1002/jcc.25869}{autocas: A
  program for fully automated multiconfigurational calculations}, Journal of
  Computational Chemistry 40~(25) (2019) 2216.
\newblock \href {https://doi.org/https://doi.org/10.1002/jcc.25869}
  {\path{doi:https://doi.org/10.1002/jcc.25869}}.
\newline\urlprefix\url{https://onlinelibrary.wiley.com/doi/abs/10.1002/jcc.25869}

\bibitem{McCaskey2019}
A.~J. McCaskey, Z.~P. Parks, J.~Jakowski, S.~V. Moore, T.~D. Morris, T.~S.
  Humble, R.~C. Pooser, V.~Moore, T.~D. Morris, T.~S. Humble, R.~C. Pooser,
  {\it Quantum chemistry as a benchmark for near-term quantum computers}, npj
  Quantum Inf. 5 (2019) 1.
\newblock \href {https://doi.org/10.1038/s41534-019-0209-0}
  {\path{doi:10.1038/s41534-019-0209-0}}.

\bibitem{Reiher2016}
M.~Reiher, N.~Wiebe, K.~M. Svore, D.~Wecker, M.~Troyer, {\it Elucidating
  reaction mechanisms on quantum computers}, Proceedings of the National
  Academy of Sciences 114 (2017) 7555.
\newblock \href {https://doi.org/10.1073/pnas.1619152114}
  {\path{doi:10.1073/pnas.1619152114}}.

\bibitem{Li2018a}
Z.~Li, J.~Li, N.~S. Dattani, C.~J. Umrigar, G.~K.-L. Chan, {\it The electronic
  complexity of the ground-state of the FeMo cofactor of nitrogenase as
  relevant to quantum simulations}, J. Chem. Phys. 150~(2) (2018) 024302.
\newblock \href {https://doi.org/10.1063/1.5063376}
  {\path{doi:10.1063/1.5063376}}.

\bibitem{Louvet2023}
T.~Louvet, T.~Ayral, X.~Waintal, {Go-No go criteria for performing quantum
  chemistry calculations on quantum computers} (2023).
\newblock \href {http://arxiv.org/abs/2306.02620v1}
  {\path{arXiv:2306.02620v1}}.

\bibitem{berry_2006}
D.~W. Berry, G.~Ahokas, R.~Cleve, B.~C. Sanders, {\it Efficient quantum
  algorithms for simulating sparse Hamiltonians}, Commun. Math. Phys. 270
  (2006) 359.
\newblock \href {https://doi.org/10.1007/s00220-006-0150-x}
  {\path{doi:10.1007/s00220-006-0150-x}}.

\bibitem{Farhi2000}
E.~Farhi, J.~Goldstone, S.~Gutmann, M.~Sipser, {\it Quantum computation by
  adiabatic evolution} (2000).
\newblock \href {https://doi.org/10.48550/arXiv.quant-ph/0001106}
  {\path{doi:10.48550/arXiv.quant-ph/0001106}}.

\bibitem{Brassard2000}
G.~Brassard, P.~Hoyer, M.~Mosca, A.~Tapp, {\it Quantum amplitude amplification
  and estimation}, AMS Contemporary Mathematics Series 305 (2000).
\newblock \href {https://doi.org/10.1090/conm/305/05215}
  {\path{doi:10.1090/conm/305/05215}}.

\bibitem{Huggins2021}
W.~J. Huggins, B.~A. O'Gorman, N.~C. Rubin, D.~R. Reichman, R.~Babbush, J.~Lee,
  {\it Unbiasing fermionic quantum Monte Carlo with a quantum computer}, Nature
  603 (2022) 416.
\newblock \href {https://doi.org/10.1038/s41586-021-04351-z}
  {\path{doi:10.1038/s41586-021-04351-z}}.

\bibitem{Mazzola2022}
G.~Mazzola, G.~Carleo, {\it Exponential challenges in unbiasing quantum Monte
  Carlo algorithms with quantum computers}, arXiv:2205.09203 (2022).
\newblock \href {https://doi.org/10.48550/arXiv.2205.09203}
  {\path{doi:10.48550/arXiv.2205.09203}}.

\bibitem{Zhang2022a}
Y.~Zhang, Y.~Huang, J.~Sun, D.~Lv, X.~Yuan, {\it Quantum Computing Quantum
  Monte Carlo}, arXiv:2206.10431 (2022).
\newblock \href {https://doi.org/10.48550/arXiv.2206.10431}
  {\path{doi:10.48550/arXiv.2206.10431}}.

\bibitem{Wecker2014a}
D.~Wecker, B.~Bauer, B.~K. Clark, M.~B. Hastings, M.~Troyer, {\it Gate-count
  estimates for performing quantum chemistry on small quantum computers}, Phys.
  Rev. A 90 (2014) 022305.
\newblock \href {https://doi.org/10.1103/PhysRevA.90.022305}
  {\path{doi:10.1103/PhysRevA.90.022305}}.

\bibitem{Cirstoiu2020}
C.~Cirstoiu, Z.~Holmes, J.~Iosue, L.~Cincio, P.~J. Coles, A.~Sornborger, {\it
  Variational fast forwarding for quantum simulation beyond the coherence
  time}, npj Quantum Inf. 6~(82) (2020).
\newblock \href {https://doi.org/10.1038/s41534-020-00302-0}
  {\path{doi:10.1038/s41534-020-00302-0}}.

\bibitem{Gibbs2021}
J.~Gibbs, K.~Gili, Z.~Holmes, B.~Commeau, A.~Arrasmith, L.~Cincio, P.~J. Coles,
  A.~Sornborger, {\it Long-time simulations with high fidelity on quantum
  hardware}, arXiv:2102.04313 (2021).
\newblock \href {https://doi.org/10.48550/ARXIV.2102.04313}
  {\path{doi:10.48550/ARXIV.2102.04313}}.

\bibitem{Jaderberg2020}
B.~Jaderberg, A.~Agarwal, K.~Leonhardt, M.~Kiffner, D.~Jaksch, {\it Minimum
  hardware requirements for hybrid quantum--classical DMFT}, Quantum Sci.
  Technol. 5~(3) (2020) 034015.
\newblock \href {https://doi.org/10.1088/2058-9565/ab972b}
  {\path{doi:10.1088/2058-9565/ab972b}}.

\bibitem{Zhang2020}
Z.-J. Zhang, J.~Sun, X.~Yuan, M.-H. Yung, {\it Low-depth hamiltonian simulation
  by adaptive product formula}, arXiv:2011.05283 (2020).
\newblock \href {https://doi.org/10.48550/arXiv.2011.05283}
  {\path{doi:10.48550/arXiv.2011.05283}}.

\bibitem{Klymko2021}
K.~Klymko, C.~Mejuto-Zaera, S.~J. Cotton, F.~Wudarski, M.~Urbanek, D.~Hait,
  M.~Head-Gordon, K.~B. Whaley, J.~Moussa, N.~Wiebe, W.~A. de~Jong, N.~M.
  Tubman, {\it Real time evolution for ultracompact Hamiltonian eigenstates on
  quantum hardware}, arxiv:103.08563 (2021).
\newblock \href {https://doi.org/10.48550/arxiv.2103.08563}
  {\path{doi:10.48550/arxiv.2103.08563}}.

\bibitem{Kokail2018}
C.~Kokail, C.~Maier, R.~van Bijnen, T.~Brydges, M.~K. Joshi, P.~Jurcevic, C.~A.
  Muschik, P.~Silvi, R.~Blatt, C.~F. Roos, P.~Zoller, {\it Self-verifying
  variational quantum simulation of lattice models}, Nature 569 (2019) 355.
\newblock \href {https://doi.org/10.1038/s41586-019-1177-4}
  {\path{doi:10.1038/s41586-019-1177-4}}.

\bibitem{Peruzzo2014}
A.~Peruzzo, et~al, {\it \it A variational eigenvalue solver on a photonic
  quantum processor }, Nat. Commun 5 (2014) 4213.
\newblock \href {https://doi.org/10.1038/ncomms5213}
  {\path{doi:10.1038/ncomms5213}}.

\bibitem{Huggins2019}
W.~J. Huggins, J.~R. McClean, N.~C. Rubin, Z.~Jiang, N.~Wiebe, K.~B. Whaley,
  R.~Babbush, {\it Efficient and noise resilient measurements for quantum
  chemistry on near-term quantum computers}, npj Quantum Inf. 7 (2021) 23.
\newblock \href {https://doi.org/10.1038/s41534-020-00341-7}
  {\path{doi:10.1038/s41534-020-00341-7}}.

\bibitem{Mitarai2018}
K.~Mitarai, M.~Negoro, M.~Kitagawa, K.~Fujii, {\it Quantum circuit learning},
  Phys. Rev. A 98 (2018) 032309.
\newblock \href {https://doi.org/10.1103/PhysRevA.98.032309}
  {\path{doi:10.1103/PhysRevA.98.032309}}.

\bibitem{Schuld2019}
M.~Schuld, V.~Bergholm, C.~Gogolin, J.~A. Izaac, N.~Killoran, {\it Evaluating
  analytic gradients on quantum hardware}, Phys. Rev. A 99 (2019) 032331.
\newblock \href {https://doi.org/10.1103/PhysRevA.99.032331}
  {\path{doi:10.1103/PhysRevA.99.032331}}.

\bibitem{izmaylov_analytic_2021}
A.~F. Izmaylov, R.~A. Lang, T.-C. Yen, {\it Analytic gradients in variational
  quantum algorithms: Algebraic extensions of the parameter-shift rule to
  general unitary transformations}, Phys. Rev. A 104 (2021) 062443.
\newblock \href {https://doi.org/10.1103/PhysRevA.104.062443}
  {\path{doi:10.1103/PhysRevA.104.062443}}.

\bibitem{Nakanishi2019}
K.~M. Nakanishi, K.~Fujii, S.~Todo, {\it Sequential minimal optimization for
  quantum-classical hybrid algorithms}, Phys. Rev. Research 2 (2020) 043158.
\newblock \href {https://doi.org/10.1103/PhysRevResearch.2.043158}
  {\path{doi:10.1103/PhysRevResearch.2.043158}}.

\bibitem{Ostaszewski2021}
M.~Ostaszewski, E.~Grant, M.~Benedetti,
  \href{https://quantum-journal.org/papers/q-2021-01-28-391/}{Structure
  optimization for parameterized quantum circuits} (Jan 2021).
\newline\urlprefix\url{https://quantum-journal.org/papers/q-2021-01-28-391/}

\bibitem{Michel2023}
A.~Michel, S.~Grijalva, L.~Henriet, C.~Domain, A.~Browaeys, {\it A blueprint
  for a Digital-Analog Variational Quantum Eigensolver using Rydberg atom
  arrays}, arxiv:2301.06453 (2023).
\newblock \href {https://doi.org/10.48550/ARXIV.2301.06453}
  {\path{doi:10.48550/ARXIV.2301.06453}}.

\bibitem{higgott_wang_brierley_2019}
O.~Higgott, D.~Wang, S.~Brierley, {\it Variational quantum computation of
  excited states}, Quantum 3 (2019) 156.
\newblock \href {https://doi.org/10.22331/q-2019-07-01-156}
  {\path{doi:10.22331/q-2019-07-01-156}}.

\bibitem{ryabinkin_genin_izmaylov_2018}
I.~G. Ryabinkin, S.~N. Genin, A.~F. Izmaylov, {\it Constrained variational
  quantum eigensolver: Quantum computer search engine in the Fock space}, J.
  Chem. Theory Comput. 15 (2018) 249.
\newblock \href {https://doi.org/10.1021/acs.jctc.8b00943}
  {\path{doi:10.1021/acs.jctc.8b00943}}.

\bibitem{nakanishi_mitarai_fujii_2019}
K.~M. Nakanishi, K.~Mitarai, K.~Fujii, {\it Subspace-search variational quantum
  eigensolver for excited states}, Phys. Rev. Research 1 (2019) 033062.
\newblock \href {https://doi.org/10.1103/physrevresearch.1.033062}
  {\path{doi:10.1103/physrevresearch.1.033062}}.

\bibitem{meitei_2021}
O.~R. Meitei, B.~T. Gard, G.~S. Barron, D.~P. Pappas, S.~E. Economou,
  E.~Barnes, N.~J. Mayhall, {\it Gate-free state preparation for fast
  variational quantum Eigensolver simulations}, npj Quantum Inf. 7 (2021) 155.
\newblock \href {https://doi.org/10.1038/s41534-021-00493-0}
  {\path{doi:10.1038/s41534-021-00493-0}}.

\bibitem{sokolov_2020}
I.~O. Sokolov, P.~K. Barkoutsos, P.~J. Ollitrault, D.~Greenberg, J.~Rice,
  M.~Pistoia, I.~Tavernelli, {\it Quantum orbital-optimized unitary coupled
  cluster methods in the strongly correlated regime: Can quantum algorithms
  outperform their classical equivalents?}, J. Chem. Phys. 152 (2020) 124107.
\newblock \href {https://doi.org/10.1063/1.5141835}
  {\path{doi:10.1063/1.5141835}}.

\bibitem{koridon_2021}
E.~Koridon, S.~Yalouz, B.~Senjean, F.~Buda, T.~E. O'Brien, L.~Visscher, {\it
  Orbital transformations to reduce the 1-norm of the electronic structure
  hamiltonian for quantum computing applications}, Phys. Rev. Research 3 (2021)
  033127.
\newblock \href {https://doi.org/10.1103/physrevresearch.3.033127}
  {\path{doi:10.1103/physrevresearch.3.033127}}.

\bibitem{tkachenko_2021}
N.~V. Tkachenko, J.~Sud, Y.~Zhang, S.~Tretiak, P.~M. Anisimov, A.~T. Arrasmith,
  P.~J. Coles, L.~Cincio, P.~A. Dub, {\it Correlation-informed permutation of
  qubits for reducing ansatz depth in the variational quantum eigensolver}, PRX
  Quantum 2 (2021) 020337.
\newblock \href {https://doi.org/10.1103/prxquantum.2.020337}
  {\path{doi:10.1103/prxquantum.2.020337}}.

\bibitem{Stair2021a}
N.~H. Stair, F.~A. Evangelista, {\it Simulating Many-Body Systems with a
  Projective Quantum Eigensolver}, PRX Quantum 2 (2021) 030301.
\newblock \href {https://doi.org/10.1103/PRXQuantum.2.030301}
  {\path{doi:10.1103/PRXQuantum.2.030301}}.

\bibitem{Siegbahn1981}
P.~E.~M. Siegbahn, J.~Almlöf, A.~Heiberg, B.~O. Roos,
  \href{https://doi.org/10.1063/1.441359}{{The complete active space SCF
  (CASSCF) method in a Newton–Raphson formulation with application to the HNO
  molecule}}, The Journal of Chemical Physics 74~(4) (1981) 2384.
\newblock \href {http://arxiv.org/abs/https://pubs.aip.org/aip/jcp/article-pdf
  74/4/2384/11249618/2384\_1\_online.pdf}
  {\path{arXiv:https://pubs.aip.org/aip/jcp/article-pdf
  74/4/2384/11249618/2384\_1\_online.pdf}}, \href
  {https://doi.org/10.1063/1.441359} {\path{doi:10.1063/1.441359}}.
\newline\urlprefix\url{https://doi.org/10.1063/1.441359}

\bibitem{Yalouz2021}
S.~Yalouz, B.~Senjean, J.~Günther, F.~Buda, T.~E. O’Brien, L.~Visscher,
  \href{https://dx.doi.org/10.1088/2058-9565/abd334}{{A state-averaged
  orbital-optimized hybrid quantum–classical algorithm for a democratic
  description of ground and excited states}} 6~(2) (2021) 024004.
\newblock \href {https://doi.org/10.1088/2058-9565/abd334}
  {\path{doi:10.1088/2058-9565/abd334}}.
\newline\urlprefix\url{https://dx.doi.org/10.1088/2058-9565/abd334}

\bibitem{McLachlan64}
A.~McLachlan, {\it A variational solution of the time-dependent Schrodinger
  equation}, Mol. Phys. 8 (1964) 39.
\newblock \href {https://doi.org/10.1080/00268976400100041}
  {\path{doi:10.1080/00268976400100041}}.

\bibitem{yuan_theory_2019}
X.~Yuan, S.~Endo, Q.~Zhao, Y.~Li, S.~Benjamin, {\it Theory of variational
  quantum simulation}, Quantum 3 (2019) 191.
\newblock \href {https://doi.org/10.22331/q-2019-10-07-191}
  {\path{doi:10.22331/q-2019-10-07-191}}.

\bibitem{Endo2020}
S.~Endo, I.~Kurata, Y.~O. Nakagawa, {\it Calculation of the Green's function on
  near-term quantum computers}, Phys. Rev. Research 2 (2020) 033281.
\newblock \href {https://doi.org/10.1103/PhysRevResearch.2.033281}
  {\path{doi:10.1103/PhysRevResearch.2.033281}}.

\bibitem{mcardle_variational_2019}
S.~McArdle, T.~Jones, S.~Endo, Y.~Li, S.~C. Benjamin, X.~Yuan, {\it Variational
  ansatz-based quantum simulation of imaginary time evolution}, npj Quantum
  Inf. 5 (2019) 75.
\newblock \href {https://doi.org/10.1038/s41534-019-0187-2}
  {\path{doi:10.1038/s41534-019-0187-2}}.

\bibitem{Motta2019}
M.~Motta, C.~Sun, A.~T.~K. Tan, M.~J. O'Rourke, E.~Ye, A.~J. Minnich, F.~G.
  S.~L. Brand\~ao, G.~K.-l. Chan, {\it Determining eigenstates and thermal
  states on a quantum computer using quantum imaginary time evolution}, Nat.
  Phys. 16 (2020) 205.
\newblock \href {https://doi.org/10.1038/s41567-019-0704-4}
  {\path{doi:10.1038/s41567-019-0704-4}}.

\bibitem{Schuld2018}
M.~Schuld, F.~Petruccione, {\it Supervised learning with quantum computers},
  Vol.~17, Springer Cham, 2018.
\newblock \href {https://doi.org/10.1007/978-3-319-96424-9}
  {\path{doi:10.1007/978-3-319-96424-9}}.

\bibitem{Benedetti2019}
M.~Benedetti, E.~Lloyd, S.~Sack, M.~Fiorentini, {\it Parameterized quantum
  circuits as machine learning models}, Quantum Sci. Technol. 4 (2019) 043001.
\newblock \href {https://doi.org/10.1088/2058-9565/ab4eb5}
  {\path{doi:10.1088/2058-9565/ab4eb5}}.

\bibitem{Biamonte2021}
J.~Biamonte, {\it Universal variational quantum computation}, Phys. Rev. A 103
  (2021) L030401.
\newblock \href {https://doi.org/10.1103/PhysRevA.103.L030401}
  {\path{doi:10.1103/PhysRevA.103.L030401}}.

\bibitem{Schuld2021}
M.~Schuld, R.~Sweke, J.~J. Meyer, {\it Effect of data encoding on the
  expressive power of variational quantum-machine-learning models}, Phys. Rev.
  A 103 (2021) 032430.
\newblock \href {https://doi.org/10.1103/PhysRevA.103.032430}
  {\path{doi:10.1103/PhysRevA.103.032430}}.

\bibitem{Goto2021}
T.~Goto, Q.~H. Tran, K.~Nakajima, {\it Universal Approximation Property of
  Quantum Machine Learning Models in Quantum-Enhanced Feature Spaces}, Phys.
  Rev. Lett. 127 (2021) 090506.
\newblock \href {https://doi.org/10.1103/PhysRevLett.127.090506}
  {\path{doi:10.1103/PhysRevLett.127.090506}}.

\bibitem{Cerezo2021}
M.~Cerezo, A.~Sone, T.~Volkoff, L.~Cincio, P.~J. Coles, {\it Cost function
  dependent barren plateaus in shallow parametrized quantum circuits}, Nat.
  Commun 12 (2021) 1791.
\newblock \href {https://doi.org/10.1038/s41467-021-21728-w}
  {\path{doi:10.1038/s41467-021-21728-w}}.

\bibitem{Gell-Mann1951}
M.~Gell-Mann, F.~Low, {\it Bound States in Quantum Field Theory}, Phys. Rev. 84
  (1951) 350.
\newblock \href {https://doi.org/10.1103/PhysRev.84.350}
  {\path{doi:10.1103/PhysRev.84.350}}.

\bibitem{Fetter2012}
A.~L. Fetter, J.~D. Walecka, {\it Quantum theory of many-particle systems},
  Courier Corporation, 2012.

\bibitem{Romero2018}
J.~Romero, R.~Babbush, J.~R. McClean, C.~Hempel, P.~J. Love, A.~Aspuru-Guzik,
  {\it Strategies for quantum computing molecular energies using the unitary
  coupled cluster ansatz}, Quantum Sci. Technol. 4 (2018) 014008.
\newblock \href {https://doi.org/10.1088/2058-9565/aad3e4}
  {\path{doi:10.1088/2058-9565/aad3e4}}.

\bibitem{Arute2020}
F.~Arute, K.~Arya, R.~Babbush, D.~Bacon, J.~C. Bardin, R.~Barends, S.~Boixo,
  M.~Broughton, B.~B. Buckley, D.~A. Buell, et~al., {\it Hartree-Fock on a
  superconducting qubit quantum computer}, Science 369 (2020) 1084.
\newblock \href {https://doi.org/10.1126/science.abb9811}
  {\path{doi:10.1126/science.abb9811}}.

\bibitem{Thouless1960}
D.~J. Thouless, {\it Stability conditions and nuclear rotations in the
  Hartree-Fock theory}, Nucl. Phys. 21 (1960) 225.
\newblock \href {https://doi.org/10.1016/0029-5582(60)90048-1}
  {\path{doi:10.1016/0029-5582(60)90048-1}}.

\bibitem{brink_broglia_2005}
D.~M. Brink, R.~A. Broglia, {\it Nuclear Superfluidity: Pairing in Finite
  Systems}, Cambridge University Press, 2005.
\newblock \href {https://doi.org/10.1017/CBO9780511534911}
  {\path{doi:10.1017/CBO9780511534911}}.

\bibitem{Dallaire-Demers2018}
P.-L. Dallaire-Demers, J.~Romero, L.~Veis, S.~Sim, A.~Aspuru-Guzik, {\it
  Low-depth circuit ansatz for preparing correlated fermionic states on a
  quantum computer}, Quantum Sci. Technol. 4 (2018) 045005.
\newblock \href {https://doi.org/10.1088/2058-9565/ab3951}
  {\path{doi:10.1088/2058-9565/ab3951}}.

\bibitem{hoffman_1972}
D.~K. Hoffman, R.~C. Raffenetti, K.~Ruedenberg, {\it Generalization of Euler
  angles to n‐dimensional orthogonal matrices}, J. Math. Phys. 13 (1972) 528.
\newblock \href {https://doi.org/10.1063/1.1666011}
  {\path{doi:10.1063/1.1666011}}.

\bibitem{valiant_2001}
L.~G. Valiant, {\it Quantum computers that can be simulated classically in
  polynomial time}, Proceedings of the thirty-third annual ACM symposium on
  Theory of computing - STOC '01 (2001) 114\href
  {https://doi.org/10.1145/380752.380785} {\path{doi:10.1145/380752.380785}}.

\bibitem{jozsa_miyake_2008}
R.~Jozsa, A.~Miyake, {\it Matchgates and classical simulation of quantum
  circuits}, Proc. R. Soc. A 464 (2008) 3089.
\newblock \href {https://doi.org/10.1098/rspa.2008.0189}
  {\path{doi:10.1098/rspa.2008.0189}}.

\bibitem{Verstraete2000}
F.~Verstraete, J.~I. Cirac, J.~I. Latorre, {\it Quantum circuits for strongly
  correlated quantum systems}, Phys. Rev. A 79 (2009) 032316.
\newblock \href {https://doi.org/10.1103/PhysRevA.79.032316}
  {\path{doi:10.1103/PhysRevA.79.032316}}.

\bibitem{ovrum2007quantum}
E.~Ovrum, M.~Hjorth-Jensen, {\it Quantum computation algorithm for many-body
  studies}, arXiv:0705.1928 (2007).
\newblock \href {https://doi.org/10.48550/ARXIV.0705.1928}
  {\path{doi:10.48550/ARXIV.0705.1928}}.

\bibitem{Jiang2018}
Z.~Jiang, K.~J. Sung, K.~Kechedzhi, V.~N. Smelyanskiy, S.~Boixo, {\it Quantum
  Algorithms to Simulate Many-Body Physics of Correlated Fermions}, Phys. Rev.
  A 9 (2018) 044036.
\newblock \href {https://doi.org/10.1103/PhysRevApplied.9.044036}
  {\path{doi:10.1103/PhysRevApplied.9.044036}}.

\bibitem{Khamoshi2020}
A.~Khamoshi, F.~Evangelista, G.~Scuseria, {\it Correlating AGP on a quantum
  computer}, Quantum Sci. Technol. 6 (2020) 014004.
\newblock \href {https://doi.org/10.1088/2058-9565/abc1bb}
  {\path{doi:10.1088/2058-9565/abc1bb}}.

\bibitem{Khamoshi2022}
A.~Khamoshi, G.~P. Chen, F.~A. Evangelista, G.~E. Scuseria, {\it AGP-based
  unitary coupled cluster theory for quantum computers}, Quantum Sci. Technol.
  8 (2023) 015006.
\newblock \href {https://doi.org/10.1088/2058-9565/ac93ae}
  {\path{doi:10.1088/2058-9565/ac93ae}}.

\bibitem{Wecker2015a}
D.~Wecker, M.~B. Hastings, N.~Wiebe, B.~K. Clark, C.~Nayak, M.~Troyer, {\it
  Solving strongly correlated electron models on a quantum computer}, Phys.
  Rev. A 92 (2015) 062318.
\newblock \href {https://doi.org/10.1103/PhysRevA.92.062318}
  {\path{doi:10.1103/PhysRevA.92.062318}}.

\bibitem{Kivlichan2018}
I.~D. Kivlichan, J.~McClean, N.~Wiebe, C.~Gidney, A.~Aspuru-Guzik, G.~K.-l.
  Chan, R.~Babbush, {\it Quantum Simulation of Electronic Structure with Linear
  Depth and Connectivity}, Phys. Rev. Lett. 120 (2018) 110501.
\newblock \href {https://doi.org/10.1103/PhysRevLett.120.110501}
  {\path{doi:10.1103/PhysRevLett.120.110501}}.

\bibitem{Farhi_2014}
E.~Farhi, J.~Goldstone, S.~Gutmann, {\it A Quantum Approximate Optimization
  Algorithm} (2014).
\newblock \href {https://doi.org/10.48550/arXiv.quant-ph/0001106}
  {\path{doi:10.48550/arXiv.quant-ph/0001106}}.

\bibitem{wecker_2015}
D.~Wecker, M.~B. Hastings, M.~Troyer, {\it Progress towards practical quantum
  variational algorithms}, Phys. Rev. A 92 (2015) 042303.
\newblock \href {https://doi.org/10.1103/PhysRevA.92.042303}
  {\path{doi:10.1103/PhysRevA.92.042303}}.

\bibitem{anselme_martin_2022}
B.~Anselme~Martin, P.~Simon, M.~J. Rančić, {\it Simulating strongly
  interacting hubbard chains with the variational hamiltonian ansatz on a
  quantum computer}, Phys. Rev. Research 4 (2022) 023190.
\newblock \href {https://doi.org/10.1103/physrevresearch.4.023190}
  {\path{doi:10.1103/physrevresearch.4.023190}}.

\bibitem{Kandala2017}
A.~Kandala, A.~Mezzacapo, K.~Temme, M.~Takita, M.~Brink, J.~M. Chow, J.~M.
  Gambetta, {\it Hardware-efficient variational quantum eigensolver for Small
  Mol. and quantum magnets}, Nature 549 (2017) 242--246.
\newblock \href {https://doi.org/10.1038/nature23879}
  {\path{doi:10.1038/nature23879}}.

\bibitem{Barkoutsos2018}
P.~K. Barkoutsos, J.~F. Gonthier, I.~Sokolov, N.~Moll, G.~Salis, A.~Fuhrer,
  M.~Ganzhorn, D.~J. Egger, M.~Troyer, A.~Mezzacapo, S.~Filipp, I.~Tavernelli,
  {\it Quantum algorithms for electronic structure calculations: Particle-hole
  Hamiltonian and optimized wave-function expansions}, Phys. Rev. A 98 (2018)
  022322.
\newblock \href {https://doi.org/10.1103/PhysRevA.98.022322}
  {\path{doi:10.1103/PhysRevA.98.022322}}.

\bibitem{McClean2018}
J.~R. McClean, S.~Boixo, V.~N. Smelyanskiy, R.~Babbush, H.~Neven, {\it Barren
  plateaus in quantum neural network training landscapes}, Nat Commun 9 (2018)
  4812.
\newblock \href {https://doi.org/10.1038/s41467-018-07090-4}
  {\path{doi:10.1038/s41467-018-07090-4}}.

\bibitem{Kim_2021}
J.~Kim, J.~Kim, D.~Rosa, Universal effectiveness of high-depth circuits in
  variational eigenproblems, Physical Review Research 3~(2) (2021).
\newblock \href {https://doi.org/10.1103/physrevresearch.3.023203}
  {\path{doi:10.1103/physrevresearch.3.023203}}.

\bibitem{Larocca_2023}
M.~Larocca, N.~Ju, D.~García-Martín, P.~J. Coles, M.~Cerezo, Theory of
  overparametrization in quantum neural networks, Nature Computational Science
  3~(6) (2023) 542–551.
\newblock \href {https://doi.org/10.1038/s43588-023-00467-6}
  {\path{doi:10.1038/s43588-023-00467-6}}.

\bibitem{Abhinav2022}
A.~Anand, P.~Schleich, S.~Alperin-Lea, P.~W.~K. Jensen, S.~Sim,
  M.~Díaz-Tinoco, J.~S. Kottmann, M.~Degroote, A.~F. Izmaylov,
  A.~Aspuru-Guzik, {\it A quantum computing view on unitary coupled cluster
  theory}, Chem. Soc. Rev. 51 (2022) 1659.
\newblock \href {https://doi.org/10.1039/D1CS00932J}
  {\path{doi:10.1039/D1CS00932J}}.

\bibitem{Qiu2019}
Y.~Qiu, T.~M. Henderson, T.~Duguet, G.~E. Scuseria, {\it Particle-number
  projected Bogoliubov-coupled-cluster theory: Application to the pairing
  Hamiltonian}, Phys. Rev. C 99 (2019) 044301.
\newblock \href {https://doi.org/10.1103/PhysRevC.99.044301}
  {\path{doi:10.1103/PhysRevC.99.044301}}.

\bibitem{Lacroix2022}
P.~S. Denis~Lacroix, Edgar Andres Ruiz~Guzman, {\it Symmetry breaking/symmetry
  preserving circuits and symmetry restoration on quantum computers: A quantum
  many-body perspective}, Eur. Phys. J. A 59 (2023) 3.
\newblock \href {https://doi.org/10.1140/epja/s10050-022-00911-7}
  {\path{doi:10.1140/epja/s10050-022-00911-7}}.

\bibitem{grimsley_economou_barnes_mayhall_2019}
H.~R. Grimsley, S.~E. Economou, E.~Barnes, N.~J. Mayhall, {\it An adaptive
  variational algorithm for exact Mol. Simul.s on a quantum computer}, Nat.
  Commun. 10 (2019) 3007.
\newblock \href {https://doi.org/10.1038/s41467-019-10988-2}
  {\path{doi:10.1038/s41467-019-10988-2}}.

\bibitem{tang_shkolnikov_2021}
H.~L. Tang, V.~Shkolnikov, G.~S. Barron, H.~R. Grimsley, N.~J. Mayhall,
  E.~Barnes, S.~E. Economou, {\it Qubit-ADAPT-VQE: An Adaptive Algorithm for
  Constructing Hardware-Efficient Ans\"atze on a Quantum Processor}, PRX
  Quantum 2 (2021) 020310.
\newblock \href {https://doi.org/10.1103/PRXQuantum.2.020310}
  {\path{doi:10.1103/PRXQuantum.2.020310}}.

\bibitem{Haidar2022}
M.~Haidar, M.~J. Ran{\v{c}}i{\'{c}}, T.~Ayral, Y.~Maday, J.-P. Piquemal, {\it
  Open Source Variational Quantum Eigensolver Extension of the Quantum Learning
  Machine (QLM) for Quantum Chemistry}, WIREs Comput. Mol. Sci., in press
  (2022).
\newblock \href {https://doi.org/10.48550/arXiv.2206.08798}
  {\path{doi:10.48550/arXiv.2206.08798}}.

\bibitem{Rudolph2022}
M.~S. Rudolph, J.~Chen, J.~Miller, A.~Acharya, A.~Perdomo-ortiz, {\it
  Decomposition of Matrix Product States into Shallow Quantum Circuits} (2022).
\newblock \href {https://doi.org/10.48550/ARXIV.2209.00595}
  {\path{doi:10.48550/ARXIV.2209.00595}}.

\bibitem{Rudolph2022a}
M.~S. Rudolph, J.~Miller, J.~Chen, A.~Acharya, A.~Perdomo-ortiz, {\it Synergy
  Between Quantum Circuits and Tensor Networks: Short-cutting the Race to
  Practical Quantum Advantage} (2022) arXiv:2208.13673.\href
  {https://doi.org/10.48550/ARXIV.2208.13673}
  {\path{doi:10.48550/ARXIV.2208.13673}}.

\bibitem{Yuan2021a}
X.~Yuan, J.~Sun, J.~Liu, Q.~Zhao, Y.~Zhou, {\it Quantum Simulation with Hybrid
  Tensor Networks}, Phys. Rev. Lett. 127 (2021) 40501.
\newblock \href {https://doi.org/10.1103/PhysRevLett.127.040501}
  {\path{doi:10.1103/PhysRevLett.127.040501}}.

\bibitem{Miao2021}
Q.~Miao, T.~Barthel, {\it A quantum-classical eigensolver using multiscale
  entanglement renormalization}, arXiv:2108.13401. (2021).
\newblock \href {https://doi.org/10.48550/ARXIV.2108.13401}
  {\path{doi:10.48550/ARXIV.2108.13401}}.

\bibitem{McClean2017a}
J.~R. McClean, M.~E. Kimchi-Schwartz, J.~Carter, W.~A. de~Jong, {\it Hybrid
  quantum-classical hierarchy for mitigation of decoherence and determination
  of excited states}, Phys. Rev. A 95 (2017) 042308.
\newblock \href {https://doi.org/10.1103/PhysRevA.95.042308}
  {\path{doi:10.1103/PhysRevA.95.042308}}.

\bibitem{Seki2021}
K.~Seki, S.~Yunoki, {\it Quantum Power Method by a Superposition of
  Time-Evolved States}, PRX Quantum 2 (2021) 010333.
\newblock \href {https://doi.org/10.1103/PRXQuantum.2.010333}
  {\path{doi:10.1103/PRXQuantum.2.010333}}.

\bibitem{Parrish2019}
R.~M. Parrish, P.~L. McMahon, {\it Quantum Filter Diagonalization: Quantum
  Eigendecomposition without Full Quantum Phase Estimation} (2019)
  arXiv:1909.08925.\href {https://doi.org/10.48550/ARXIV.1909.08925}
  {\path{doi:10.48550/ARXIV.1909.08925}}.

\bibitem{Stair2020}
N.~H. Stair, R.~Huang, F.~A. Evangelista, {\it A Multireference Quantum Krylov
  Algorithm for Strongly Correlated Electrons}, J. Chem. Theory Comput. 16
  (2020) 2236.
\newblock \href {https://doi.org/10.1021/acs.jctc.9b01125}
  {\path{doi:10.1021/acs.jctc.9b01125}}.

\bibitem{Bharti2021b}
K.~Bharti, T.~Haug, {\it Quantum-assisted simulator}, Phys. Rev. A 104 (2021)
  042418.
\newblock \href {https://doi.org/10.1103/PhysRevA.104.042418}
  {\path{doi:10.1103/PhysRevA.104.042418}}.

\bibitem{Bharti2021c}
K.~Bharti, T.~Haug, {\it Iterative quantum-assisted eigensolver}, Phys. Rev. A
  104 (2021) L050401.
\newblock \href {https://doi.org/10.1103/PhysRevA.104.L050401}
  {\path{doi:10.1103/PhysRevA.104.L050401}}.

\bibitem{Bespalova2021}
T.~A. Bespalova, O.~Kyriienko, {\it Hamiltonian Operator Approximation for
  Energy Measurement and Ground-State Preparation}, PRX Quantum 2 (2021)
  030318.
\newblock \href {https://doi.org/10.1103/PRXQuantum.2.030318}
  {\path{doi:10.1103/PRXQuantum.2.030318}}.

\bibitem{Jamet2021}
F.~F. Jamet, A.~Agarwal, C.~Lupo, D.~E. Browne, C.~Weber, I.~Rungger, {\it
  Krylov variational quantum algorithm for first principles materials
  simulations}~(1) (2021).
\newblock \href {http://arxiv.org/abs/2105.13298v2}
  {\path{arXiv:2105.13298v2}}, \href
  {https://doi.org/10.48550/ARXIV.2105.13298}
  {\path{doi:10.48550/ARXIV.2105.13298}}.

\bibitem{Cortes2022}
C.~L. Cortes, S.~K. Gray, {\it Quantum Krylov subspace algorithms for ground-
  and excited-state energy estimation}, Phys. Rev. A 105 (2022) 022417.
\newblock \href {https://doi.org/10.1103/PhysRevA.105.022417}
  {\path{doi:10.1103/PhysRevA.105.022417}}.

\bibitem{Haug2022}
T.~Haug, K.~Bharti, {\it Generalized quantum assisted simulator}, Quantum Sci.
  Technol. 7 (2022) 045019.
\newblock \href {https://doi.org/10.1088/2058-9565/ac83e7}
  {\path{doi:10.1088/2058-9565/ac83e7}}.

\bibitem{Wei2022}
J.~W.~Z. Lau, T.~Haug, L.~C. Kwek, K.~Bharti, {\it NISQ Algorithm for
  Hamiltonian simulation via truncated Taylor series}, SciPost Phys. 12 (2022)
  122.
\newblock \href {https://doi.org/10.21468/SciPostPhys.12.4.122}
  {\path{doi:10.21468/SciPostPhys.12.4.122}}.

\bibitem{Aulicino2022}
J.~C. Aulicino, T.~Keen, B.~Peng, {\it State preparation and evolution in
  quantum computing: A perspective from Hamiltonian moments}, Int. J. of Quant.
  Chem. 122 (2022) e26853.
\newblock \href {https://doi.org/10.1002/qua.26853}
  {\path{doi:10.1002/qua.26853}}.

\bibitem{Kitaev1996}
A.~Y. Kitaev, {\it Quantum measurements and the Abelian Stabilizer Problem},
  Electron. Colloquium Comput. Complex. TR96-003 (1996).
\newblock \href {https://doi.org/10.48550/arXiv.quant-ph/9511026}
  {\path{doi:10.48550/arXiv.quant-ph/9511026}}.

\bibitem{Griffiths1996}
R.~B. Griffiths, C.-S. Niu, {\it Semiclassical Fourier Transform for Quantum
  Computation}, Phys. Rev. Lett. 76 (1996) 3228.
\newblock \href {https://doi.org/10.1103/PhysRevLett.76.3228}
  {\path{doi:10.1103/PhysRevLett.76.3228}}.

\bibitem{Dobsicek2007}
M.~Dob\ifmmode \check{s}\else \v{s}\fi{}\'{\i}\ifmmode~\check{c}\else
  \v{c}\fi{}ek, G.~Johansson, V.~Shumeiko, G.~Wendin, {\it Arbitrary accuracy
  iterative quantum phase estimation algorithm using a single ancillary qubit:
  A two-qubit benchmark}, Phys. Rev. A 76 (2007) 030306.
\newblock \href {https://doi.org/10.1103/PhysRevA.76.030306}
  {\path{doi:10.1103/PhysRevA.76.030306}}.

\bibitem{Svore2013}
K.~M. Svore, M.~B. Hastings, M.~Freedman, {\it Faster phase estimation},
  Quantum Inf. Comput. 14 (2014) 306.
\newblock \href {https://doi.org/10.26421/QIC14.3-4-7}
  {\path{doi:10.26421/QIC14.3-4-7}}.

\bibitem{Lin2022}
L.~Lin, {\it Lecture Notes on Quantum Algorithms for Scientific Computation}
  (2022).
\newblock \href {https://doi.org/10.48550/ARXIV.2201.08309}
  {\path{doi:10.48550/ARXIV.2201.08309}}.

\bibitem{Choi2021}
K.~Choi, D.~Lee, J.~Bonitati, Z.~Qian, J.~Watkins, {\it Rodeo Algorithm for
  Quantum Computing}, Phys. Rev. Lett. 127 (2021) 040505.
\newblock \href {https://doi.org/10.1103/PhysRevLett.127.040505}
  {\path{doi:10.1103/PhysRevLett.127.040505}}.

\bibitem{Qian2021}
Z.~Qian, J.~Watkins, G.~Given, J.~Bonitati, K.~Choi, D.~Lee, {\it Demonstration
  of the Rodeo Algorithm on a Quantum Computer}, arXiv:2110.07747. (2021).
\newblock \href {https://doi.org/10.48550/ARXIV.2110.07747}
  {\path{doi:10.48550/ARXIV.2110.07747}}.

\bibitem{Bee-Lindgren2022}
M.~Bee-Lindgren, Z.~Qian, M.~DeCross, N.~C. Brown, C.~N. Gilbreth, J.~Watkins,
  X.~Zhang, D.~Lee, {\it Rodeo Algorithm with Controlled Reversal Gates},
  arXiv:2208.13557. (2022).
\newblock \href {https://doi.org/10.48550/ARXIV.2208.13557}
  {\path{doi:10.48550/ARXIV.2208.13557}}.

\bibitem{wang_2019}
D.~Wang, O.~Higgott, S.~Brierley, {\it Accelerated variational quantum
  eigensolver}, Phys. Rev. Lett. 122 (2019) 140504.
\newblock \href {https://doi.org/10.1103/physrevlett.122.140504}
  {\path{doi:10.1103/physrevlett.122.140504}}.

\bibitem{Somma2019}
R.~D. Somma, {\it \it Quantum eigenvalue estimation via time series analysis},
  New J. Phys. 21 (2019) 123025.
\newblock \href {https://doi.org/10.1088/1367-2630/ab5c60}
  {\path{doi:10.1088/1367-2630/ab5c60}}.

\bibitem{Endo2020b}
S.~Endo, J.~Sun, Y.~Li, S.~C. Benjamin, X.~Yuan, {\it \it Variational Quantum
  Simulation of General Processes }, Phys. Rev. Lett. 125 (2020) 6.
\newblock \href {https://doi.org/10.1103/PhysRevLett.125.010501}
  {\path{doi:10.1103/PhysRevLett.125.010501}}.

\bibitem{Baker2021}
T.~E. Baker, {\it Lanczos recursion on a quantum computer for the Green's
  function and ground state}, Phys. Rev. A 103 (2021) 032404.
\newblock \href {https://doi.org/10.1103/PhysRevA.103.032404}
  {\path{doi:10.1103/PhysRevA.103.032404}}.

\bibitem{Tong2021}
Y.~Tong, D.~An, N.~Wiebe, L.~Lin, {\it Fast inversion, preconditioned quantum
  linear system solvers, fast Green's-function computation, and fast evaluation
  of matrix functions}, Phys. Rev. A 104 (2021) 032422.
\newblock \href {https://doi.org/10.1103/PhysRevA.104.032422}
  {\path{doi:10.1103/PhysRevA.104.032422}}.

\bibitem{Stenger2022}
J.~P.~T. Stenger, G.~Ben-Shach, D.~Pekker, N.~T. Bronn, {\it Simulating
  spectroscopy experiments with a superconducting quantum computer}, Phys. Rev.
  Res. 4 (2022) 043106.
\newblock \href {http://arxiv.org/abs/2202.12910} {\path{arXiv:2202.12910}},
  \href {https://doi.org/10.1103/PhysRevResearch.4.043106}
  {\path{doi:10.1103/PhysRevResearch.4.043106}}.

\bibitem{Rizzo2022}
J.~Rizzo, F.~Libbi, F.~Tacchino, P.~J. Ollitrault, N.~Marzari, I.~Tavernelli,
  {\it One-particle Green's functions from the quantum equation of motion
  algorithm}, Phys. Rev. Research 4 (2022) 043011.
\newblock \href {https://doi.org/10.1103/PhysRevResearch.4.043011}
  {\path{doi:10.1103/PhysRevResearch.4.043011}}.

\bibitem{Wilde2017}
M.~M. Wilde, {\it Quantum information theory}, Cambridge University Press,
  2013.
\newblock \href {https://doi.org/10.1017/CBO9781139525343}
  {\path{doi:10.1017/CBO9781139525343}}.

\bibitem{Breuer2002}
H.-P. Breuer, F.~Petruccione, {\it The theory of open quantum systems}, Oxford
  University Press, 2007.
\newblock \href {https://doi.org/10.1093/acprof:oso/9780199213900.001.0001}
  {\path{doi:10.1093/acprof:oso/9780199213900.001.0001}}.

\bibitem{Rissler2006}
J.~Rissler, R.~M. Noack, S.~R. White, {\it Measuring orbital interaction using
  quantum information theory}, Chem. Phys. 323 (2006) 519.
\newblock \href {https://doi.org/10.1016/j.chemphys.2005.10.018}
  {\path{doi:10.1016/j.chemphys.2005.10.018}}.

\bibitem{Boguslawski2015}
K.~Boguslawski, P.~Tecmer, {\it Orbital entanglement in quantum chemistry},
  Int. J. Quantum Chem. 115 (2015) 1289.
\newblock \href {https://doi.org/10.1002/qua.24832}
  {\path{doi:10.1002/qua.24832}}.

\bibitem{Robin2021}
C.~Robin, M.~J. Savage, N.~Pillet, {\it Entanglement rearrangement in
  self-consistent nuclear structure calculations}, Phys. Rev. C 103 (2021)
  034325.
\newblock \href {https://doi.org/10.1103/PhysRevC.103.034325}
  {\path{doi:10.1103/PhysRevC.103.034325}}.

\bibitem{Lacroix2022b}
D.~Lacroix, A.~B. Balantekin, M.~J. Cervia, A.~V. Patwardhan, P.~Siwach, {\it
  Role of non-gaussian quantum fluctuations in neutrino entanglement}, Phys.
  Rev. D 106 (2022) 123006.
\newblock \href {https://doi.org/10.1103/PhysRevD.106.123006}
  {\path{doi:10.1103/PhysRevD.106.123006}}.

\bibitem{nachtergaele_2010}
B.~Nachtergaele, R.~Sims, {\it Lieb-Robinson bounds in quantum many-body
  physics}, Entropy and the Quantum 529 (2010) 141.
\newblock \href {https://doi.org/10.1090/conm/529/10429}
  {\path{doi:10.1090/conm/529/10429}}.

\bibitem{orus_2014}
R.~Orús, {\it A practical introduction to tensor networks: Matrix product
  states and projected entangled pair states}, Ann. Phys. 349 (2014) 117.
\newblock \href {https://doi.org/10.1016/j.aop.2014.06.013}
  {\path{doi:10.1016/j.aop.2014.06.013}}.

\bibitem{ran_2020}
S.-J. Ran, {\it Encoding of matrix product states into quantum circuits of one-
  and two-qubit gates}, Phys. Rev. A 101 (2020) 032310.
\newblock \href {https://doi.org/10.1103/physreva.101.032310}
  {\path{doi:10.1103/physreva.101.032310}}.

\bibitem{Srednicki_1993}
M.~Srednicki, {\it Entropy and area}, Phys. Rev. Lett. 71 (1993) 666.
\newblock \href {https://doi.org/10.1103/physrevlett.71.666}
  {\path{doi:10.1103/physrevlett.71.666}}.

\bibitem{Vidal2003}
G.~Vidal, {\it Efficient Classical Simulation of Slightly Entangled Quantum
  Computations}, Phys. Rev. Lett. 91 (2003) 147902.
\newblock \href {https://doi.org/10.1103/PhysRevLett.91.147902}
  {\path{doi:10.1103/PhysRevLett.91.147902}}.

\bibitem{Zhou2020}
Y.~Zhou, E.~M. Stoudenmire, X.~Waintal, {\it What limits the simulation of
  quantum computers?}, Phys. Rev. X 10 (2020) 041038.
\newblock \href {https://doi.org/10.1103/PhysRevX.10.041038}
  {\path{doi:10.1103/PhysRevX.10.041038}}.

\bibitem{Ayral2022}
T.~Ayral, T.~Louvet, Y.~Zhou, C.~Lambert, E.~M. Stoudenmire, X.~Waintal, {\it A
  density-matrix renormalisation group algorithm for simulating quantum
  circuits with a finite fidelity} (2022).
\newblock \href {https://doi.org/10.48550/ARXIV.2207.05612}
  {\path{doi:10.48550/ARXIV.2207.05612}}.

\bibitem{Cabello2002}
A.~Cabello, {\it Bell's theorem with and without inequalities for the
  three-qubit Greenberger-Horne-Zeilinger and W states}, Phys. Rev. A 65 (2002)
  032108.
\newblock \href {https://doi.org/10.1103/PhysRevA.65.032108}
  {\path{doi:10.1103/PhysRevA.65.032108}}.

\bibitem{Greenberger2007}
D.~M. Greenberger, M.~A. Horne, A.~Zeilinger, {\it Going Beyond Bell's Theorem.
  In: Kafatos, M. (eds) Bell’s Theorem, Quantum Theory and Conceptions of the
  Universe. Fundamental Theories of Physics}, Springer, Dordrecht 37 (1989) 69.
\newblock \href {https://doi.org/10.1007/978-94-017-0849-4_10}
  {\path{doi:10.1007/978-94-017-0849-4_10}}.

\bibitem{Paeckel2019}
S.~Paeckel, T.~Köhler, A.~Swoboda, S.~R. Manmana, U.~Schollwöck, C.~Hubig,
  {\it Time-evolution methods for matrix-product states}, Ann. Phys. 411 (2019)
  167998.
\newblock \href {https://doi.org/10.1016/j.aop.2019.167998}
  {\path{doi:10.1016/j.aop.2019.167998}}.

\bibitem{bravo_2020}
C.~Bravo-Prieto, J.~Lumbreras-Zarapico, L.~Tagliacozzo, J.~I. Latorre, {\it
  Scaling of variational quantum circuit depth for Condensed Matter Systems},
  Quantum 4 (2020) 272.
\newblock \href {https://doi.org/10.22331/q-2020-05-28-272}
  {\path{doi:10.22331/q-2020-05-28-272}}.

\bibitem{Schlosshauer2019}
M.~Schlosshauer, {\it Quantum decoherence}, Phys. Rep. 831 (2019) 1.
\newblock \href {https://doi.org/10.1016/j.physrep.2019.10.001}
  {\path{doi:10.1016/j.physrep.2019.10.001}}.

\bibitem{Watrous2018}
J.~Watrous, {\it The Theory of Quantum Information}, Cambridge University
  Press, 2018.
\newblock \href {https://doi.org/10.1017/9781316848142}
  {\path{doi:10.1017/9781316848142}}.

\bibitem{Oliver2013}
W.~D. Oliver,
  \href{https://equs.mit.edu/wp-content/uploads/2016/11/SC_qubits_Oliver_IFF_Spring_School_20140330.pdf}{{\it
  Superconducting Qubits}}, Vol.~52, Forschungszentrum J\"ulich, 2013, p. 1000.
\newline\urlprefix\url{https://equs.mit.edu/wp-content/uploads/2016/11/SC_qubits_Oliver_IFF_Spring_School_20140330.pdf}

\bibitem{Krantz2019}
P.~Krantz, M.~Kjaergaard, F.~Yan, T.~P. Orlando, S.~Gustavsson, W.~D. Oliver,
  {\it A Quantum Engineer's Guide to Superconducting Qubits}, Applied Physics
  Reviews 6 (2019) 021318.
\newblock \href {http://arxiv.org/abs/1904.06560} {\path{arXiv:1904.06560}},
  \href {https://doi.org/10.1063/1.5089550} {\path{doi:10.1063/1.5089550}}.

\bibitem{Paladino2014}
E.~Paladino, Y.~Galperin, G.~Falci, B.~L. Altshuler, {\it 1/ f noise:
  Implications for solid-state quantum information}, Rev. Mod. Phys. 86 (2014)
  361.
\newblock \href {https://doi.org/10.1103/RevModPhys.86.361}
  {\path{doi:10.1103/RevModPhys.86.361}}.

\bibitem{Breuer2016}
H.-P. Breuer, E.-M. Laine, J.~Piilo, B.~Vacchini, {\it Colloquium:
  Non-Markovian dynamics in open quantum systems}, Rev. Mod. Phys. 88 (2016)
  021002.
\newblock \href {https://doi.org/10.1103/RevModPhys.88.021002}
  {\path{doi:10.1103/RevModPhys.88.021002}}.

\bibitem{deVega2017}
I.~de~Vega, D.~Alonso, {\it Dynamics of non-Markovian open quantum systems},
  Rev. Mod. Phys. 89 (2017) 015001.
\newblock \href {https://doi.org/10.1103/RevModPhys.89.015001}
  {\path{doi:10.1103/RevModPhys.89.015001}}.

\bibitem{Diosi1998}
L.~Di\'osi, N.~Gisin, W.~T. Strunz, {\it Non-Markovian quantum state
  diffusion}, Phys. Rev. A 58 (1998) 1699.
\newblock \href {https://doi.org/10.1103/PhysRevA.58.1699}
  {\path{doi:10.1103/PhysRevA.58.1699}}.

\bibitem{Breuer2004}
H.-P. Breuer, D.~Burgarth, F.~Petruccione, {\it Non-Markovian dynamics in a
  spin star system: Exact solution and approximation techniques}, Phys. Rev. B
  70 (2004) 045323.
\newblock \href {https://doi.org/10.1103/PhysRevB.70.045323}
  {\path{doi:10.1103/PhysRevB.70.045323}}.

\bibitem{Sargsyan2017}
V.~V. Sargsyan, D.~Lacroix, G.~G. Adamian, N.~V. Antonenko, {\it Non-Markovian
  dynamics of fully coupled fermionic and bosonic oscillators}, Phys. Rev. A 95
  (2017) 032119.
\newblock \href {https://doi.org/10.1103/PhysRevA.95.032119}
  {\path{doi:10.1103/PhysRevA.95.032119}}.

\bibitem{Lacroix2020_oqs}
D.~Lacroix, V.~V. Sargsyan, G.~G. Adamian, N.~V. Antonenko, A.~A. Hovhannisyan,
  {\it Non-Markovian modeling of Fermi-Bose systems coupled to one or several
  Fermi-Bose thermal baths}, Phys. Rev. A 102 (2020) 022209.
\newblock \href {https://doi.org/10.1103/PhysRevA.102.022209}
  {\path{doi:10.1103/PhysRevA.102.022209}}.

\bibitem{Kraus1971}
K.~Kraus, {\it General State Changes in Quantum Theory}, Ann. Phys. 64 (1971)
  311.
\newblock \href {https://doi.org/10.1016/0003-4916(71)90108-4}
  {\path{doi:10.1016/0003-4916(71)90108-4}}.

\bibitem{Kraus1983}
K.~Kraus, A.~B\"ohm, J.~D. Dollard, W.~Wootters, {\it States, Effects, and
  Operations Fundamental Notions of Quantum Theory: Lectures in Mathematical
  Physics at the University of Texas at Austin}, Springer, 1983.
\newblock \href {https://doi.org/10.1007/3-540-12732-1}
  {\path{doi:10.1007/3-540-12732-1}}.

\bibitem{Chow2012}
J.~M. Chow, J.~M. Gambetta, A.~D. C\'orcoles, S.~T. Merkel, J.~A. Smolin,
  C.~Rigetti, S.~Poletto, G.~A. Keefe, M.~B. Rothwell, J.~R. Rozen, M.~B.
  Ketchen, M.~Steffen, {\it Universal Quantum Gate Set Approaching
  Fault-Tolerant Thresholds with Superconducting Qubits}, Phys. Rev. Lett. 109
  (2012) 060501.
\newblock \href {https://doi.org/10.1103/PhysRevLett.109.060501}
  {\path{doi:10.1103/PhysRevLett.109.060501}}.

\bibitem{Chuang1997}
I.~L. Chuang, M.~A. Nielsen, {\it Prescription for experimental determination
  of the dynamics of a quantum black box}, J. Mod. Opt. 44 (1997) 2455.
\newblock \href {https://doi.org/10.1080/09500349708231894}
  {\path{doi:10.1080/09500349708231894}}.

\bibitem{Choi1975a}
M.-D. Choi, {\it Positive Linear Maps on Complex Matrices}, Linear Algebra
  Appl. 10 (1975) 285.
\newblock \href {https://doi.org/10.1016/0024-3795(75)90075-0}
  {\path{doi:10.1016/0024-3795(75)90075-0}}.

\bibitem{Jamiolkowski1972}
A.~Jamiolkowski, {\it Linear Transformations which preserve trace and positive
  semidefiniteness of operators}, Rep. Math. Phys. 3 (1972) 275.
\newblock \href {https://doi.org/10.1016/0034-4877(72)90011-0}
  {\path{doi:10.1016/0034-4877(72)90011-0}}.

\bibitem{Stinespring1955}
F.~Stinespring, {\it Positive Functions on C*-algebras}, Proc. Amer. Math. Soc.
  6 (1955) 211.
\newblock \href {https://doi.org/10.2307/2032342} {\path{doi:10.2307/2032342}}.

\bibitem{Wood2015}
C.~J. Wood, J.~D. Biamonte, D.~G. Cory, {\it Tensor networks and graphical
  calculus for open quantum systems}, Quantum Inf Comput 15 (2015) 759.
\newblock \href {https://doi.org/10.26421/QIC15.9-10-3}
  {\path{doi:10.26421/QIC15.9-10-3}}.

\bibitem{Preskill2015}
J.~Preskill, \href{http://theory.caltech.edu/~preskill/ph219/chap2_13.pdf}{{\it
  Lecture Notes for Ph219/CS219: Quantum Information}}, 2015.
\newline\urlprefix\url{http://theory.caltech.edu/~preskill/ph219/chap2_13.pdf}

\bibitem{Aharonov1998}
D.~Aharonov, A.~Kitaev, N.~Nisan, {\it Quantum circuits with mixed states},
  Vol.~1, ACM Press, 1998, p.~20.
\newblock \href {https://doi.org/10.1145/276698.276708}
  {\path{doi:10.1145/276698.276708}}.

\bibitem{Blume-Kohout2013}
R.~Blume-Kohout, J.~K. Gamble, E.~Nielsen, J.~Mizrahi, J.~D. Sterk, P.~Maunz,
  {\it Robust, self-consistent, closed-form tomography of quantum logic gates
  on a trapped ion qubit}, arxiv:1310.4492. (2013).
\newblock \href {https://doi.org/10.48550/arXiv.1310.4492}
  {\path{doi:10.48550/arXiv.1310.4492}}.

\bibitem{Merkel2013}
S.~T. Merkel, J.~M. Gambetta, J.~A. Smolin, S.~Poletto, A.~D. C\'orcoles, B.~R.
  Johnson, C.~A. Ryan, M.~Steffen, {\it Self-consistent quantum process
  tomography}, Phys. Rev. A 87 (2013) 062119.
\newblock \href {https://doi.org/10.1103/PhysRevA.87.062119}
  {\path{doi:10.1103/PhysRevA.87.062119}}.

\bibitem{greenbaum_2015}
D.~Greenbaum, {\it Introduction to Quantum Gate Set Tomography},
  arXiv:1509.02921. (2015).
\newblock \href {https://doi.org/10.48550/arXiv.1509.02921}
  {\path{doi:10.48550/arXiv.1509.02921}}.

\bibitem{Carignan-dugas2019}
A.~Carignan-Dugas, J.~J. Wallman, J.~Emerson, {\it Bounding the average gate
  fidelity of composite channels using the unitarity}, New J. Phys. 21 (2019)
  053016.
\newblock \href {https://doi.org/10.1088/1367-2630/ab1800}
  {\path{doi:10.1088/1367-2630/ab1800}}.

\bibitem{Cai2022}
Z.~Cai, R.~Babbush, S.~C. Benjamin, S.~Endo, W.~J. Huggins, Y.~Li, R.~Jarrod,
  T.~E.~O. Brien, {\it Quantum Error Mitigation}, arXiv:2210.00921. (2022).
\newblock \href {https://doi.org/10.48550/arXiv.2210.00921}
  {\path{doi:10.48550/arXiv.2210.00921}}.

\bibitem{rubin_2018}
N.~C. Rubin, R.~Babbush, J.~McClean, {\it Application of fermionic marginal
  constraints to hybrid quantum algorithms}, New J. Phys. 20 (2018) 053020.
\newblock \href {https://doi.org/10.1088/1367-2630/aab919}
  {\path{doi:10.1088/1367-2630/aab919}}.

\bibitem{mcweeny_1960}
R.~McWeeny, {\it Some recent advances in density matrix theory}, Rev. Mod.
  Phys. 32 (1960) 335.
\newblock \href {https://doi.org/10.1103/revmodphys.32.335}
  {\path{doi:10.1103/revmodphys.32.335}}.

\bibitem{he_zero-noise_2020}
A.~He, B.~Nachman, W.~A. de~Jong, C.~W. Bauer, {\it Zero-noise extrapolation
  for quantum-gate error mitigation with identity insertions}, Phys. Rev. A 102
  (2020) 012426.
\newblock \href {https://doi.org/10.1103/PhysRevA.102.012426}
  {\path{doi:10.1103/PhysRevA.102.012426}}.

\bibitem{kandala_error_2019}
A.~Kandala, K.~Temme, A.~D. Córcoles, A.~Mezzacapo, J.~M. Chow, J.~M.
  Gambetta, {\it Error mitigation extends the computational reach of a noisy
  quantum processor}, Nature 567 (2019) 491.
\newblock \href {https://doi.org/10.1038/s41586-019-1040-7}
  {\path{doi:10.1038/s41586-019-1040-7}}.

\bibitem{czarnik_error_2021}
P.~Czarnik, A.~Arrasmith, P.~J. Coles, L.~Cincio, {\it Error mitigation with
  Clifford quantum-circuit data}, Quantum 5 (2021) 592.
\newblock \href {https://doi.org/10.22331/q-2021-11-26-592}
  {\path{doi:10.22331/q-2021-11-26-592}}.

\bibitem{strikis_learning-based_2021}
A.~Strikis, D.~Qin, Y.~Chen, S.~C. Benjamin, Y.~Li, {\it Learning-based quantum
  error mitigation}, PRX Quantum 2 (2021) 040330.
\newblock \href {https://doi.org/10.1103/PRXQuantum.2.040330}
  {\path{doi:10.1103/PRXQuantum.2.040330}}.

\bibitem{montanaro_2021}
A.~Montanaro, S.~Stanisic, {\it Error mitigation by training with fermionic
  linear optics}, arXiv.2102.02120 (2021).
\newblock \href {https://doi.org/10.48550/arXiv.2102.02120}
  {\path{doi:10.48550/arXiv.2102.02120}}.

\bibitem{huggins_2021}
W.~J. Huggins, S.~McArdle, T.~E. O’Brien, J.~Lee, N.~C. Rubin, S.~Boixo,
  K.~B. Whaley, R.~Babbush, J.~R. McClean, {\it Virtual distillation for
  quantum error mitigation}, Phys. Rev. X 11 (2021) 041036.
\newblock \href {https://doi.org/10.1103/physrevx.11.041036}
  {\path{doi:10.1103/physrevx.11.041036}}.

\bibitem{bultrini_2021}
D.~Bultrini, M.~H. Gordon, P.~Czarnik, A.~Arrasmith, P.~J. Coles, L.~Cincio,
  {\it Unifying and benchmarking state-of-the-art quantum error mitigation
  techniques}, arxiv:2107.13470. (2021).
\newblock \href {https://doi.org/10.48550/ARXIV.2107.13470}
  {\path{doi:10.48550/ARXIV.2107.13470}}.

\bibitem{temme_error_2017}
K.~Temme, S.~Bravyi, J.~M. Gambetta, {\it Error mitigation for short-depth
  quantum circuits}, Phys. Rev. Lett. 119 (2017) 180509.
\newblock \href {https://doi.org/10.1103/PhysRevLett.119.180509}
  {\path{doi:10.1103/PhysRevLett.119.180509}}.

\bibitem{piveteau_quasiprobability_2022}
C.~Piveteau, D.~Sutter, S.~Woerner, {\it Quasiprobability decompositions with
  reduced sampling overhead}, Npj Quantum Inf. 8 (2022) 12.
\newblock \href {https://doi.org/10.1038/s41534-022-00517-3}
  {\path{doi:10.1038/s41534-022-00517-3}}.

\bibitem{piveteau_2021}
C.~Piveteau, {\it Advanced methods for quasiprobabilistic quantum error
  mitigation}, Master thesis (2021).
\newblock \href {https://doi.org/10.3929/ethz-b-000504508}
  {\path{doi:10.3929/ethz-b-000504508}}.

\bibitem{nielsen_gamble_2021}
E.~Nielsen, J.~K. Gamble, K.~Rudinger, T.~Scholten, K.~Young, R.~Blume-Kohout,
  {\it Gate set tomography}, Quantum 5 (2021) 557.
\newblock \href {https://doi.org/10.22331/q-2021-10-05-557}
  {\path{doi:10.22331/q-2021-10-05-557}}.

\bibitem{endo_practical_2018}
S.~Endo, S.~C. Benjamin, Y.~Li, {\it Practical Quantum Error Mitigation for
  Near-Future Applications}, Phys. Rev. X 8 (2018) 031027.
\newblock \href {https://doi.org/10.1103/PhysRevX.8.031027}
  {\path{doi:10.1103/PhysRevX.8.031027}}.

\bibitem{Girvin2022}
S.~M. Girvin, {\it Introduction to Quantum Error Correction and Fault
  Tolerance}, arXiv.2111.08894 (2022).
\newblock \href {https://doi.org/10.48550/arXiv.2111.08894}
  {\path{doi:10.48550/arXiv.2111.08894}}.

\bibitem{Knill1997}
E.~Knill, R.~Laflamme, {\it Theory of quantum error-correcting codes}, Phys.
  Rev. A 55 (1997) 900.
\newblock \href {https://doi.org/10.1103/PhysRevA.55.900}
  {\path{doi:10.1103/PhysRevA.55.900}}.

\bibitem{Gottesman1997}
D.~Gottesman, {\it Stabilizer codes and quantum error correction}, Ph.D. thesis
  arXiv:quant-ph/9705052 (1997).
\newblock \href {https://doi.org/10.48550/arXiv.quant-ph/9705052}
  {\path{doi:10.48550/arXiv.quant-ph/9705052}}.

\bibitem{Shor1996}
P.~Shor, {\it Fault-tolerant quantum computation}, IEEE Comput. Soc. Press,
  1996, p.~56.
\newblock \href {https://doi.org/10.1109/SFCS.1996.548464}
  {\path{doi:10.1109/SFCS.1996.548464}}.

\bibitem{Aharonov2008}
D.~Aharonov, M.~Ben-Or, {\it Fault-Tolerant Quantum Computation With Constant
  Error Rate}, SIAM J. Comput. 38 (1999) 1207.
\newblock \href {https://doi.org/10.1137/S0097539799359385}
  {\path{doi:10.1137/S0097539799359385}}.

\bibitem{Kitaev2003}
A.~Y. Kitaev, {\it Fault-tolerant quantum computation by anyons}, Ann. Phys.
  303 (1997) 2.
\newblock \href {https://doi.org/10.1016/S0003-4916(02)00018-0}
  {\path{doi:10.1016/S0003-4916(02)00018-0}}.

\bibitem{Dennis2002}
E.~Dennis, A.~Kitaev, A.~Landahl, J.~Preskill, {\it Topological quantum
  memory}, J. Math. Phys. 43 (2002) 4452.
\newblock \href {https://doi.org/10.1063/1.1499754}
  {\path{doi:10.1063/1.1499754}}.

\bibitem{Fowler2012}
A.~G. Fowler, M.~Mariantoni, J.~M. Martinis, A.~N. Cleland, {\it Surface codes:
  Towards practical large-scale quantum computation}, Phys. Rev. A 86 (2012)
  032324.
\newblock \href {https://doi.org/10.1103/PhysRevA.86.032324}
  {\path{doi:10.1103/PhysRevA.86.032324}}.

\bibitem{Bombin2006}
H.~Bombin, M.~A. Martin-Delgado, {\it Topological Quantum Distillation}, Phys.
  Rev. Lett. 97 (2006) 180501.
\newblock \href {https://doi.org/10.1103/PhysRevLett.97.180501}
  {\path{doi:10.1103/PhysRevLett.97.180501}}.

\bibitem{Landahl2011}
A.~J. Landahl, J.~T. Anderson, P.~R. Rice, {\it Fault-tolerant quantum
  computing with color codes}, arXiv:1108.5738. (2011).
\newblock \href {https://doi.org/10.48550/arXiv.1108.5738}
  {\path{doi:10.48550/arXiv.1108.5738}}.

\bibitem{Krinner2021}
S.~Krinner, N.~Lacroix, A.~Remm, A.~D. Paolo, E.~Genois, C.~Leroux,
  C.~Hellings, S.~Lazar, F.~Swiadek, J.~Herrmann, G.~J. Norris, K.~Andersen,
  M.~Markus, A.~Blais, C.~Eichler, A.~Wallraff, {\it Realizing Repeated Quantum
  Error Correction in a Distance-Three Surface Code}, Nature 605 (2021) 669.
\newblock \href {https://doi.org/10.1038/s41586-022-04566-8}
  {\path{doi:10.1038/s41586-022-04566-8}}.

\bibitem{Acharya2022}
R.~Acharya, et~al., {\it Suppressing quantum errors by scaling a surface code
  logical qubit}, arXiv:2207.06431. (2022).
\newblock \href {https://doi.org/10.48550/arXiv.2207.06431}
  {\path{doi:10.48550/arXiv.2207.06431}}.

\bibitem{Ryan-Anderson2022}
C.~Ryan-Anderson, J.~G. Bohnet, K.~Lee, D.~Gresh, A.~Hankin, J.~P. Gaebler,
  D.~Francois, A.~Chernoguzov, D.~Lucchetti, N.~C. Brown, T.~M. Gatterman,
  S.~K. Halit, K.~Gilmore, J.~A. Gerber, B.~Neyenhuis, D.~Hayes, R.~P. Stutz,
  {\it Realization of Real-Time Fault-Tolerant Quantum Error Correction}, Phys.
  Rev. X 11 (2022) 41058.
\newblock \href {https://doi.org/10.1103/PhysRevX.11.041058}
  {\path{doi:10.1103/PhysRevX.11.041058}}.

\bibitem{Guillaud2022}
J.~Guillaud, J.~Cohen, M.~Mirrahimi, {\it Quantum computation with cat qubits},
  arXiv.2203.03222 (2022).
\newblock \href {https://doi.org/10.48550/arXiv.2203.03222}
  {\path{doi:10.48550/arXiv.2203.03222}}.

\end{thebibliography}

\end{document}